\documentclass[reviewcopy]{elsarticle}

\usepackage[reviewcopy]{adndt}
\usepackage{longtable}
\usepackage{multirow}

\usepackage{amsmath}

\usepackage{amssymb}

\biboptions{square,sort&compress}
\bibpunct[]{[}{]}{,}{n}{}{;}
\begin{document}

\begin{frontmatter}

\journal{Atomic Data and Nuclear Data Tables}


\title{Tables of E2 Transition Probabilities from the first $2^{+}$ States in Even-Even Nuclei
}

  \author[One]{B. Pritychenko\corref{cor1}}
  \ead{E-mail: pritychenko@bnl.gov}

  \author[Two]{M. Birch}
  \author[Two]{B. Singh}
  \author[Three] {M. Horoi}

  \cortext[cor1]{Corresponding author.}

  \address[One]{National Nuclear Data Center, Brookhaven National Laboratory, Upton, NY 11973-5000, USA}
  \address[Two]{Department of Physics \& Astronomy, McMaster University, Hamilton, Ontario L8S 4M1, Canada}
  \address[Three]{Department of Physics, Central Michigan University,
Mount Pleasant, MI 48859, USA}

\date{13.12.2013} 
\begin{abstract}
Experimental results of E2 transition probabilities or B(E2) values for the known first 2$^{+}$ states in 447 even-even nuclei
have been compiled and evaluated. The evaluation policies for the analysis of experimental data have been described and
new results are discussed. The recommended B(E2) values have been compared with comprehensive shell model calculations 
for a selected set of nuclei, where such theoretical procedures are amenable. 
The present work was motivated by a rapid increase in the number of new B(E2) measurements for the first 2$^{+}$ states
since the previous evaluation of such data by S. Raman {\it et al.} published in 2001. Future plans to investigate the
systematics of B(E2)$\uparrow$ values, and intercomparison of different experimental techniques to obtain these data are outlined.
\end{abstract}

\end{frontmatter}




\newpage

\tableofcontents
\listofDtables
\listofDfigures
\vskip5pc


\section{Introduction}

Measurements of the quadrupole collectivity of atomic nuclei started in the early 1950s.
Such collectivity was extensively studied along the N$\sim$Z or ``valley of stability" region.
In this region, many nuclear structure phenomena, such as nuclear shell closure,
were identified and explained in the framework of the nuclear shell model
\cite{1950Go80,1990Wa02}. With the advent of radioactive beam and isotope production
techniques, scientists were presented with a unique opportunity to test nuclear models for neutron- and
proton-rich nuclei. This approach produced many interesting and often unexpected results on the 
evolution of nuclear properties near the neutron and proton driplines \cite{2006XxZZ}.

To demonstrate the importance of B(E2) values in nuclear physics research and model development, we will consider nuclear
``magic numbers" and their evolution along the nuclear chart. In stable nuclei, large gaps exist between nuclear shells
when the proton or neutron number is equal to 2, 8, 20, 28, 50, 82, and 126 \cite{1950Go80,2008So14}. These gaps result
in large transition energy values between the ground and first excited states, relatively low quadrupole collectivities and
small neutron capture cross sections. The ``magic numbers" and their values are not preserved; they evolve for unstable nuclei
due to nuclear structure effects. Therefore, nuclear properties of the first excited 2$^{+}_{1}$ states in even-even nuclei
provide important information on the evolution of nuclear properties and shell model studies. Accurate knowledge of these 
properties is necessary for continuing the development of nuclear model calculations and theoretical understanding of many 
interesting phenomena in the quantum world.

Another significant application of B(E2) evaluated data is for nuclear reaction model calculations.
Precise values of quadrupole deformation parameters are essential for the Reference Input Parameter Library (RIPL)
\cite{2009RIPL} and nuclear reaction model codes such as EMPIRE and TALYS \cite{2007EMPIRE,2012TALYS}, which are
extensively used for ENDF evaluations \cite{11Chad,11Kon,11Shi}. Such evaluations are critically important for applications
of nuclear data as the ENDF/B-VII.1 library \cite{11Chad} provides evaluated
neutron cross sections for frequently-used nuclear science and technology codes GEANT and MCNP.

The importance of compilation and evaluation of E2 transition probabilities for even-even nuclei was
recognized in the 1960s by P.H. Stelson and L. Grodzins at Oak Ridge National Laboratory. They produced the first
compilation of B(E2) values  for 2$_{1}^{+}$ states \cite{1965St16}, which was then continued by S. Raman {\it et al.}
\cite{1987Ra01,2001Ra27}. Presently, this work proceeds under the auspices of the U.S. Nuclear Data Program (USNDP).
It began as periodic update of B(E2) values for the mass regions, where a large number of new experimental results became available.
The first update of B(E2) values for Cr, Fe, Ni and Zn isotopes (Z$\sim$28 region) has been recently published by the joint
effort of NNDC, Brookhaven National Laboratory (BNL), McMaster and Central Michigan Universities \cite{2012Pr08}.
This update supplied valuable feedback to our collaboration from the research community \cite{2014Pr06}, which
has helped to improve the quality of the present work. The detailed description of compilation and evaluation tools, 
procedures and results is given in the following sections.

\section{Nuclear Databases used in the present work}
Nuclear Science References (NSR) \cite{2011Pr03}, Evaluated Nuclear Structure Data File (ENSDF) \cite{1990Bu35,1996Tu01} and
Experimental Unevaluated Nuclear Data List (XUNDL) \cite{XUNDL}  databases each played a crucial role in this project.
 A short description of the databases is presented below.

The NSR database \cite{2011Pr03} is the most comprehensive source of low- and intermediate-energy nuclear physics bibliographical information,
containing more than 219000 articles, mostly in peer-reviewed journals, since the beginning of nuclear science. It consists of primary (journals)
and secondary (proceedings, lab reports, theses, private communications) references.
The main goal of the NSR is to provide bookmarks for experimental and
theoretical articles in nuclear science using keywords. NSR keywords are assigned to articles that contain results on atomic nuclei and masses, nuclear decays,
nuclear reactions and other properties. Keywords are also used to build author and subject indexes, which allow users to search for 
articles by subject (Coulomb excitation, $\sigma$, B(E2), T$_{1/2}$, etc.) or author.
This database is updated weekly and serves as a primary source of bibliographical information for the ENSDF database.

The ENSDF database \cite{1990Bu35,1996Tu01} contains evaluated nuclear structure and decay data. An international network of evaluators \cite{2007Ni15}
contributes to the database. For each nuclide, all known experimental data used to deduce nuclear structure information are included.
Each type of experiment is presented as a separate dataset. In addition, there is a dataset of ``adopted" level
and $\gamma$-ray transition properties, which represent the evaluator's determination of the recommended values for these properties,
based on all the available experimental data. Information in the database is regularly
updated and most of this information is also published in Elsevier Nuclear Data Sheets journal as
A-chain evaluations. Due to the large scope of the database, evaluation updates
are often conducted on an $\sim$10-year basis, with some nuclides updated more frequently.

The XUNDL database \cite{XUNDL} contains compiled experimental nuclear structure data from current
publications in the ``ENSDF" format. In general, the information
in a given XUNDL dataset comes from a single journal article, or from a set of closely-related articles
by one group of authors. The information in the XUNDL database is often used in the updated ENSDF evaluations.

We primarily used NSR and XUNDL database content for the experimental data search. These searches were verified using the ENSDF database,
previous evaluation of S. Raman {\it et al.} \cite{2001Ra27} and references from the original experimental papers.

\section{Experimental B(E2) values}

Experimental values of B(E2), $\tau$ and $\beta_{2}$ are compiled in Table \ref{tableExp}.
This Table extends the list of the previously reported quantities by S. Raman {\it et al.} \cite{2001Ra27},
and  includes target, beam, beam energy and annotation where the beam energy exceeds the Coulomb
barrier \cite{1999ZaZY}. 
In general, Coulomb excitation and nuclear resonance fluorescence measurements list 0$^{+}_{1} {\rightarrow} 2^{+}_{1}$ transitions,
while lifetime measurements list 2$^{+}_{1} {\rightarrow} 0^{+}_{1}$ transitions.
A short review of the recent experimental results that  motivated the current evaluation is presented below
and provides summary of experimental activities in the last 10-15 years.
It lists new nuclides, nuclear physics rationale, experimental techniques, theoretical calculations,
laboratories, references, etc.
The following data indicate strong international collaborations and broad popularity of quadrupole collectivity studies worldwide.

\subsection{$^6$He}
A neutron-$\alpha$-particle coincidence experiment was performed at Notre Dame University to study breakup of $^{6}$He \cite{2007Ko23},
and a B(E2) value of 0.00054(7) e$^2$b$^2$ was deduced for breakup via the 2$^{+}$ excited state reaction channel.
The measured collectivity is for the particle unbound state. These data are also model dependent due to Coulomb-nuclear interference effects.

\subsection{$^{10,12}$Be}
Lifetimes 0.205$\pm$5 (stat) $\pm$7 (syst) ps and 2.5$\pm$7 (stat) $\pm$3 (syst) ps of the first
2$^{+}$ states in $^{10}$Be and $^{12}$Be have
been measured using the Doppler shift attenuation method at Argonne National Laboratory \cite{2009Mc02}
and inelastic scattering at RIKEN \cite{2009Im01}, respectively. The former measurement provides a discriminating test
of {\it ab initio} calculations of light nuclei. While the later result shows a large quadrupole strength in the ground state
transition, providing further evidence on the disappearance of the $N=8$ ``magic number".

\subsection{$^{10,16,18,20}$C}
To further test  {\it ab initio} model predictions, the lifetime of the 2$^{+}_{1}$ state in $^{10}$C has been precisely re-measured  at
Argonne using the Doppler shift attenuation method \cite{2012Mc03} to be 0.219(12) ps. Four different measurements \cite{2004Im01,2008Wi04,2008On02}
helped to pin-point the recommended B(E2) value for $^{16}$C at 0.00179(20) $e^2b^2$.
The electric quadrupole transition from the first 2$^{+}$ state to the ground 0$^{+}$ state in $^{18}$C was studied through a
lifetime measurement by an upgraded recoil shadow method  at RIKEN \cite{2008On02}.  The lifetime of the 2$^{+}_{1}$ state
in the near-dripline nucleus $^{20}$C  was recently measured to be 9.8($^{+28}_{-30}$) ps at the
Michigan State University (MSU) Cyclotron Laboratory \cite{2011Pe21}. That measurement is consistent with the previous limit \cite{2009El03}.

\subsection{$^{22,24}$O}
Recent inelastic scattering experiments provided model-dependent data on quadrupole deformation parameter values,
$\beta_2$ of 0.24 and 0.26 for $^{22}$O \cite{2006El05,2006Be04}, and 0.15(4) for $^{24}$O \cite{2012Ts03}.
These data provide complementary experimental evidence for a  new ``magic number"  $N=14$.

\subsection{$^{18,26,28,30}$Ne}
Quadrupole collectivity of neon isotopes  has been extensively studied at RIKEN.
$^{18}$Ne collectivity was verified in a Coulomb excitation experiment (Coulex) \cite{2006YaZV}.
The measurement of the 2$^{+}_{1} {\rightarrow} 0^{+}_{1}$ transition in $^{30}$Ne \cite{2003Ya05}  and  confirmation of
B(E2) values  in $^{26,28}$Ne  \cite{2007Gi06,2005Iw02} have helped to pin-point the
boundary of the ``island of inversion" or nuclear shell ordering along $Z$=10. Finally, $^{28,30}$Ne deformation lengths and parameters were extracted
from the angle-integrated cross sections using distorted-wave calculations \cite{2014Mi09}.

\subsection{$^{20,30,32,34}$Mg}
Investigation of nuclear shell closure effects in the ``island of inversion" region served as
an additional motivation for study of magnesium isotopes.
Coulomb excitation  of  $^{20,34}$Mg was performed at the RIKEN cyclotron facility \cite{2008Iw04,2001Iw07}.
The deduced B(E2)$\uparrow$  value for $^{34}$Mg  of 0.0631(126) e$^2$b$^2$  is in agreement with
the MSU measurements \cite{1999Pr09,2005Ch66}. In addition, REX-ISOLDE Coulex B(E2)$\uparrow$ values
of 0.0241(31) and 0.0434(52) e$^2$b$^2$ in $^{30}$Mg and $^{32}$Mg \cite{2005Ni11,2005NiZS}, respectively,
confirmed the previous MSU and RIKEN results \cite{1999Pr09,1995Mo16}.
Recently, complementary values of deformation lengths and parameters for the first
2$^{+}$ states in $^{32,34,36}$Mg were measured at RIKEN  \cite{2014Mi09}.
These deformation parameters provide a glimpse of quadrupole collectivity in the vicinity of $^{32}$Mg.
Finally, in 2015, the RIKEN group demonstrated that the electromagnetic properties of $^{32}$Mg can be studied with
Coulomb excitation at beam energies of a few hundreds of MeV/nucleon, where a thicker target can be used to increase
the excitation yields \cite{2015Li28}.

\subsection{$^{24,36,38,40}$Si}
The Coulomb excitation technique was used for the study of collectivity in the proton-rich nucleus $^{24}$Si. The reduced transition probability
from its 2$^{+}_{1}$ state was probed using a radioactive beam of $^{24}$Si at 57.9 MeV/nucleon bombarding a $^{208}$Pb target \cite{2002Ka80}.
This B(E2)$\downarrow$ value of 19.1$\pm$5.9 e$^{2}$fm$^{4}$ is smaller than that of the mirror nucleus $^{24}$Ne.
$\beta_2$ values of 0.25(4), 0.36(3), and 0.37(5) have been deduced for $^{36,38,40}$Si, respectively,
using inelastic proton-scattering cross sections at MSU \cite{2007Ca35}.
Enhanced collectivity at $N$=26 indicates a reduced $N$=28 shell gap at large neutron excess in this chain of isotopes.

\subsection{$^{28}$S}
$^{28}$S was measured at RIKEN with the Coulomb excitation technique \cite{2012To06}.  The
resulting B(E2) value of 181(31) e$^{2}$fm$^{4}$ is smaller than the expected value
based on empirical B(E2) systematics. These results indicate the emergence of the
``magic number" $Z$=16 in the $\left\vert T_z \right\vert$=2 nucleus $^{28}$S.

\subsection{$^{32,44,46,48}$Ar}
The collective strengths of the 0$_{1}^{gs} {\rightarrow} 2_{1}^{+}$ excitations in $^{32,48}$Ar
were measured at MSU using NaI- and SeGA- gamma detector arrays \cite{2002Co09,2012Wi05}.  The $^{32}$Ar measurement, taken together with previously existing Coulomb
excitation data for $^{32}$Si \cite{1998Ib01}, yields the isoscalar and isovector 
multipole matrix elements for the transition between
$T$=2 states in the $A$=32 system. Complementary  Coulex and RDM measurements of  $^{44,46}$Ar were conducted at GANIL, France
and Legnaro, Italy facilities \cite{2009Zi01,2010Me07}, respectively. These experiments addressed the development of deformation and
shape coexistence in the vicinity of the doubly magic $^{48}$Ca, related to the weakening of the $N$=28 shell closure.

\subsection{$^{40,42,44,46,48,50}$Ca}
Electric quadrupole strength distributions in doubly magic nuclei $^{40,48}$Ca were studied using the resonance
fluorescence technique at Darmstadt, Germany \cite{2002Ha13}.
The transient field technique was employed to study collectivity in $^{42,44,46}$Ca at the Cologne tandem accelerator \cite{2003Sc21,2003Sp04}.
The $^{46}$Ca $g$-factor \cite{2003Sp04} is in disagreement with the large positive value predicted
by the large scale shell model (LSSM) calculations which included $sd$ shell core excitations
into the $fp$ shell and accounted well for the corresponding $^{42,44}$Ca results \cite{2003Sc21}.
Both $g$(2$^{+}_{1}$) and B(E2) in $^{46}$Ca can be explained by full $fp$-shell model calculations using the
FPD6 interaction without invoking core excitations.
Lifetimes of the first excited state in $^{46,48}$Ca were measured with the recoil distance method using
the PRISMA-CLARA setup at Legnaro \cite{2009Me23,2012Mo11}.  The same facility has been employed to measure
a lifetime of the first excited state of the $N$=30 isotone, $^{50}$Ca \cite{2009Va06}. This extends the lifetime
knowledge beyond the f$_{7/2}$-shell closure  near the doubly magic nucleus $^{48}$Ca.

\subsection{$^{52,54,56,58}$Ti}
The even-even $^{52,54,56,58}$Ti isotopes have been studied with 
intermediate-energy Coulex experiments at MSU and absolute B(E2)
transition rates have been deduced \cite{2005Di05}. These data 
confirm the presence of a subshell closure at neutron number $N$=32 in
neutron-rich nuclei above the doubly magic nucleus $^{48}$Ca and do not support the predicted $N$=34 closure.
$^{52}$Ti low-level structure properties were verified using an inverse kinematics reaction
Doppler shift attenuation method at K{\"o}ln (Cologne), Germany \cite{2006Sp02}.
Finally, the $^{58}$Ti deformation length was recently probed at RIKEN, Japan \cite{2013Su20}. The energy of the first 2$^{+}$ state and
the deformation length value are comparable to the ones of $^{56}$Ti, which indicates that the collectivity of the Ti isotopes
does not increase significantly with neutron number until $N$=36.

\subsection{$^{46,56,58,60,62,64}$Cr}
To complete the systematics in the $N = Z = 28$ region, a B(E2;$0_{1}^{+} \rightarrow 2_{1}^{+}$) value of 0.093(20)  e$^2$b$^2$  has been
reported from the intermediate-energy Coulex of $^{46}$Cr \cite{2005Ya26}.
Coulomb excitation B(E2) values of 8.7(30) and 14.8(42) W.u. for $^{56,58}$Cr, respectively, have been measured by the
RISING collaboration \cite{2005Bu29}. These results agree well with the shell model calculation based
on GXPF1A and GXPF1 effective interactions \cite{2005Ho32,2004Ho08}.  B(E2) and lifetime values for the first
2$^{+}$ states of $^{58,60,62,64}$Cr have been recently measured at MSU \cite{2012Ba31,2013Cr02,2015Br10}.  The deformation length and quadrupole deformation parameter have been
studied in the inelastic scattering of  $^{60,62}$Cr on hydrogen \cite{2009Ao01}.
These data provide evidence for enhanced collectivity in chromium nuclei. Recently, quadrupole collectivity of neutron-rich $^{64}$Cr
was measured with the Coulex technique \cite{2013Cr02}. Its deformation has been interpreted with shell-model calculations
using the state-of-the-art LNPS  effective interaction.

\subsection{$^{50,52,62,64,66,68}$Fe}
A B(E2;$0_{1}^{+} \rightarrow 2_{1}^{+}$) value of  0.140(30) e$^2$b$^2$ in  $^{50}$Fe has been reported from an
intermediate-energy Coulex experiment \cite{2005Ya26}. A Coulex measurement at MSU \cite{2004Yu07}  has produced a B(E2;$0_{1}^{+} \rightarrow 2_{1}^{+}$)  value of 0.082(10) e$^2$b$^2$ for $^{52}$Fe. The increase in  B(E2) strength with respect to the even-mass neighbor $^{54}$Fe agrees
with shell model calculations as the ``magic number" $N$=28 is approached.   $^{62,64}$Fe lifetimes of 7.4(9) and 7.4(26) ps \cite{2010Lj01}, have been originally reported  by the GANIL group using the recoil-distance Doppler shift method after multinucleon transfer reactions in inverse kinematics.
A recent MSU measurement of lifetimes provides the following results: 8.0(10), 10.3(10), 39.0(40) ps for $^{62,64,66}$Fe \cite{2011Ro02}, respectively.
The deduced B(E2) strengths demonstrate the enhanced collectivity of the neutron-rich Fe isotopes up to $N=40$. Note that both groups used the plunger method.
B(E2) values of 0.1445(124) and 0.1777(216) e$^2$b$^2$ for $^{66,68}$Fe, respectively, have also been recently measured at MSU \cite{2013Cr02}.

\subsection{$^{54,58,60,62,64,70,74}$Ni}
The Coulex technique was employed to deduce the B(E2) value for $^{54}$Ni \cite{2004Yu10,2005Ya26}.
High-precision reduced electric-quadrupole transition probabilities have been measured from the single-step Coulomb
excitation of semi-magic $^{58,60,62,64}$Ni beams at 1.8 MeV per nucleon on a natural carbon target at the
Holifield Radioactive Ion Beam Facility \cite{2014Al20}.
A reduced transition probability, B(E2;$0_{1}^{+} \rightarrow 2_{1}^{+}$), of  0.086(14) e$^2$b$^2$ \cite{2006Pe13}
has been measured by Coulex for the neutron-rich nucleus $^{70}$Ni in a $^{208}$Pb target at intermediate energy.
The current B(E2) value for $^{70}$Ni is unexpectedly large, which may indicate that neutrons added above
$N=40$ strongly polarize the $Z=28$ proton core. The deformation length and quadrupole deformation parameter have been
measured by inelastic scattering of $^{74}$Ni on hydrogen \cite{2010Ao01}.  Results of this experiment indicate that the
magic character of $Z=28$ or $N=50$ is weakened in $^{74}$Ni. The precision of this measurements was improved with Coulomb excitation techniques at MSU \cite{2014Ma85}.

\subsection{$^{72,74,76,78,80}$Zn}
A reduced transition probability, B(E2;$0_{1}^{+} \rightarrow 2_{1}^{+}$), of  0.174(21) e$^2$b$^2$ \cite{2002Le17}
for the $^{72}$Zn nucleus has been measured by the Coulex technique at intermediate energy.
This result is consistent with the expected values from the neighboring nucleus $^{73}$Zn and
indicates that the behavior of B(E2) strengths around the $N = 40$ sub-shell closure in Zn is very different from the Ni isotopic chains.
A reduced transition probability B(E2;$0_{1}^{+} \rightarrow 2_{1}^{+}$) of  0.204(15) e$^2$b$^2$ \cite{2006Pe13}
for the neutron-rich  $^{74}$Zn nucleus has been measured by Coulomb excitation on a $^{208}$Pb target at intermediate energy.
This result agrees well with 0.201(16) e$^2$b$^2$ which was measured at REX-ISOLDE \cite{2007Va20}.
Recent B(E2) measurements at GANIL and Legnaro facilities \cite{2013Ce01,2013Lo04} highlight needs for additional systematics.
The reduced transition probabilities, B(E2;$0_{1}^{+} \rightarrow 2_{1}^{+}$), of  0.145(18), 0.077(19) and 0.073(9) e$^2$b$^2$
for $^{76,78,80}$Zn have been reported by the REX-ISOLDE group \cite{2007Va20,2009Va01}  using the Coulex method at low-energy.
Lifetimes of $^{70,72,74}$Zn were deduced with the AGATA spectrometer demonstrator \cite{2013Lo04}. These data are consistent with shell model
predictions using JUN45 and LNPS  effective interactions.

\subsection{$^{64,66,70,76,78,80,82}$Ge}
Collectivity of germanium isotopes was extensively studied in the last ten years. A lifetime value of 3.3(5) ps for the $N=Z$ nucleus $^{64}$Ge was
measured with the recoil distance method at MSU \cite{2007St16}. The last result is in excellent agreement with large-scale shell-model calculations applying the GXPF1A interactions.
Recent lifetime measurements in $^{66}$Ge \cite{2013Co23,2012Lu03} indicate potential problems with the original measurement \cite{1979Wa23}. A low-level structure of $^{70}$Ge was investigated at Munich tandem with the
Doppler shift attenuation technique \cite{2006Le31}. Complementary B(E2) values for a $^{76}$Ge primary beam were deduced and used for calibration of secondary fragment
values at MSU, GANIL, and Legnaro \cite{2005Di05,2006Pe13,2013Lo04}. Reduced transition probabilities in $^{78,80,82}$Ge were investigated at RIKEN, Oak Ridge,
and MSU \cite{2005Iw03,2005Pa23,2010Ga14}. The B(E2) systematic trend approaching $N$=50 indicates strong sensitivity of its values to the effective interaction.

\subsection{$^{68,70,72,82,84}$Se}
Recently B(E2) values have been deduced from the Coulomb excitation of $^{68,82,84}$Se at MSU and RIKEN \cite{2009Ob02,2005Iw03,2010Ga14}.
It was found that the $^{68}$Se transition strength is similar to that of the triaxial $^{64}$Ge nucleus \cite{2007St16};
in sharp contrast to the much stronger collectivity observed for the oblate $^{72}$Kr nucleus \cite{2005Ga22}.
Meanwhile, a $^{84}$Se measurement \cite{2010Ga14} has helped to complete the B(E2) systematics for the $N$=50 isotones from zinc to molybdenum.
$^{70,72}$Se lifetimes were measured with the recoil-distance method at Legnaro \cite{2008Lj01}. The Legnaro results reveal considerable
discrepancies with the literature values \cite{1986He17}. The HFB-based configuration-mixing calculations indicate an oblate rotational ground-state band in
$^{68}$Se \cite{2008Lj01}. The collectivity in  $^{68,70}$Se was recently verified at MSU using the recoil distance Doppler shift technique \cite{2014Ni09}.
This trend is consistent with shell model calculations
using the GXPF1A interaction in an $fp$-model space including the Coulomb, spin-orbit and isospin non-conserving interactions.

\subsection{$^{72,74,76,78,88,90,92,94,96}$Kr}
A B(E2)$\uparrow$ value of 0.4997(647) e$^2$b$^2$ for the $N=Z$ nucleus $^{72}$Kr was measured with the Coulex technique at MSU \cite{2005Ga22}.
This value is in agreement with the self-consistent models that predict an oblate shape for the ground state of $^{72}$Kr.
Quadrupole collectivity of $^{74,76,78}$Kr was  studied with the Coulomb excitation and recoil distance methods.
The $^{74,76}$Kr results \cite{2005Go43} resolve discrepancies between lifetime and Coulomb excitation measurements.
A series of $^{78}$Kr measurements \cite{2005Ga22,2006Be18,2009Ob02} agree well with one another.  A GRETINA array lifetime measurement
of  $^{72,74}$Kr  \cite{2014Iw01} agrees with the previous  values \cite{2005Ga22,2005Go43} and  indicates a future potential use of this detector.
Results for $^{88,92}$Kr were reported at a recent conference \cite{2009MuZW}, and may require further experimental work.
The $^{90}$Kr lifetime was measured by cold-neutron-induced fission of $^{235}$U \cite{2014Re15}.
B(E2) values of  0.247(28) and 0.436(93) e$^2$b$^2$ for $^{94,96}$Kr, respectively, were measured with Coulex at the CERN REX-ISOLDE facility \cite{2012Al03}.
This measurement helped to clarify energies of the first excited states and the erroneous statements on sudden onset of deformation in $^{96}$Kr.

\subsection{$^{76,78,84,86,88,96}$Sr}
The lifetimes of the first excited states in $^{76,78}$Sr and $^{84,86,88}$Sr  were measured with the Doppler shift
attenuation technique at MSU \cite{2012Le05} and Yale \cite{2012Ku14}, respectively. The former results  highlight the importance of
the mixing of coexisting shapes for the description of well-deformed nuclei, and the latter data are consistent with
the large-scale shell-model calculations using the JUN45 interaction.
The B(E2) value of $^{96}$Sr was measured in a Coulex experiment at the REX-ISOLDE facility \cite{2011Cl03}. The combination of a rather large B(E2) value
with a large spectroscopic quadrupole moment in $^{96}$Sr suggests a quasi vibrator character
and excludes static quadrupole deformation. These results are reproduced with Gogny D1S force calculations.

\subsection{$^{88,92,96,104,106}$Zr}
Lifetimes of 3.6(4) and 0.82(10) ps for the first 2$^{+}$ states in $^{88,96}$Zr, respectively,
were measured at Yale with the Doppler shift attenuation method \cite{2012Ku14,2003Ku11}. These
data are in fair agreement with shell-model calculations.
$^{92}$Zr collectivity was investigated at Darmstadt and K{\"o}ln \cite{2013Sc01,2002We15}. A combination of experimental data
and shell model calculations shows that both, single particle and collective degrees of freedom are present in  $^{92}$Zr.
The first excited state lifetime in $^{104}$Zr  was deduced to be 2885(435) ps at Argonne by employing
a $^{252}$Cf(SF) decay technique  \cite{2006Hw01}.  This experiment indicates that $^{104}$Zr has one of the most
deformed 2$^{+}$ state among medium and heavy even-even nuclei. The deformation can also be reproduced with HFB calculations. 
Fast timing in beta decay results were published recently for $^{104,106}$Zr \cite{2015BrP}.

\subsection{$^{106,108}$Mo}
The $^{106}$Mo lifetime of 173(14) ps was deduced at Lawrence Berkeley National Laboratory with a delayed coincidence (DC) technique \cite{2006Hw01}.
This result helped to resolve ambiguities in previous measurements. The previously discussed preliminary results have helped to
shed more light on $^{106,108}$Mo \cite{2015Br03}.

\subsection{$^{96,98,106}$Ru}
Lifetimes of $^{96,98}$Ru were extracted with the Doppler shift attenuation technique at K{\"o}ln   \cite{2002Kl07} and  Yale \cite{2012To01,2012Ra03}.
The K{\"o}ln measurement is well reproduced with shell model calculations, and the Yale experimental ratio of 4$^{+}$/2$^{+}$ transition strengths
agrees well with the vibrational character of the low-energy excitations in $^{98}$Ru.
A time-delayed technique at Jyvaskyla, Finland was used to deduce a lifetime of 249(5) ps in $^{106}$Ru \cite{2008Sa05}. This value is consistent with the
General Collective Model and Interacting Boson Approximation (IBA) calculations.

\subsection{$^{98,100,110,114}$Pd}
The recoil distance method (RDM) was employed at K{\"o}ln to set a lifetime limit  of $<$16.3 ps for $^{98}$Pd \cite{2009FrZZ}
and a value of 9.0(4) ps for $^{100}$Pd \cite{2009Ra28}.
The $^{98}$Pd nucleus is not very collective due to  its closeness to doubly-magic $^{100}$Sn, and $^{100}$Pd is well reproduced by shell model calculations.
The Yale RDM value \cite{2012An17} is rather preliminary in nature, and was excluded from the evaluation process.  The identical technique was used at MSU
to deduce lifetimes of 67(8) and 118(20) ps for $^{110,114}$Pd nuclei, respectively \cite{2008De30}. The new B(E2) values 
are described  in the framework of the Interacting Boson Model (IBM),
and  $^{114}$Pd data fit nicely into the systematic trends deduced from the lighter Pd isotopes.

\subsection{$^{100,102,104,110,122,124,126}$Cd}
Coulomb excitation and recoil distance techniques were used to measure excitation strength
in $^{100,102,104}$Cd \cite{2009Ek01,2007Bo17,2001Li24}. These data could be described within
the shell-model using realistic matrix elements obtained from a G-matrix renormalized CD-Bonn interaction.
The recoil distance Doppler shift technique  was used to deduce a
lifetime of 8.7(12) ps in $^{110}$Cd at K{\"o}ln \cite{2001Ha09}. The E2-transition probabilities in $^{110}$Cd are
in rather good agreement with the predictions of the U(5)-limit of the IBM-1.
The REX-ISOLDE collaboration employed the Coulex technique  to deduce B(E2) values in $^{122,124,126}$Cd \cite{2014Il01}.
These data agree well with other preliminary results \cite{2006KrZV,2008KrZZ},
and clarify the onset of collectivity in the vicinity of the $Z$=50 and $N$=82 shell closures.

\subsection{$^{104,106,108,110,112,114,116,118,120,122,124,126,128,130,132,134}$Sn}
Quadrupole collectivity of even-A tin isotopes was extensively studied during the last decade.
The intermediate-energy Coulomb excitation technique was used to
deduce the B(E2) value in $^{104}$Sn \cite{2013Ba57,2014Do19}.
Both results are consistent, and show the enhanced collectivity below the midshell, approaching $N$=$Z$=50. These results
disagree with the modern many-body calculations.
The same technique was used to measure B(E2) values in $^{106,108,110}$Sn  at GSI RISING, MSU,
and REX-ISOLDE at CERN \cite{2005Bb09,2007Va22,2008Ek01,2008EkZZ}.  These results show that the transition
strengths for these nuclei are larger than predicted by current state-of-the-art shell-model calculations.
For spectroscopic purposes, $^{112,114}$Sn nuclei were re-measured at MSU, and IUAC in New Delhi, India \cite{2007Va22,2010Ku07,2011Ku05}.
Precise measurements  of the first 2$^{+}$ excited states lifetimes in $^{112,114,116,122}$Sn and 
B(E2) values in $^{116,118,120,124}$Sn were conducted with the Doppler shift attenuation technique
at GSI and Australian National University, respectively  \cite{2011Ju01}.  For the isotopes $^{112,114,116}$Sn, 
the E2 transition strengths, deduced from the measured lifetimes, are in disagreement
with the previous values and indicate a shallow minimum at $N$=66.
A series of Coulomb excitation and  Doppler shift attenuation measurements were conducted at Oak Ridge National Laboratory
to measure collectivity in $^{124,126,128,130,132,134}$Sn, employing carbon and titanium targets \cite{2012Ku24,2011Al25,2004Ra27,2005Va31}.
The Oak Ridge data were compared to large-scale shell-model and quasiparticle random-phase calculations. The shell model predictions  are consistent
with a generalized-seniority scheme, which predicts relatively constant 2$^{+}_{1}$ energies and a parabolic trend in the matrix elements for  $A$=102-130.

\subsection{$^{108,112,114,118,120,122,128,130,132,134,136}$Te}
The lifetime of the first excited 2$^{+}$ state in $^{108}$Te has been measured, using a combined recoil decay tagging and
recoil distance Doppler shift technique at  Jyvaskyla (JYFL), Finland \cite{2011Ba37}.  In contrast to the earlier results for the light tin isotopes,
$^{108}$Te does not show any enhanced transition probability with respect to the theoretical predictions and the tellurium systematics.
The lifetime in the neutron-deficicient nucleus $^{112}$Te has been measured  using the DPUNS plunger and the recoil distance Doppler shift technique \cite{2015Do04}.
$^{114}$Te lifetimes were determined using the recoil distance Doppler-shift technique
with a plunger device at K{\"o}ln \cite{2005Mo20}. The energy spectrum of $^{114}$Te is a slightly anharmonic vibrator, however, the obtained B(E2) values are
in strong contradiction with the theoretical predictions of the U(5) limit of IBM.
Lifetimes of excited states in $^{118}$Te have been measured using the Doppler Shift Attenuation method
(DSAM) and Recoil Distance method (RDM) at the Niels Bohr Institute in Denmark \cite{2002Pa19}.  The excitation
energies and B(E2) values are satisfactorily interpreted in the framework of IBM.
$^{120}$Te was recently studied at Yale with a plunger device and inverse kinematics Coulomb excitation
with heavy beams \cite{2010We12}, and at IUAC, New Delhi by DSAM \cite{2014Sa49}. $^{122}$Te excitations have been investigated
using $\gamma$-ray spectroscopy following inelastic neutron scattering at Kentucky \cite{2005Hi04}. The energies of low-lying levels of
tellurium are described by the IBM.
$^{126,128,130,132}$Te collectivities were measured with time correlation between fission fragments
and $\gamma$-rays at Grenoble \cite{2001Ge07}.
Independently, B(E2) values of $^{132,134,136}$Te were studied with Coulomb excitation at Oak Ridge \cite{2002Ra21}, and explained within the shell model formalism.
The results of this measurement were further re-analyzed by the Oak Ridge group and updated values were published in a subsequent measurement publication \cite{2011Da21}.
Complementary B(E2) values in $^{130,132,134}$Te were determined through Coulomb excitation in inverse kinematics \cite{2003Ba01}.
This led to the extension of  systematics of experimental quadrupole collectivities from the ground state to the first excited state  to the $N$=82 shell closure.

\subsection{$^{114,124,126,128,130,132,134,138,140,142,144}$Xe}
Quadrupole collectivity in $^{114}$Xe  was studied  using the 4$\pi$ spectrometer, EUROBALL IV and Cologne plunger device \cite{2002De26}, then explained
with a total Routhian surfaces calculation. As a first test of SeGA Ge-array, the MSU group has conducted a Coulex  experiment at the
Argonne tandem \cite{2006Mu04} to study  $^{124,126,128,130,132,134}$Xe. These results agree well with the previously published data.
Preliminary values for the B(E2) values in $^{138,140,142,144}$Xe were deduced using the
Coulex technique and MINIBALL Ge-array \cite{2007Kr12,2007Kr19,2008KrZZ}.
The $^{140}$Xe value agrees well with a previously published measurement \cite{1999Li18},
while the $^{138}$Xe experimental value exceeds that predicted by the quasiparticle random phase approximation.

\subsection{$^{122,136,140}$Ba}
$^{122}$Ba lifetime was studied with RDM using the Cologne plunger device \cite{2010Bi11}.
The corresponding B(E2) value agreed with the predictions of the X(5) model and calculations performed in the framework of the IBA-1 and IBA-2 models.
A $^{136}$Ba stable beam Coulex measurement  at Oak Ridge yielded a reduced transition probability of 0.46(4) e$^2$b$^2$ \cite{2002Ra21}.
The transition probability is in agreement with the adopted value \cite{2001Ra27}.
A B(E2) value and lifetime of 0.484($^{+38}_{-101}$) e$^2$b$^2$ and 10.4($^{+22}_{-8}$) ps, respectively,
were measured at REX-ISOLDE and MINIBALL setup at CERN using $^{140}$Ba  particle beams \cite{2012Ba40}.
The present result agrees with predictions of Monte Carlo shell-model and energy density functional  calculations.

\subsection{$^{148,152}$Ce}
A $^{252}$Cf(SF) radioactive source and the Gammasphere array were employed to measure lifetimes
of  130(43) and 360(24) ps in $^{148,152}$Ce \cite{2006Hw01,2005Fo17}, respectively.
The $^{148}$Ce lifetime is marginally lower but still consistent with the previously reported values, while $^{152}$Ce was measured for the first time.

\subsection{$^{136,140}$Nd}
A relativistic Coulex technique was employed to deduce a B(E2) strength of 80(11) W.u. in $^{136}$Nd at Darmstadt \cite{2008Sa35}.
The comparison with the asymmetric rotor  and the Geometrical Collective Model (GCM) yields information on the nuclear shape,
quadrupole deformation parameters, and indicates $\gamma$-softness of the $N$=76 isotone.
A low-energy Coulex experiment was used to deduced a B(E2)$\uparrow$ value of 0.72(5) e$^2$b$^2$ in $^{140}$Nd
using the Miniball array at the REX-ISOLDE-CERN facility \cite{2013Ba38}.
The quasiparticle phonon and large-scale shell model calculations of $N$=80 isotones
could not reproduce an E2 strength enhancement in $^{140}$Nd.

\subsection{$^{140,142,152}$Sm}
Lifetime of the first 2$^{+}$ excited state at 530.7 keV was measured from recoil-distance Doppler shift 
method \cite{2015Be25} at the Heavy Ion Laboratory of the University of Warsaw. 
The Coulomb excitation technique was used to investigate evolution of quadrupole collectivity in $^{142}$Sm \cite{2015St08}.
A recent in-flight fast-timing measurement of the $^{152}$Sm lifetime \cite{2014Pl01} agrees well with the ENSDF recommendation.

\subsection{$^{138,148,160,162,164}$Gd}
The first excited state lifetime, 308(17) ps, for $^{138}$Gd was confirmed using RDM with the Cologne plunger at the 
Javaskyla facility \cite{2011Pr10}. The excitation energies in $^{138}$Gd can be reproduced with X(5) critical-point calculations, however,
large experimental B(E2) uncertainties  cannot rule out contributions from rotational and vibrational modes of excitation.
The same technique was applied to measure a lifetime of 6.0(19) ps in $^{148}$Gd using the EUROBALL array \cite{2003Po02}, and results were reproduced
with shell model calculations. $^{148}$Gd has the smallest B(E2) value among the $N{>}82$ nuclei in the region.
Lifetimes in $^{160,162,164}$Gd  were recently measured using a $\beta$-$\gamma$ timing technique at JAEA \cite{2010NaZY}.
These results suggest that the deformation of nuclear shape would be enhanced at $N$=98.

\subsection{$^{156}$Dy}
To test the X(5) model,  the lifetime for $^{156}$Dy was measured to be 106(15) ps with the recoil
distance Doppler-shift method using  the Cologne coincidence plunger apparatus at Legnaro \cite{2006Mo22}.
A fit of the data using the general collective model suggests contribution of a deeper collective potential.

\subsection{$^{158,170}$Er}
A lifetime of  341(10) ps for $^{158}$Er was measured with the recoil distance technique and the Gammasphere array \cite{2002Sh09}.
This result is consistent with the previous measurement \cite{1986Os02} and was reproduced using Ultimate Cranker model calculations.
The Coulex technique was used to deduce properties of $^{170}$Er at Legnaro \cite{2011Di07}.
The reduced matrix elements extracted with the Coulomb excitation code GOSIA  are in agreement with collective model predictions.

\subsection{$^{168,170,172,174,176}$Hf}
Preliminary values of the lifetimes for the first 2$^{+}$ states
in $^{168,172}$Hf, were measured using the delayed-coincidence technique at Yale \cite{2011We08,2010We12} 
to be 1237(10) and 2655(79) ps, respectively. The results for $^{168}$Hf and  $^{172}$Hf are
in agreement and slightly higher than ENSDF adopted values, respectively. These results and the 
transition strengths in $^{174,176}$Hf were tested at the university of Cologne \cite{2011ReZZ,2015Ru03}.
A lifetime of 1740(60) ps for $^{170}$Hf was deduced at the Stony Brook University TANDEM-LINAC facility
with the help of a pulsed beam technique  \cite{2006Co20}. The corresponding E2 transition rate follows the expected trend
and empirically confirms the correlation between deformation and the filling of major shells.
An extended e$^{-}$-e$^{-}$ lifetime measurement of $^{174}$Hf  has been performed at the
Cologne Tandem Van-de-Graaff accelerator \cite{2009Re20}. This measurement suggests a value lower than previously reported.

\subsection{$^{172,174,176,178,188}$W}
The first excited 2$^{+}$ state lifetimes of 970(29), 1431(9), and 1642(21) ps in $^{172,176,178}$W, respectively,
were measured  in fast timing experiments using conversion electron spectroscopy at K{\"o}ln \cite{2010Ru12,2009Re20}.
IBM calculations reproduce systematics of energy levels  for the tungsten isotopes, however, transition rates could only be satisfactorily
reproduced with individual adjustments of the effective charge.
The preliminary value of the $^{174}$W lifetime was deduced using a DC  technique  at Yale \cite{2011We08}.
The IFIN-HH, Romania facility was used to measure a lifetime of 1255(173) ps in $^{188}$W with a fast-timing technique \cite{2013Ma66}.
This result, in combination with systematics  and Woods-Saxon potential energy surface calculations, suggests  a prolate deformed minima with rapidly
increasing $\gamma$-softness for tungsten isotopes.

\subsection{$^{174,176,178,180,188,190}$Os}
The first excited 2$^{+}$ state lifetime of 513(20) ps in $^{174}$Os was measured with a DC technique
at the China Institute of Atomic Energy \cite{2012Li50}. The low uncertainty makes this value sufficiently precise
to serve as a normalization parameter for meaningful tests of nuclear models.
The DC technique was also employed to deduce the lifetimes  of  $^{176,178,180}$Os at K{\"o}ln \cite{2005Mo33}.
Data for the even-even osmium isotopes transition strengths show a maximum value at
the $N$=104 midshell that corresponds to the simple expectation of the N$_{\pi}$N$_{\nu}$ rule of the IBA.
Lifetimes of 930(140) and 540(36) ps in $^{188,190}$Os, respectively,  were measured with the recoil distance technique at Yale \cite{2001Wu03}.
The measured lifetimes confirm the E2 properties derived from prior heavy-ion induced Coulomb excitation experiments \cite{1996Wu07}.
The previously known $^{190}$Os lifetime was verified at the IFIN-HH facility \cite{2012MaZP}.

\subsection{$^{178,182,186,196}$Pt}
The lifetime of the 2$^{+}_{1}$ state in $^{178}$Pt was measured by using fast-timing techniques with the high-purity Ge and
LaBr$_{3}$ scintillator at the China Institute of Atomic Energy \cite{2014Li45}. The first excited 2$^{+}$ state lifetime of 590(102) ps   in $^{182}$Pt was measured
with RDM at K{\"o}ln \cite{2012Gl01}.  Calculations within the IBM and the
GCM indicate shape coexistence in $^{182}$Pt.
This is consistent with the previous measurement lifetimes of 709(43) and 318(24) ps
in $^{182,186}$Pt, which were deduced using the same  method at the ATLAS facility  \cite{2012Wa16}. The experimental lifetime value
in $^{196}$Pt has been revisited recently \cite{2015Jo01}.

\subsection{$^{180,182,184,186}$Hg}
The first excited 2$^{+}$ state lifetimes of 17.5(25) and 41(3) ps   in $^{180,182}$Hg were measured with RDM
at Jyvaskyla \cite{2009Gr09}. These results support the shape coexistence of weak prolate and intruding prolate
structures in neutron-deficient Hg nuclei.
$^{184,186}$Hg lifetimes were measured using the recoil distance Doppler-shift method using the   K{\"o}ln plunger device \cite{2014Ga04}.  
These more precise lifetime values have been used in the analysis of
Coulomb excitation of $^{182,184,186,188}$Hg measurements at the REX-ISOLDE facility \cite{2014Br05}.
Further analysis of properties of the low-lying states in $^{182-188}$Hg indicates a partial agreement with beyond mean field and
IBM-based models and a possible interpretation within a two-state mixing model.

\subsection{$^{186,188,208}$Pb}
Lifetimes of prolate intruder states in  $^{186,188}$Pb were measured with RDM
at Jyvaskyla \cite{2008Gr04}. Reduced transition probabilities, derived from the measured lifetimes
confirm the high collectivity of the intruder states in this region, and shed more light on shape coexistence
typical for the nuclei near $Z$=82 and $N$=104.
A lifetime of 0.00147(10) ps and B(E2)$\uparrow$  value of 0.25(6) e$^2$b$^2$  in $^{208}$Pb were deduced
with nuclear resonance fluorescence technique at Darmstadt \cite{2003En07} and the NIAIS, Japan \cite{2008Sh23}, respectively.
The latter result was compared with an estimation of self-consistent random phase approximation using a semi-realistic interaction.

\subsection{$^{194,196,198,200,202}$Po}
The first excited 2$^{+}$ state lifetimes, 37(7) and 11.6(15) ps, in $^{194,196}$Po, respectively,  were measured
with RDM at Jyvaskyla \cite{2008Gr04,2009Gr08}. Self-consistent mean-field calculations suggest that
oblate intruder states in $^{194}$Po could dominate the ground state. A calculated collectivity in $^{196}$Po,
considerably smaller than the experimental value of 47(6) W.u., indicates a contribution from the intruder structures. 
E2 matrix elements for $^{196,198,200,202}$Po have been extracted at Leuven with GOSIA analysis \cite{2015KeZZ}. The values of  
nuclear matrix elements hint towards mixing of a spherical structure with a weakly-deformed rotational structure.

\subsection{$^{202,204,220}$Rn}
Shape coexistence in $^{202,204}$Rn \cite{2015Ga19} has been studied at CERN. The same facility also measured a B(E2)$\uparrow$  value of 
1.88(11) e$^2$b$^2$ in the 'octupole deformed' or distorted pear-shaped nucleus $^{220}$Rn \cite{2013Ga23}.

\subsection{$^{224}$Ra}
In another pear-shaped nucleus, $^{224}$Ra, quadrupole collectivity  was investigated in the same work at CERN \cite{2013Ga23}.
Its B(E2)$\uparrow$  value of 3.96(12) e$^2$b$^2$ provides evidence for a stronger octupole deformation than in $^{220}$Rn.

\section{B(E2)$\uparrow$ Evaluation Policies}

The current evaluation represents the recommended values of B(E2)$\uparrow$ in $e^{2}b^{2}$, mean lifetimes ($\tau$) in picoseconds (ps)
and deformation parameters ($\beta_{2}$) for the first 2$^{+}$ states in $Z$=2-104, even $N$ nuclei. These quantities are mutually related:
\begin{equation}
\label{eq0}
\tau =  40.81 \times 10^{13} E^{-5}_{\gamma} [B(E2)\uparrow / e^{2}b^{2}]^{-1} (1+\alpha_{T})^{-1}
\end{equation}
\begin{equation}
\label{eq1}
\beta_{2} = (4\pi / 3 ZR_0^2)[B(E2)\uparrow/ e^{2}]^{1/2},
\end{equation}
where $E_{\gamma}$ and $\alpha_{T}$ are the $\gamma$-ray energy in keV and the total conversion electron coefficient, 
respectively, and $R_{0}^{2} = (1.2 {\times} 10^{-13} A^{1/3}$cm)$^2$. To introduce an additional measure of collectivity
for nuclear excitations, Weisskopf units (W.u.) are added.
Transition quadrupole moment values $Q_0$ in barns (b) are not included in the current evaluation, however can
be deduced from the presented work
\begin{equation}
\label{eq2}
Q_{0} = [16 \pi B(E2)\uparrow / 5 e^2]^{1/2}.
\end{equation}
All the measured values can be organized using three classes of experimental techniques:
\begin{itemize}
\item Model-independent or traditional types of measurements \cite{2001Ra27}: transmission Doppler-shift attenuation lifetime (TDSA), recoil distance Doppler-shift (RDM or RDDS), delayed coincidences (DC or TCS), low-energy and intermediate-energy Coulomb  excitation (CE) and nuclear resonance fluorescence ($\gamma$,$\gamma^{\prime}$).
\item Low model-dependent: electron scattering (E,E$^{\prime}$), hyperfine splitting.
\item Model-dependent: inelastic scattering of light and heavy ions (IN-EL).
\end{itemize}

\subsection{B(E2)$\uparrow$ Evaluation Procedure}
This evaluation is based on the analysis of results from 2579 quadrupole collectivity measurements and 1273 experimental references. The literature cut-off date is September 2015. This number includes
120  pre-2000 experimental references that were not listed in the previous compilation of S. Raman \cite{2001Ra27}.
These data span more than sixty years, and experimental techniques have evolved over time. It is worth noting that
in the older measurements results may have been affected by the lack of side-feeding corrections, and the newer measurements should take precedence.
The evaluation procedure for deducing the adopted (recommended) B(E2)$\uparrow$ values is presented below:
\begin{itemize}
\item Compile a list of experimental B(E2)$\uparrow$,$\downarrow$ or W.u., $\tau$ and $\beta_2$ values as reported in the original papers. Reported values depend on the measured quantities and are deduced from experimental data in the offline analysis.
\item Convert experimental  values into B(E2)$\uparrow$ in  $e^{2}b^{2}$.
\item Analyze B(E2)$\uparrow$ values. In a few of the older results, where uncertainties were not quoted by the authors, we have taken
the values as adopted by Raman {\it et al.} \cite{2001Ra27}.
The experimental values listed in Table 3, with the original uncertainties as quoted by the authors (not adjusted for evaluation purposes).
\item Round uncertainties to two (rarely three) significant digits.
\item Accept asymmetric uncertainties, if necessary.
\item Direct communication with authors in discrepant cases, if possible.
\item Deduce B(E2)$\uparrow$ recommended values using model-independent or traditional, combined (model-independent and low model-dependent)
and model-dependent data sets with the visual averaging library software package \cite{vavelib} using the selected data sets.
\end{itemize}

\subsection{Asymmetric Uncertainties}
In this work, evaluated B(E2)$\uparrow$ values are deduced from the measured values of B(E2), mean lifetime, 
and, in rare cases, deformation parameters. Note that the first two quantities are inversely dependent.
Previous evaluations of S. Raman {\it et al.} \cite{1987Ra01,2001Ra27} contain two different treatments 
of central values and uncertainties. Originally, S. Raman {\it et al.} \cite{1987Ra01} used a standard mathematical 
procedure  to convert a particular $\tau$ value to the corresponding B(E2)$\downarrow$ value. Later, this procedure was changed in favor of
converting the central $\tau$ value \cite{2001Ra27} to a value between the upper and lower bounds,
by extracting the mean of the two values and assigning an uncertainty so that the value overlapped the two bounds. This treatment produced symmetric uncertainties,
however the original lifetime values could not be directly reproduced from the modified B(E2)$\uparrow$ central values.
To resolve this discrepancy, we kept central values and accepted asymmetric uncertainties that arise from the inverse dependence described above.
In addition, original measurements may contain asymmetric uncertainties due to particular experimental conditions and analyses.

\subsection{A Brief Review of the Previous Results}
Consistency between the present results and the work of S. Raman et {\it al.} \cite{2001Ra27} is an important issue.
The authors of the Ref. \cite{2001Ra27} indicate: ``Where several B(E2)$\uparrow$ values are available for a given nuclide, we have generally used weighting values that
are inversely proportional to the quoted uncertainty rather than inversely proportional to the square of the quoted uncertainty, which would be the correct procedure if the
uncertainties were purely statistical. We believe that our weighting procedure results in a more reliable average value. We did not, however, adhere religiously to the weighting
procedure outlined above in all cases."

We do not know the exact course of action taken by S. Raman and his collaborators for the evaluation of B(E2) values in each particular case.
However, in the present work we rely on the general statistical and uncertainty handling procedures employed in nuclear data evaluations such as in the ENSDF database, or in
Particle Data Project \cite{pdg2014}, and employ
the Visual Averaging Library code  \cite{vavelib} where weighting values are inversely proportional to the square of
the quoted uncertainty. To check the validity of S. Raman's claim to have used an inverse weighting procedure, we developed a custom extension for the 
Visual Averaging Library that produces values using averaging weights which are inversely proportional to the quoted uncertainty. This approach
is also used to make an overall consistency check as described below between our results and those in Raman et {\it al.} \cite{2001Ra27} where no new experimental
values are available.

\begin{figure}[!htb]
\includegraphics[height=12cm,angle=0]{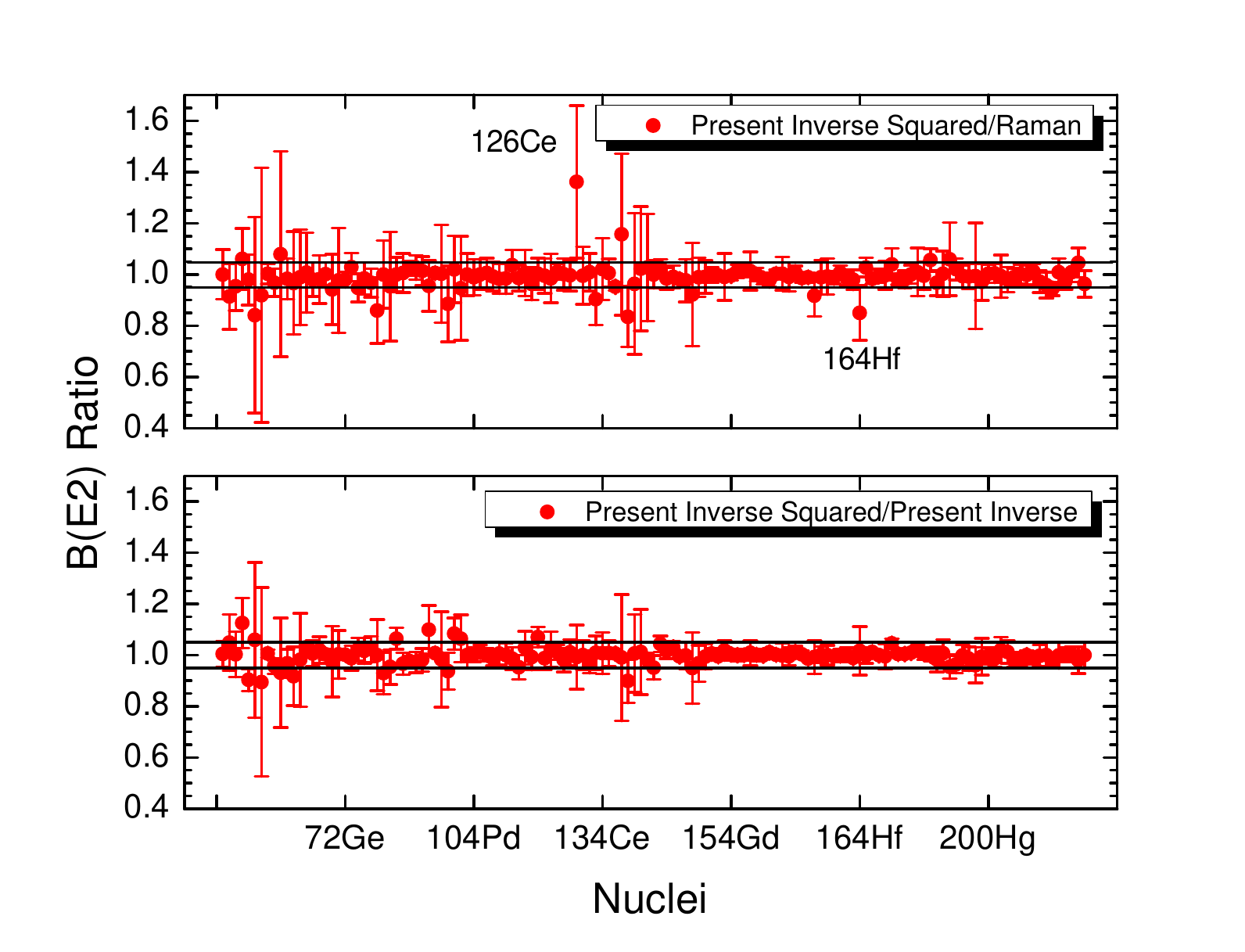}
\caption{The ratio of the present B(E2) values to Raman's evaluation and inverse squared to inverse (linear) averages 
for 135 nuclides are shown in the upper and lower panels, respectively. The majority of calculated values lie within a $\pm$5$\%$ band near unity.}
\label{figPreRaman}
\end{figure}

In this work, we have selected 135 nuclides where no new measurements have been reported since the previous evaluation,
and B(E2)$\uparrow$ value for each nuclide value was measured at least twice.  The ratio of these data are shown in the upper panel of Fig.~\ref{figPreRaman}.
The majority of the B(E2)$\uparrow$ values are within 5$\%$ agreement. Notable deviations from unity in  $^{126}$Ce and $^{164}$Hf are
due to missing data and adoption by Raman of the earliest results, respectively. 

To extend this analysis we calculate both the inverse squared and inverse B(E2)$\uparrow$ averages  for these nuclides.
A comparative analysis of the current inverse squared, current inverse,  and Raman's B(E2)$\uparrow$ results for $Z$=2-104 isotopes
is shown in Table \ref{tableResults} and  Fig.~\ref{figPreRaman}. 
The  data analysis indicates that we have a good agreement between the present inverse squared averages and Raman's values,
and the inverse averaging often results in comparable values with the corresponding inverse squared averages values. 
These facts and comments in the table clearly indicate that S. Raman et {\it al.} \cite{2001Ra27} were not following their practice of linear weighting consistently.

\section{Adopted  values}

The recommended values for $Z$=2-104 isotopes from this work are shown in Table \ref{tableAdopted}.
These data extend the previous work of S. Raman {\it et al.} \cite{2001Ra27} with 119 new B(E2)$\uparrow$ values as well as a large number of updated values. 
The current work also contains 646 $\gamma$-ray energies for the first 2$^{+}$ states in even-even nuclei. A comparative analysis of the two 
evaluations is presented below.

In the present evaluation, we  used the latest version of the visual averaging library \cite{vavelib}, Band-Raman  calculation of Internal conversion coefficients ($\alpha_{T}$) \cite{2008Ki07} and presently available data.
The visual averaging library program includes unweighted and weighted averages as well as the limitation of relative statistical weights (LWM) \cite{2004Ni03}, 
normalized residual (NRM) \cite{1992Ja06}, Rajeval technique (RT) \cite{1992Ra08}, the Expected Value (EVM) \cite{2014Bi13}, bootstrap and
Mandel-Paule (MP) \cite{1998RuAA} statistical methods to calculate averages of experimental data with uncertainties. In our evaluation, 
we generally adopted the weighted average, using NRM in some discrepant cases.
We accepted reduced $\chi^{2}<2$ as a reasonable fit for available data sets.
Previously, S. Raman {\it et al.} \cite{2001Ra27} used an averaging procedure based on the inverse of the
quoted uncertainties, while current evaluation uses statistical methods  based on the inverse squared value of the quoted uncertainties. Our procedure, in addition to being mathematically justifiable, is consistent with 
the general methodology used in treatment of data for ENSDF database and  horizontal evaluations.

\begin{figure}[!htb]
\includegraphics[height=12cm, angle=0]{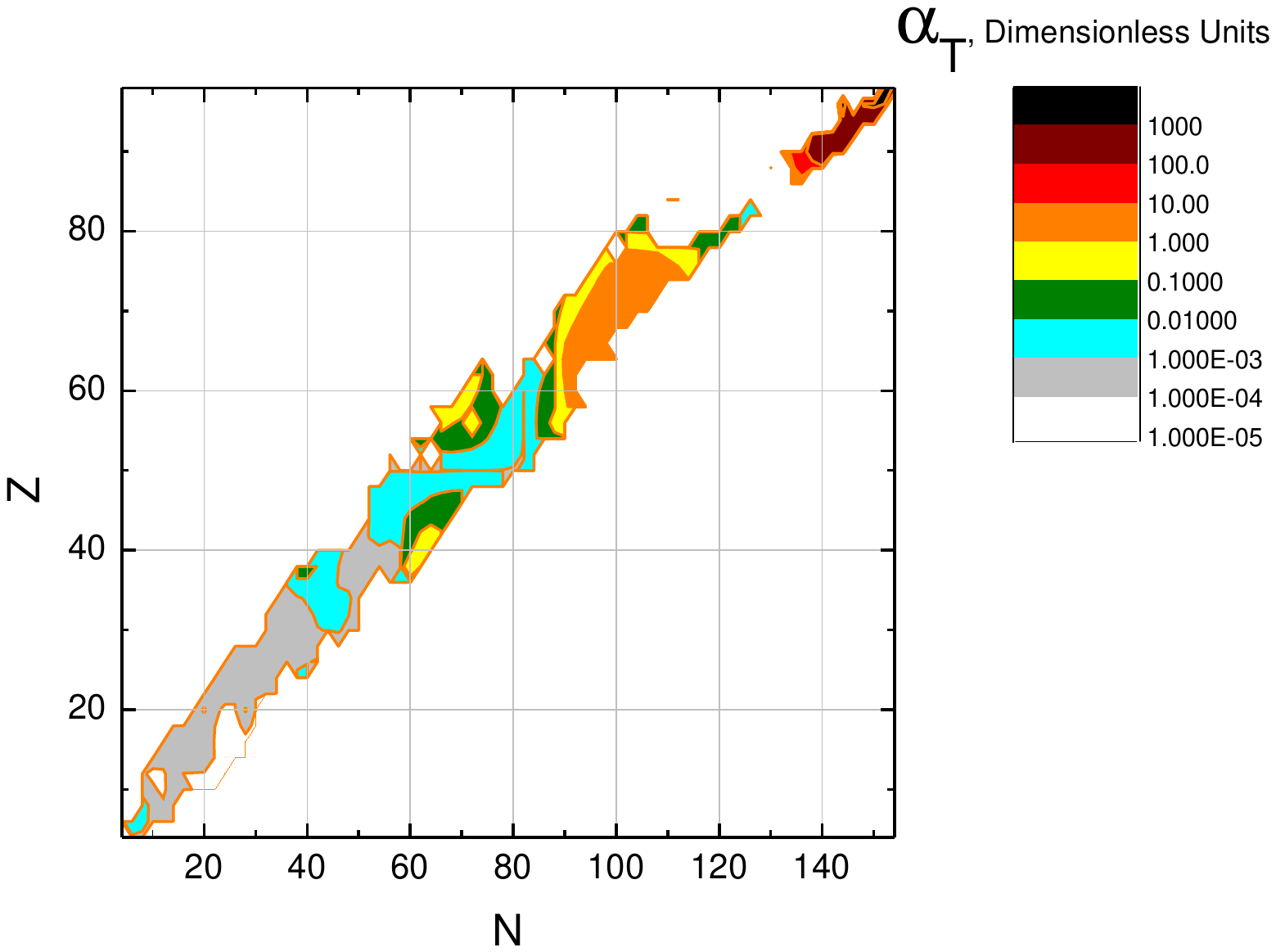}
\caption{Total Conversion Coefficients ($\alpha_{T}$) for even-even $Z$=2-104 nuclei. The coefficients have been deduced using the frozen orbital (FO) version of the BrIcc code.}
\label{figICC}
\end{figure}

The Band-Raman method \cite{2008Ki07} was used in this work, while the previous evaluation \cite{2001Ra27} employed the internal conversion coefficients code (ICCDF)  \cite{1993Ba60}.
The former code incorporates the Dirac-Fock atomic model with the exchange interaction between atomic
electrons  and the free electron receding to infinity during the conversion process.
Total conversion coefficients for the  $Z$=2-104 - region were calculated using the Australian National University BrIcc code  {\it http://bricc.anu.edu.au/}
and shown in Fig. \ref{figICC}. The coefficient values increase over eight orders of magnitude across the nuclear chart,
and reach maximum values in the actinide region.

For low $Z$ values and relatively high 2$^{+}_{1}{\rightarrow}0^{+}_{1}$ transition energies, the
total E2 conversion coefficients are relatively small ($\alpha_{T} <$ 0.002) and do not substantially affect the adopted values.
A complementary comparison between the present model-independent and the previous evaluation
adopted values for $^{14}$C,  $^{28,34,36,38}$Si, $^{38,40,42}$S and $^{38}$Ca, where no new data were added, shows good agreement.
Consequently, the differences between the current work and  S. Raman {\it et al.} \cite{2001Ra27} evaluation for light and medium nuclei are mainly due to  the addition of new  experimental results.

We recommend using model-independent or traditional B(E2)$\uparrow$ adopted values as the most reliable. If a model-independent value is not available,
a low model-dependent value should be used. Finally, a model-dependent value can be used if no other values are available. Table \ref{tableAdopted} recommended values for
Coulomb excitation and in-elastic scattering measurements in $^{28}$Ne and $^{30,32,34}$Mg isotopes support this conclusion.
This is consistent with the previous evaluation of Raman {\it et al.} \cite{2001Ra27},
who treated data as follows: ``However, our adopted B(E2)$\uparrow$ values are based only on the traditional types of
measurements because these are more direct and involve essentially model-independent analyses."
Our new recommended values are interpreted within the scope of large-scale shell-model calculations which are presented in the following sections.

\subsection{Analysis of Adopted Values}

\begin{figure}
     \begin{minipage}{0.78\textwidth}
      \includegraphics[width=\textwidth,height=\textheight,keepaspectratio]{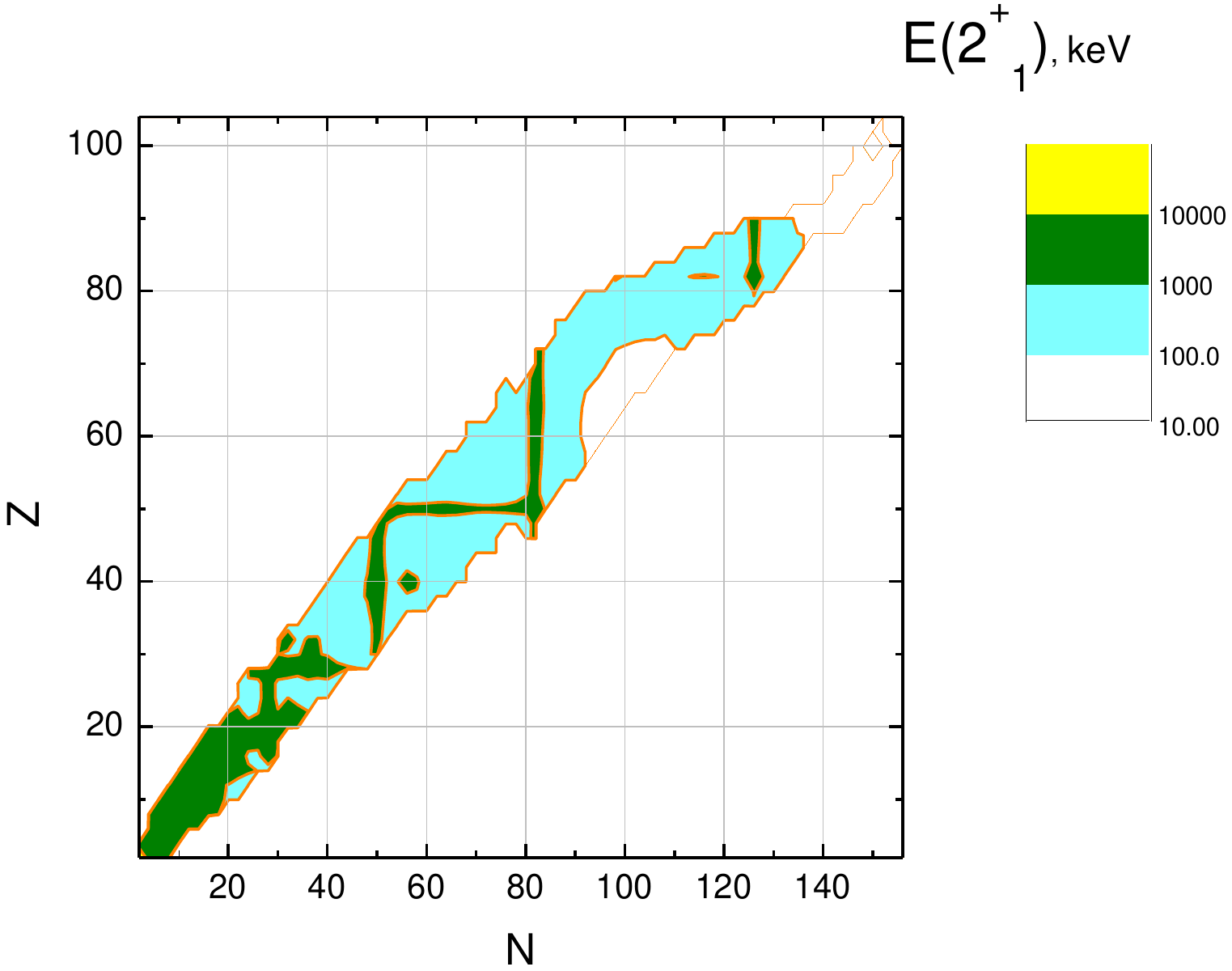}
      \caption{Energies of 2$^{+}_{1}$ states (E($2^{+}_{1}$)) for even-even $Z$=2-104 nuclei, in keV.}
      \label{figE2plus}
     \end{minipage}
     \begin{minipage}{0.78\textwidth}
      \includegraphics[width=\textwidth,height=\textheight,keepaspectratio]{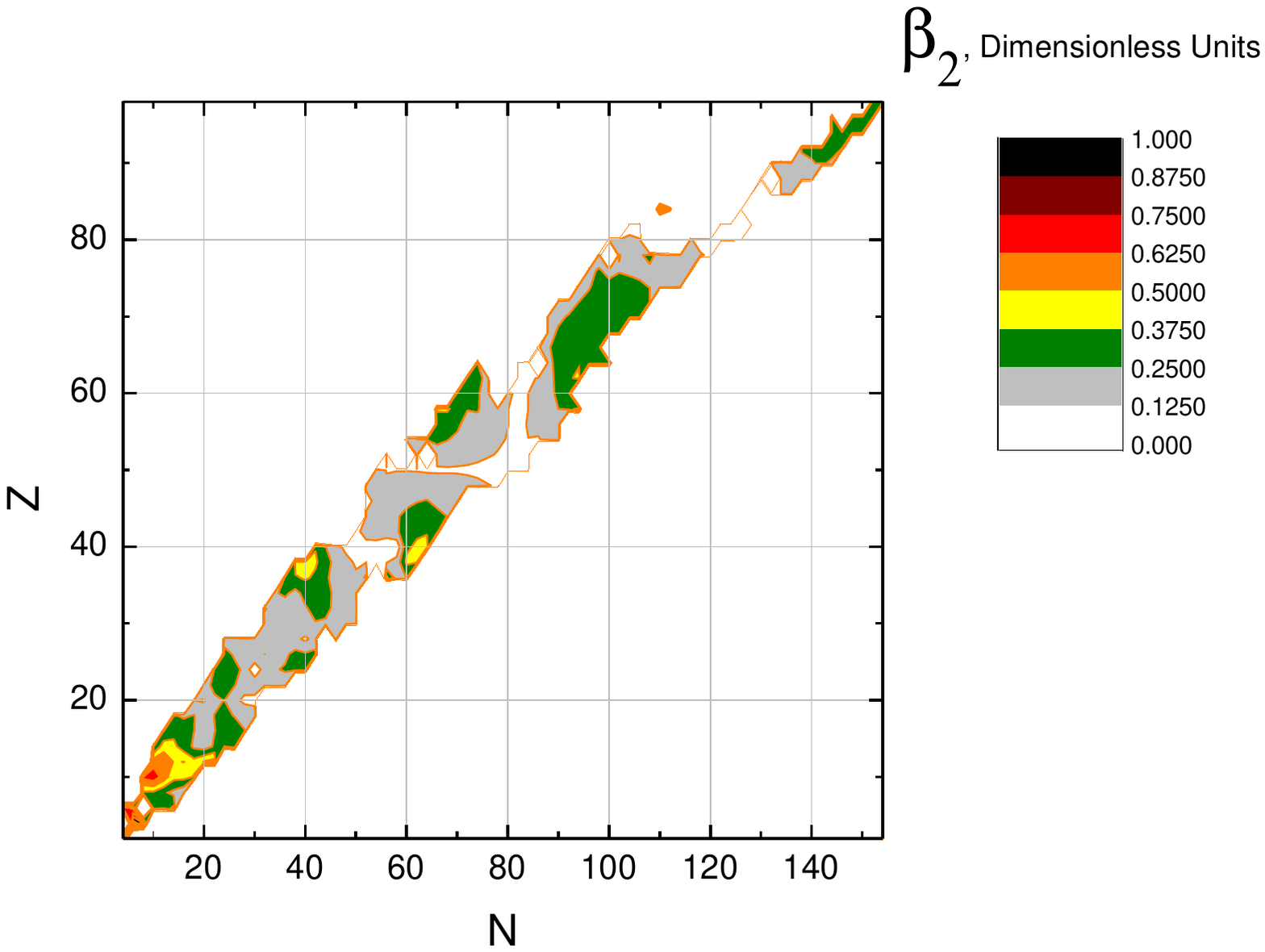}
      \caption{Quadrupole deformation parameter ($\beta_2$) values for even-even $Z$=2-104 nuclei.}
     \label{figBeta2}
     \end{minipage}
    \end{figure}

\begin{figure}[!htb]
\includegraphics[height=18cm, angle=0]{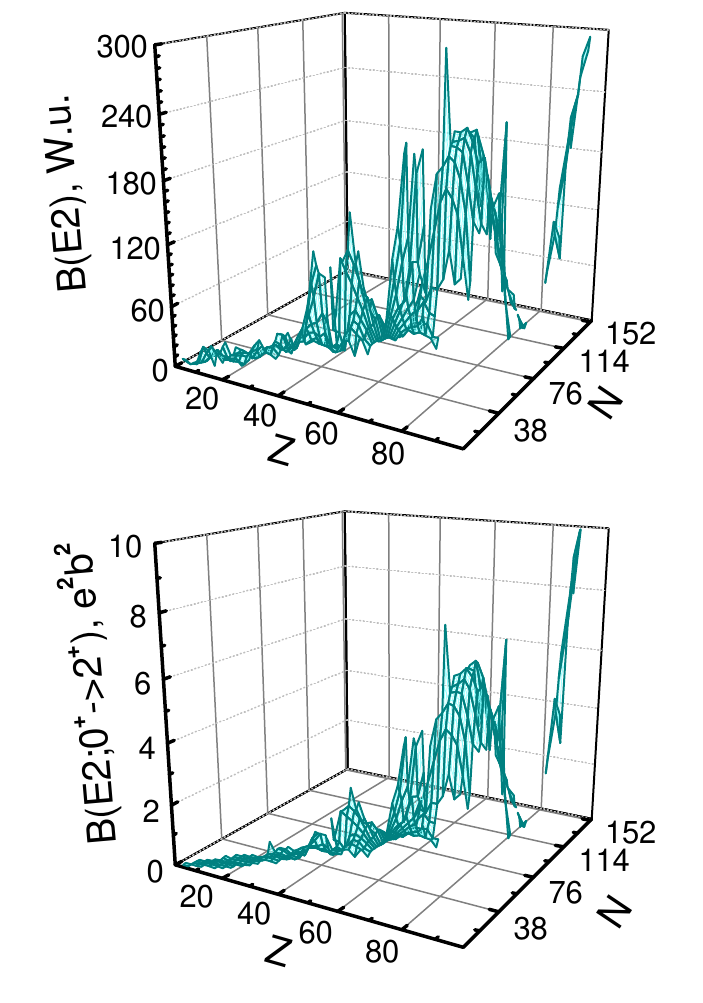}
\caption{ B(E2)$\uparrow$ in e$^{2}$b$^{2}$ and W.u. for even-even $Z$=2-104 nuclei.}
\label{figBE2}
\end{figure}

Evaluated values are traditionally given in a tabular format as in the Table \ref{tableAdopted}. In addition, we will also show these data in the
two-, and three-dimensional graphic form and  conduct a brief ``visual inspection". Plots of evaluated 2$_{1}^{+}$ state energies, quadrupole
deformation parameters, and quadrupole collectivities in $e^{2}b^{2}$ and W.u. units as functions of $N$ and $Z$ are shown in
Figs. \ref{figE2plus},\ref{figBeta2},  and \ref{figBE2}, respectively.
Fig. \ref{figE2plus} shows that energies of the 2$^{+}_{1}$ states are relatively high near the closed shells
at $Z$=20, $Z$=28 and $N$=28, $Z$=50 and $N$=50, $Z$=82 and $N$=82, and $N$=126.
However,  2$^{+}_{1}{\rightarrow}0^{+}_{1}$ transition energies are not sufficient for the
understanding of nuclear structure effects across the nuclear chart.

Furthermore, a combination of the transition energy and quadrupole deformation plots supplies a more compelling picture of nuclear shell closure
 of atomic nuclei. The deformation parameter chart indicates an anti-correlation effect between its values and  the 2$^{+}_{1}$ state energies,
 as shown in Figs. \ref{figBeta2} and \ref{figE2plus}, respectively.
The nuclear shell closure effects result in small deformation parameter values and relatively large first excited state energies.
These effects  near  $Z$=$N$=8, $Z$=20, $Z$=40 and $N$=50, $Z$=50 and $N$=82, and $Z$=82, and the deformation regions
are shown in Fig. \ref{figBeta2} using a vertical-line pattern.
To gain additional insights on nuclear collectivity a complementary analysis of the B(E2)$\uparrow$ adopted values has
been conducted in Figs.  \ref{figBE2}. The last Figure clearly demonstrates distinct nuclear properties for
light (Z$<$30), medium (Z$<$50), heavy (Z$<$84), and actinide (Z$>$88) nuclides.
In addition, systematic trends of evaluated B(E2)$\uparrow$ and E$_{2^+_1}$ values are shown in Graphs \ref{fig:graph2}-\ref{fig:graph104}. These Graphs demonstrate  
the evolution of nuclear properties of even-even nuclei and could motivate new measurements.

\section{Shell Model Calculations}

The experimental data presented in this paper covers regions of the nuclear chart that are best
treated by a diversity of nuclear structure models, including {\it ab initio} models such as Green's Function Monte Carlo
(GFMC) \cite{2004Pi09}, No-Core Shell Model (NCSM) \cite{2007Na16}, and Coupled-Cluster model \cite{2011Ja06},
but also effective models such as the traditional shell model with effective interactions, Quasi-Particle Random
Phase Approximation \cite{2008Te08}, Generator-Coordinate Method \cite{2012Ro39}, etc. Attempting to describe the data using
all these models is clearly a tall goal. Therefore, we confine ourselves to the description of a limited amount of data using
the traditional shell model with effective interactions. This model seems to have a wide range of applicability,
from light $p$-shell nuclei to nuclei around $^{208}$Pb, provided that good effective interactions are available.
Here, we only use the shell model to give some examples for $p$-shell nuclei, $sd$-shell nuclei, $sdpf$-shell nuclei,
and $pf$-shell nuclei, for which effective shell model interactions are established. In that vein, we avoided cases
where protons or neutrons are near closed shells or in the ``island of inversion".

For the $p$-shell examples we used the CKIHE interaction \cite{1988St06} for ${^6}$He and PWT interaction \cite{1992Wa22}
for Be and C. For the $sd$-shell nuclei we used the USDB interaction \cite{2006Br18} and for cases with protons in $sd$-shell and
neutrons in the $pf$-shell we used the SDPFU interaction \cite{2009No01}. Finally, for the few cases of $pf$-shell
nuclei we used the GXPF1A interaction \cite{2004Ho08,2005Ho32}. A few other cases for the $A$=60-100 region could be
also considered using the JUN45 interaction \cite{2009Ho14} in a model space that includes
the $1p_{3/2},\ 1p_{1/2},\ 0f_{5/2},\ 0g_{9/2}$ orbitals, but more insight into this region of nuclei
is required for higher reliability. Isolated cases of Sn, Te, Xe, and Ba isotopes could be considered,
but the effective interactions need to be further refined to show consistent reliability. Results of these
calculations  are shown in Table \ref{tableShell}, and complementary details of shell model calculations and analysis
of Cr, Fe, Ni and Zn nuclei could be found in our previous publication \cite{2012Pr08}. These results
were produced with ``canonical" effective charges: 0.5e for neutrons and 1.5e for protons.

Finally, the shell model and evaluated 2$_{1}^{+}$ state energies and quadrupole collectivities, are plotted in Graphs:
\ref{fig:graph2} --- \ref{fig:graph104}
for $Z$=2, 8, 20, and 28, respectively. These values and their mutual correlations provide strong evidence
for nuclear shell model across the nuclear chart. The strong correlations between transition energies and
quadrupole collectivities are, particularly,  evident for doubly-magic nuclei $^{40,48}$Ca, $^{56}$Ni, $^{132}$Sn, and $^{208}$Pb.
In addition, analysis of Graph \ref{fig:graph2} data indicates ``magic" properties for neutron-rich nucleus $^{24}$O.
There are other theoretical calculations of the B(E2)$\uparrow$ values and first excited states
in even-even nuclei \cite{2012Ro40,2013Sc21}.  These calculations could be used for nuclei where present
shell model calculations are missing.

\section{Future Plans \& Complementary Analyses}
There is a large volume of B(E2) experimental activities worldwide; a new nucleus has been measured every month in the last 10-15 years.
In such an active field of experimental work, constant compilation and evaluation work is required.
A compilation of the latest experimental results will be posted
on the B(E2) project website ({\it http://www.nndc.bnl.gov/be2}), and the next evaluation published in about ten years.

Due to space limitation, Grodzins systematics \cite{1962Gr38} and comparison of evaluated values
based on the different types of measurements \cite{2006Co03} are not presented here.
These analyses will be addressed in subsequent publications.

\section{Conclusions}
A new B(E2;$0_{1}^{+} \rightarrow 2_{1}^{+}$) compilation and evaluation of even-even nuclei has been
performed under the auspices of the USNDP.  It is a continuation of the nuclear data work by P.H. Stelson and L. Grodzins, and S. Raman {\it et al.}
on quadrupole transition probabilities  \cite{1965St16,1987Ra01,2001Ra27}.  The current evaluation literature cut-off date is September 2015,  it includes experimental B(E2)$\uparrow$ values for 119 new nuclei, a large number of updates 
and extends the previous evaluation  to 447 nuclei. The evaluation incorporates many features requested by nuclear data
users and broadens the list of compiled experimental quantities. The present evaluated results are compared with the
 previous evaluation, and large-scale shell model calculations, where available.

\section*{Acknowledgments}
The authors are grateful to Dr. M. Herman (BNL) for his constant support of this project,  J. Choquette and B. Karamy (McMaster University) for the
help with compilation work in early stages of this project. We acknowledge Profs. T. Glasmacher, V.G. Zelevinsky (MSU), T. Motobayashi (RIKEN),
V. Denisov (KINR) and J. Totans and M. Blennau (BNL) for  productive discussions, help with the references  and
careful reading of the manuscript and useful suggestions, respectively. Finally, we would like to thank many authors,
who provided valuable insights on their data. These communications have helped to resolve many experimental discrepancies.
This work was funded by the Office of Nuclear Physics, Office of Science of
the U.S. Department of Energy,  under Contract No. DE-AC02-98CH10886 with Brookhaven Science Associates, LLC.
Work at McMaster University was also supported by DOE and NSERC of Canada.
MH acknowledges support from DOE grant DE-FC02-09ER41584 (UNEDF SciDAC Collaboration).

\newpage

\TableExplanation



\bigskip
\renewcommand{\arraystretch}{1.0}

\section*{Table 1.\label{tbl3te} Experimental B(E2)$\uparrow$-, $\tau$- and $\beta_{2}$-values for Z=2-104 nuclei.}
\begin{tabular*}{0.95\textwidth}{@{}@{\extracolsep{\fill}}lp{5.5in}@{}}
\multicolumn{2}{p{0.95\textwidth}}{(Throughout this table,
        bracketed numbers refer to the uncertainties in the last
        digits of the quoted values.)}\\

Nuclide & The even $Z$, even $N$ nuclide studied\\

$B(E2)\!\!\uparrow$
        & Reduced electric quadrupole transition rate for the ground
        state to $2^+$ state transition in units of ${\textrm{e}^2}$b$^2$\\

$\tau$  & Mean lifetime of the state in ps\\

$\beta_{2}$
        & Quadrupole deformation parameter or $\delta$ deformation length\\

Target & Target nuclide\\
Beam & Incident beam \\
Energy & Incident beam energy\\
Method & CE: Coulomb excitation\\
       & CE$^{*}$: Coulomb excitation with beam energy above the Coulomb barrier\\
       & EE (e,e$^{\prime}$): Inelastic electron scattering\\
       & DC: Delayed Coincidence\\
       & GG ($\gamma$,$\gamma^{\prime}$): Resonance fluorescence\\
       & IN-EL: Inelastic scattering of light and heavy ions\\
       & PB: Pulsed beam \\
       & RDM, RDDS, TRDM: Recoil distance method \\
       & RSM: Recoil shadow method \\
       & SCATT: Neutron scattering \\
       & TDSA, DSAM: Doppler shift attenuation\\
       & TCS: Time coincidences\\

Reference & NSR database \cite{2011Pr03} keynumber\\
Comments  & ENSDF: ENSDF analysis \\
          & Ex: excluded \\
          & Gos: GOSIA code \\
          & MD: model dependent \\
          & NR: not in Raman \\
          & Rad: Raman adjusted \\
          & Su: superseded \\
          & Un: uncertainty introduced by Raman \\

\end{tabular*}
\label{tableI}

\section*{Table 2.\label{tbl3comp}  Comparative analysis of the present and S. Raman et {\it al.} \cite{2001Ra27} results.}
\begin{tabular*}{0.95\textwidth}{@{}@{\extracolsep{\fill}}lp{5.5in}@{}}

Nuclide & The even $Z$, even $N$ nuclide studied\\

Inverse Squared $B(E2)\!\!\uparrow$
        & Reduced electric quadrupole transition rate for the ground
        state to $2^+$ state transition in units of ${\textrm{e}^2}$b$^2$\\

Inverse $B(E2)\!\!\uparrow$
        & Reduced electric quadrupole transition rate for the ground
        state to $2^+$ state transition in units of ${\textrm{e}^2}$b$^2$\\

Raman's $B(E2)\!\!\uparrow$ \cite{2001Ra27}
        & Reduced electric quadrupole transition rate for the ground
        state to $2^+$ state transition in units of ${\textrm{e}^2}$b$^2$\\

Comments on Raman's values \cite{2001Ra27} & Description \\

\end{tabular*}
\label{tableII}

\section*{Table 3.\label{tbl1te} Adopted (recommended) B(E2)$\uparrow$-, $\tau$- and $\beta_2$-values for Z=2-104 nuclei.}
\begin{tabular*}{0.95\textwidth}{@{}@{\extracolsep{\fill}}lp{5.5in}@{}}
\multicolumn{2}{p{0.95\textwidth}}{(Throughout this table,
        bracketed  numbers refer to the uncertainties in the last
        digits of the quoted values.)}\\

Nuclide & The even $Z$, even $N$ nuclide studied\\

$E$(level)
        & Energy of the first excited 2$^+$ state in keV either from a
        compilation or from current literature\\

$B(E2)\!\!\uparrow$
        & Reduced electric quadrupole transition rate for the ground
        state to $2^+$ state transition in units of ${\textrm{e}^2}$b$^2$\\

        & Reduced electric quadrupole transition rate for the ground
        state to $2^+$ state transition in W.u. (Weisskopf units);
        the Weisskopf single-particle  value is $B(E2)\uparrow_{(sp)}$ = 2.97$\times$10$^{-5}$A$^{4/3}$  e$^2$b$^2$ \cite{2001Ra27}
         \\
        & {\bf Comments}: Multiply $B(E2)\uparrow$ by a factor of 0.2 to convert it to $B(E2)\downarrow$,\\
        & ($B(E2)\downarrow$ = $\frac{(2J_{i} + 1)}{(2J_{f} + 1)}B(E2)\uparrow$, where $J_{i}$=0 and $J_{f}$=2); \\
        & use formula (\ref{eq2}) to extract transition quadrupole moment values \\

$\tau$  & Mean lifetime of the state in ps\\
        & \quad$\tau =
  40.81\times10^{13}E^{-5}[B(E2)\!\!\uparrow/{\textrm{e}^2}\textrm{b}^2]^{-1}
                (1+\alpha)^{-1}$, where $\alpha$ - Band-Raman internal conversion coefficients \\

$\beta_{2}$
        & Quadrupole deformation parameter\\
        & \quad\begin{array}[t]{l@{=}l}
          \beta_{2} & (4\pi/3 ZR_0^2)[B(E2)\!\!\uparrow/{\textrm{e}^2}]^{1/2},\
    \textrm{where}\\
                R_{0}^{2} & (1.2\times10^{-13}A^{1/3}\textrm{cm})^2\\
                          &0.0144 A^{2/3}\textrm{b}
          \end{array}\\

\end{tabular*}
\label{tableIII}

\section*{Table 4.\label{tbl2te} Shell model E(2$_1^+$)-, B(E2$\uparrow$)-values for even-even nuclei.}
\begin{tabular*}{0.95\textwidth}{@{}@{\extracolsep{\fill}}lp{5.5in}@{}}

Nuclide & The even $Z$, even $N$ nuclide studied\\

$E$(level)
        & Energy of the first excited 2$^+$ state in MeV \\

$B(E2)\!\!\uparrow$
        & Reduced electric quadrupole transition rate for the ground
        state to $2^+$ state transition in units of ${\textrm{e}^2}$b$^2$\\

Model Space &  Description\\
Effective Interaction &  Description \\
Comments     & Description \\
\end{tabular*}
\label{tableIV}

\newpage

\GraphExplanation

\section*{Graph 1. Evaluated and shell model calculated energies, E($2^{+}_{1}$), and B(E; $0_{1}^{+} \rightarrow 2_{1}^{+}$) values for He nuclei.}
\begin{tabular}{@{}p{1in}p{6in}@{}}
Theory  &  CKIHE \cite{1988St06} \\
Evaluation      &   \\
\end{tabular}

\section*{Graph 2. Evaluated and shell model calculated energies, E($2^{+}_{1}$), and B(E; $0_{1}^{+} \rightarrow 2_{1}^{+}$) values for Be nuclei.}
\begin{tabular}{@{}p{1in}p{6in}@{}}
Theory  &  PWT \cite{1992Wa22} \\
Evaluation      &   \\
\end{tabular}

\section*{Graph 3. Evaluated and shell model calculated energies, E($2^{+}_{1}$), and B(E; $0_{1}^{+} \rightarrow 2_{1}^{+}$) values for C nuclei.}
\begin{tabular}{@{}p{1in}p{6in}@{}}
Theory  &  PWT \cite{1992Wa22}, USDB \cite{2006Br18}  \\
Evaluation      &   \\
\end{tabular}

\section*{Graph 4. Evaluated and shell model calculated energies, E($2^{+}_{1}$), and B(E; $0_{1}^{+} \rightarrow 2_{1}^{+}$) values for O nuclei.}
\begin{tabular}{@{}p{1in}p{6in}@{}}
Theory          &  USDB \cite{2006Br18} \\
Evaluation      &  \\
\end{tabular}

\section*{Graph 5. Evaluated and shell model calculated energies, E($2^{+}_{1}$), and B(E; $0_{1}^{+} \rightarrow 2_{1}^{+}$) values for Ne nuclei.}
\begin{tabular}{@{}p{1in}p{6in}@{}}
Theory          &  USDB \cite{2006Br18} \\
Evaluation      &  \\
\end{tabular}

\section*{Graph 6. Evaluated and shell model calculated energies, E($2^{+}_{1}$), and B(E; $0_{1}^{+} \rightarrow 2_{1}^{+}$) values for Mg nuclei.}
\begin{tabular}{@{}p{1in}p{6in}@{}}
Theory          &  USDB \cite{2006Br18} \\
Evaluation      &  \\
\end{tabular}

\section*{Graph 7. Evaluated and shell model calculated energies, E($2^{+}_{1}$), and B(E; $0_{1}^{+} \rightarrow 2_{1}^{+}$) values for Si nuclei.}
\begin{tabular}{@{}p{1in}p{6in}@{}}
Theory          &  SDPFU \cite{2009No01} \\
Evaluation      &  \\
\end{tabular}

\section*{Graph 8. Evaluated and shell model calculated energies, E($2^{+}_{1}$), and B(E; $0_{1}^{+} \rightarrow 2_{1}^{+}$) values for S nuclei.}
\begin{tabular}{@{}p{1in}p{6in}@{}}
Theory          &  SDPFU \cite{2009No01} \\
Evaluation      &  \\
\end{tabular}

\section*{Graph 9. Evaluated and shell model calculated energies, E($2^{+}_{1}$), and B(E; $0_{1}^{+} \rightarrow 2_{1}^{+}$) values for Ar nuclei.}
\begin{tabular}{@{}p{1in}p{6in}@{}}
Theory          &   \\
Evaluation      &  \\
\end{tabular}

\section*{Graph 10. Evaluated and shell model calculated energies, E($2^{+}_{1}$), and B(E; $0_{1}^{+} \rightarrow 2_{1}^{+}$) values for Ca nuclei.}
\begin{tabular}{@{}p{1in}p{6in}@{}}
Theory  &  GXPF1A \cite{2004Ho08,2005Ho32} \\
Evaluation      &  \\
\end{tabular}

\section*{Graph 11. Evaluated and shell model calculated energies, E($2^{+}_{1}$), and B(E; $0_{1}^{+} \rightarrow 2_{1}^{+}$) values for Ti nuclei.}
\begin{tabular}{@{}p{1in}p{6in}@{}}
Theory  &  GXPF1A \cite{2004Ho08,2005Ho32} \\
Evaluation      &  \\
\end{tabular}

\section*{Graph 12. Evaluated and shell model calculated energies, E($2^{+}_{1}$), and B(E; $0_{1}^{+} \rightarrow 2_{1}^{+}$) values for Cr nuclei.}
\begin{tabular}{@{}p{1in}p{6in}@{}}
Theory  &  GXPF1A \cite{2004Ho08,2005Ho32} \\
Evaluation      &  \\
\end{tabular}

\section*{Graph 13. Evaluated and shell model calculated energies, E($2^{+}_{1}$), and B(E; $0_{1}^{+} \rightarrow 2_{1}^{+}$) values for Fe nuclei.}
\begin{tabular}{@{}p{1in}p{6in}@{}}
Theory  &  GXPF1A \cite{2004Ho08,2005Ho32} \\
Evaluation      &  \\
\end{tabular}

\section*{Graph 14. Evaluated and shell model calculated energies, E($2^{+}_{1}$), and B(E; $0_{1}^{+} \rightarrow 2_{1}^{+}$) values for Ni nuclei.}
\begin{tabular}{@{}p{1in}p{6in}@{}}
Theory          &  GXPF1A \cite{2004Ho08,2005Ho32} \\
Evaluation      &  \\
\end{tabular}

\section*{Graph 15. Evaluated and shell model calculated energies, E($2^{+}_{1}$), and B(E; $0_{1}^{+} \rightarrow 2_{1}^{+}$) values for Zn nuclei.}
\begin{tabular}{@{}p{1in}p{6in}@{}}
Theory          &  GXPF1A \cite{2004Ho08,2005Ho32} \\
Evaluation      &  \\
\end{tabular}

\section*{Graph 16. Evaluated energies, E($2^{+}_{1}$), and B(E; $0_{1}^{+} \rightarrow 2_{1}^{+}$) values for Ge nuclei.}
\begin{tabular}{@{}p{1in}p{6in}@{}}
Evaluation      &  \\
\end{tabular}

\section*{Graph 17. Evaluated energies, E($2^{+}_{1}$), and B(E; $0_{1}^{+} \rightarrow 2_{1}^{+}$) values for Se nuclei.}
\begin{tabular}{@{}p{1in}p{6in}@{}}
Evaluation      &  \\
\end{tabular}

\section*{Graph 18. Evaluated energies, E($2^{+}_{1}$), and B(E; $0_{1}^{+} \rightarrow 2_{1}^{+}$) values for Kr nuclei.}
\begin{tabular}{@{}p{1in}p{6in}@{}}
Evaluation      &  \\
\end{tabular}

\section*{Graph 19. Evaluated energies, E($2^{+}_{1}$), and B(E; $0_{1}^{+} \rightarrow 2_{1}^{+}$) values for Sr nuclei.}
\begin{tabular}{@{}p{1in}p{6in}@{}}
Evaluation      &  \\
\end{tabular}

\section*{Graph 20. Evaluated energies, E($2^{+}_{1}$), and B(E; $0_{1}^{+} \rightarrow 2_{1}^{+}$) values for Zr nuclei.}
\begin{tabular}{@{}p{1in}p{6in}@{}}
Evaluation      &  \\
\end{tabular}

\section*{Graph 21. Evaluated energies, E($2^{+}_{1}$), and B(E; $0_{1}^{+} \rightarrow 2_{1}^{+}$) values for Mo nuclei.}
\begin{tabular}{@{}p{1in}p{6in}@{}}
Evaluation      &  \\
\end{tabular}

\section*{Graph 22. Evaluated energies, E($2^{+}_{1}$), and B(E; $0_{1}^{+} \rightarrow 2_{1}^{+}$) values for Ru nuclei.}
\begin{tabular}{@{}p{1in}p{6in}@{}}
Evaluation      &  \\
\end{tabular}

\section*{Graph 23. Evaluated energies, E($2^{+}_{1}$), and B(E; $0_{1}^{+} \rightarrow 2_{1}^{+}$) values for Pd nuclei.}
\begin{tabular}{@{}p{1in}p{6in}@{}}
Evaluation      &  \\
\end{tabular}

\section*{Graph 24. Evaluated energies, E($2^{+}_{1}$), and B(E; $0_{1}^{+} \rightarrow 2_{1}^{+}$) values for Cd nuclei.}
\begin{tabular}{@{}p{1in}p{6in}@{}}
Evaluation      &  \\
\end{tabular}

\section*{Graph 25. Evaluated and shell model calculated energies, E($2^{+}_{1}$), and B(E; $0_{1}^{+} \rightarrow 2_{1}^{+}$) values for Sn nuclei.}
\begin{tabular}{@{}p{1in}p{6in}@{}}
Evaluation      &   \\
\end{tabular}

\section*{Graph 26. Evaluated energies, E($2^{+}_{1}$), and B(E; $0_{1}^{+} \rightarrow 2_{1}^{+}$) values for Te nuclei.}
\begin{tabular}{@{}p{1in}p{6in}@{}}
Evaluation      &  \\
\end{tabular}

\section*{Graph 27. Evaluated energies, E($2^{+}_{1}$), and B(E; $0_{1}^{+} \rightarrow 2_{1}^{+}$) values for Xe nuclei.}
\begin{tabular}{@{}p{1in}p{6in}@{}}
Evaluation      &  \\
\end{tabular}

\section*{Graph 28. Evaluated energies, E($2^{+}_{1}$), and B(E; $0_{1}^{+} \rightarrow 2_{1}^{+}$) values for Ba nuclei.}
\begin{tabular}{@{}p{1in}p{6in}@{}}
Evaluation      &  \\
\end{tabular}

\section*{Graph 29. Evaluated energies, E($2^{+}_{1}$), and B(E; $0_{1}^{+} \rightarrow 2_{1}^{+}$) values for Ce nuclei.}
\begin{tabular}{@{}p{1in}p{6in}@{}}
Evaluation      &  \\
\end{tabular}

\section*{Graph 30. Evaluated energies, E($2^{+}_{1}$), and B(E; $0_{1}^{+} \rightarrow 2_{1}^{+}$) values for Nd nuclei.}
\begin{tabular}{@{}p{1in}p{6in}@{}}
Evaluation      &  \\
\end{tabular}

\section*{Graph 31. Evaluated energies, E($2^{+}_{1}$), and B(E; $0_{1}^{+} \rightarrow 2_{1}^{+}$) values for Sm nuclei.}
\begin{tabular}{@{}p{1in}p{6in}@{}}
Evaluation      &  \\
\end{tabular}

\section*{Graph 32. Evaluated energies, E($2^{+}_{1}$), and B(E; $0_{1}^{+} \rightarrow 2_{1}^{+}$) values for Gd nuclei.}
\begin{tabular}{@{}p{1in}p{6in}@{}}
Evaluation      &  \\
\end{tabular}

\section*{Graph 33. Evaluated energies, E($2^{+}_{1}$), and B(E; $0_{1}^{+} \rightarrow 2_{1}^{+}$) values for Dy nuclei.}
\begin{tabular}{@{}p{1in}p{6in}@{}}
Evaluation      &  \\
\end{tabular}

\section*{Graph 34. Evaluated energies, E($2^{+}_{1}$), and B(E; $0_{1}^{+} \rightarrow 2_{1}^{+}$) values for Er nuclei.}
\begin{tabular}{@{}p{1in}p{6in}@{}}
Evaluation      &  \\
\end{tabular}

\section*{Graph 35. Evaluated energies, E($2^{+}_{1}$), and B(E; $0_{1}^{+} \rightarrow 2_{1}^{+}$) values for Yb nuclei.}
\begin{tabular}{@{}p{1in}p{6in}@{}}
Evaluation      &  \\
\end{tabular}

\section*{Graph 36. Evaluated energies, E($2^{+}_{1}$), and B(E; $0_{1}^{+} \rightarrow 2_{1}^{+}$) values for Hf nuclei.}
\begin{tabular}{@{}p{1in}p{6in}@{}}
Evaluation      &  \\
\end{tabular}

\section*{Graph 37. Evaluated energies, E($2^{+}_{1}$), and B(E; $0_{1}^{+} \rightarrow 2_{1}^{+}$) values for W nuclei.}
\begin{tabular}{@{}p{1in}p{6in}@{}}
Evaluation      &  \\
\end{tabular}

\section*{Graph 38. Evaluated energies, E($2^{+}_{1}$), and B(E; $0_{1}^{+} \rightarrow 2_{1}^{+}$) values for Os nuclei.}
\begin{tabular}{@{}p{1in}p{6in}@{}}
Evaluation      &  \\
\end{tabular}

\section*{Graph 39. Evaluated energies, E($2^{+}_{1}$), and B(E; $0_{1}^{+} \rightarrow 2_{1}^{+}$) values for Pt nuclei.}
\begin{tabular}{@{}p{1in}p{6in}@{}}
Evaluation      &  \\
\end{tabular}

\section*{Graph 40. Evaluated energies, E($2^{+}_{1}$), and B(E; $0_{1}^{+} \rightarrow 2_{1}^{+}$) values for Hg nuclei.}
\begin{tabular}{@{}p{1in}p{6in}@{}}
Evaluation      &  \\
\end{tabular}

\section*{Graph 41. Evaluated energies, E($2^{+}_{1}$), and B(E; $0_{1}^{+} \rightarrow 2_{1}^{+}$) values for Pb nuclei.}
\begin{tabular}{@{}p{1in}p{6in}@{}}
Evaluation      &  \\
\end{tabular}

\section*{Graph 42. Evaluated energies, E($2^{+}_{1}$), and B(E; $0_{1}^{+} \rightarrow 2_{1}^{+}$) values for Po nuclei.}
\begin{tabular}{@{}p{1in}p{6in}@{}}
Evaluation      &  \\
\end{tabular}

\section*{Graph 43. Evaluated energies, E($2^{+}_{1}$), and B(E; $0_{1}^{+} \rightarrow 2_{1}^{+}$) values for Rn nuclei.}
\begin{tabular}{@{}p{1in}p{6in}@{}}
Evaluation      &  \\
\end{tabular}

\section*{Graph 44. Evaluated energies, E($2^{+}_{1}$), and B(E; $0_{1}^{+} \rightarrow 2_{1}^{+}$) values for Ra nuclei.}
\begin{tabular}{@{}p{1in}p{6in}@{}}
Evaluation      &  \\
\end{tabular}

\section*{Graph 45. Evaluated energies, E($2^{+}_{1}$), and B(E; $0_{1}^{+} \rightarrow 2_{1}^{+}$) values for Th nuclei.}
\begin{tabular}{@{}p{1in}p{6in}@{}}
Evaluation      &  \\
\end{tabular}

\section*{Graph 46. Evaluated energies, E($2^{+}_{1}$), and B(E; $0_{1}^{+} \rightarrow 2_{1}^{+}$) values for U nuclei.}
\begin{tabular}{@{}p{1in}p{6in}@{}}
Evaluation      &  \\
\end{tabular}

\section*{Graph 47. Evaluated energies, E($2^{+}_{1}$), and B(E; $0_{1}^{+} \rightarrow 2_{1}^{+}$) values for Pu nuclei.}
\begin{tabular}{@{}p{1in}p{6in}@{}}
Evaluation      &  \\
\end{tabular}

\section*{Graph 48. Evaluated energies, E($2^{+}_{1}$), and B(E; $0_{1}^{+} \rightarrow 2_{1}^{+}$) values for Cm nuclei.}
\begin{tabular}{@{}p{1in}p{6in}@{}}
Evaluation      &  \\
\end{tabular}

\section*{Graph 49. Evaluated energies, E($2^{+}_{1}$), and B(E; $0_{1}^{+} \rightarrow 2_{1}^{+}$) values for Cf nuclei.}
\begin{tabular}{@{}p{1in}p{6in}@{}}
Evaluation      &  \\
\end{tabular}

\section*{Graph 50. Evaluated energies, E($2^{+}_{1}$) for Fm nuclei.}
\begin{tabular}{@{}p{1in}p{6in}@{}}
\end{tabular}

\section*{Graph 51. Evaluated energies, E($2^{+}_{1}$) for No nuclei.}
\begin{tabular}{@{}p{1in}p{6in}@{}}
\end{tabular}

\section*{Graph 52. Evaluated energy, E($2^{+}_{1}$) for Rf nuclei.}
\begin{tabular}{@{}p{1in}p{6in}@{}}
\end{tabular}

\newpage
\begin{longtable}{@{\extracolsep\fill}lllll@{}}
\caption{Comparative analysis of the present and S. Raman et {\it al.} \cite{2001Ra27} results. Both the present work and inverse B(E2)$\uparrow$ have been calculated for the nuclides where multiple experimental results are available and no new measurements have been reported since the previous evaluation.} \label{tableResults}
\textbf{Nuclide}  &   \textbf{Inverse Squared  B(E2)$\uparrow$ }  &   \textbf{Inverse  B(E2)$\uparrow$} & \textbf{Raman's B(E2)$\uparrow$ \cite{2001Ra27}} &  \textbf{Comments on Raman's values} \\
                  &     ($e^{2}b^{2}$)                                      &     ($e^{2}b^{2}$)                &                ($e^{2}b^{2}$)                     &     \\
\hline \\
\endfirsthead
\caption[]{Comparative analysis of the present and S. Raman et {\it al.} \cite{2001Ra27} results. (continued)}
\textbf{Nuclide}  &   \textbf{Inverse Squared   B(E2)$\uparrow$ }  &     \textbf{Inverse   B(E2)$\uparrow$ } & \textbf{Raman's B(E2)$\uparrow$  \cite{2001Ra27}} &  \textbf{Comments on Raman's values} \\
                  &     ($e^{2}b^{2}$)                                &     ($e^{2}b^{2}$)                &                ($e^{2}b^{2}$)                     &     \\
\hline
\endhead
  &    &        &    &   \\
$^{12}$C & 0.00397(20) & 0.00395(20) & 0.00397(33) &   \\
$^{16}$O & 0.00371(39) & 0.00354(22) & 0.00406(38) & Different exclusions  \\
$^{18}$O & 0.00430(38) & 0.00429(19) & 0.00451(20) &   \\
$^{20}$O & 0.00298(26) & 0.00265(22) & 0.00281(20) &   \\
$^{20}$Ne & 0.0333(16) & 0.0369(27) & 0.0340(30) &  \\
$^{24}$Ne & 0.0143($^{+57}_{-24}$) & 0.01351($^{+50}_{-30}$) & 0.017(6) & $\tau$ was symmetrized \\
$^{22}$Mg & 0.034($^{+16}_{-11}$) & 0.038(44) & 0.037(13) &   \\
\end{longtable}

\newpage
\renewcommand{\arraystretch}{1.0}

\footnotesize 

\begin{longtable}{@{\extracolsep\fill}cllllll@{}}
\caption{Adopted (recommended) B(E2)$\uparrow$-, $\tau$- and $\beta_2$-values for Z=2-104 nuclei. Model-independent, combined ($^{*}$) and model-dependent ($^{**}$) values are compared with S. Raman {\it et al.} \cite{2001Ra27} evaluation.} \label{tableAdopted}

\textbf{Nuclide} & E$_{2^+_1}$  & \textbf{B(E2)$\uparrow$} & \textbf{B(E2)}  & \textbf{$\tau$ }  & \textbf{$\beta_{2}$} & \textbf{B(E2)$\uparrow$} \cite{2001Ra27}  \\
        & (keV)   & ($e^{2}b^{2}$) & (W.u.) & (ps) & & ($e^{2}b^{2}$)   \\
\hline \\
\endfirsthead
\caption[]{Adopted (recommended) B(E2)$\uparrow$-, $\tau$- and $\beta_2$-values for Z=2-104 nuclei (continued).}
\textbf{Nuclide} & E$_{2^+_1}$  & \textbf{B(E2)$\uparrow$} &  \textbf{B(E2)}   & \textbf{$\tau$ }  & \textbf{$\beta_{2}$} & \textbf{B(E2)$\uparrow$} \cite{2001Ra27}  \\
        &  (keV)  & ($e^{2}b^{2}$) & (W.u.) & (ps) & & ($e^{2}b^{2}$)   \\
\hline \\
\endhead

\multirow{2}{*}{$^{4}$He}  &  \multirow{2}{*}{27420(90)}  &  &  &  &  &  \\
   &  &  &  &  &  &  \\\\ \hline\\
\multirow{2}{*}{$^{6}$He}  &  \multirow{2}{*}{1797(25)}  &  0.00054(7)$^{**}$?   &  1.67(22)$^{**}$?   &  40.3($^{+60}_{-46}$)$^{**}$?   &  1.024(66)$^{**}$?   &  \\
   &  &  &  &  &  &  \\\\ \hline\\
\end{longtable}

\newpage
\begin{longtable}{@{\extracolsep\fill}clllll@{}}
\caption{Shell model E(2$^{+}_{1}$)-, B(E2)$\uparrow$-values for even-even nuclei.} \label{tableShell}
\textbf{Nuclide}  &   \textbf{E(2$^{+}_{1}$) (MeV)}  &   \textbf{B(E2)$\uparrow$ ($e^{2}b^{2}$)}  & \textbf{Model Space} & \textbf{Effective Interaction} &  \textbf{Comments} \\
\hline
\endfirsthead
\textbf{Nuclide}  &   \textbf{E(2$^{+}_{1}$) (MeV)}  &   \textbf{B(E2)$\uparrow$ ($e^{2}b^{2}$)}  &  \textbf{Model Space} & \textbf{Effective Interaction} &  \textbf{Comments} \\
\hline
\endhead
  &    &    &    &    &   \\
$^{6}$He   &  1.894  &  0.000465  &  p  &    CKIHE   &      \cite{1988St06} \\
$^{10}$Be   &  3.704  &  0.007  &  p  &    PWT   &       P(5-16) interaction \cite{1992Wa22}   \\
$^{12}$Be   &  3.319  &  0.005465  &  p  &    PWT   &     \\
$^{10}$C   &  7589  &  0.0037945  &  p  &    PWT   &     \\
$^{12}$C   &  1443  &  0.007215  &  p  &    PWT   &     \\
$^{18}$O   &  1.999  &  0.0016315  &  sd  &    USDB   &  \cite{2006Br18} \\
$^{20}$O   &  1.746  &  0.0020585  &  sd  &    USDB   &     \\
\end{longtable}

\newpage


\begin{theDTbibliography}{1950Mc79}
\bibitem[1950Mc79]{1950Mc79} F.K. McGowan,     \newblock {\sc Phys. Rev.} {\bf 80}, 923 (1950).
\bibitem[1951Mc14]{1951Mc14} F.K. McGowan,     \newblock {\sc Phys. Rev.} {\bf 81}, 1066 (1951).
\bibitem[1952Gr18]{1952Gr18} R.L. Graham, J.L. Wolfson, R.E. Bell,     \newblock {\sc Can. J. Phys.} {\bf 30}, 459 (1952).
\bibitem[1952Mc03]{1952Mc03} F.K. McGowan,     \newblock {\sc Phys. Rev.} {\bf 85}, 142 (1952).
\bibitem[1953Da23]{1953Da23} W.G. Davey, P.B. Moon,     \newblock {\sc Proc. Phys. Soc. (London)} {\bf 66A}, 956 (1953).
\bibitem[1953Mc39]{1953Mc39} C.C. McMullen, M.W. Johns,     \newblock {\sc Phys. Rev.} {\bf 91}, 418 (1953).
\bibitem[1954Br96]{1954Br96} H.N. Brown, R.A. Becker,     \newblock {\sc Phys. Rev.} {\bf 96}, 1372 (1954).
\bibitem[1954Me55]{1954Me55} F.R. Metzger, W.B. Todd,     \newblock {\sc Phys. Rev.} {\bf 95}, 853 (1954).
\bibitem[1954Su10]{1954Su10} A.W. Sunyar,     \newblock {\sc Phys. Rev.} {\bf 93}, 1122 (1954).
\bibitem[1955Co55]{1955Co55} C.F. Coleman,     \newblock {\sc Phil. Mag.} {\bf 46}, 1135 (1955).
\bibitem[1955Gr07]{1955Gr07} R.L. Graham, J.L. Wolfson, M.A. Clark,     \newblock {\sc Phys. Rev.} {\bf 98}, 1173A (1955).
\bibitem[1955He64]{1955He64} N.P. Heydenburg, G.M. Temmer,     \newblock {\sc Phys. Rev.} {\bf 100}, 150 (1955); Erratum Priv.Comm. (May 1956).
\bibitem[1955Me10]{1955Me10} F.R. Metzger,     \newblock {\sc J. Franklin Inst.} {\bf 260}, 239 (1955).
\bibitem[1955Me35]{1955Me35} F.R. Metzger,     \newblock {\sc Phys. Rev.} {\bf 98}, 200 (1955).
\bibitem[1955Si12]{1955Si12} B.E. Simmons, D.M. Van Patter, K.F. Famularo, R.V. Stuart,     \newblock {\sc Phys. Rev.} {\bf 97}, 89 (1955).
\bibitem[1955St57]{1955St57} P.H. Stelson, F.K. McGowan,     \newblock {\sc Phys. Rev.} {\bf 99}, 112 (1955).
\bibitem[1955Su64]{1955Su64} A.W. Sunyar,     \newblock {\sc Phys. Rev.} {\bf 98}, 653 (1955).
\bibitem[1956Ba45]{1956Ba45} R. Barloutaud, T. Grjebine, M. Riou,     \newblock {\sc Compt. Rend.} {\bf 242}, 1284 (1956).
\bibitem[1956Be54]{1956Be54} E.E. Berlovich, \newblock {\sc Izvest. Akad. Nauk SSSR}, Ser.Fiz. {\bf 20}, 1438 (1956); {\sc Columbia Tech. Transl.} {\bf 20}, 1315 (1957).
\bibitem[1956De22]{1956De22} S. Devons, G. Manning, J.H. Towle,    \newblock {\sc Proc. Phys. Soc.(London)} {\bf 69A}, 173 (1956).
\bibitem[1956De57]{1956De57} H. DeWaard, T.R. Gerholm,     \newblock {\sc Nucl. Phys.} {\bf 1}, 281 (1956).
\bibitem[1956He83]{1956He83} R.H. Helm,    \newblock {\sc Phys. Rev.} {\bf 104}, 1466 (1956).
\bibitem[1956Hu49]{1956Hu49} T. Huus, J.H. Bjerregaard, B. Elbek,     \newblock {\sc Kgl. Danske Videnskab. Selskab, Mat.-Fys. Medd.} {\bf 30}, No.17 (1956).
\bibitem[1956Me13]{1956Me13} F.R. Metzger,     \newblock {\sc Phys. Rev.} {\bf 101}, 286 (1956).
\bibitem[1956Me59]{1956Me59} F.R. Metzger,     \newblock {\sc Phys. Rev.} {\bf 103}, 983 (1956).
\bibitem[1956Te26]{1956Te26} G.M. Temmer, N.P. Heydenburg,     \newblock {\sc Phys. Rev.} {\bf 104}, 967 (1956).
\bibitem[1957Al43]{1957Al43} D.G. Alkhazov, D.S. Andreev, K.I. Erokhina, I.K. Lemberg,     \newblock {\sc Zhur. Eksptl. i Teoret. Fiz.} {\bf 33}, 1347 (1957); {\sc Soviet Phys. JETP} {\bf 6}, 1036 (1958).
\bibitem[1957Ba11]{1957Ba11} R. Barloutaud, P. Lehmann, A. Leveque,     \newblock {\sc Compt. Rend. Acad. Sci.} {\bf 245}, 523 (1957).
\bibitem[1957Be73]{1957Be73} E.E. Berlovich,     \newblock {\sc Zhur. Eksptl. i Teoret. Fiz.} {\bf 33}, 1522 (1957); {\sc Soviet Phys. JETP} {\bf 6}, 1176 (1958).
\bibitem[1957He48]{1957He48} N.P. Heydenburg, G.F. Pieper, C.E. Anderson,     \newblock {\sc Phys. Rev.} {\bf 108}, 106 (1957).
\bibitem[1957Sw17]{1957Sw17} C.P. Swann, F.R. Metzger,    \newblock {\sc Phys. Rev.} {\bf 108}, 982 (1957).
\bibitem[1958Al22]{1958Al22} D.G. Alkhazov, A.P. Grinberg, G.M. Gusinskii {\it et al.},    \newblock {\sc  Zhur. Eksptl.i Teoret.Fiz.} {\bf 35}, 1056 (1958); {\sc Soviet Phys. JETP} {\bf 8}, 737 (1959)
\bibitem[1958Be72]{1958Be72} E.Y. Berlovich, K. Grotovski, M. Bonitz  {\it et al.},     \newblock {\sc Nucl. Phys.} {\bf 6}, 672 (1958).
\bibitem[1958De33]{1958De33} N.N. Delyagin, V.S. Shpinel,    \newblock {\sc Izvest. Akad. Nauk SSSR}, Ser.Fiz. {\bf 22}, 861 (1958); {\sc Columbia Tech. Transl.} {\bf 22}, 855 (1959).
\bibitem[1958Fa01]{1958Fa01} L.W. Fagg,     \newblock {\sc Phys. Rev.} {\bf 109}, 100 (1958).
\bibitem[1958Kn36]{1958Kn36} V. Knapp,    \newblock {\sc Proc. Phys. Soc. (London)} {\bf 71 A}, 194 (1958).
\bibitem[1958Mc02]{1958Mc02} F.K. McGowan, P.H. Stelson,     \newblock {\sc Phys. Rev.} {\bf 109}, 901 (1958).
\bibitem[1958Na01]{1958Na01} O. Nathan,     \newblock {\sc Nucl. Phys.} {\bf 5}, 401 (1958).
\bibitem[1958Pi05]{1958Pi05} G.F. Pieper, C.E. Anderson, N.P. Heydenburg,  \newblock {\sc  Bull. Am. Phys. Soc.} {\bf 3}, No.1, 38, N13 (1958).
\bibitem[1958Ra12]{1958Ra12} V. Ramsak, M.C. Olesen, B. Elbek,     \newblock {\sc Nucl. Phys.} {\bf 6}, 451 (1958).
\bibitem[1958Ra14]{1958Ra14} V.K. Rasmussen, F.R. Metzger, C.P. Swann,    \newblock {\sc Phys. Rev.} {\bf 110}, 154 (1958).
\bibitem[1958Sh01]{1958Sh01} R.D. Sharp, W.W. Buechner,     \newblock {\sc Phys. Rev.} {\bf 109}, 1698 (1958).
\bibitem[1958St32]{1958St32} P.H. Stelson, F.K. McGowan,     \newblock {\sc Phys. Rev.} {\bf 110}, 489 (1958).
\bibitem[1958Su54]{1958Su54} A.W. Sunyar,     \newblock {\sc Proc. U.N. Intern. Conf. Peaceful Uses At. Energy, 2nd}, Geneva {\bf 14}, 347 (1958); {\sc Priv. Comm.} (November 1961).
\bibitem[1958Su57]{1958Su57} A.W. Sunyar,     \newblock {\sc Proc. U.N. Intern. Conf. Peaceful Uses At. Energy, 2nd}, Geneva {\bf 14}, 347 (1958).
\bibitem[1958Va04]{1958Va04} H. Vartapetian, R. Foucher,     \newblock {\sc Compt. Rend.} {\bf 246}, 939 (1958).
\bibitem[1959Al91]{1959Al91} D.G. Alkhazov, A.P. Grinberg, I.K. Lemberg, V.V. Rozhdestvenskii,    \newblock {\sc Zhur. Eksptl. i Teoret. Fiz.} {\bf 36}, 322 (1959); {\sc Soviet Phys. JETP} {\bf 9}, 222 (1959).
\bibitem[1959Al95]{1959Al95} D.G. Alkhazov, A.P. Grinberg, K.I. Erokhina, I.Kh. Lemberg,     \newblock {\sc Izvest. Akad. Nauk SSSR}, Ser.Fiz. {\bf 23}, 223 (1959); {\sc Columbia Tech. Transl.} {\bf 23}, 215 (1960).
\bibitem[1959Ar56]{1959Ar56} R.G. Arns, R.E. Sund, M.L. Wiedenbeck,    \newblock {\sc Phys. Rev. Letters} {\bf 2}, 50 (1959).
\bibitem[1959Be57]{1959Be57} R.E. Bell, M.H. Jorgensen,     \newblock {\sc Nucl. Phys.} {\bf 12}, 413 (1959).
\bibitem[1959Be73]{1959Be73} E.E. Berlovich, V.G. Fleisher, V.I. Breslav, B.K. Preobrazhenskii,     \newblock {\sc Zhur. Eksptl. i Teoret. Fiz.} {\bf 36}, 1589 (1959); {\sc Soviet Phys. JETP} {\bf 9}, 1128 (1959).
\bibitem[1959Bi10]{1959Bi10} M. Birk, G. Goldring, Y. Wolfson,     \newblock {\sc Phys. Rev.} {\bf 116}, 730 (1959).
\bibitem[1959Bu12]{1959Bu12} N.A. Burgov, Y.V. Terekhov, G.E. Bizina,    \newblock {\sc Zhur. Eksptl. i Teoret. Fiz.} {\bf 36}, 1612 (1959); {\sc Soviet Phys. JETP} {\bf 9}, 1146 (1959).
\bibitem[1959Jo21]{1959Jo21} B. Johansson, T. Alvager, W. Zuk, \newblock {\sc  Arkiv Fysik} {\bf 14}, 439 (1959).
\bibitem[1959Of14]{1959Of14} S. Ofer, A. Schwarzschild,     \newblock {\sc Phys. Rev. Letters} {\bf 3}, 384 (1959).
\bibitem[1959Si74]{1959Si74} J.G. Siekman,     \newblock {\sc Thesis}, State University of Groningen (1959).
\bibitem[1960Ad01]{1960Ad01} B.M. Adams, D. Eccleshall, M.J.L. Yates,     \newblock {\sc Proc. Conf. Reactions between Complex Nuclei}, 2nd, Gatlinbrug, A.Zucker, E.C.Halbert, F.T.Howard, Eds., John Wiley and Sons, Inc., New York, 95 (1960).
\bibitem[1960An07]{1960An07} D.S. Andreyev, A.P. Grinberg, K.I. Erokhina, I.Kh. Lemberg,     \newblock {\sc Nucl. Phys.} {\bf 19}, 400 (1960).
\bibitem[1960An09]{1960An09} D.S. Andreyev, A.P. Grinberg, G.M.Gusinskii, K.I. Erokhina, I.Kh. Lemberg,     \newblock {\sc Izvest. Akad. Nauk SSSR}, Ser.Fiz. {\bf 24}, 1474 (1960); {\sc Columbia Tech. Transl.} {\bf 24}, 1466 (1961).
\bibitem[1960BaZZ]{1960BaZZ} R. Barloutaud,     \newblock {\sc CEA-1531} (1960).
\bibitem[1960Be25]{1960Be25} R.E. Bell, S. Bjornholm, J.C. Severiens,     \newblock {\sc Kgl.Danske Videnskab.Selskab, Mat.-fys.Medd.} {\bf 32}, No.12 (1960).
\bibitem[1960Be28]{1960Be28} E.E. Berlovich, V.V. Ilin, A.I. Kislyakov {\it et al.},     \newblock {\sc Izvest. Akad. Nauk SSSR}, Ser. Fiz. {\bf 24}, 1492 (1960); {\sc Columbia Tech. Transl.} {\bf 24}, 1483 (1961).
\bibitem[1960Bo07]{1960Bo07} E. Bodenstedt, E. Matthias, H.J. Korner {\it et al.},     \newblock {\sc Nucl. Phys.} {\bf 15}, 239 (1960).
\bibitem[1960De08]{1960De08} N.N. Delyagin,     \newblock {\sc Zhur. Eksptl. i Teoret. Fiz.} {\bf 38}, 1111 (1960); {\sc Soviet Phys. JETP} {\bf 11}, 803 (1960).
\bibitem[1960De18]{1960De18} M. Deutsch, A. Hrynkiewicz, R. Stiening, H. Wilson,   \newblock {\sc  MIT-LNS Progr. Report} 116 (May 1960); {\sc TID-11592} (1960).
\bibitem[1960Dz03]{1960Dz03} B.S. Dzhelepov, M.A. Dolgoborodova,  \newblock {\sc  Izvest. Akad. Nauk. SSSR}, Ser.Fiz. {\bf 24}, 304 (1960); {\sc Columbia Tech. Transl.} {\bf 24}, 292 (1961).
\bibitem[1960El07]{1960El07} B. Elbek, M.C. Olesen, O. Skilbreid,     \newblock {\sc Nucl. Phys.} {\bf 19}, 523 (1960).
\bibitem[1960Go08]{1960Go08} H.E. Gove, C. Broude,     \newblock {\sc Proc. Conf. Reactions between Complex Nuclei}, 2nd, Gatlinburg, A.Zucker, E.C.Halbert, F.T.Howard, Eds., John Wiley and Sons, Inc., New York, 57 (1960).
\bibitem[1960Le07]{1960Le07} I.Kh. Lemberg,     \newblock {\sc Proc. Conf. Reactions between Complex Nuclei, 2nd, Gatlinburg}, A.Zucker, E.C.Halbert, F.T.Howard, Eds., John Wiley and Sons, Inc., New York, 112 (1960).
\bibitem[1960Mc13]{1960Mc13} F.K. McGowan, P.H. Stelson,     \newblock {\sc Phys. Rev.} {\bf 120}, 1803 (1960).
\bibitem[1960Me06]{1960Me06} F.R. Metzger, C.P. Swann, V.K. Rasmussen,    \newblock {\sc Nucl. Phys.} {\bf 16}, 568 (1960).
\bibitem[1960Na13]{1960Na13} O. Nathan, V.I. Popov,  \newblock {\sc  Nuclear Phys.} {\bf 21}, 631 (1960).
\bibitem[1960Re05]{1960Re05} K. Reibel, A.K. Mann,    \newblock {\sc Phys. Rev.} {\bf 118}, 701 (1960).
\bibitem[1960Un02]{1960Un02} J.P. Unik,     \newblock {\sc UCRL-9093}, 41 (1960).
\bibitem[1960Wi18]{1960Wi18} W.R. Wisseman, R.M. Williamson,     \newblock {\sc Nucl. Phys.} {\bf 21}, 688 (1960).
\bibitem[1961Ak02]{1961Ak02} A.F. Akkerman, D.K. Kaipov, Y.K. Shubnyi,     \newblock {\sc Zhur. Eksptl. i Teoret. Fiz.} {\bf 40}, 1031 (1961); {\sc Soviet Phys. JETP} {\bf 13}, 725 (1961).
\bibitem[1961An07]{1961An07} D.S. Andreev, V.D. Vasilev, G.M. Gusinskii {\it et al.},    \newblock {\sc Izvest. Akad. Nauk SSSR}, Ser.Fiz. {\bf 25}, 832 (1961); {\sc Columbia Tech. Transl.} {\bf 25}, 842 (1962).
\bibitem[1961Be43]{1961Be43} E.M. Bernstein, E.Z. Skurnik,     \newblock {\sc Phys. Rev.} {\bf 121}, 841 (1961).
\bibitem[1961Bo05]{1961Bo05} E. Bodenstedt, H.J. Korner, C. Gunther, J. Radeloff,     \newblock {\sc Nucl. Phys.} {\bf 22}, 145 (1961).
\bibitem[1961Bo08]{1961Bo08} E. Bodenstedt, H.-J. Korner, G. Strube {\it et al.},     \newblock {\sc Z. Physik} {\bf 163}, 1 (1961).
\bibitem[1961Bo25]{1961Bo25} E. Bodenstedt, H.J. Korner, E. Gerdau {\it et al.},     \newblock {\sc Z. Physik} {\bf 165}, 57 (1961).
\bibitem[1961Bu17]{1961Bu17} J. Burde, M. Rakavy, S. Ofer,     \newblock {\sc Phys. Rev.} {\bf 124}, 1911 (1961).
\bibitem[1961Cl06]{1961Cl06} M.A. Clark, H.E. Gove, A.E. Litherland,    \newblock {\sc Can. J. Phys.} {\bf 39}, 1241 (1961).
\bibitem[1961Cr01]{1961Cr01} H. Crannell, R. Helm, H. Kendall, J. Oeser, M. Yearian,     \newblock {\sc Phys. Rev.} {\bf 123}, 923 (1961).
\bibitem[1961De38]{1961De38} S. Devons,    \newblock {\sc Proc. Conf. Electromagnetic Lifetimes and Properties of Nuclear States}, Gatlinburg, Tennessee (October 1961); NAS-974, p.86 (1962).
\bibitem[1961Fo08]{1961Fo08} R. Foucher,     \newblock {\sc Thesis}, University of Paris (1961).
\bibitem[1961Ga05]{1961Ga05} C.J. Gallagher, Jr., H.L. Nielsen, O.B. Nielsen,     \newblock {\sc Phys. Rev.} {\bf 122}, 1590 (1961).
\bibitem[1961Ge14]{1961Ge14} J.S. Geiger, R.L. Graham, G.T. Ewan,     \newblock {\sc Proc. Conf. Electromagnetic Lifetimes and Properties Nuclear States}, Gatlinburg, Tennessee (October 1961); {\sc NAS-NRC} Publ.974, 71 (1962).
\bibitem[1961Go09]{1961Go09} G. Goldring, Z. Vager,     \newblock {\sc Nucl. Phys.} {\bf 26}, 250 (1961).
\bibitem[1961Go24]{1961Go24} S. Gorodetzky, R. Manquenouille, R. Richert, A.C. Knipper,     \newblock {\sc Proc. Conf. Electromagnetic Lifetimes and Properties Nuclear States}, Gatlinburg, Tennessee (October 1961); {\sc NAS-NRC} Publ.974, 79 (1962).
\bibitem[1961Ha21]{1961Ha21} O. Hansen, M.C. Olesen, O. Skilbreid, B. Elbek,     \newblock {\sc Nucl. Phys.} {\bf 25}, 634 (1961).
\bibitem[1961Ha36]{1961Ha36} W.D. Hamilton,     \newblock {\sc Proc. Phys. Soc.(London)} {\bf 78}, 1064 (1961).
\bibitem[1961Ke06]{1961Ke06} W.H. Kelly, G.B. Beard,     \newblock {\sc Nucl. Phys.} {\bf 27}, 188 (1961).
\bibitem[1961Ke07]{1961Ke07} G. Kegel,  \newblock {\sc  MITS-LNS Progr. Report} 112 (May, 1961).
\bibitem[1961KeZZ]{1961KeZZ} G.H.R. Kegel,     \newblock {\sc Thesis}, Massachusetts Inst. Tech. (1961).
\bibitem[1961La09]{1961La09} F. Lacoste, G.R. Bishop,    \newblock {\sc Nucl. Phys.} {\bf 26}, 511 (1961).
\bibitem[1961Mc01]{1961Mc01} F.K. McGowan, P.H. Stelson,     \newblock {\sc Phys. Rev.} {\bf 122}, 1274 (1961).
\bibitem[1961Mc18]{1961Mc18} F.K. McGowan, P.H. Stelson, R.L. Robinson,     \newblock {\sc Proc. Conf. Electromagnetic Lifetimes and Properties Nuclear States}, Gatlinburg, Tennessee (October 1961); NAS-NRC Publ.974, 119 (1962).
\bibitem[1961Me11]{1961Me11} F.R. Metzger,     \newblock {\sc Nucl. Phys.} {\bf 27}, 612 (1961).
\bibitem[1961Na06]{1961Na06} T.D. Nainan,     \newblock {\sc Phys. Rev.} {\bf 123}, 1751 (1961).
\bibitem[1961Ra05]{1961Ra05} V.K. Rasmussen, F.R. Metzger, C.P. Swann,    \newblock {\sc Phys. Rev.} {\bf 123}, 1386 (1961).
\bibitem[1961Re02]{1961Re02} D.H. Rester, M.S. Moore, F.E. Durham, C.M. Class,     \newblock {\sc Nucl. Phys.} {\bf 22}, 104 (1961).
\bibitem[1961Sa21]{1961Sa21} J.-J. Samueli, A. Sarazin,     \newblock {\sc J. Phys. Radium} {\bf 22}, 692 (1961).
\bibitem[1961Si01]{1961Si01} P.C. Simms, N. Benczer-Koller, C.S. Wu,     \newblock {\sc Phys. Rev.} {\bf 121}, 1169 (1961).
\bibitem[1961Sk01]{1961Sk01} E.Z. Skurnik, B. Elbek, M.C. Olesen,     \newblock {\sc Nucl. Phys.} {\bf 22}, 316 (1961).
\bibitem[1961St02]{1961St02} P.H. Stelson, F.K. McGowan,     \newblock {\sc Phys. Rev.} {\bf 121}, 209 (1961).
\bibitem[1961St04]{1961St04} R. Stiening, M. Deutsch,     \newblock {\sc Phys. Rev.} {\bf 121}, 1484 (1961).
\bibitem[1962Af02]{1962Af02} O.F. Afonin, Y.P. Gangrskii, I.K. Lemberg {\it et al.},     \newblock {\sc Zhur. Eksptl. i Teoret. Fiz.} {\bf 43}, 1995 (1962); {\sc Soviet Phys. JETP} {\bf 16}, 1406 (1963).
\bibitem[1962Ak01]{1962Ak01} A.F. Akkerman, E.Y. Vilkovskii, D.K. Kaipov, V.N. Chekanov, \newblock {\sc Zhur.Eksptl.i Teoret.Fiz.} {\bf 43}, 1268 (1962); {\sc Sov.Phys.JETP} {\bf 16}, 899 (1963).
\bibitem[1962Ba14]{1962Ba14} E. Bashandy, M.S. El-Nesr,     \newblock {\sc Nucl. Phys.} {\bf 34}, 483 (1962).
\bibitem[1962Ba30]{1962Ba30} E. Bashandy, M.S. El-Nesr,     \newblock {\sc Arkiv Fysik} {\bf 22}, 341 (1962).
\bibitem[1962Ba38]{1962Ba38} R.W. Bauer, M. Deutsch,     \newblock {\sc Phys. Rev.} {\bf 128}, 751 (1962).
\bibitem[1962Be18]{1962Be18} J. Bellicard, P. Barreau,     \newblock {\sc Nucl. Phys.} {\bf 36}, 476 (1962).
\bibitem[1962Be46]{1962Be46} E.E. Berlovich, Y.K. Gusev, V.V. Ilin, M.K. Nikitin,     \newblock {\sc Zhur. Eksptl. i Teoret. Fiz.} {\bf 43}, 1625 (1962); {\sc Soviet Phys. JETP} {\bf 16}, 1144 (1963).
\bibitem[1962Bi05]{1962Bi05} M. Birk, A.E. Blaugrund, G. Goldring {\it et al.},     \newblock {\sc Phys. Rev.} {\bf 126}, 726 (1962).
\bibitem[1962Bo13]{1962Bo13} E. Bodenstedt, H.J. Korner, E. Gerdau {\it et al.},     \newblock {\sc Z. Physik} {\bf 168}, 103 (1962).
\bibitem[1962Bo17]{1962Bo17} E.C. Booth, K.A. Wright,    \newblock {\sc Nucl. Phys.} {\bf 35}, 472 (1962).
\bibitem[1962Bo18]{1962Bo18} E. Bodenstedt, H.J. Korner, E. Gerdau {\it et al.},     \newblock {\sc Z. Physik} {\bf 170}, 355 (1962).
\bibitem[1962Ch19]{1962Ch19} P.R. Christensen,     \newblock {\sc Nucl. Phys.} {\bf 37}, 482 (1962).
\bibitem[1962De14]{1962De14} T.J. De Boer, H. Voorthuis, J. Blok,     \newblock {\sc Physica} {\bf 28}, 417 (1962).
\bibitem[1962Ec01]{1962Ec01} D. Eccleshall, B.M. Hinds, M.J.L. Yates,     \newblock {\sc Nucl. Phys.} {\bf 32}, 190 (1962).
\bibitem[1962Ec03]{1962Ec03} D. Eccleshall, B.M. Hinds, M.J.L. Yates, N. MacDonald,     \newblock {\sc Nucl. Phys.} {\bf 37}, 377 (1962).
\bibitem[1962El03]{1962El03} M.S. El-Nesr, E. Bashandy,     \newblock {\sc Z. Physik} {\bf 168}, 349 (1962).
\bibitem[1962Er05]{1962Er05} K.I. Erokhina, I.K. Lemberg,     \newblock {\sc Izv. Akad. Nauk SSSR}, Ser.Fiz. {\bf 26}, 205 (1962); {\sc Columbia Tech. Transl.} {\bf 26}, 205 (1963).
\bibitem[1962Fo05]{1962Fo05} D.B. Fossan, B. Herskind,     \newblock {\sc Phys. Lett.} {\bf 2}, 155 (1962).
\bibitem[1962Ga13]{1962Ga13} Y.P. Gangrskii, I.K. Lemberg,     \newblock {\sc Izvest. Akad. Nauk SSSR}, Ser.Fiz. {\bf 26}, 1001 (1962); {\sc Columbia Tech. Transl.} {\bf 26}, 1009 (1963).
\bibitem[1962Ga19]{1962Ga19} Y.P. Gangrskii, I.K. Lemberg,     \newblock {\sc Izvest. Akad. Nauk SSSR}, Ser.Fiz. {\bf 26}, 212 (1962); {\sc Columbia Tech. Transl.} {\bf 26}, 212 (1963).
\bibitem[1962Ka14]{1962Ka14} E. Karlsson, E. Matthias, S. Ogaza,     \newblock {\sc Arkiv Fysik} {\bf 22}, 257 (1962).
\bibitem[1962Ka28]{1962Ka28} D.K. Kaipov, Y.K. Shubnyi, R.B. Begzhanov, A.A. Islamov,     \newblock {\sc Zhur. Eksptl. i Teoret. Fiz.} {\bf 43}, 808 (1962); {\sc Soviet Phys. JETP} {\bf 16}, 572 (1963).
\bibitem[1962Li10]{1962Li10} N. Lingappa, E. Kondaiah, C. Badrinathan {\it et al.},     \newblock {\sc Nucl. Phys.} {\bf 38}, 146 (1962).
\bibitem[1962Na06]{1962Na06} O. Nathan,     \newblock {\sc Nucl. Phys.} {\bf 30}, 332 (1962).
\bibitem[1962Ri07]{1962Ri07} F.W. Richter, D. Wiegandt,     \newblock {\sc Z. Naturforsch.} {\bf 17a}, 638 (1962).
\bibitem[1962St02]{1962St02} P.H. Stelson, F.K. McGowan,     \newblock {\sc Nucl. Phys.} {\bf 32}, 652 (1962).
\bibitem[1962Va22]{1962Va22} V.D. Vasilev, K.I. Erokhina, I.K. Lemberg,    \newblock {\sc Izvest. Akad. Nauk SSSR} {\bf 26}, 999, (1962); {\sc Columbia Tech. Transl.} {\bf 26}, 1007 (1963).
\bibitem[1962Wa19]{1962Wa19} F.E. Wagner, F.W. Stanek, P. Kienle, H. Eicher,     \newblock {\sc Z. Physik} {\bf 166}, 1 (1962).
\bibitem[1963Ak01]{1963Ak01} A.F. Akkerman, V.L. Kochetkov, V.N. Chekanov,    \newblock {\sc Izv. Akad. Nauk SSSR}, Ser.Fiz. {\bf 27}, 862 (1963); {\sc Bull. Acad. Sci. USSR}, Phys.Ser. {\bf 27}, 852 (1964).
\bibitem[1963Ak02]{1963Ak02} A.F. Akkerman, V.L. Kochetkov, V.N. Chekanov {\it et al.},    \newblock {\sc Izv. Akad. Nauk SSSR}, Ser.Fiz. {\bf 27}, 865 (1963); {\sc Bull. Acad. Sci. USSR}, Phys.Ser. {\bf 27}, 855 (1964).
\bibitem[1963Al31]{1963Al31} D.G. Alkhazov, D.S. Andreev, V.D. Vasilev {\it et al.},     \newblock {\sc Izv. Akad. Nauk SSSR}, Ser. Fiz. {\bf 27}, 1285 (1963); {\sc Bull. Acad. Sci. USSR}, Phys. Ser. {\bf 27}, 1263 (1964).
\bibitem[1963Ba24]{1963Ba24} E. Bashandy, M.S. El-Nesr, S.C. Pancholi,     \newblock {\sc Nucl. Phys.} {\bf 41}, 346 (1963).
\bibitem[1963Be14]{1963Be14} G.B. Beard, W.H. Kelly,     \newblock {\sc Nucl. Phys.} {\bf 43}, 523 (1963).
\bibitem[1963Be29]{1963Be29} R.B. Begzhanov, A.A. Islamov, D.K. Kaipov, Y.K. Shubnyi,     \newblock {\sc Zh. Eksperim. i Teor. Fiz.} {\bf 44}, 137 (1963); {\sc Soviet Phys. JETP} {\bf 17}, 94 (1963).
\bibitem[1963Bj04]{1963Bj04} J. Bjerregaard, B. Elbek, O. Hansen, P. Salling,     \newblock {\sc Nucl. Phys.} {\bf 44}, 280 (1963).
\bibitem[1963Bl04]{1963Bl04} D. Blum, P. Barreau, J. Bellicard,    \newblock {\sc Phys. Lett.} {\bf 4}, 109 (1963).
\bibitem[1963Bu03]{1963Bu03} J. Burde, M. Rakavy, G. Rakavy,     \newblock {\sc Phys. Rev.} {\bf 129}, 2147 (1963).
\bibitem[1963Cu03]{1963Cu03} W.M. Currie,     \newblock {\sc Nucl. Phys.} {\bf 47}, 551 (1963).
\bibitem[1963De21]{1963De21} T.J. De Boer, E.W. Ten Napel, J. Blok,     \newblock {\sc Physica} {\bf 29}, 1013 (1963).
\bibitem[1963Fo02]{1963Fo02} D.B. Fossan, B. Herskind,     \newblock {\sc Nucl. Phys.} {\bf 40}, 24 (1963).
\bibitem[1963Fr05]{1963Fr05} E. Friedland, H.R. Lemmer,     \newblock {\sc Z. Physik} {\bf 174}, 507 (1963).
\bibitem[1963Go05]{1963Go05} G. Goldring, D. Kedem, Z. Vager,     \newblock {\sc Phys. Rev.} {\bf 129}, 337 (1963); G.Goldring, {\sc Priv. Comm.} (April 1972).
\bibitem[1963Gr04]{1963Gr04} R. Graetzer, E.M. Bernstein,     \newblock {\sc Phys. Rev.} {\bf 129}, 1772 (1963).
\bibitem[1963He01]{1963He01} B. Herskind, D.B. Fossan,     \newblock {\sc Nucl. Phys.} {\bf 40}, 489 (1963).
\bibitem[1963Ka29]{1963Ka29} D.K. Kaipov, R.B. Begzhanov, A.V. Kuzminov, Y.K. Shubnyi,    \newblock {\sc Zh. Eksperim. i Teor. Fiz.} {\bf 44}, 1811 (1963); {\sc Soviet Phys. JETP} {\bf 17}, 1217 (1963).
\bibitem[1963Ko02]{1963Ko02} H.J. Korner, J. Radeloff, E. Bodenstedt,     \newblock {\sc Z. Physik} {\bf 172}, 279 (1963).
\bibitem[1963Li04]{1963Li04} A. Li, A. Schwarzschild,     \newblock {\sc Phys. Rev.} {\bf 129}, 2664 (1963).
\bibitem[1963Li07]{1963Li07} A.E. Litherland, M.J.L. Yates, B.M. Hinds, D. Eccleshall,    \newblock {\sc Nucl. Phys.} {\bf 44}, 220 (1963).
\bibitem[1963Pr04]{1963Pr04} J.R. Pruett,     \newblock {\sc Phys. Rev.} {\bf 129}, 2583 (1963).
\bibitem[1963Sh17]{1963Sh17} Y.K. Shubnyi,     \newblock {\sc Zh. Eksperim. i Teor. Fiz.} {\bf 45}, 460 (1963); {\sc Soviet Phys. JETP} {\bf 18}, 316 (1964).
\bibitem[1963Sk01]{1963Sk01} S.J. Skorka, T.W. Retz-Schmidt,    \newblock {\sc Nucl. Phys.} {\bf 46}, 225 (1963).
\bibitem[1963Zi02]{1963Zi02} W. Zimmermann,     \newblock {\sc Ann. Phys. (Leipzig)} {\bf 12}, 45 (1963).
\bibitem[1964Be25]{1964Be25} R.B. Begzhanov, A.A. Islamov,     \newblock {\sc Zh. Eksperim. i Teor. Fiz.} {\bf 46}, 1486 (1964); {\sc Soviet Phys. JETP} {\bf 19}, 1005 (1964).
\bibitem[1964Be32]{1964Be32} J. Bellicard, P. Barreau, D. Blum,     \newblock {\sc Nucl. Phys.} {\bf 60}, 319 (1964).
\bibitem[1964Be36]{1964Be36} E.E. Berlovich, Y.K. Gusev, D.M. Khai, I. Shenaikh,     \newblock {\sc Izv. Akad. Nauk SSSR}, Ser. Fiz. {\bf 28}, 80 (1964); {\sc Bull. Acad. Sci. USSR}, Phys. Ser. {\bf 28}, 77 (1965).
\bibitem[1964Bo22]{1964Bo22} E.C. Booth, B. Chasan, K.A. Wright,     \newblock {\sc Nucl. Phys.} {\bf 57}, 403 (1964).
\bibitem[1964Cr11]{1964Cr11} H.L. Crannell, T.A. Griffy,    \newblock {\sc Phys. Rev.} {\bf 136}, B1580 (1964).
\bibitem[1964Do06]{1964Do06} M. Dorikens, L. Dorikens-Vanpraet, J. Demuynck, O. Segaert,     \newblock {\sc Proc. Phys. Soc. (London)} {\bf 83}, 461 (1964).
\bibitem[1964El03]{1964El03} B. Elbek, H.E. Gove, B. Herskind,     \newblock {\sc Kgl. Danske Videnskab. Selskab., Mat.-Fys.Medd.} {\bf 34}, No.8 (1964).
\bibitem[1964Es02]{1964Es02} M.A. Eswaran, C. Broude,    \newblock {\sc Can. J. Phys.} {\bf 42}, 1311 (1964).
\bibitem[1964Gu01]{1964Gu01} C. Gunther, W. Engels, E. Bodenstedt,     \newblock {\sc Phys. Lett.} {\bf 10}, 77 (1964).
\bibitem[1964Ho25]{1964Ho25} B.W. Hooton,     \newblock {\sc Nucl. Phys.} {\bf 59}, 332 (1964).
\bibitem[1964Ja09]{1964Ja09} J. Jastrzebski, M. Moszynski, K. Pawlak, K. Stryczniewicz,     \newblock {\sc Compt. Rend. Congr. Intern. Phys. Nucl.}, Paris, P.Gugenberger, Ed., Centre National de la Recherche Scientifique, Paris, {\bf II}, 573 (1964).
\bibitem[1964Ko13]{1964Ko13} H.J. Korner, E. Gerdau, J. Heisenberg, J. Braunsfurth,     \newblock {\sc Perturbed Angular Correlations}, E.Karlsson, E.Matthias, K.Siegbahn, Ed., North-Holland Publishing Co., 200 (1964).
\bibitem[1964Lo08]{1964Lo08} R. Lombard, P. Kossanyi-Demay, G.R. Bishop,    \newblock {\sc Nucl. Phys.} {\bf 59}, 398 (1964).
\bibitem[1964Ma01]{1964Ma01} D.L. Malaker, L. Schaller, W.C. Miller,    \newblock {\sc Bull. Am. Phys. Soc.} {\bf 9}, No.1, 9, AB7 (1964).
\bibitem[1964No01]{1964No01} T. Novakov, J.M. Hollander, R.L. Graham,     \newblock {\sc Bull. Am. Phys. Soc.} {\bf 9}, No.1, 9, AB6 (1964).
\bibitem[1964Pa17]{1964Pa17} J.C. Palathingal,     \newblock {\sc Phys. Rev.} {\bf 136}, B1553 (1964).
\bibitem[1964Ro19]{1964Ro19} R. Rougny, J.J. Samueli, A. Sarazin,     \newblock {\sc J. Phys. (Paris)} {\bf 25}, 989 (1964).
\bibitem[1964Sc21]{1964Sc21} R.P. Scharenberg, J.D. Kurfess, G. Schilling {\it et al.},     \newblock {\sc Nucl. Phys.} {\bf 58}, 658 (1964); R.P.Scharenberg, {\sc Priv. Comm.} (April 1972).
\bibitem[1964St04]{1964St04} P.H. Stelson, F.K. McGowan, R.L. Robinson, J.L.C. Ford, Jr.,     \newblock {\sc Bull. Am. Phys. Soc.} {\bf 9}, No.4, 484, JA8 (1964).
\bibitem[1964Sy01]{1964Sy01} G.D. Symons, J. De Boer,     \newblock {\sc Bull. Am. Phys. Soc.} {\bf 9}, No.5, 554, N9 (1964).
\bibitem[1965Ab02]{1965Ab02} H. Abou-Leila, N.N. Perrin, J. Valentin,     \newblock {\sc Arkiv Fysik} {\bf 29}, 53 (1965).
\bibitem[1965Do02]{1965Do02} M. Dorikens, O. Segaert, J. Demuynck {\it et al.},     \newblock {\sc Nucl. Phys.} {\bf 61}, 33 (1965).
\bibitem[1965Es01]{1965Es01} M.A. Eswaran, H.E. Gove, A.E. Litherland, C. Broude,     \newblock {\sc Nucl. Phys.} {\bf 66}, 401 (1965).
\bibitem[1965Ev03]{1965Ev03} H.C. Evans, M.A. Eswaran, H.E. Gove,     \newblock {\sc Can. J. Phys.} {\bf 43}, 82 (1965).
\bibitem[1965Fr11]{1965Fr11} A.M. Friedman, J.R. Erskine, T.H. Braid,     \newblock {\sc Bull. Am. Phys. Soc.} {\bf 10}, No.4, 540, KA13 (1965).
\bibitem[1965Ga05]{1965Ga05} Y.P. Gangrskii, I.K. Lemberg,     \newblock {\sc Yadern. Fiz.} {\bf 1}, 1025 (1965); {\sc Soviet J. Nucl. Phys.} {\bf 1}, 731 (1965).
\bibitem[1965Gu02]{1965Gu02} C. Gunther, G. Strube, U. Wehmann {\it et al.},     \newblock {\sc Z. Physik} {\bf 183}, 472 (1965).
\bibitem[1965Gu10]{1965Gu10} G.M. Gusinskii, K.I. Erokhina, I.K. Lemberg,    \newblock {\sc Yadern. Fiz.} {\bf 2}, 794 (1965); {\sc Soviet J. Nucl. Phys.} {\bf 2}, 567 (1966)
\bibitem[1965Hu02]{1965Hu02} A. Hubner,     \newblock {\sc Z. Physik} {\bf 183}, 25 (1965).
\bibitem[1965Ka15]{1965Ka15} D.K. Kaipov, Y.K. Shubnyi, V.M. Amerbaev {\it et al.},    \newblock {\sc Zh. Eksperim. i Teor. Fiz.} {\bf 48}, 1221 (1965); {\sc Soviet Phys. JETP} {\bf 21}, 815 (1965).
\bibitem[1965Mc05]{1965Mc05} F.K. McGowan, R.L. Robinson, P.H. Stelson, J.L.C. Ford, Jr., \newblock {\sc Nucl. Phys.} {\bf 66}, 97 (1965).
\bibitem[1965Me08]{1965Me08} W. Meiling, F. Stary,     \newblock {\sc Nucl. Phys.} {\bf 74}, 113 (1965).
\bibitem[1965Ne03]{1965Ne03} W.R. Neal, H.W. Kraner,     \newblock {\sc Phys.Rev.} {\bf 137}, B1164 (1965).
\bibitem[1965Ro17]{1965Ro17} R. Rougny, J.J. Samueli, A. Sarazin,     \newblock {\sc J. Phys. (Paris)} {\bf 26}, 63 (1965).
\bibitem[1965Sc05]{1965Sc05} R.P. Scharenberg, J.D. Kurfess, G. Schilling {\it et al.},     \newblock {\sc Phys. Rev.} {\bf 137}, B26 (1965).
\bibitem[1965Si02]{1965Si02} J.J. Simpson, J.A. Cookson, D. Eccleshall, M.J.L. Yates,     \newblock {\sc Nucl. Phys.} {\bf 62}, 385 (1965).
\bibitem[1965Ta13]{1965Ta13} G.K. Tandon,     \newblock {\sc Thesis}, Yale University (1965).
\bibitem[1965Ti02]{1965Ti02} J.W. Tippie, R.P. Scharenberg,     \newblock {\sc Phys.Lett.} {\bf 16}, 154 (1965); R.P. Scharenberg, {\sc Priv.Comm.} (April 1972).
\bibitem[1966Ab02]{1966Ab02} H. Abou-Leila, J. Treherne,     \newblock {\sc J. Phys. (Paris)} {\bf 27}, 5 (1966).
\bibitem[1966As03]{1966As03} D. Ashery, N. Assaf, G. Goldring  {\it et al.},     \newblock {\sc Nucl. Phys.} {\bf 77}, 650 (1966).
\bibitem[1966Be16]{1966Be16} G.B. Beard,     \newblock {\sc Phys. Rev.} {\bf 145}, 862 (1966).
\bibitem[1966Be53]{1966Be53} R.B. Begzhanov, A.A. Islamov,     \newblock {\sc Program and Theses, Proc. 16th All-Union Conf. Nucl. Spectroscopy and Struct. Of At. Nuclei}, Moscow, 21 (1966).
\bibitem[1966Bl08]{1966Bl08} D. Bloess, A. Krusche, F. Munnich,     \newblock {\sc Z. Physik} {\bf 192}, 358 (1966).
\bibitem[1966Ec02]{1966Ec02} D. Eccleshall, M.J.L. Yates, J.J. Simpson,     \newblock {\sc Nucl. Phys.} {\bf 78}, 481 (1966).
\bibitem[1966Fu03]{1966Fu03} E.G. Funk, H.J. Prask, J.W. Mihelich,     \newblock {\sc Phys. Rev.} {\bf 141}, 1200 (1966).
\bibitem[1966Go20]{1966Go20} S. Gorodetzky, N. Schulz, E. Bozek, A.C. Knipper,     \newblock {\sc Nucl. Phys.} {\bf 85}, 529 (1966).
\bibitem[1966Gr20]{1966Gr20} L. Grodzins, R.R. Borchers, G.B. Hagemann,     \newblock {\sc Nucl. Phys.} {\bf 88}, 474 (1966).
\bibitem[1966Hr01]{1966Hr01} A.Z. Hrynkiewicz, S. Kopta, S. Szymczyk, T. Walczak,     \newblock {\sc Nucl. Phys.} {\bf 79}, 495 (1966).
\bibitem[1966Hr03]{1966Hr03} B. Hrastnik, V. Knapp, M. Vlatkovic,     \newblock {\sc Nucl. Phys.} {\bf 89}, 412 (1966).
\bibitem[1966Ja16]{1966Ja16} J. Jastrzebski, M. Moszynski, A. Zglinski,  \newblock {\sc  Nukleonika} {\bf 11}, 471 (1966).
\bibitem[1966Li07]{1966Li07} K.P. Lieb,     \newblock {\sc Nucl. Phys.} {\bf 85}, 461(1966).
\bibitem[1966Li08]{1966Li08} H. Liesem,    \newblock {\sc Z. Phys.} {\bf 196}, 174 (1966).
\bibitem[1966Mc07]{1966Mc07} R.E. McAdams, E.N. Hatch,     \newblock {\sc Nucl. Phys.} {\bf 82}, 372 (1966).
\bibitem[1966Mc18]{1966Mc18} F.K. McGowan, P.H. Stelson, R.L. Robinson {\it et al.},     \newblock {\sc Proc.Conf.Bases for Nucl.Spin-Parity Assignments}, Gatlinburg, Tenn. (1966), N.B. Gove, R.L. Robinson, Eds., Academic Press, New York, 222 (1966)
\bibitem[1966Me11]{1966Me11} F.R. Metzger, G.K. Tandon,    \newblock {\sc Phys. Rev.} {\bf 148}, 1133 (1966).
\bibitem[1966Ra04]{1966Ra04} B.V.N. Rao, S. Jnanananda,     \newblock {\sc Nucl. Phys.} {\bf 75}, 109 (1966).
\bibitem[1966Sc06]{1966Sc06} A. Schwarzschild,     \newblock {\sc Phys. Rev.} {\bf 141}, 1206 (1966).
\bibitem[1966Se06]{1966Se06} G.G. Seaman, J.S. Greenberg, D.A. Bromley, F.K. McGowan,     \newblock {\sc Phys. Rev.} {\bf 149}, 925 (1966).
\bibitem[1966Sk01]{1966Sk01} S.J. Skorka, D. Evers, J. Hertel {\it et al.},    \newblock {\sc Nucl. Phys.} {\bf 81}, 370 (1966).
\bibitem[1966Ti01]{1966Ti01} J.W. Tippie, R.P. Scharenberg,     \newblock {\sc Phys. Rev.} {\bf 141}, 1062 (1966).
\bibitem[1966Wa10]{1966Wa10} E.K. Warburton, J.W. Olness, K.W. Jones {\it et al.},    \newblock {\sc Phys. Rev.} {\bf 148}, 1072 (1966)
\bibitem[1967Ab06]{1967Ab06} H. Abou-Leila,     \newblock {\sc Ann. Phys. (Paris)} {\bf 2}, 181 (1967).
\bibitem[1967Af03]{1967Af03} O.F. Afonin, A.P. Grinberg, I.K. Lemberg, I.N. Chugunov,     \newblock {\sc Yadern. Fiz.} {\bf 6}, 219 (1967); {\sc Soviet J. Nucl. Phys.} {\bf 6}, 160 (1968).
\bibitem[1967As03]{1967As03} D. Ashery, N. Bahcall, G. Goldring {\it et al.},     \newblock {\sc Nucl. Phys.} {\bf A 101}, 51 (1967).
\bibitem[1967Ba27]{1967Ba27} A. Backlin, S.G. Malmskog, H. Solhed,     \newblock {\sc Arkiv Fysik} {\bf 34}, 495 (1967).
\bibitem[1967Ba52]{1967Ba52} P. Barreau, J.B. Bellicard,     \newblock {\sc Phys. Rev. Lett.} {\bf 19}, 1444 (1967).
\bibitem[1967Be39]{1967Be39} R.B. Begzhanov, A.A. Islamov,     \newblock {\sc Yadern. Fiz.} {\bf 5}, 483 (1967); {\sc Soviet J. Nucl. Phys.} {\bf 5}, 339 (1967).
\bibitem[1967Be62]{1967Be62} E.E. Berlovich, V.V. Lukashevich,     \newblock {\sc Izv. Akad. Nauk SSSR}, Ser. Fiz. {\bf 31}, 1603 (1967); {\sc Bull. Acad. Sci. USSR}, Phys. Ser. {\bf 31}, 1643 (1968).
\bibitem[1967Br01]{1967Br01} C. Broude, P.J.M. Smulders, T.K. Alexander,    \newblock {\sc Nucl. Phys.} {\bf A 90}, 321 (1967).
\bibitem[1967BuZX]{1967BuZX} G.A. Burginyon, J.S. Greenberg, R.F. Casten, D.A. Bromley,     \newblock {\sc Contrib. Intern. Conf. Nucl. Struct.}, Tokyo, 155 (1967).
\bibitem[1967Ca02]{1967Ca02} A.L. Catz, S. Amiel,    \newblock {\sc Nucl. Phys.} {\bf A 92}, 222 (1967).
\bibitem[1967Ca08]{1967Ca08} R.F. Casten, J.S. Greenberg, G.A. Burginyon, D.A. Bromley,     \newblock {\sc Phys. Rev. Lett.} {\bf 18}, 912 (1967).
\bibitem[1967Cl02]{1967Cl02} J.E. Clarkson, R.M. Diamond, F.S. Stephens, I. Perlman,     \newblock {\sc Nucl. Phys.} {\bf A 93}, 272 (1967).
\bibitem[1967Cr01]{1967Cr01} H. Crannell, T.A. Griffy, L.R. Suelzle, M.R. Yearian,    \newblock {\sc Nucl. Phys.} {\bf A 90}, 152 (1967).
\bibitem[1967DeZW]{1967DeZW} J.de Boer, A.M. Kleinfeld, R. Covello-Moro, H.P. Lie,    \newblock {\sc Bull. Am. Phys. Soc.} {\bf 12}, No.4, 535, GD8 (1967).
\bibitem[1967Du07]{1967Du07} M.A. Duguay, C.K. Bockelman, T.H. Curtis, R.A. Eisenstein,     \newblock {\sc Phys. Rev.} {\bf 163}, 1259 (1967).
\bibitem[1967Gi02]{1967Gi02} P. Gilad, G. Goldring, R. Herber, R. Kalish,     \newblock {\sc Nucl. Phys.} {\bf A 91}, 85 (1967); G.Goldring, {\sc Priv. Comm.} (April 1972).
\bibitem[1967Gl02]{1967Gl02} J.E. Glenn, J.X. Saladin,     \newblock {\sc Phys. Rev. Lett.} {\bf 19}, 33 (1967).
\bibitem[1967Ka16]{1967Ka16} R. Kalish, L. Grodzins, R.R. Borchers {\it et al.},     \newblock {\sc Phys. Rev.} {\bf 161}, 1196 (1967); R.R.Borchers, {\sc Priv. Comm.} (May 1972).
\bibitem[1967Ku07]{1967Ku07} J.D. Kurfess, R.P. Scharenberg,     \newblock {\sc Phys. Rev.} {\bf 161}, 1185 (1967).
\bibitem[1967Li05]{1967Li05} K.P. Lieb, H. Grawe, H. Ropke,    \newblock {\sc Nucl. Phys.} {\bf A 98}, 145 (1967).
\bibitem[1967Si03]{1967Si03} J.J. Simpson, D. Eccleshall, M.J.L. Yates, N.J. Freeman,     \newblock {\sc Nucl. Phys.} {\bf A 94}, 177 (1967).
\bibitem[1967St03]{1967St03} R.G. Stokstad, I. Hall, G.D. Symons, J. de Boer,     \newblock {\sc Nucl. Phys.} {\bf A 92}, 319 (1967).
\bibitem[1967St16]{1967St16} R.G. Stokstad, I. Hall,     \newblock {\sc Nucl. Phys.} {\bf A 99}, 507 (1967).
\bibitem[1967TaZZ]{1967TaZZ} G.K. Tandon,    \newblock {\sc Bull. Am. Phys. Soc.} {\bf 12}, No.5, 683, DE14 (1967).
\bibitem[1967Wo06]{1967Wo06} P.J. Wolfe, R.P. Scharenberg,     \newblock {\sc Phys. Rev.} {\bf 160}, 866 (1967).
\bibitem[1968An20]{1968An20} D.S. Andreev, O.F. Afonin, V.K. Bondarev {\it et al.},    \newblock {\sc Izv. Akad. Nauk SSSR}, Ser.Fiz. {\bf 32}, 1671 (1968); {\sc Bull. Acad. Sci. USSR}, Phys.Ser. {\bf 32}, 1543 (1969).
\bibitem[1968Cr07]{1968Cr07} W.L. Creten, R.J. Jacobs, H.M. Ferdinande,    \newblock {\sc Nucl. Phys.} {\bf A 120}, 126 (1968).
\bibitem[1968Cu05]{1968Cu05} W.M. Currie, C.H. Johnson,    \newblock {\sc Nucl. Instr. Methods} {\bf 63}, 221 (1968).
\bibitem[1968Do12]{1968Do12} K.W. Dolan, D.K. McDaniels,    \newblock {\sc Phys. Rev.} {\bf 175}, 1446 (1968).
\bibitem[1968Ev03]{1968Ev03} D. Evers, G. Flugge, J. Morgenstern {\it et al.},    \newblock {\sc Phys. Letters} {\bf B 27}, 423 (1968).
\bibitem[1968Fi09]{1968Fi09} T.R. Fisher, S.S. Hanna, D.C. Healey, P. Paul,    \newblock {\sc Phys. Rev.} {\bf 176}, 1130(1968)
\bibitem[1968Gi05]{1968Gi05} E.F. Gibson, K. Battleson, D.K. McDaniels,    \newblock {\sc Phys. Rev.} {\bf 172}, 1004 (1968).
\bibitem[1968Ha18]{1968Ha18} O. Hausser, T.K. Alexander, C. Broude,    \newblock {\sc Can. J. Phys.} {\bf 46}, 1035 (1968).
\bibitem[1968Ka01]{1968Ka01} G. Kaye,     \newblock {\sc Nucl. Phys.} {\bf A 108}, 625 (1968).
\bibitem[1968Ke04]{1968Ke04} R.J. Keddy, Y. Yoshizawa, B. Elbek  {\it et al.},     \newblock {\sc Nucl. Phys.} {\bf A 113}, 676 (1968) .
\bibitem[1968Ku03]{1968Ku03} H.W. Kugel, E.G. Funk, J.W. Mihelich,     \newblock {\sc Phys. Rev.} {\bf 165}, 1352 (1968).
\bibitem[1968La26]{1968La26} J.M. Lagrange, G. Albouy, M. Pautrat {\it et al.},     \newblock {\sc J. Phys. (Paris)} {\bf 29}, Suppl.No.1, Colloq.C1-191 (1968).
\bibitem[1968LaZZ]{1968LaZZ} P.G. Lawson,    \newblock {\sc Thesis}, Oxford Univ. (1968); Quoted by 1976As04.
\bibitem[1968Li04]{1968Li04} K.P. Lieb, H. Ropke, H. Grawe {\it et al.},    \newblock {\sc Nucl. Phys.} {\bf A 108}, 233(1968).
\bibitem[1968Li12]{1968Li12} H. Lindeman, G.A.P. Engelbertink, M.W. Ockeloen, H.S. Pruys,    \newblock {\sc Nucl. Phys.} {\bf A 122}, 373(1968).
\bibitem[1968Ma05]{1968Ma05} J.R. MacDonald, D.F.H. Start, R. Anderson {\it et al.},    \newblock {\sc Nucl. Phys.} {\bf A 108}, 6 (1968).
\bibitem[1968Ma14]{1968Ma14} S.G. Malmskog, M. Hojeberg,     \newblock {\sc Arkiv Fysik} {\bf 35}, 229 (1968).
\bibitem[1968Mc08]{1968Mc08} F.K. McGowan, R.L. Robinson, P.H. Stelson, W.T. Milner,     \newblock {\sc Nucl. Phys.} {\bf A 113}, 529 (1968).
\bibitem[1968MiZZ]{1968MiZZ} W.T. Milner,     \newblock {\sc Thesis}, Univ. Tennessee (1968); ORNL-TM-2121 (1968).
\bibitem[1968Pe02]{1968Pe02} G.A. Peterson, J. Alster,     \newblock {\sc Phys. Rev.} {\bf 166}, 1136 (1968).
\bibitem[1968Ra32]{1968Ra32} B.V.N. Rao, K.M.M.S. Ayyangar, S. Jnanananda,     \newblock {\sc Indian J. Pure Appl. Phys.} {\bf 6}, 358 (1968).
\bibitem[1968Ri09]{1968Ri09} F.W. Richter, J. Schutt, D. Wiegandt,     \newblock {\sc Z. Physik} {\bf 213}, 202 (1968).
\bibitem[1968Ri16]{1968Ri16} F. Riess, P. Paul, J.B. Thomas, S.S. Hanna,    \newblock {\sc Phys. Rev.} {\bf 176}, 1140 (1968).
\bibitem[1968Ro05]{1968Ro05} S.W. Robinson, R.D. Bent,     \newblock {\sc Phys. Rev.} {\bf 168}, 1266 (1968).
\bibitem[1968Sc04]{1968Sc04} D. Schroeer, P.S. Jastram,     \newblock {\sc Phys. Rev.} {\bf 166}, 1212 (1968).
\bibitem[1968Sc13]{1968Sc13} M. Schumacher,     \newblock {\sc Phys. Rev.} {\bf 171}, 1279 (1968).
\bibitem[1968Se02]{1968Se02} B. Sethi, S.K. Mukherjee,     \newblock {\sc Phys. Rev.} {\bf 166}, 1227 (1968).
\bibitem[1968SeZZ]{1968SeZZ} G.G. Seaman, M.C. Bertin, J.W. Tape {\it et al.},    \newblock {\sc Bull. Am. Phys. Soc.} {\bf 13}, No.11, 1384, BD15 (1968); {\sc Priv.Comm.} (1969).
\bibitem[1968Si05]{1968Si05} J.J. Simpson, U. Smilansky, J.P. Wurm,     \newblock {\sc Phys. Lett.} {\bf B 27}, 633 (1968); Erratum Phys.Letters {\bf B 28}, 422 (1969).
\bibitem[1968St04]{1968St04} M. Stroetzel,    \newblock {\sc Phys. Letters} {\bf B 26}, 376 (1968).
\bibitem[1968St13]{1968St13} R.G. Stokstad, B. Persson,     \newblock {\sc Phys. Rev.} {\bf 170}, 1072 (1968).
\bibitem[1968Ve01]{1968Ve01} E. Veje, B. Elbek, B. Herskind, M.C. Olesen,     \newblock {\sc Nucl. Phys.} {\bf A 109}, 489 (1968).
\bibitem[1968Wa08]{1968Wa08} H.K. Walter, A. Weitsch, \newblock {\sc Z. Physik} {\bf 211}, 304 (1968).
\bibitem[1968Zi02]{1968Zi02} J.F. Ziegler, G.A. Peterson,     \newblock {\sc Phys. Rev.} {\bf 165}, 1337 (1968).
\bibitem[1969Af01]{1969Af01} V.D. Afanasev, N.G. Afanasev, I.S. Gulkarov {\it et al.},     \newblock {\sc Yadern. Fiz.} {\bf 10}, 33 (1969); {\sc Soviet J. Nucl. Phys.} {\bf 10}, 18 (1970).
\bibitem[1969An08]{1969An08} J.H. Anderson, R.C. Ritter,    \newblock {\sc Nucl. Phys.} {\bf A 128}, 305 (1969).
\bibitem[1969Av01]{1969Av01} R. Avida, Y. Dar, P. Gilad {\it et al.},     \newblock {\sc Nucl. Phys.} {\bf A 127}, 412 (1969).
\bibitem[1969Be31]{1969Be31} R.A.L. Bell, J. L'Ecuyer, R.D. Gill {\it et al.},    \newblock {\sc Nucl. Phys.} {\bf A 133}, 337 (1969).
\bibitem[1969Be34]{1969Be34} R.A. Belt, H.W. Kugel, J.M. Jaklevich, E.G. Funk,     \newblock {\sc Nucl. Phys.} {\bf A 134}, 225 (1969).
\bibitem[1969Be48]{1969Be48} M.C. Bertin, N. Benczer-Koller, G.G. Seaman, J.R. MacDonald,     \newblock {\sc Phys. Rev.} {\bf 183}, 964 (1969).
\bibitem[1969Bh01]{1969Bh01} K. Bharuth-Ram, K.P. Jackson, K.W. Jones, E.K. Warburton,    \newblock {\sc Nucl. Phys.} {\bf A 137}, 262 (1969).
\bibitem[1969Bi09]{1969Bi09} M. Bister, A. Anttila, J. Rasanen,    \newblock {\sc Can. J. Phys.} {\bf 47}, 2539 (1969).
\bibitem[1969Bi11]{1969Bi11} P.G. Bizzeti, A.M. Bizzeti-Sona, A. Cambi {\it et al.},    \newblock {\sc Nuovo Cimento Lett.} {\bf 2}, 775 (1969).
\bibitem[1969Ca19]{1969Ca19} R.F. Casten, J.S. Greenberg, S.H. Sie {\it et al.},    \newblock {\sc  Phys. Rev.} {\bf 187}, 1532 (1969).
\bibitem[1969Ca24]{1969Ca24} L.E. Carlson,    \newblock {\sc Priv. Comm.}, quoted by 1969Fl03 (1969).
\bibitem[1969Cl05]{1969Cl05} D.Cline, H.S.Gertzman, H.E.Gove {\it et al.},     \newblock {\sc Nucl. Phys.} {\bf A 133}, 445 (1969).
\bibitem[1969Cu06]{1969Cu06} T.H. Curtis, R.A. Eisenstein, D.W. Madsen, C.K. Bockelman,     \newblock {\sc Phys. Rev.} {\bf 184}, 1162 (1969).
\bibitem[1969Di02]{1969Di02} R.M. Diamond, F.S. Stephens, W.H. Kelly, D. Ward,     \newblock {\sc Phys. Rev. Letters} {\bf 22}, 546 (1969).
\bibitem[1969Ei03]{1969Ei03} R.A. Eisenstein, D.W. Madsen, H. Theissen {\it et al.},    \newblock {\sc Phys. Rev.} {\bf 188}, 1815 (1969).
\bibitem[1969En04]{1969En04} G.A.P. Engelbertink, G. van Middelkoop,    \newblock {\sc Nucl. Phys.} {\bf A 138}, 588 (1969).
\bibitem[1969Fo07]{1969Fo07} M. Forker, H.F. Wagner,     \newblock {\sc Nucl. Phys.} {\bf A 138}, 13 (1969).
\bibitem[1969Fo08]{1969Fo08} M. Forker, H.F. Wagner, G. Schmidt,     \newblock {\sc Nucl. Phys.} {\bf A 138}, 97 (1969).
\bibitem[1969FuZX]{1969FuZX} E.G. Funk,     \newblock {\sc Priv. Comm.}, quoted by 1970Ra18 (1969).
\bibitem[1969Ga25]{1969Ga25} L.N. Galperin, A.Z. Ilyasov, I.K. Lemberg, G.A. Firsanov,     \newblock {\sc Yad. Fiz.} {\bf 9}, 225 (1969); {\sc Sov. J. Nucl. Phys.} {\bf 9}, 133 (1969).
\bibitem[1969Gl08]{1969Gl08} J.E. Glenn, R.J. Pryor, J.X. Saladin,     \newblock {\sc Phys. Rev.} {\bf 188}, 1905 (1969).
\bibitem[1969GlZY]{1969GlZY} J.E. Glenn,     \newblock {\sc COO-535-603}, 28 (1969).
\bibitem[1969Gr03]{1969Gr03} H.Grawe, K.P.Lieb,     \newblock {\sc Nucl. Phys.} {\bf A 127}, 13 (1969).
\bibitem[1969Ha02]{1969Ha02} R. Hartmann, K.P. Lieb, H. Ropke,    \newblock {\sc Nucl. Phys.} {\bf A 123}, 437 (1969).
\bibitem[1969Ha31]{1969Ha31} O. Hausser, T.K. Alexander, D. Pelte {\it et al.},     \newblock {\sc Phys. Rev. Lett.} {\bf 23}, 320 (1969).
\bibitem[1969Jo10]{1969Jo10} K.W. Jones, A.Z. Schwarzschild, E.K. Warburton, D.B. Fossan,     \newblock {\sc Phys. Rev.} {\bf 178}, 1773 (1969).
\bibitem[1969Ka10]{1969Ka10} C.D. Kavaloski, W.J. Kossler,    \newblock {\sc Phys. Rev.} {\bf 180}, 971 (1969).
\bibitem[1969KeZX]{1969KeZX} J.R. Kerns,     \newblock {\sc Thesis}, Univ.Pittsburgh (1969); {\sc Diss. Abstr. Int.} {\bf 31B}, 336 (1970).
\bibitem[1969Ko03]{1969Ko03} W.J. Kossler, J. Winkler, C.D. Kavaloski,    \newblock {\sc Phys. Rev.} {\bf 177}, 1725(1969).
\bibitem[1969Me14]{1969Me14} M.A. Meyer, N.S. Wolmarans,    \newblock {\sc Nucl. Phys.} {\bf A 136}, 663 (1969).
\bibitem[1969Mi07]{1969Mi07} W.T. Milner, F.K. McGowan, P.H. Stelson {\it et al.},     \newblock {\sc Nucl. Phys.} {\bf A 129}, 687 (1969).
\bibitem[1969Ni09]{1969Ni09} R.J. Nickles,    \newblock {\sc Nucl. Phys.} {\bf A 134}, 308 (1969).
\bibitem[1969Pe08]{1969Pe08} D. Pelte, O. Hausser, T.K. Alexander {\it et al.},    \newblock {\sc Phys. Lett.} {\bf B 29}, 660 (1969).
\bibitem[1969Pe11]{1969Pe11} D. Pelte, O. Hausser, T.K. Alexander, H.C. Evans,    \newblock {\sc Can. J. Phys.} {\bf 47}, 1929 (1969).
\bibitem[1969Po04]{1969Po04} A.R. Poletti, A.D.W. Jones, J.A. Becker, R.E. McDonald,    \newblock {\sc Phys. Rev.} {\bf 181}, 1606 (1969).
\bibitem[1969Ro05]{1969Ro05} R.L. Robinson, F.K. McGowan, P.H. Stelson {\it et al.},     \newblock {\sc Nucl. Phys.} {\bf A 124}, 553 (1969).
\bibitem[1969Ro08]{1969Ro08} B.C. Robertson, R.A.I. Bell, J. L'Ecuyer {\it et al.},    \newblock {\sc Nucl. Phys.} {\bf A 126}, 431 (1969).
\bibitem[1969Sa14]{1969Sa14} G.A. Savitskii, N.G. Afanasev, I.V. Andreeva {\it et al.},    \newblock {\sc Izv. Akad. Nauk SSSR}, Ser.Fiz. {\bf 33}, 53 (1969); {\sc Bull. Acad. Sci. USSR}, Phys.Ser. {\bf 33}, 50 (1970).
\bibitem[1969Sa27]{1969Sa27} J.X. Saladin, J.E. Glenn, R.J. Pryor,     \newblock {\sc Phys. Rev.} {\bf 186}, 1241 (1969).
\bibitem[1969ScZV]{1969ScZV} D. Schwalm, B. Povh,     \newblock {\sc Contrib. Intern. Conf. Properties Nucl. States}, Suppl., Montreal, Canada, 15 (1969).
\bibitem[1969Si15]{1969Si15} J.J. Simpson, U. Smilansky, D. Ashery,     \newblock {\sc Nucl. Phys.} {\bf A 138}, 529 (1969).
\bibitem[1969Sp05]{1969Sp05} G.D. Sprouse, S.S. Hanna,     \newblock {\sc Nucl. Phys.} {\bf A137}, 658 (1969).
\bibitem[1969Th01]{1969Th01} M.J. Throop,     \newblock {\sc Phys. Rev.} {\bf 179}, 1011 (1969).
\bibitem[1969Th03]{1969Th03} J.P. Thibaud, M.M. Aleonard, D. Castera {\it et al.},    \newblock {\sc Nucl. Phys.} {\bf A 135}, 281 (1969).
\bibitem[1969Ti01]{1969Ti01} O. Titze,    \newblock {\sc Z. Physik} {\bf 220}, 66 (1969).
\bibitem[1969To08]{1969To08} Y. Torizuka, Y. Kojima, M. Oyamada {\it et al.},     \newblock {\sc Phys. Rev.} {\bf 185}, 1499 (1969).
\bibitem[1970Ab14]{1970Ab14} H. Abou-Leila, S.M. Darwish, A. Abd El-Haliem, Z. Awwad,     \newblock {\sc Nucl. Phys.} {\bf A 158}, 568 (1970).
\bibitem[1970Af04]{1970Af04} V.D. Afanasev, N.G. Afanasev, A.Y. Buki {\it et al.},     \newblock {\sc Yad. Fiz.} {\bf 12}, 885 (1970); {\sc Sov. J. Nucl. Phys.} {\bf 12}, 480 (1971).
\bibitem[1970AgZV]{1970AgZV} A.P. Agnihotry, K.P. Gopinathan, M.C. Joshi, K.G. Prasad,     \newblock {\sc Proc. Nucl. Phys. And Solid State Phys. Symp.}, Nucl. Phys., Madurai, 161 (1970).
\bibitem[1970Al05]{1970Al05} M.M. Aleonard, D. Castera, P. Hubert {\it et al.},    \newblock {\sc Nucl. Phys.} {\bf A 146}, 90 (1970).
\bibitem[1970Al10]{1970Al10} T.K. Alexander, A. Bell,    \newblock {\sc Nucl. Instrum. Methods} {\bf 81}, 22 (1970).
\bibitem[1970BaYH]{1970BaYH} R.D. Barton, J.S. Wadden, V.K. Carriere,     \newblock {\sc Bull. Amer. Phys. Soc.} {\bf 15}, No.6, 806, EE6 (1970).
\bibitem[1970Be07]{1970Be07} J. Bellicard, P. Leconte, T.H. Curtis {\it et al.},     \newblock {\sc Nucl. Phys.} {\bf A 143}, 213 (1970).
\bibitem[1970Be08]{1970Be08} R. Beraud, I. Berkes, R. Chery {\it et al.},     \newblock {\sc Phys. Rev.} {\bf C 1}, 303 (1970).
\bibitem[1970Be18]{1970Be18} T. Bedike, N.G. Zaitseva, V.A. Morozov {\it et al.},     \newblock {\sc Yad. Fiz.} {\bf 11}, 481 (1970); {\sc Sov. J. Nucl. Phys.} {\bf 11}, 269 (1970).
\bibitem[1970Be36]{1970Be36} I. Ben-Zvi, P. Gilad, M.B. Goldberg {\it et al.},     \newblock {\sc Nucl. Phys.} {\bf A 151}, 401 (1970); G.Goldring, {\sc Priv.Comm.} (April 1972).
\bibitem[1970Be39]{1970Be39} N. Benczer-Koller, G.G. Seaman, M.C. Bertin {\it et al.},    \newblock {\sc Phys. Rev.} {\bf C 2}, 1037 (1970).
\bibitem[1970Bi08]{1970Bi08} M. Bini, P.G. Bizzeti, A.M. Bizzeti-Sona {\it et al.},    \newblock {\sc Lett. Nuovo Cimento} {\bf 3}, 235 (1970).
\bibitem[1970Br18]{1970Br18} F. Brandolini, C. Signorini,    \newblock {\sc Nuovo Cimento} {\bf 67 A}, 247 (1970).
\bibitem[1970Br26]{1970Br26} E.J. Bruton, J.A. Cameron, A.W. Gibb {\it et al.},     \newblock {\sc Nucl. Phys.} {\bf A 152}, 495 (1970).
\bibitem[1970BrZP]{1970BrZP} R. Broda, J. Golczewski, A.Z. Hrynkiewicz {\it et al.},     \newblock {\sc JINR-E6-5070} (1970).
\bibitem[1970Ch01]{1970Ch01} A. Christy, I. Hall, R.P. Harper {\it et al.},     \newblock {\sc Nucl. Phys.} {\bf A 142}, 591 (1970).
\bibitem[1970Ch11]{1970Ch11} E. Cheifetz, R.C. Jared, S.G. Thompson, J.B. Wilhelmy,     \newblock {\sc Phys. Rev. Lett.} {\bf 25}, 38 (1970).
\bibitem[1970Ch14]{1970Ch14} P.R. Christensen, G. Lovhoiden, J. Rasmussen,     \newblock {\sc Nucl. Phys.} {\bf A 149}, 302 (1970).
\bibitem[1970ChZH]{1970ChZH} E. Cheifetz, R.C. Jared, S.G. Thompson, J.B. Wilhelmy,     \newblock {\sc Proc. Int. Conf. Prop. Nuclei Far from Region of Beta-Stability}, Leysin, Switzerland, {\bf 2}, 883 (1970); {\sc CERN 70-30} (1970).
\bibitem[1970Co09]{1970Co09} P.M. Cockburn, W.J. Stark, R.W. Krone,    \newblock {\sc Phys. Rev.} {\bf C 1}, 1757 (1970).
\bibitem[1970Cu02]{1970Cu02} W.M. Currie, L.G. Earwaker, J. Martin, A.K. Sen Gupta,    \newblock {\sc J. Phys.} {\bf A 3}, 73 (1970).
\bibitem[1970De01]{1970De01} P.R. de Kock, J.W. Koen, W.L. Mouton,    \newblock {\sc Nucl. Phys.} {\bf A 140}, 190 (1970).
\bibitem[1970En01]{1970En01} G. Engler,     \newblock {\sc Phys. Rev.} {\bf C 1}, 734 (1970).
\bibitem[1970ErZY]{1970ErZY} B.R. Erdal, M. Finger, R. Foucher {\it et al.},     \newblock {\sc Proc. Int. Conf. Prop. Nuclei Far from Region of Beta-Stability}, Leysin, Switzerland, {\bf 2}, 1031 (1970); {\sc CERN-70-30} (1970).
\bibitem[1970Ge07]{1970Ge07} H.S. Gertzman, D. Cline, H.E. Gove {\it et al.},     \newblock {\sc Nucl. Phys.} {\bf A 151}, 273 (1970).
\bibitem[1970Gr11]{1970Gr11} M.W. Greene, P.R. Alderson, D.C. Bailey {\it et al.},    \newblock {\sc Nucl. Phys.} {\bf A 148}, 351 (1970).
\bibitem[1970Ha04]{1970Ha04} O. Hausser, B.W. Hooton, D. Pelte {\it et al.},    \newblock {\sc Can. J. Phys.} {\bf 48}, 35 (1970).
\bibitem[1970Ha24]{1970Ha24} O. Hausser, D. Pelte, T.K. Alexander, H.C. Evans,    \newblock {\sc Nucl.Phys.} {\bf A 150}, 417 (1970).
\bibitem[1970He01]{1970He01} D. Herrmann, J. Kalus,    \newblock {\sc Nucl. Phys.} {\bf A 140}, 257 (1970).
\bibitem[1970Hi03]{1970Hi03} D. Hitlin, S. Bernow, S. Devons {\it et al.},     \newblock {\sc Phys. Rev.} {\bf C 1}, 1184 (1970).
\bibitem[1970Hu14]{1970Hu14} F.C.P. Huang, D.K. McDaniels,    \newblock {\sc Phys. Rev.} {\bf C 2}, 1342 (1970).
\bibitem[1970It01]{1970It01} K. Itoh, M. Oyamada, Y. Torizuka,    \newblock {\sc Phys. Rev.} {\bf C 2}, 2181 (1970).
\bibitem[1970Jo20]{1970Jo20} W. John, F.W. Guy, J.J. Wesolowski,    \newblock {\sc   Phys. Rev.} {\bf C 2}, 1451 (1970).
\bibitem[1970Ka09]{1970Ka09} R. Kalish, R.R. Borchers, H.W. Kugel,     \newblock {\sc Nucl. Phys.} {\bf A 147}, 161 (1970); R.R.Borchers, {\sc Priv. Comm.} (May 1972).
\bibitem[1970KaZK]{1970KaZK} G. Kaspar, W. Knupfer, W. Ebert  {\it et al.},     \newblock {\sc Proc. Intern. Conf. Nucl. Reactions Induced by Heavy Ions}, Heidelberg, Germany (1969); R.Bock, W.R.Hering, Eds, North-Holland Publishing Co., Amsterdam, 471 (1970).
\bibitem[1970Ke15]{1970Ke15} P.F. Kenealy, G.B. Beard, K. Parsons,     \newblock {\sc Phys. Rev.} {\bf C 2}, 2009 (1970).
\bibitem[1970Kh05]{1970Kh05} V.M. Khvastunov, N.G. Afanasev, V.D. Afanasev {\it et al.},    \newblock {\sc Yad. Fiz.} {\bf 12}, 9 (1970); {\sc Sov. J. Nucl. Phys.} {\bf 12}, 5 (1971).
\bibitem[1970Kl06]{1970Kl06} A.M. Kleinfeld, R. Covello-Moro, H. Ogata {\it et al.},     \newblock {\sc Nucl. Phys.} {\bf A 154}, 499 (1970).
\bibitem[1970Kl12]{1970Kl12} A.M. Kleinfeld, J.D. Rogers, J. Gastebois {\it et al.},     \newblock {\sc Nucl. Phys.} {\bf A 158}, 81 (1970).
\bibitem[1970LaZM]{1970LaZM} J.M. Lagrange,     \newblock {\sc Thesis}, Univ.Paris (1970); {\sc NP-18447} (1970); {\sc Priv. Comm.} (1973).
\bibitem[1970Le17]{1970Le17} P.M.S. Lesser, D. Cline, J.D. Purvis,     \newblock {\sc Nucl. Phys.} {\bf A 151}, 257 (1970).
\bibitem[1970Me08]{1970Me08} F.R. Metzger,     \newblock {\sc Nucl. Phys.} {\bf A 158}, 88 (1970).
\bibitem[1970Me09]{1970Me09} M.W. Mekshes, N. Hershkowitz {\it et al.},     \newblock {\sc Phys. Rev.} {\bf C 2}, 289 (1970).
\bibitem[1970Me18]{1970Me18} F.R. Metzger,     \newblock {\sc Nucl. Phys.} {\bf A 148}, 362 (1970).
\bibitem[1970MiZQ]{1970MiZQ} W.T. Milner, F.K. McGowan, P.H. Stelson, R.L. Robinson,     \newblock {\sc Bull. Amer. Phys. Soc.} {\bf 15}, No.11, 1358, DF11 (1970).
\bibitem[1970Mo39]{1970Mo39} V.A. Morozov, T.M. Muminov, A.B. Khalikulov,     \newblock {\sc JINR-P6-5201} (1970).
\bibitem[1970Na05]{1970Na05} K. Nakai, J.L. Quebert, F.S. Stephens, R.M. Diamond,    \newblock {\sc Phys. Rev. Lett.} {\bf 24}, 903 (1970).
\bibitem[1970Na07]{1970Na07} K. Nakai, F.S. Stephens, R.M. Diamond,    \newblock {\sc Nucl. Phys.} {\bf A 150}, 114 (1970).
\bibitem[1970Pe15]{1970Pe15} R.J. Peterson, H. Theissen, W.J. Alston,     \newblock {\sc Nucl. Phys.} {\bf A153}, 610 (1970).
\bibitem[1970Pr07]{1970Pr07} R.J. Pryor, F. Rosel, J.X. Saladin, K. Alder,     \newblock {\sc Phys. Lett.} {\bf B 32}, 26 (1970).
\bibitem[1970Pr09]{1970Pr09} R.J. Pryor, J.X. Saladin,     \newblock {\sc Phys. Rev.} {\bf C 1}, 1573 (1970).
\bibitem[1970Qu02]{1970Qu02} J.L. Quebert, K. Nakai, R.M. Diamond, F.S. Stephens,     \newblock {\sc Nucl. Phys.} {\bf A 150}, 68 (1970).
\bibitem[1970Ra17]{1970Ra17} C.E. Ragan III, R.V. Poore, N.R. Roberson {\it et al.},    \newblock {\sc Phys. Rev.} {\bf C 1}, 2012 (1970).
\bibitem[1970Ra18]{1970Ra18} R.L. Rasera, A. Li-Scholz,     \newblock {\sc Phys. Rev.} {\bf B 1}, 1995 (1970).
\bibitem[1970RaZC]{1970RaZC} V.K. Rasmussen,    \newblock {\sc Priv. Comm.}, quoted by 1971Ma03 (1970).
\bibitem[1970Sa09]{1970Sa09} R.O. Sayer, P.H. Stelson, F.K. McGowan  {\it et al.},     \newblock {\sc Phys. Rev.} {\bf C 1}, 1525 (1970).
\bibitem[1970St10]{1970St10} P. Strehl,    \newblock {\sc Z. Phys.} {\bf 234}, 416 (1970).
\bibitem[1970St17]{1970St17} S.G. Steadman, A.M. Kleinfeld, G.G. Seaman {\it et al.},     \newblock {\sc Nucl. Phys.} {\bf A 155}, 1 (1970).
\bibitem[1970St20]{1970St20} P.H. Stelson, F.K. McGowan, R.L. Robinson, W.T. Milner,     \newblock {\sc Phys. Rev.} {\bf C 2}, 2015 (1970).
\bibitem[1970StZP]{1970StZP} D.F.H. Start,    \newblock {\sc Priv. Comm.}, quoted by 1971Ma03 (1970).
\bibitem[1970Sw03]{1970Sw03} C.P. Swann,    \newblock {\sc Nucl. Phys.} {\bf A 150}, 300 (1970).
\bibitem[1970Th04]{1970Th04} J.P. Thibaud, M.M. Aleonard, D. Castera {\it et al.},    \newblock {\sc J. Phys. (Paris)} {\bf 31}, 131 (1970).
\bibitem[1970To08]{1970To08} H. Ton, W. Beens, S. Roodbergen, J. Blok,     \newblock {\sc Nucl. Phys.} {\bf A 155}, 235 (1970).
\bibitem[1970Wa04]{1970Wa04} B. Wakefield, I.M. Naqib, R.P. Harper {\it et al.},     \newblock {\sc Phys. Lett.} {\bf B 31}, 56 (1970).
\bibitem[1970Wa05]{1970Wa05} R.L. Watson, J.B. Wilhelmy, R.C. Jared {\it et al.},     \newblock {\sc Nucl. Phys.} {\bf A 141}, 449 (1970).
\bibitem[1971Ab05]{1971Ab05} H. Abou-Leila, A. Abd El-Haliem, S.M. Darwish,     \newblock {\sc Nucl. Phys.} {\bf A 175}, 663 (1971).
\bibitem[1971Ba59]{1971Ba59} J. Barrette, M. Barrette, A. Boutard {\it et al.},     \newblock {\sc  Nucl.Phys.} {\bf A 172}, 41 (1971).
\bibitem[1971Bb09]{1971Bb09} T. Badica,     \newblock {\sc Stud. Cercet. Fiz.} {\bf 23}, 877 (1971).
\bibitem[1971Bo08]{1971Bo08} H.H. Bolotin, D.A. McClure,     \newblock {\sc Phys. Rev.} {\bf C 3}, 797 (1971).
\bibitem[1971Bo13]{1971Bo13} P.D. Bond, J.D. McGervey, S. Jha,     \newblock {\sc Nucl. Phys.} {\bf A 163}, 571 (1971).
\bibitem[1971Bo23]{1971Bo23} M. Bocciolini, P. Sona, N. Taccetti,    \newblock {\sc Lett. Nuovo Cim.} {\bf 1}, 695 (1971).
\bibitem[1971BrYK]{1971BrYK} B.A. Brown, M. Marmor, D.B. Fossan,    \newblock {\sc Proc. Topical Conf. Struct. of 1f7/2 Nuclei}, R.A.Ricci, Ed., Editrice Compositori, Bologna, p.123 (1971).
\bibitem[1971Ca35]{1971Ca35} J.A. Cameron, Z. Zamori,     \newblock {\sc Can. J. Phys.} {\bf 49}, 2690 (1971) .
\bibitem[1971Ch26]{1971Ch26} A. Charvet, Do Huu Phuoc, R. Duffait {\it et al.},     \newblock {\sc J. Phys. (Paris)} {\bf 32}, 359 (1971).
\bibitem[1971ChZF]{1971ChZF} J. Charbonneau, N.V.De Castro Faria, J. L'Ecuyer, D. Vitoux,     \newblock {\sc Bull. Amer. Phys. Soc.} {\bf 16}, No.4, 625, JH2 (1971).
\bibitem[1971Cr01]{1971Cr01} P.A. Crowley, J.R. Kerns, J.X. Saladin,     \newblock {\sc Phys. Rev.} {\bf C 3}, 2049 (1971).
\bibitem[1971DaZM]{1971DaZM} W.G. Davies, J.S. Forster, I.M. Szoghy, D. Ward,     \newblock {\sc AECL-3996}, 16 (1971).
\bibitem[1971De29]{1971De29} N.V. de Castro Faria, J. Charbonneau, J. L'Ecuyer, R.J.A. Levesque,    \newblock {\sc Nucl. Phys.} {\bf A 174}, 37 (1971).
\bibitem[1971Di02]{1971Di02} R.M. Diamond, F.S. Stephens, K. Nakai, R. Nordhagen,     \newblock {\sc Phys. Rev.} {\bf C 3}, 344 (1971).
\bibitem[1971Ej01]{1971Ej01} H. Ejiri, G.B. Hagemann,     \newblock {\sc Nucl. Phys.} {\bf A 161}, 449 (1971).
\bibitem[1971El03]{1971El03} C. Ellegaard, P.D. Barnes, E.R. Flynn, G.J. Igo,     \newblock {\sc Nucl. Phys.} {\bf A 162}, 1 (1971).
\bibitem[1971Fa14]{1971Fa14} J. Fagot, R. Lucas, H. Nifenecker, M. Schneeberger,    \newblock {\sc Nucl. Instrum. Methods} {\bf 95}, 421 (1971).
\bibitem[1971Fo17]{1971Fo17} J.L.C. Ford, Jr., P.H. Stelson, C.E. Bemis {\it et al.},     \newblock {\sc Phys. Rev. Lett.} {\bf 27}, 1232 (1971).
\bibitem[1971FoZV]{1971FoZV} J.S. Forster, G.C. Ball, W.G. Davies,    \newblock {\sc Bull. Amer. Phys. Soc.} {\bf 16}, No.4, 555, EE8 (1971).
\bibitem[1971FoZW]{1971FoZW} J.L.C. Ford, Jr., P.H. Stelson, R.L. Robinson {\it et al.},     \newblock {\sc Bull. Amer. Phys. Soc.} {\bf 16}, No.4, 515, BG11 (1971); {\sc Priv. Comm.} (July 1971).
\bibitem[1971Ga01]{1971Ga01} G.T. Garvey, K.W. Jones, L.E. Carlson {\it et al.},    \newblock {\sc Nucl. Phys.} {\bf A 160}, 25 (1971).
\bibitem[1971Gr31]{1971Gr31} E. Grosse, M. Dost, K. Haberkant {\it et al.},     \newblock {\sc Nucl. Phys.} {\bf A 174}, 525 (1971).
\bibitem[1971Ha08]{1971Ha08} R.P. Harper, A. Christy, I. Hall {\it et al.},     \newblock {\sc Nucl. Phys.} {\bf A 162}, 161 (1971).
\bibitem[1971Ha12]{1971Ha12} R. Hartmann, H. Grawe,    \newblock {\sc Nucl. Phys.} {\bf A 164}, 209 (1971).
\bibitem[1971Ha26]{1971Ha26} O. Hausser, T.K. Alexander, A.B. McDonald {\it et al.},    \newblock {\sc Nucl. Phys.} {\bf A 168}, 17 (1971).
\bibitem[1971Ha47]{1971Ha47} O. Hausser, T.K. Alexander, A.B. McDonald, W.T. Diamond,    \newblock {\sc Nucl. Phys.} {\bf A 175}, 593 (1971).
\bibitem[1971HaXH]{1971HaXH} O. Hausser, D.L. Disdier, A.J. Ferguson, T.K. Alexander,    \newblock {\sc AECL-4068}, 12 (1971).
\bibitem[1971He08]{1971He08} J. Heisenberg, J.S. McCarthy, I. Sick,     \newblock {\sc Nucl. Phys.} {\bf A164}, 353 (1971).
\bibitem[1971Ho14]{1971Ho14} A. Hoglund, S.G. Malmskog, A. Marelius {\it et al.},     \newblock {\sc Nucl. Phys.} {\bf A 169}, 49 (1971).
\bibitem[1971HuZR]{1971HuZR} R.B. Huber, W. Kutschera, C. Signorini,    \newblock {\sc J. Phys. (Paris)} Suppl. C6-207 (1971).
\bibitem[1971In02]{1971In02} F. Ingebretsen, B.W. Sargent, A.J. Ferguson {\it et al.},    \newblock {\sc Nucl. Phys.} {\bf A 161}, 433 (1971).
\bibitem[1971ImZY]{1971ImZY} H. Imada,   \newblock {\sc Thesis, Texas A and M Univ.} (1971); {\sc Diss.Abstr.Int.} {\bf B 32}, 1772 (1971).
\bibitem[1971Ja10]{1971Ja10} A.N. James, P.R. Alderson, D.C. Bailey {\it et al.},    \newblock {\sc Nucl. Phys.} {\bf A 168}, 56 (1971).
\bibitem[1971Ja15]{1971Ja15} A.N.James, P.R.Alderson, C.D.Bailey {\it et al.},    \newblock {\sc Nucl. Phys.} {\bf A 172}, 401 (1971).
\bibitem[1971Ka03]{1971Ka03} R. Kalish, R.R. Borchers, H.W. Kugel,     \newblock {\sc Nucl. Phys.} {\bf A 161}, 637 (1971).
\bibitem[1971Ma03]{1971Ma03} J.R. MacDonald, D.H. Wilkinson, D.E. Alburger,    \newblock {\sc Phys. Rev.} {\bf C 3}, 219 (1971).
\bibitem[1971Ma27]{1971Ma27} D.W. Madsen, L.S. Cardman, J.R. Legg, C.K. Bockelman,     \newblock {\sc Nucl. Phys.} {\bf A 168}, 97 (1971).
\bibitem[1971Mc20]{1971Mc20} A.B. McDonald, T.K. Alexander, O. Hausser, G.T. Ewan,    \newblock {\sc Can. J. Phys.} {\bf 49}, 2886 (1971).
\bibitem[1971Mi08]{1971Mi08} W.T. Milner, F.K. McGowan, R.L. Robinson {\it et al.},     \newblock {\sc Nucl. Phys.} {\bf A 177}, 1 (1971).
\bibitem[1971MiZK]{1971MiZK} W.T. Mills,     \newblock {\sc Priv. Comm.}, quoted by 1972Me04 (1971).
\bibitem[1971Na06]{1971Na06} K. Nakai, F.S. Stephens, R.M. Diamond,    \newblock {\sc Phys. Lett.} {\bf B 34}, 389 (1971).
\bibitem[1971NoZT]{1971NoZT} R.H. Nord,     \newblock {\sc Thesis}, Univ.Wisconsin (1971); Diss.Abstr. 32B, No.6, 3572 (1971).
\bibitem[1971Ob02]{1971Ob02} L.W. Oberley, N. Hershkowitz, S.A. Wender, A.B. Carpenter,     \newblock {\sc Phys. Rev.} {\bf C 3}, 1585 (1971).
\bibitem[1971Pe11]{1971Pe11} R.J. Peterson,     \newblock {\sc Ann. Phys. (N.Y.)} {\bf 65}, 125 (1971).
\bibitem[1971Re15]{1971Re15} M.J. Renan, R.J. Keddy,    \newblock {\sc Nuovo Cimento} {\bf 3 A}, 347 (1971).
\bibitem[1971RiZJ]{1971RiZJ} L.L. Riedinger, G. Schilling, A.E. Rainis {\it et al.},     \newblock {\sc Nucl. Struct. Lab., Univ. Notre Dame, Ann. Rep. 1971}, 121 (1971).
\bibitem[1971Sh11]{1971Sh11} J.F. Sharpey-Schafer, P.R. Alderson, D.C. Bailey {\it et al.},    \newblock {\sc Nucl. Phys.} {\bf A 167}, 602 (1971).
\bibitem[1971SiYA]{1971SiYA} S.H. Sie, J.S. Geiger, R.L. Graham {\it et al.},     \newblock {\sc AECL-4068}, 50 (1971).
\bibitem[1971Sp06]{1971Sp06} H. Spehl, N. Wertz,     \newblock {\sc Z. Phys.} {\bf 243}, 431 (1971).
\bibitem[1971Sp12]{1971Sp12} S.W. Sprague, R.G. Arns, B.J. Brunner {\it et al.},     \newblock {\sc Phys. Rev.} {\bf C4}, 2074 (1971).
\bibitem[1971Sw07]{1971Sw07} C.P. Swann,    \newblock {\sc Phys. Rev.} {\bf C 4}, 1489 (1971).
\bibitem[1971Vi01]{1971Vi01} D. Vitoux, R.C. Haight, J.X. Saladin,    \newblock {\sc Phys. Rev.} {\bf C 3}, 718 (1971).
\bibitem[1971WaZP]{1971WaZP} D. Ward, A. Christy, J.S. Geiger, R.L. Graham,     \newblock {\sc AECL-3912}, 17 (1971).
\bibitem[1971Yo02]{1971Yo02} D.H. Youngblood, R.L. Kozub, J.C. Hill,     \newblock {\sc Nucl. Phys.} {\bf A 166}, 198 (1971).
\bibitem[1972ArZD]{1972ArZD} R.G. Arnold,     \newblock {\sc Thesis}, Univ. Boston (1972); {\sc Diss. Abst. Int.} {\bf 33B}, 1723 (1972).
\bibitem[1972Aw04]{1972Aw04} Z. Awwad, O.E. Badawy, M.R. El-Aasser, A.H. El-Farrash,     \newblock {\sc Indian J. Pure Appl. Phys.} {\bf 10}, 870 (1972).
\bibitem[1972Ba93]{1972Ba93} S.I. Baker, C.R. Gossett, P.A. Treado {\it et al.},    \newblock {\sc Nucl. Phys.} {\bf A 196}, 197 (1972).
\bibitem[1972Be53]{1972Be53} I. Berkes, R. Rougny, M. Meyer-Levy {\it et al.},     \newblock {\sc Phys. Rev.} {\bf C 6}, 1098 (1972).
\bibitem[1972Be66]{1972Be66} Z. Berant, R.A. Eisenstein, Y. Horowitz {\it et al.},     \newblock {\sc Nucl. Phys.} {\bf A 196}, 312 (1972).
\bibitem[1972BeVM]{1972BeVM} R.B. Begzhanov, T. Boranov, D.A. Gladyshev {\it et al.},     \newblock {\sc Program and Theses, Proc. 22nd Ann. Conf. Nucl. Spectrosc. Struct. At. Nuclei}, Kiev, 128 (1972).
\bibitem[1972Bi17]{1972Bi17} M. Bini, P.G. Bizzeti, A.M. Bizzeti-Sona {\it et al.},     \newblock {\sc Lett. Nuovo Cim.} {\bf 5}, 913 (1972).
\bibitem[1972Bo04]{1972Bo04} B. Bochev, S.A. Karamian, T. Kutsarova {\it et al.},     \newblock {\sc Compt. Rend. Acad. Bulg. Sci.} {\bf 25}, 905 (1972).
\bibitem[1972BrYV]{1972BrYV} W. Bruckner, D. Pelte, B. Povh {\it et al.},     \newblock {\sc Contrib. Int. Conf. Nuclear Moments and Nuclear Structure}, Osaka, Japan, 107 (1972).
\bibitem[1972Ca05]{1972Ca05} J.A. Cameron, A.W. Gibb, T. Taylor, Z. Zamori,     \newblock {\sc Can. J. Phys.} {\bf 50}, 475 (1972).
\bibitem[1972Ca22]{1972Ca22} J.M.G. Caraca, R.D. Gill, A.J. Cox, H.J. Rose,    \newblock {\sc Nucl. Phys.} {\bf A 193}, 1 (1972).
\bibitem[1972ClZN]{1972ClZN} R.G. Clark,     \newblock {\sc   Thesis, Iowa State Univ.} (1972); See Also 1974ClZX.
\bibitem[1972Co13]{1972Co13} W.F. Coetzee, M.A. Meyer, D. Reitmann,    \newblock {\sc Nucl. Phys.} {\bf A 185}, 644 (1972).
\bibitem[1972CrZN]{1972CrZN} H. Crannell, P.L. Hallowell, J.T. O'Brien {\it et al.},    \newblock {\sc Proc. Int. Conf. Nucl. Struct. Studies Using Electron Scattering and Photoreaction}, Sendai, Japan, K.Shoda, H.Ui, Eds., Res.Rep.Lab.Nucl.Sci.Tohoku Univ. 5, Suppl. p.375 (1972).
\bibitem[1972Du05]{1972Du05} J.L. Durell, P.R. Alderson, D.C. Bailey {\it et al.},    \newblock {\sc J. Phys. (London)} {\bf A 5}, 302 (1972).
\bibitem[1972El08]{1972El08} T.W. Elze, J.R. Huizenga,     \newblock {\sc Nucl. Phys.} {\bf A 187}, 545 (1972).
\bibitem[1972El20]{1972El20} M.R. El-Aasser, O.E. Badawy, Z. Awwad, A.H. El-Farrash,     \newblock {\sc Z. Naturforsch.} {\bf 27a}, 1229 (1972).
\bibitem[1972Er04]{1972Er04} K.A. Erb, J.E. Holden, I.Y. Lee {\it et al.},     \newblock {\sc Phys. Rev. Lett.} {\bf 29}, 1010 (1972).
\bibitem[1972Fi12]{1972Fi12} M. Finger, R. Foucher, J.P. Husson {\it et al.},     \newblock {\sc Nucl. Phys.} {\bf A 188}, 369 (1972).
\bibitem[1972Ga05]{1972Ga05} A. Gallmann, F. Haas, M. Toulemonde,    \newblock {\sc Can. J. Phys.} {\bf 50}, 278 (1972).
\bibitem[1972Gr04]{1972Gr04} H. Gruppelaar, P.J.M. Smulders,    \newblock {\sc Nucl. Phys.} {\bf A 179}, 737 (1972).
\bibitem[1972Gr05]{1972Gr05} M.I. Green, P.F. Kenealy, G.B. Beard,     \newblock {\sc Nucl. Instrum. Methods} {\bf 99}, 445 (1972).
\bibitem[1972GrYQ]{1972GrYQ} J.S. Greenberg, A.H. Shaw,     \newblock {\sc Contrib. Int. Conf. Nuclear Moments and Nuclear Structure}, Osaka, Japan, 113 (1972).
\bibitem[1972Gu03]{1972Gu03} D.K. Gupta, G.N. Rao,     \newblock {\sc Nucl. Phys.} {\bf A 182}, 669 (1972).
\bibitem[1972Ha59]{1972Ha59} O. Hausser, F.C. Khanna, D. Ward,     \newblock {\sc Nucl. Phys.} {\bf A 194}, 113 (1972).
\bibitem[1972HaYA]{1972HaYA} O. Hausser, A.J. Ferguson, D.L. Disdier,    \newblock {\sc AECL-4205}, 10 (1972).
\bibitem[1972Ho40]{1972Ho40} G.A. Hokken, J.A.J.G. Hendricx, J. de Kogel,    \newblock {\sc Nucl. Phys.} {\bf A 194}, 481 (1972).
\bibitem[1972HuZL]{1972HuZL} J.P. Husson, \newblock {\sc  Thesis}, Univ.Paris (1972); {\sc FRNC-TH-355} (1972).
\bibitem[1972Ka22]{1972Ka22} D.K. Kaipov, Y.G. Kosyak, L.N. Smirin, Y.K. Shubnyi,     \newblock {\sc Izv. Akad. Nauk SSSR}, Ser.Fiz. {\bf 36}, 137 (1972); {\sc Bull. Acad. Sci. USSR}, Phys.Ser. {\bf 36}, 128 (1973).
\bibitem[1972KaXR]{1972KaXR} D.K. Kaipov, Y.G. Kosyak, D.N. Smirin, Y.K. Shubny,    \newblock {\sc Program and Theses, Proc. 22nd Ann. Conf. Nucl. Spectrosc. Struct. At. Nuclei}, Kiev, 48 (1972).
\bibitem[1972Ke16]{1972Ke16} J.R. Kerns, J.X. Saladin,     \newblock {\sc Phys. Rev.} {\bf C 6}, 1016 (1972).
\bibitem[1972Ku14]{1972Ku14} W. Kutschera, W. Dehnhardt, O.C. Kistner {\it et al.},     \newblock {\sc Phys. Rev.} {\bf C 5}, 1658 (1972).
\bibitem[1972La16]{1972La16} S.A. Lane, J.X. Saladin,     \newblock {\sc Phys. Rev.} {\bf C 6}, 613 (1972).
\bibitem[1972Le19]{1972Le19} P.M.S. Lesser, D. Cline, P. Goode, R.N. Horoshko,     \newblock {\sc Nucl. Phys.} {\bf A190}, 597 (1972).
\bibitem[1972LeYB]{1972LeYB} J.R. Legg,     \newblock {\sc Thesis}, Yale Univ. (1972).
\bibitem[1972Li28]{1972Li28} A.S. Litvinenko, N.G. Shevchenko, O.Y. Buki {\it et al.},     \newblock {\sc Ukr. Fiz. Zh.} {\bf 17}, 1197 (1972).
\bibitem[1972Lo01]{1972Lo01} P.C. Lopiparo, R.L. Rasera, M.E. Caspari,     \newblock {\sc Nucl. Phys.} {\bf A 178}, 577 (1972).
\bibitem[1972Me04]{1972Me04} F.R. Metzger,     \newblock {\sc Nucl. Phys.} {\bf A 182}, 213 (1972).
\bibitem[1972Me09]{1972Me09} M.A. Meyer, J.P. Reinecke, D. Reitmann,    \newblock {\sc Nucl. Phys.} {\bf A 185}, 625 (1972); {\sc Erratum Nucl. Phys.} {\bf A 196}, 635 (1972).
\bibitem[1972Mo31]{1972Mo31} J.M. Moss, D.L. Hendrie, C. Glashausser, J. Thirion,     \newblock {\sc Nucl. Phys.} {\bf A194}, 12 (1972).
\bibitem[1972Na06]{1972Na06} A. Nakada, Y. Torizuka,    \newblock {\sc J. Phys. Soc. Jap.} {\bf 32}, 1 (1972).
\bibitem[1972Pr18]{1972Pr18} J.G. Pronko, R.E.McDonald,    \newblock {\sc Phys. Rev.} {\bf C 6}, 2065 (1972).
\bibitem[1972Ra14]{1972Ra14} S. Raman, R.L. Auble, W.T. Milner {\it et al.},     \newblock {\sc AECL-4314}, 9 (1972).
\bibitem[1972Ro20]{1972Ro20} C. Rolfs, R. Kraemer, F. Riess, E. Kuhlmann,     \newblock {\sc Nucl. Phys.} {\bf A 191}, 209 (1972).
\bibitem[1972Sa09]{1972Sa09} E.A. Samworth, J.W. Olness,    \newblock {\sc Phys. Rev.} {\bf C 5}, 1238 (1972).
\bibitem[1972Sa27]{1972Sa27} G.C. Salzman, A. Goswami, D.K. McDaniels,     \newblock {\sc Nucl. Phys.} {\bf A 192}, 312 (1972).
\bibitem[1972Sa42]{1972Sa42} T.K. Saylor, J.X. Saladin, I.Y. Lee, K.A. Erb,     \newblock {\sc Phys. Lett.} {\bf B 42}, 51 (1972).
\bibitem[1972Sh38]{1972Sh38} Y.K. Shubnyi, Y.A. Lysikov,     \newblock {\sc Izv. Akad. Nauk SSSR}, Ser. Fiz. {\bf 36}, 2531 (1972); {\sc Bull. Acad. Sci. USSR}, Phys. Ser. {\bf 36}, 2199 (1973).
\bibitem[1972Si01]{1972Si01} C.H. Sinex, R.S. Cox, C.M. Class,    \newblock {\sc Nucl. Phys.} {\bf A 178}, 612 (1972).
\bibitem[1972SiZI]{1972SiZI} S.H. Sie, R.L. Graham, J.S. Geiger {\it et al.},     \newblock {\sc AECL-4205}, 12 (1972).
\bibitem[1972SiZP]{1972SiZP} S.H. Sie, J.S. Geiger, R.L. Graham {\it et al.},     \newblock {\sc AECL-4147}, 14 (1972).
\bibitem[1972Sn01]{1972Sn01} F.D. Snyder,     \newblock {\sc Phys. Rev.} {\bf C 6}, 204 (1972).
\bibitem[1972Sz05]{1972Sz05} H. Sztark, J.L. Quebert, P. Gil, L. Marquez,     \newblock {\sc J. Phys. (Paris)} {\bf 33}, 841 (1972).
\bibitem[1972WaYZ]{1972WaYZ} D. Ward, I.M. Szoghy, J.S. Forster, W.G. Davies,     \newblock {\sc AECL-4314}, 9 (1972).
\bibitem[1972Yo01]{1972Yo01} D.H. Youngblood, R.L. Kozub, J.C. Hill,     \newblock {\sc Nucl. Phys.} {\bf A183}, 197 (1972).
\bibitem[1973An01]{1973An01} N. Anyas-Weiss, R. Griffiths, N.A. Jelley {\it et al.},    \newblock {\sc Nucl. Phys.} {\bf A 201}, 513 (1973).
\bibitem[1973Ba02]{1973Ba02} T.T. Bardin, J.A. Becker, T.R. Fisher,    \newblock {\sc Phys. Rev.} {\bf C 7}, 190 (1973).
\bibitem[1973Be40]{1973Be40} C.E. Bemis, Jr., P.H. Stelson, F.K. McGowan {\it et al.},     \newblock {\sc Phys. Rev.} {\bf C 8}, 1934 (1973).
\bibitem[1973Be44]{1973Be44} C.E. Bemis, Jr., F.K. McGowan, J.L.C. Ford {\it et al.},     \newblock {\sc Phys. Rev.} {\bf C 8}, 1466 (1973).
\bibitem[1973Be49]{1973Be49} J.C. Bergstrom, I.P. Auer, F.J. Kline, H.S. Caplan    \newblock {\sc Nucl. Phys.} {\bf A 213}, 609 (1973).
\bibitem[1973BeYD]{1973BeYD} W. Beens,     \newblock {\sc Thesis}, Vrije Univ., Amsterdam (1973).
\bibitem[1973Br02]{1973Br02} W. Bruckner, J.G. Merdinger, D. Pelte  {\it et al.},     \newblock {\sc Phys. Rev. Lett.} {\bf 30}, 57 (1973).
\bibitem[1973Br33]{1973Br33} C. Broude, F.A. Beck, P. Engelstein,     \newblock {\sc Nucl. Phys.} {\bf A 216}, 603 (1973).
\bibitem[1973Ca29]{1973Ca29} A.B. Carpenter, N. Hershkowitz,     \newblock {\sc Phys. Rev.} {\bf C 8}, 2302 (1973).
\bibitem[1973Ch13]{1973Ch13} P.R. Christensen, I. Chernov, E.E. Gross {\it et al.},     \newblock {\sc Nucl. Phys.} {\bf A 207}, 433 (1973).
\bibitem[1973Ch26]{1973Ch26} R.C. Chopra, P.N. Tandon, S.H. Devare, H.G. Devare,     \newblock {\sc Nucl. Phys.} {\bf A 209}, 461 (1973).
\bibitem[1973Ch28]{1973Ch28} A. Charvet, R. Chery, R. Duffait {\it et al.},     \newblock {\sc Nucl. Phys.} {\bf A 213}, 117 (1973).
\bibitem[1973ClZF]{1973ClZF} D. Cline, P. Jennens, C.W. Towsley, H.S. Gertzman,     \newblock {\sc Proc. Int. Conf. Nucl. Moments and Nucl. Struct.}, Osaka, Japan (1972), H.Horie, K.Sugimoto, Eds., 443 (1973); {\sc J. Phys. Soc. Japan} {\bf 34} Suppl. (1973).
\bibitem[1973Co38]{1973Co38} A.J. Cox, J.M.G. Caraca, B. Schlenk {\it et al.},    \newblock {\sc Nucl. Phys.} {\bf A 217}, 400 (1973).
\bibitem[1973De09]{1973De09} W. Dehnhardt, O.C. Kistner, W. Kutschera, H.J. Sann,     \newblock {\sc Phys. Rev.} {\bf C7}, 1471 (1973).
\bibitem[1973Di04]{1973Di04} W.R. Dixon, R.S. Storey, J.J. Simpson,    \newblock {\sc Nucl. Phys.} {\bf A 202}, 579 (1973).
\bibitem[1973DoZB]{1973DoZB} R. Doerr, E. Obst, F. Rauch {\it et al.},     \newblock {\sc IKF-32}, 22 (1973).
\bibitem[1973El06]{1973El06} C. Ellegaard, P.D. Barnes, R. Eisenstein {\it et al.},     \newblock {\sc Nucl. Phys.} {\bf A 206}, 83 (1973).
\bibitem[1973Fi03]{1973Fi03} T.R. Fisher, T.T. Bardin, J.A. Becker {\it et al.},    \newblock {\sc Phys. Rev.} {\bf C 7}, 1878 (1973).
\bibitem[1973Fi15]{1973Fi15} T.R. Fisher, P.D. Bond,     \newblock {\sc Part. Nucl.} {\bf 6}, 119 (1973).
\bibitem[1973FrZN]{1973FrZN} I. Fraser, J.S. Greenberg,     \newblock {\sc Priv. Comm.} (September 1973).
\bibitem[1973Gr05]{1973Gr05} J.S. Greenberg,     \newblock {\sc Priv. Comm.} (March 1973).
\bibitem[1973GrXX]{1973GrXX} M.I. Green,     \newblock {\sc Thesis}, Wayne State Univ. (1973); Diss.Abstr.Int. {\bf 34B}, 2845 (1973).
\bibitem[1973Ha07]{1973Ha07} T. Hammer, H. Ejiri, G.B. Hagemann,     \newblock {\sc Nucl. Phys.} {\bf A 202}, 321 (1973).
\bibitem[1973Ha10]{1973Ha10} R. Hartmann, H. Grawe, K. Kandler,    \newblock {\sc Nucl. Phys.} {\bf A 203}, 401 (1973).
\bibitem[1973Ha13]{1973Ha13} P.L. Hallowell, W. Bertozzi, J. Heisenberg {\it et al.},    \newblock {\sc Phys. Rev.} {\bf C 7}, 1396 (1973).
\bibitem[1973Ho05]{1973Ho05} K. Hosoyama, Y. Torizuka, Y. Kawazoe, H. Ui,     \newblock {\sc Phys. Rev. Lett.} {\bf 30}, 388 (1973).
\bibitem[1973KaZV]{1973KaZV} D.K. Kaipov, Y.A. Lysikov, Y.K. Shubnyi,     \newblock {\sc Program and Theses, Proc. 23rd Ann. Conf. Nucl. Spectrosc. Struct. At. Nuclei}, Tbilisi, 46 (1973).
\bibitem[1973Ku10]{1973Ku10} W. Kutschera, R.B. Huber, C. Signorini, P. Blasi,     \newblock {\sc Nucl. Phys.} {\bf A210}, 531 (1973).
\bibitem[1973Ku15]{1973Ku15} E. Kuhlmann, W. Albrecht, A. Hofmann,    \newblock {\sc Nucl. Phys.} {\bf A 213}, 82 (1973).
\bibitem[1973Le15]{1973Le15} F. Leccia, M.M. Aleonard, D. Castera {\it et al.},    \newblock {\sc J.Phys.(Paris)} {\bf 34}, 147 (1973).
\bibitem[1973Le17]{1973Le17} E.W. Lees, A. Johnston, S.W. Brain {\it et al.},    \newblock {\sc J. Phys. (London)} {\bf A 6}, L116 (1973).
\bibitem[1973Li24]{1973Li24} A.S. Litvinenko, N.G. Shevchenko, N.G. Afanasev {\it et al.},     \newblock {\sc Yad. Fiz.} {\bf 18}, 250 (1973); {\sc Sov. J. Nucl. Phys.} {\bf 18}, 128 (1974).
\bibitem[1973Mc16]{1973Mc16} J.D. McCullen, D.J. Donahue,    \newblock {\sc Phys. Rev.} {\bf C 8}, 1406 (1973).
\bibitem[1973Ol02]{1973Ol02} J.W. Olness, E.K. Warburton, J.A. Becker,    \newblock {\sc Phys. Rev.} {\bf C 7}, 2239 (1973).
\bibitem[1973Pi04]{1973Pi04} R. Pitthan,     \newblock {\sc Z. Phys.} {\bf 260}, 283 (1973).
\bibitem[1973RaWV]{1973RaWV} F. Rauch,     \newblock {\sc Priv. Comm.}, quoted by 1973BeYD (1973).
\bibitem[1973Ro20]{1973Ro20} H. Ronsin, P. Beuzit, J. Delaunay {\it et al.},    \newblock {\sc  Nucl. Phys.} {\bf A 207}, 577 (1973).
\bibitem[1973Ru08]{1973Ru08} N. Rud, D. Ward, H.R. Andrews {\it et al.},     \newblock {\sc Phys. Rev. Lett.} {\bf 31}, 1421 (1973).
\bibitem[1973Sc28]{1973Sc28} N. Schulz, J. Chevallier, B. Haas {\it et al.},     \newblock {\sc Phys. Rev.} {\bf C 8}, 1779 (1973).
\bibitem[1973ScWZ]{1973ScWZ} D. Schwalm,     \newblock {\sc Thesis}, Univ.Heidelberg (1973).
\bibitem[1973Si31]{1973Si31} R.P. Singhal, H.S. Caplan, J.R. Moreira, T.E. Drake,    \newblock {\sc Can. J. Phys.} {\bf 51}, 2125 (1973).
\bibitem[1973Sm01]{1973Sm01} G.J. Smith, P.C. Simms,     \newblock {\sc Nucl. Phys.} {\bf A 202}, 409 (1973).
\bibitem[1973To07]{1973To07} C.W. Towsley, D. Cline, R.N. Horoshko,    \newblock {\sc Nucl. Phys.} {\bf A 204}, 574 (1973).
\bibitem[1973ToXW]{1973ToXW} C.W. Towsley, R. Cook, D. Cline, R.N. Horoshko,     \newblock {\sc Proc. Int. Conf. Nucl. Moments and Nucl. Struct.}, Osaka, Japan (1972), H.Horie, K.Sugimoto, Eds., 442 (1973); {\sc J. Phys. Soc. Jap.} {\bf 34} Suppl. (1973).
\bibitem[1973WeYO]{1973WeYO} D. Werdecker, A.M. Kleinfeld, J.S. Greenberg,     \newblock {\sc Proc. Int. Conf. Nucl. Moments and Nucl. Struct.}, Osaka, Japan (1972), H.Horie, K.Sugimoto, Eds., 195 (1973); {\sc J. Phys. Soc. Jap.} 34 Suppl. (1973).
\bibitem[1974Aw03]{1974Aw03} Z. Awwad, A. Abdel-Haliem, M.R. El-Aasser,     \newblock {\sc Acta Phys.} {\bf 37}, 141 (1974).
\bibitem[1974Ba43]{1974Ba43} C. Baktash, J.X. Saladin,     \newblock {\sc Phys. Rev.} {\bf C 10}, 1136 (1974).
\bibitem[1974Ba45]{1974Ba45} J. Barrette, M. Barrette, R. Haroutunian {\it et al.},     \newblock {\sc Phys. Rev.} {\bf C 10}, 1166 (1974).
\bibitem[1974Ba80]{1974Ba80} J. Barrette, M. Barrette, G. Lamoureux {\it et al.},     \newblock {\sc Nucl. Phys.} {\bf A 235}, 154 (1974).
\bibitem[1974Be18]{1974Be18} R.A.I. Bell, J.V. Thompson, I.G. Graham, L.E. Carlson,    \newblock {\sc Nucl. Phys.} {\bf A 222}, 477 (1974).
\bibitem[1974Be25]{1974Be25} Z. Berant, C. Broude, G. Engler, D.F.H. Start,    \newblock {\sc Nucl. Phys.} {\bf A 225}, 55 (1974).
\bibitem[1974Br04]{1974Br04} B.A. Brown, D.B. Fossan, J.M. McDonald, K.A. Snover,     \newblock {\sc Phys. Rev.} {\bf C9}, 1033 (1974).
\bibitem[1974Br31]{1974Br31} W. Bruckner, D. Husar, D. Pelte  {\it et al.},     \newblock {\sc Nucl. Phys.} {\bf A 231}, 159 (1974).
\bibitem[1974Bu13]{1974Bu13} J. Burde, S. Eshhar, A. Ginzburg, E. Navon,     \newblock {\sc Nucl. Phys.} {\bf A 229}, 387 (1974).
\bibitem[1974Ch09]{1974Ch09} Y.T. Cheng, A. Goswami, M.J. Throop, D.K. McDaniels,    \newblock {\sc Phys. Rev.} {\bf C 9}, 1192 (1974).
\bibitem[1974De12]{1974De12} W. Dehnhardt, S.J. Mills, M. Muller-Veggian {\it et al.},     \newblock {\sc Nucl. Phys.} {\bf A 225}, 1 (1974).
\bibitem[1974El03]{1974El03} M.R. El-Aasser, Z. Awwad,     \newblock {\sc ATOMKI Kozlem.} {\bf 16}, 141 (1974).
\bibitem[1974Fi01]{1974Fi01} T.R. Fisher, T.T. Bardin, J.A. Becker, B.A. Watson,    \newblock {\sc Phys. Rev.} {\bf C 9}, 598 (1974).
\bibitem[1974Fi05]{1974Fi05} S.P. Fivozinsky, S. Penner, J.W. Lightbody, Jr., D. Blum,     \newblock {\sc Phys. Rev.} {\bf C 9}, 1533 (1974).
\bibitem[1974Fo13]{1974Fo13} J.S. Forster, D. Ward, G.J. Costa {\it et al.},    \newblock {\sc Phys. Lett.} {\bf B 51}, 133 (1974).
\bibitem[1974Gr06]{1974Gr06} H. Grawe, R. Konig,    \newblock {\sc Z. Phys.} {\bf 266}, 41 (1974).
\bibitem[1974Gr19]{1974Gr19} H. Grawe, U. Lohle, R. Konig,    \newblock {\sc Z. Phys.} {\bf 268}, 419 (1974).
\bibitem[1974Gu11]{1974Gu11} G. Guillaume, B. Rastegar, P. Fintz, A. Gallmann,    \newblock {\sc Nucl. Phys.} {\bf A 227}, 284 (1974).
\bibitem[1974Iv01]{1974Iv01} M. Ivascu, D. Popescu, E. Dragulescu {\it et al.},     \newblock {\sc Nucl. Phys.} {\bf A 218}, 104 (1974).
\bibitem[1974JaZN]{1974JaZN} R.C. Jared, H. Nifenecker, S.G. Thompson,  LBL-2366, 38 (1974).
\bibitem[1974Jo02]{1974Jo02} P.M. Johnson, M.A. Meyer, D. Reitmann,    \newblock {\sc Nucl. Phys.} {\bf A 218}, 333 (1974).
\bibitem[1974Jo10]{1974Jo10} A. Johnston, T.E. Drake,    \newblock {\sc J. Phys. (London)} {\bf A 7}, 898 (1974).
\bibitem[1974Le13]{1974Le13} P.M.S. Lesser, D. Cline, C. Kalbach-Cline, A. Bahnsen, \newblock {\sc  Nucl. Phys.} {\bf A 223}, 563 (1974).
\bibitem[1974Le17]{1974Le17} E.W. Lees, A. Johnston, S.W. Brain {\it et al.},    \newblock {\sc J. Phys. (London)} {\bf A 7}, 936 (1974).
\bibitem[1974MaYP]{1974MaYP} R. Maas, C.W. de Jager,     \newblock {\sc Proc. Int. Conf. Nucl. Struct. and Spectrosc.}, Amsterdam, H.P.Blok, A.E.L.Dieperink, Eds., Scholar's Press, Amsterdam {\bf 1}, 204 (1974).
\bibitem[1974Mc17]{1974Mc17} A.B. McDonald, T.K. Alexander, O. Hausser {\it et al.},    \newblock {\sc Can. J. Phys.} {\bf 52}, 1381 (1974).
\bibitem[1974Me13]{1974Me13} F.R. Metzger,     \newblock {\sc Phys. Rev.} {\bf C 9}, 1525 (1974).
\bibitem[1974No08]{1974No08} E. Nolte, Y. Shida, W. Kutschera {\it et al.}, \newblock {\sc Z. Phys.} {\bf 268}, 267 (1974).
\bibitem[1974Ol01]{1974Ol01} D.K. Olsen, A.R. Barnett, S.F. Biagi {\it et al.}, \newblock {\sc  Nucl.Phys.} {\bf A 220}, 541 (1974).
\bibitem[1974Ol02]{1974Ol02} A. Olin, O. Hausser, T.K. Alexander {\it et al.}, \newblock {\sc Nucl. Phys.} {\bf A 221}, 555 (1974).
\bibitem[1974Po15]{1974Po15} A.R. Poletti, B.A. Brown, D.B. Fossan, E.K. Warburton, \newblock {\sc Phys. Rev.} {\bf C10}, 2329 (1974).
\bibitem[1974Pr02]{1974Pr02} D. Proetel, R.M. Diamond, F.S. Stephens,  \newblock {\sc Phys. Lett.} {\bf B 48}, 102 (1974).
\bibitem[1974Pr04]{1974Pr04} J.G. Pronko, T.T. Bardin, J.A. Becker {\it et al.},  \newblock {\sc Phys. Rev.} {\bf C 9}, 1430 (1974); {\sc Erratum Phys. Rev.} {\bf C 10}, 1249 (1974).
\bibitem[1974Ra15]{1974Ra15} B. Rastegar, G. Guillaume, P. Fintz, A. Gallmann,  \newblock {\sc Nucl. Phys.} {\bf A 225}, 80 (1974).
\bibitem[1974SaZH]{1974SaZH} R.O. Sayer, R.L. Robinson, N.C. Singhal {\it et al.},     \newblock {\sc ORNL-4937}, 117 (1974).
\bibitem[1974Sh12]{1974Sh12} A.H. Shaw, J.S. Greenberg,     \newblock {\sc Phys. Rev.} {\bf C 10}, 263 (1974).
\bibitem[1974Si01]{1974Si01} R.P. Singhal, S.W. Brain, W.A. Gillespie {\it et al.},     \newblock {\sc Nucl. Phys.} {\bf A 218}, 189 (1974).
\bibitem[1974Sw05]{1974Sw05} C.P. Swann,     \newblock {\sc J. Franklin Inst.} {\bf 298}, 321 (1974).
\bibitem[1974ThZG]{1974ThZG} T.F. Thorsteinsen, F. Videbaek,     \newblock {\sc BUP-65} (1974).
\bibitem[1974ToZJ]{1974ToZJ} C.W. Towsley,     \newblock {\sc Thesis}, Univ.Rochester (1974); Diss.Abstr.Int. {\bf 35B}, 1864 (1974).
\bibitem[1974Wa04]{1974Wa04} B.A. Watson, J.A. Becker, T.R. Fisher,     \newblock {\sc Phys. Rev.} {\bf C9}, 1200 (1974).
\bibitem[1974Wo01]{1974Wo01} H.J. Wollersheim, W. Wilcke, T.W. Elze, D. Pelte,     \newblock {\sc Phys. Lett.} {\bf B 48}, 323 (1974).
\bibitem[1974Ye01]{1974Ye01} R. Yen, L.S. Cardman, D. Kalinsky {\it et al.},     \newblock {\sc Nucl. Phys.} {\bf A 235}, 135 (1974).
\bibitem[1975Ba72]{1975Ba72} P. Barreau, P. Bertin, B. Bihoreau  {\it et al.},     \newblock {\sc J. Phys. (Paris)}, Colloq. {\bf 36}, C5-77 (1975).
\bibitem[1975Be15]{1975Be15} Z. Berant, C. Broude, G. Engler {\it et al.},    \newblock {\sc Nucl. Phys.} {\bf A 243}, 519 (1975).
\bibitem[1975Bi03]{1975Bi03} S.F. Biagi, W.R. Phillips, A.R. Barnett,    \newblock {\sc Nucl. Phys.} {\bf A 242}, 160 (1975).
\bibitem[1975Bo39]{1975Bo39} H. Bohn, P. Kienle, D. Proetel, R.L. Hershberger,     \newblock {\sc Z. Phys.} {\bf A 274}, 327 (1975).
\bibitem[1975Bu08]{1975Bu08} P.A. Butler, J. Meyer-Ter-Vehn, D. Ward {\it et al.},     \newblock {\sc Phys. Lett.} {\bf B 56}, 453 (1975).
\bibitem[1975DeXW]{1975DeXW} J.E.P. de Bie, C.W. de Jager, A.A.C. Klaasse {\it et al.},     \newblock {\sc IKO Progr. Rept. 1975}, 3 (1975).
\bibitem[1975Eb01]{1975Eb01} J.L. Eberhardt, R.E. Horstman, H.A. Doubt, G. Van Middelkoop,    \newblock {\sc Nucl. Phys.} {\bf A 244}, 1 (1975).
\bibitem[1975EdZY]{1975EdZY} L.-O. Edvardson, L.O. Norlin,     \newblock {\sc Priv. Comm.} (March 1975); {\sc UUIP-895} (1975).
\bibitem[1975Fr04]{1975Fr04} H.-G. Friederichs, A. Gelberg, B. Heits {\it et al.},     \newblock {\sc Phys. Rev. Lett.} {\bf 34}, 745 (1975).
\bibitem[1975Go18]{1975Go18} D.M. Gordon, L.S. Eytel, H. de Waard, D.E. Murnick,     \newblock {\sc Phys. Rev.} {\bf C 12}, 628 (1975).
\bibitem[1975Gr04]{1975Gr04} H. Grawe, K. Holzer, K. Kandler, A.A. Pilt,    \newblock {\sc Nucl. Phys.} {\bf A 237}, 18 (1975).
\bibitem[1975Gr30]{1975Gr30} R. Graetzer, S.M. Cohick, J.X. Saladin,     \newblock {\sc Phys. Rev.} {\bf C 12}, 1462 (1975).
\bibitem[1975GuYV]{1975GuYV} G.M. Gusinskii, M.A. Ivanov, A.S. Mishin,     \newblock {\sc Program and Theses, Proc. 25th Ann. Conf. Nucl. Spectrosc. Struct. At. Nuclei}, Leningrad, 380 (1975).
\bibitem[1975GuYW]{1975GuYW} G.M. Gusinskii, M.A. Ivanov, M.K. Kudoyarov {\it et al.},     \newblock {\sc Program and Theses, Proc. 25th Ann. Conf. Nucl. Spectrosc. Struct. At. Nuclei}, Leningrad, 379 (1975).
\bibitem[1975Ha04]{1975Ha04} B. Haas, P. Taras, J.C. Merdinger, R. Vaillancourt,     \newblock {\sc Nucl. Phys.} {\bf A238}, 253 (1975).
\bibitem[1975HaYU]{1975HaYU} E.C. Hagen,    \newblock {\sc Thesis}, Duke Univ. (1974); {\sc Diss. Abstr. Int.} {\sc 35B}, 4101 (1975).
\bibitem[1975He25]{1975He25} J.A.J. Hermans, G.A.P. Engelbertink, M.A. van Driel {\it et al.},    \newblock {\sc Nucl. Phys.} {\bf A 255}, 221 (1975).
\bibitem[1975Ho15]{1975Ho15} R.E. Horstman, J.L. Eberhardt, H.A. Doubt {\it et al.},    \newblock {\sc Nucl. Phys.} {\bf A 248}, 291 (1975).
\bibitem[1975JaYL]{1975JaYL} R.C. Jared, H. Nifenecker, E. Cheifetz {\it et al.},     \newblock {\sc Priv. Comm.} (1975).
\bibitem[1975Ka11]{1975Ka11} H. Karwowski, S. Majewski, B. Pietrzyk {\it et al.},     \newblock {\sc J. Phys. (Paris)} {\bf 36}, 471 (1975).
\bibitem[1975Kl07]{1975Kl07} A.M. Kleinfeld, H.G. Maggi, D. Werdecker,     \newblock {\sc Nucl. Phys.} {\bf A 248}, 342 (1975).
\bibitem[1975Kl09]{1975Kl09} A.M. Kleinfeld, K.P. Lieb, D. Werdecker, U. Smilansky,    \newblock {\sc Phys. Rev. Lett.} {\bf 35}, 1329 (1975).
\bibitem[1975Kl10]{1975Kl10} F.J. Kline, I.P. Auer, J.C. Bergstrom, H.S. Caplan,     \newblock {\sc Nucl. Phys.} {\bf A 255}, 435 (1975).
\bibitem[1975Le22]{1975Le22} I.Y. Lee, J.X. Saladin, J. Holden {\it et al.},     \newblock {\sc Phys. Rev.} {\bf C 12}, 1483 (1975).
\bibitem[1975Lo08]{1975Lo08} K.E.G. Lobner, G. Dannhauser, D.J. Donahue {\it et al.},     \newblock {\sc Z. Phys.} {\bf A 274}, 251 (1975).
\bibitem[1975Mi08]{1975Mi08} H. Miska, H.D. Graf, A. Richter {\it et al.},    \newblock {\sc Phys. Lett.} {\bf 58 B}, 155 (1975).
\bibitem[1979Sa05]{1979Sa05} S. Salem Vasconcelos, M.N. Rao, N. Ueta, C.R. Appoloni, \newblock {\sc  Nucl. Phys.} {\bf A 313}, 333 (1979).
\bibitem[1975Si21]{1975Si21} R.P. Singhal, S.W. Brain, C.S. Curran {\it et al.},     \newblock {\sc J. Phys. (London)} {\bf G 1}, 558 (1975).
\bibitem[1975Th01]{1975Th01} M.J.Throop, Y.T.Cheng, D.K.McDaniels,     \newblock {\sc Nucl. Phys.} {\bf A239}, 333 (1975).
\bibitem[1975To06]{1975To06} C.W. Towsley, D. Cline, R.N. Horoshko,     \newblock {\sc Nucl. Phys.} {\bf A250}, 381 (1975).
\bibitem[1975Tr08]{1975Tr08} W. Trautmann, D. Proetel, O. Hausser {\it et al.},     \newblock {\sc Phys. Rev. Lett.} {\bf 35}, 1694 (1975).
\bibitem[1975Wa10]{1975Wa10} P. Wagner, M.A. Ali, J.P. Coffin, A. Gallmann,    \newblock {\sc Phys. Rev.} {\bf C 11}, 1622 (1975).
\bibitem[1975Wo08]{1975Wo08} H.J. Wollersheim, W. Wilcke, T.W. Elze,     \newblock {\sc Phys. Rev.} {\bf C 11}, 2008 (1975).
\bibitem[1976AlYY]{1976AlYY} A.A. Aleksandrov, M.A. Ivanov, V.G. Kiptilyi {\it et al.},     \newblock {\sc Program and Theses, Proc. 26th Ann. Conf. Nucl. Spectrosc. At. Nuclei}, Baku, 412 (1976).
\bibitem[1976As04]{1976As04} J. Asher, M.A. Grace, P.D. Johnston {\it et al.},    \newblock {\sc J. Phys. (London)} {\bf G2}, 477 (1976).
\bibitem[1976As07]{1976As07} J. Asher, M.A. Grace, P.D. Johnston {\it et al.},    \newblock {\sc Hyperfine Interactions} {\bf 2}, 378 (1976).
\bibitem[1976Ba06]{1976Ba06} F.T. Baker, T.H. Kruse, W. Hartwig {\it et al.},     \newblock {\sc Nucl. Phys.} {\bf A 258}, 43 (1976).
\bibitem[1976Ba23]{1976Ba23} F.T. Baker, A. Scott, T.H. Kruse {\it et al.},     \newblock {\sc Phys. Rev. Lett.} {\bf 37}, 193 (1976).
\bibitem[1976Ba35]{1976Ba35} F.T. Baker, A. Scott, T.H. Kruse {\it et al.},     \newblock {\sc Nucl. Phys.} {\bf A 266}, 337 (1976).
\bibitem[1976Be64]{1976Be64} R.B. Begzhanov, F.S. Akilov, A.K. Khalikov, M.S. Rakhmankulov,    \newblock {\sc Izv. Akad. Nauk Uzb. SSR}, Ser.Fiz.-Mat.Nauk No.4, 59 (1976).
\bibitem[1976Bo12]{1976Bo12} A. Bockisch, A.M. Kleinfeld,     \newblock {\sc Nucl. Phys.} {\bf A 261}, 498 (1976).
\bibitem[1976Bo27]{1976Bo27} B. Bochev, S.A. Karamian, T. Kutsarova {\it et al.},     \newblock {\sc Nucl. Phys.} {\bf A 267}, 344 (1976).
\bibitem[1976Bu20]{1976Bu20} D.K. Butt, M.A. Raoof, S.A. Raoof,     \newblock {\sc J. Phys. (London)} {\bf G 2}, 823 (1976).
\bibitem[1976Ch11]{1976Ch11} A. Charvet, R. Duffait, T. Negadi {\it et al.},     \newblock {\sc Phys. Rev.} {\bf C13}, 2237 (1976).
\bibitem[1976Es02]{1976Es02} M.T. Esat, D.C. Kean, R.H. Spear, A.M. Baxter,     \newblock {\sc Nucl. Phys.} {\bf A 274}, 237 (1976).
\bibitem[1976Fo12]{1976Fo12} J.S. Forster, G.C. Ball, C. Broude {\it et al.},    \newblock {\sc Phys. Rev.} {\bf C 14}, 596 (1976).
\bibitem[1976Ha01]{1976Ha01} J.H. Hamilton, H.L. Crowell, R.L. Robinson {\it et al.},    \newblock {\sc Phys. Rev. Lett.} {\bf 36}, 340 (1976).
\bibitem[1976He05]{1976He05} B. Heits, H.-G. Friederichs, A. Gelberg {\it et al.},     \newblock {\sc Phys. Lett.} {\bf B 61}, 33 (1976).
\bibitem[1976KaYY]{1976KaYY} D.K. Kaipov, Yu.A. Lysikov, L.M. Dautov, N.N. Lashkul,    \newblock {\sc  Program and Theses, Proc. 26th Ann. Conf. Nucl. Spectrosc. Struct. At. Nuclei}, Baku, 62 (1976).
\bibitem[1976Kl04]{1976Kl04} A. Kluge, W. Thomas,     \newblock {\sc Nucl. Instrum. Methods} {\bf 134}, 525 (1976).
\bibitem[1976Li19]{1976Li19} J.W. Lightbody, Jr., J.W. Lightbody, S. Penner {\it et al.},     \newblock {\sc Phys. Rev.} {\bf C14}, 952 (1976).
\bibitem[1976Mc02]{1976Mc02} A.B. McDonald, T.K. Alexander, C. Broude {\it et al.},    \newblock {\sc Nucl. Phys.} {\bf A 258}, 152 (1976).
\bibitem[1976MoZB]{1976MoZB} E. Monnand, B. Fogelberg,     \newblock {\sc Proc. Int. Conf. Nuclei Far from Stability, 3rd}, Cargese, France, R.Klapisch, Ed., {\sc CERN-76-13}, 503 (1976).
\bibitem[1976Ne06]{1976Ne06} R. Neuhausen, J.W. Lightbody, Jr., J.W. Lightbody {\it et al.},     \newblock {\sc Nucl. Phys.} {\bf A263}, 249 (1976).
\bibitem[1976Pa13]{1976Pa13} P. Paradis, G. Lamoureux, R. Lecomte, S. Monaro,     \newblock {\sc Phys. Rev.} {\bf C 14}, 835 (1976).
\bibitem[1976Ra03]{1976Ra03} V.K. Rasmussen,    \newblock {\sc Phys. Rev.} {\bf C 13}, 631 (1976); {\sc Erratum Phys. Rev.} {\bf C 13}, 2596 (1976).
\bibitem[1976So03]{1976So03} J.R. Southon, L.K. Fifield, A.R. Poletti,    \newblock {\sc J. Phys. (London)} {\bf G 2}, 117 (1976).
\bibitem[1977Al14]{1977Al14} A.A.Aleksandrov, V.S.Zvonov, M.A.Ivanov {\it et al.},     \newblock {\sc Izv. Akad. Nauk SSSR}, Ser.Fiz. {\bf 41}, 49 (1977); {\sc Bull. Acad. Sci. USSR}, Phys.Ser. {\bf 41}, No.1, 39 (1977).
\bibitem[1977Ar19]{1977Ar19} A. Arnesen, K. Johansson, E. Karlsson {\it et al.},     \newblock {\sc Hyperfine Interactions} {\bf 5}, 81 (1977).
\bibitem[1977BeYM]{1977BeYM} A.M. Bergdolt, J. Chevallier, J.C. Merdinger {\it et al.},     \newblock {\sc Proc. Int. Conf. Nucl. Structure}, Tokyo, Japan, Int. Academic Printing Co., Ltd. Japan, {\bf 1}, 359 (1977).
\bibitem[1977Bo14]{1977Bo14} B. Bochev, S. Iliev, R. Kalpakchieva {\it et al.},     \newblock {\sc Nucl. Phys.} {\bf A 282}, 159 (1977).
\bibitem[1977Br16]{1977Br16} S.W. Brain, A. Johnstons, W.A. Gillespie {\it et al.},    \newblock {\sc J. Phys. (London)} {\bf G3}, 821 (1977).
\bibitem[1977BrYO]{1977BrYO} J.F. Bruandet, Tsan Ung Chan, C. Morand {\it et al.},     \newblock {\sc Int. Symp. High-Spin States}, Nucl. Struct., Dresden, L.Funke, Ed., ZfK-336, p.119 (1977).
\bibitem[1977Ca14]{1977Ca14} Y. Cauchois, H. ben Abdelaziz, Y. Heno {\it et al.},     \newblock {\sc C.R. Acad. Sci.}, Ser. {\bf B 284}, 65 (1977).
\bibitem[1977Co10]{1977Co10} D.F. Coope, L.E. Cannell, M.K. Brussel,     \newblock {\sc Phys. Rev.} {\bf C 15}, 1977 (1977).
\bibitem[1977Di07]{1977Di07} W.R. Dixon, R.S. Storey, J.J. Simpson, \newblock {\sc  Phys. Rev.} {\bf C 15}, 1896 (1977).
\bibitem[1977Es02]{1977Es02} M.T. Esat, D.C. Kean, R.H. Spear {\it et al.},     \newblock {\sc Phys. Lett.} {\bf B 72}, 49 (1977).
\bibitem[1977Fi01]{1977Fi01} H. Fischer, D. Kamke, H.J. Kittling  {\it et al.},     \newblock {\sc Phys. Rev.} {\bf C 15}, 921 (1977) .
\bibitem[1977Fi09]{1977Fi09} J.M. Finn, H. Crannell, P.L. Hallowell {\it et al.},    \newblock {\sc Nucl. Phys.} {\bf A 290}, 99 (1977).
\bibitem[1977Fl10]{1977Fl10} C. Flaum, J. Barrette, M.J. LeVine, C.E. Thorn,    \newblock {\sc Phys. Rev. Lett.} {\bf 39}, 446 (1977).
\bibitem[1977Ga06]{1977Ga06} M.G. Gavrilov, A.B. Davydov, M.M. Korotkov,     \newblock {\sc Yad. Fiz.} {\bf 25}, 240 (1977); {\sc Sov. J. Nucl. Phys.} {\bf 25}, 131 (1977).
\bibitem[1977Gi13]{1977Gi13} W.A. Gillespie, M.W.S. Macauley, A. Johnston {\it et al.},     \newblock {\sc J. Phys. (London)} {\bf G 3}, L-169 (1977).
\bibitem[1977Gu08]{1977Gu08} G.M. Gusinskii, M.A. Ivanov, I.K. Lemberg {\it et al.},     \newblock {\sc Izv. Akad. Nauk SSSR}, Ser. Fiz. {\bf 41}, 66 (1977); {\sc Bull. Acad. Sci. USSR}, Phys. Ser. {\bf 41}, No.1, 52 (1977).
\bibitem[1977He12]{1977He12} J.A.J. Hermans, G.A.P. Engelbertink, L.P. Ekstrom {\it et al.},    \newblock {\sc Nucl. Phys.} {\bf A 284}, 307 (1977).
\bibitem[1977Ho01]{1977Ho01} R.E. Horstman, J.L. Eberhardt, P.C. Zalm {\it et al.},    \newblock {\sc Nucl. Phys.} {\bf A 275}, 237 (1977).
\bibitem[1977Ho10]{1977Ho10} M. Hoshi, Y. Yoshizawa,     \newblock {\sc J. Phys. Soc. Jap.} {\bf 42}, 1106 (1977).
\bibitem[1977HoZF]{1977HoZF} R. Hofmann, H.G. Andresen, B. Dreher {\it et al.},     \newblock {\sc Proc. Int. Conf. Nucl. Structure}, Tokyo, Japan, Int. Academic Printing Co., Ltd.Japan, {\bf 1}, 387 (1977).
\bibitem[1977Hu10]{1977Hu10} D. Husar, S.J. Mills, H. Graf {\it et al.},     \newblock {\sc Nucl. Phys.} {\bf A 292}, 267 (1977).
\bibitem[1977Jo05]{1977Jo05} N.R. Johnson, P.P. Hubert, E. Eichler {\it et al.},     \newblock {\sc Phys. Rev.} {\bf C 15}, 1325 (1977).
\bibitem[1977Kl05]{1977Kl05} A.M. Kleinfeld, A. Bockisch, K.P. Lieb,     \newblock {\sc Nucl. Phys.} {\bf A 283}, 526 (1977).
\bibitem[1977La15]{1977La15} H. Lancman, A.P.M. van't Westende, H.D. Graber,    \newblock {\sc Nucl. Phys.} {\bf A 291}, 293 (1977).
\bibitem[1977La16]{1977La16} J. Lange, A.T. Kandil, J. Neuber {\it et al.},     \newblock {\sc Nucl. Phys.} {\bf A 292}, 301 (1977).
\bibitem[1977Le11]{1977Le11} R. Lecomte, P. Paradis, J. Barrette {\it et al.},     \newblock {\sc Nucl. Phys.} {\bf A 284}, 123 (1977).
\bibitem[1977LiZS]{1977LiZS} J.S. Lilley, A.R. Barnett, M. Franey {\it et al.},    \newblock {\sc Bull. Amer. Phys. Soc.} {\bf 22}, No.4, 552, BJ11 (1977).
\bibitem[1977Me01]{1977Me01} F.R. Metzger,     \newblock {\sc Phys. Rev.} {\bf C 15}, 193 (1977).
\bibitem[1977Me10]{1977Me10} F.R. Metzger,     \newblock {\sc Phys. Rev.} {\bf C 15}, 2253 (1977).
\bibitem[1977MiZM]{1977MiZM} V.K. Mittal, D.K. Avasthi, I.M. Govil,    \newblock {\sc Proc. Nucl. Phys. and Solid State Phys. Symp.}, Nucl. Phys., Pune, {\bf 20 B}, p.151 (1977).
\bibitem[1977Mo20]{1977Mo20} C. Morand, J.F. Bruandet, A. Giorni, Tsan Ung Chan,     \newblock {\sc J. Phys. (Paris)} {\bf 38}, 1319 (1977).
\bibitem[1977Na01]{1977Na01} A. Nakada, N. Haik, J. Alster {\it et al.},     \newblock {\sc Phys. Rev. Lett.} {\bf 38}, 584 (1977).
\bibitem[1977Ne05]{1977Ne05} R. Neuhausen,     \newblock {\sc Nucl. Phys.} {\bf A282}, 125 (1977).
\bibitem[1977Og03]{1977Og03} M. Ogawa, E. Arai,    \newblock {\sc J. Phys. Soc. Jap.} {\bf 42}, 376 (1977).
\bibitem[1977Ra01]{1977Ra01} D.C. Radford, A.R. Poletti {\it et al.},    \newblock {\sc Nucl. Phys.} {\bf A 275}, 141 (1977).
\bibitem[1977Ro08]{1977Ro08} R.M. Ronningen, J.H. Hamilton, A.V. Ramayya {\it et al.},     \newblock {\sc Phys. Rev.} {\bf C 15}, 1671 (1977).
\bibitem[1977Ro16]{1977Ro16} R.M. Ronningen, R.B. Piercey, A.V. Ramayya {\it et al.},     \newblock {\sc Phys. Rev.} {\bf C 16}, 571 (1977).
\bibitem[1977Ro26]{1977Ro26} R.M. Ronningen, J.H. Hamilton, L. Varnell {\it et al.},     \newblock {\sc  Phys. Rev.} {\bf C 16}, 2208 (1977).
\bibitem[1977Ro27]{1977Ro27} R.M. Ronningen, R.B. Piercey, J.H. Hamilton {\it et al.},     \newblock {\sc Phys. Rev.} {\bf C 16}, 2218 (1977).
\bibitem[1977Sa04]{1977Sa04} M. Samuel, U. Smilansky, B.A. Watson {\it et al.},     \newblock {\sc Nucl. Phys.} {\bf A 279}, 210 (1977).
\bibitem[1977Sc33]{1977Sc33} R. Schormann, H. Fischer, E. Kuhlmann,     \newblock {\sc Phys. Rev.} {\bf C 16}, 2165 (1977).
\bibitem[1977Sc36]{1977Sc36} D. Schwalm, E.K. Warburton, J.W. Olness,    \newblock {\sc Nucl. Phys.} {\bf A 293}, 425 (1977).
\bibitem[1977Sw03]{1977Sw03} C.P. Swann,     \newblock {\sc Phys. Rev.} {\bf C 15}, 1967 (1977).
\bibitem[1977Vo07]{1977Vo07} P.B. Vold, D. Cline, P. Russo {\it et al.},    \newblock {\sc Phys. Rev. Lett.} {\bf 39}, 325 (1977).
\bibitem[1977Wa10]{1977Wa10} E.K.Warburton, J.W.Olness, A.M.Nathan {\it et al.},     \newblock {\sc Phys. Rev.} {\bf C16}, 1027 (1977).
\bibitem[1977Wo02]{1977Wo02} H.J. Wollersheim, T.W. Elze,     \newblock {\sc Nucl. Phys.} {\bf A 278}, 87 (1977).
\bibitem[1978Ad03]{1978Ad03} I. Adam, W. Andrejtscheff, K.Y. Gromov {\it et al.},     \newblock {\sc Nucl. Phys.} {\bf A 311}, 188 (1978).
\bibitem[1978Ar07]{1978Ar07} R.G. Arthur, R.P. Singhal, S.W. Brain {\it et al.},     \newblock {\sc J. Phys. (London)} {\bf G 4}, 961 (1978).
\bibitem[1978Av02]{1978Av02} M. Avrigeanu, V. Avrigeanu, D. Bucurescu {\it et al.},     \newblock {\sc J. Phys. (London)} {\bf G 4}, 261 (1978).
\bibitem[1978Ba16]{1978Ba16} H. Backe, L. Richter, R. Willwater {\it et al.},     \newblock {\sc Z. Phys.} {\bf A 285}, 159 (1978).
\bibitem[1978Ba38]{1978Ba38} C. Baktash, J.X. Saladin, J.J. O'Brien, J.G. Alessi,     \newblock {\sc Phys. Rev.} {\bf C 18}, 131 (1978).
\bibitem[1978Be10]{1978Be10} M.J. Bechara, O. Dietzsch, M. Samuel, U. Smilansky,     \newblock {\sc Phys. Rev.} {\bf C 17}, 628 (1978).
\bibitem[1978Bo35]{1978Bo35} H.H. Bolotin, A.E. Stuchbery, K. Amos, I. Morrison,     \newblock {\sc Nucl. Phys.} {\bf A311}, 75 (1978).
\bibitem[1978DeYT]{1978DeYT} G. de Villiers, J.W. Koen, W.J. Naude, N.J.A. Rust,    \newblock {\sc SUNI-55, South. Univ. Nucl. Inst.}, (S.Africa), 1977 Ann.Rept., p.21 (1978).
\bibitem[1978DuZY]{1978DuZY} E.H. du Marchie Van Voorthuysen, M.J.A. de Voigt, J.F.W. Jansen,     \newblock {\sc KVI Ann. Rept. 1977}, 47 (1978).
\bibitem[1978Fa08]{1978Fa08} C. Fahlander, L. Hasselgren, G. Possnert, J.E. Thun,     \newblock {\sc Phys. Scr.} {\bf 18}, 47 (1978).
\bibitem[1978He13]{1978He13} H.P. Hellmeister, E. Schmidt, M. Uhrmacher {\it et al.},     \newblock {\sc Phys. Rev.} {\bf C 17}, 2113 (1978).
\bibitem[1978Jo04]{1978Jo04} A.M.R. Joye, A.M. Baxter, S. Hinds {\it et al.},     \newblock {\sc Phys. Lett.} {\bf B 72}, 307 (1978).
\bibitem[1978Ke11]{1978Ke11} D.L. Kennedy, H.H. Bolotin, I. Morrison, K. Amos,     \newblock {\sc Nucl. Phys.} {\bf A 308}, 14 (1978).
\bibitem[1978Ki09]{1978Ki09} G. Kindleben, T.W. Elze,     \newblock {\sc Z. Phys.} {\bf A 286}, 415 (1978).
\bibitem[1978KlZR]{1978KlZR} D.L. Kennedy, H.H. Bolotin, I. Morrison, K. Amos,     \newblock {\sc UM-P-88}, 9 (1978).
\bibitem[1978Le22]{1978Le22} R. Lecomte, S. Landsberger, P. Paradis, S. Monaro,     \newblock {\sc Phys. Rev.} {\bf C 18}, 2801 (1978).
\bibitem[1978Li13]{1978Li13} B.J. Linard, D.L. Kennedy, I. Morrison {\it et al.},     \newblock {\sc Nucl. Phys.} {\bf A 302}, 214 (1978).
\bibitem[1978Me08]{1978Me08} F.R. Metzger,     \newblock {\sc Phys. Rev.} {\bf C 17}, 939 (1978).
\bibitem[1978Me16]{1978Me16} F.R. Metzger,     \newblock {\sc Phys. Rev.} {\bf C 18}, 1603 (1978).
\bibitem[1978Og03]{1978Og03} M. Ogawa, R. Broda, K. Zell {\it et al.},     \newblock {\sc Phys. Rev. Lett.} {\bf 41}, 289 (1978).
\bibitem[1978Oh04]{1978Oh04} H. Ohnuma, J. Kasagi, Y. Iritani, N. Kishida, \newblock {\sc  J.Phys.Soc.Jpn.} {\bf 45}, 1092 (1978).
\bibitem[1978Po04]{1978Po04} V.N. Polishchuk, N.G. Shevchenko, N.G. Afanasev {\it et al.},     \newblock {\sc Yad. Fiz.} {\bf 27}, 1145 (1978); {\sc Sov. J. Nucl. Phys.} {\bf 27}, 607 (1978).
\bibitem[1978Ul01]{1978Ul01} G. Ulfert, D. Habs, V. Metag, H.J. Specht,     \newblock {\sc Nucl. Instrum. Methods} {\bf 148}, 369 (1978).
\bibitem[1978Ya02]{1978Ya02} S.W. Yates, N.R. Johnson, L.L. Riedinger, A.C. Kahler,     \newblock {\sc Phys. Rev.} {\bf C 17}, 634 (1978).
\bibitem[1978Ya11]{1978Ya11} Y. Yamazaki, E.B. Shera, M.V. Hoehn, R.M. Steffen,     \newblock {\sc Phys. Rev.} {\bf C 18}, 1474 (1978).
\bibitem[1979Az01]{1979Az01} R.E. Azuma, G.L. Borchert, L.C. Carraz {\it et al.},     \newblock {\sc Phys. Lett.} {\bf B 86}, 5 (1979).
\bibitem[1979Be41]{1979Be41} D.J. Beale, A.R. Poletti, J.R. Southon,    \newblock {\sc Aust. J. Phys.} {\bf 32}, 195 (1979).
\bibitem[1979Bo16]{1979Bo16} A. Bockisch, K. Bharuth-Ram, A.M. Kleinfeld, K.P. Lieb,     \newblock {\sc Z. Phys.} {\bf A 291}, 245 (1979).
\bibitem[1979Bo28]{1979Bo28} A. Bockisch, M. Miller, A.M. Kleinfeld {\it et al.},     \newblock {\sc Z. Phys.} {\bf A 292}, 265 (1979).
\bibitem[1979Bo29]{1979Bo29} B. Bochev, S. Iliev, R. Kalpakchieva {\it et al.},     \newblock {\sc Yad. Fiz.} {\bf 30}, 593 (1979); {\sc Sov. J. Nucl. Phys.} {\bf 30}, 305 (1979).
\bibitem[1979Bo31]{1979Bo31} H.H. Bolotin, I. Katayama, H. Sakai {\it et al.},     \newblock {\sc J. Phys. Soc. Jpn.} {\bf 47}, 1397 (1979).
\bibitem[1979DuZY]{1979DuZY} E.H. du Marchie van Voorthuysen, W. Moonen, J.F.W. Jansen, M.J.A. de Voigt,     \newblock {\sc KVI Ann.Rept.}, 1978, 63 (1979).
\bibitem[1979Ek03]{1979Ek03} L.P. Ekstrom, G.D. Jones, F. Kearns {\it et al.},     \newblock {\sc J. Phys.(London)} {\bf G5}, 803 (1979).
\bibitem[1979Fe05]{1979Fe05} M.P. Fewell, S. Hinds, D.C. Kean, T.H. Zabel,    \newblock {\sc Nucl. Phys.} {\bf A 319}, 214 (1979).
\bibitem[1979Fe06]{1979Fe06} M.P. Fewell, A.M. Baxter, D.C. Kean {\it et al.},    \newblock {\sc Nucl. Phys.} {\bf A 321}, 457 (1979).
\bibitem[1979Fe08]{1979Fe08} M.P. Fewell, D.C. Kean, R.H. Spear {\it et al.},    \newblock {\sc Phys. Rev. Lett.} {\bf 43}, 1463 (1979).
\bibitem[1979Fo02]{1979Fo02} J.S. Forster, T.K. Alexander, G.C. Ball {\it et al.},    \newblock {\sc Nucl. Phys.} {\bf A 313}, 397 (1979).
\bibitem[1979He07]{1979He07} H.P. Hellmeister, U. Kaup, J. Keinonen {\it et al.},     \newblock {\sc Phys. Lett.} {\bf B 85}, 34 (1979).
\bibitem[1979Ho23]{1979Ho23} M.V. Hoehn, E.B. Shera,   \newblock {\sc  Phys. Rev.} {\bf C 20}, 1934 (1979).
\bibitem[1979Ki17]{1979Ki17} V.G. Kiptilyi, I.Kh. Lemberg, A.S. Mishin, A.A. Pasternak,     \newblock {\sc Izv. Akad. Nauk SSSR}, Ser.Fiz. {\bf 43}, 2276 (1979); {\sc Bull. Acad. Sci. USSR}, Phys.Ser. {\bf 43}, No.11, 26 (1979).
\bibitem[1979Ma13]{1979Ma13} X.K. Maruyama, F.J. Kline, J.W. Lightbody {\it et al.},    \newblock {\sc Phys. Rev.} {\bf C 19}, 1624 (1979).
\bibitem[1979Mo01]{1979Mo01} C. Morand, J.F. Bruandet, B. Chambon {\it et al.},     \newblock {\sc Nucl. Phys.} {\bf A 313}, 45 (1979).
\bibitem[1979Pa11]{1979Pa11} P. Paradis, R. Lecomte, S. Landsberger, S. Monaro,     \newblock {\sc Phys. Rev.} {\bf C 20}, 1201 (1979).
\bibitem[1979Po01]{1979Po01} A.R. Poletti, L.K. Fifield, J. Asher, B.E. Cooke,    \newblock {\sc J. Phys. (London)} {\bf G5}, 575 (1979).
\bibitem[1979Po04]{1979Po04} R.J. Powers, P. Barreau, B. Bihoreau {\it et al.},     \newblock {\sc  Nucl. Phys.} {\bf A 316}, 295 (1979).
\bibitem[1979Ri06]{1979Ri06} L. Richter,     \newblock {\sc Z. Phys.} {\bf A 290}, 213 (1979).
\bibitem[1979Se03]{1979Se03} G. Seiler-Clark, D. Husar, R. Novotny {\it et al.},     \newblock {\sc Phys. Lett.} {\bf B 80}, 345 (1979).
\bibitem[1979Wa23]{1979Wa23} R. Wadsworth, L.P. Ekstrom, G.D. Jones {\it et al.},     \newblock {\sc J. Phys. (London)} {\bf G5}, 1761 (1979).
\bibitem[1980Ah01]{1980Ah01} I. Ahmad, A.M. Friedman, S.W. Yates,     \newblock {\sc Phys. Rev.} {\bf C 21}, 874 (1980).
\bibitem[1980Ba40]{1980Ba40} G.C. Ball, O. Hausser, T.K. Alexander {\it et al.},    \newblock {\sc Nucl. Phys.} {\bf A 349}, 271 (1980).
\bibitem[1980Bi14]{1980Bi14} E. Bitterwolf, P. Betz, A. Burkard {\it et al.},    \newblock {\sc Z. Phys.} {\bf A 298}, 279 (1980)
\bibitem[1980Ch22]{1980Ch22} T. Chapuran, R. Vodhanel, M.K. Brussel,     \newblock {\sc Phys. Rev.} {\bf C 22}, 1420 (1980).
\bibitem[1980ChZM]{1980ChZM} E. Cheifetz, H.A. Selic, A. Wolf {\it et al.},     \newblock {\sc Proc. Conf. Nucl. Spectr. Fission Products}, Grenoble, 1979, 193 (1980).
\bibitem[1980Ek03]{1980Ek03} L.P. Ekstrom, G.D. Jones, F. Kearns {\it et al.},     \newblock {\sc J. Phys. (London)} {\bf G 6}, 1415 (1980).
\bibitem[1980FaZW]{1980FaZW} C. Fahlander, A. Backlin, L. Hasselgren {\it et al.},     \newblock {\sc Proc. 6th European Phys. Soc. Nucl. Div. Conf. on Structure of Medium-Heavy Nuclei}, Rhodes, Greece, 1979, 291 (1980).
\bibitem[1980HiZV]{1980HiZV} J.H. Hirata, O. Dietzsch,     \newblock {\sc Proc. Int. Conf. on Nucl. Phys.}, Berkeley, 102 (1980).
\bibitem[1980KaZT]{1980KaZT} T.Katou, Y.Tendow, H.Kumagai {\it et al.},     \newblock {\sc Proc. Int. Conf. on Nucl. Phys.}, Berkeley, 751 (1980).
\bibitem[1980Ke04]{1980Ke04} D.L. Kennedy, A.E. Stuchbery, H.H. Bolotin,     \newblock {\sc Nucl. Instrum. Methods} {\bf 171}, 361 (1980).
\bibitem[1980La01]{1980La01} S. Landsberger, R. Lecomte, P. Paradis, S. Monaro,     \newblock {\sc Phys. Rev.} {\bf C 21}, 588 (1980).
\bibitem[1980Le16]{1980Le16} R. Lecomte, M. Irshad, S. Landsberger {\it et al.},     \newblock {\sc Phys. Rev.} {\bf C 22}, 1530 (1980).
\bibitem[1980Li14]{1980Li14} B.J. Lieb, H.S. Plendl, H.O. Funsten {\it et al.},    \newblock {\sc Phys. Rev.} {\bf C 22}, 1612 (1980).
\bibitem[1980LuZT]{1980LuZT} M. Luontama, A. Backlin, L. Westerberg {\it et al.},     \newblock {\sc JYFL Ann. Rept.}, 1980, 44 (1980).
\bibitem[1980Mi16]{1980Mi16} C. Michel, Y. El Masri, R. Holzmann {\it et al.},     \newblock {\sc Z. Phys.} {\bf A 298}, 213 (1980).
\bibitem[1980Ru01]{1980Ru01} A.J. Rutten, A. Holthuizen, J.A.G. De Raedt {\it et al.},    \newblock {\sc Nucl. Phys.} {\bf A 344}, 294 (1980).
\bibitem[1980Sc13]{1980Sc13} F. Schussler, J.A. Pinston, E. Monnand {\it et al.},     \newblock {\sc Nucl. Phys.} {\bf A 339}, 415 (1980).
\bibitem[1980Sc25]{1980Sc25} D.E.C. Scherpenzeel, G.A.P. Engelbertink, H.J.M. Aarts {\it et al.},    \newblock {\sc Nucl. Phys.} {\bf A 349}, 513 (1980).
\bibitem[1980Sp05]{1980Sp05} R.H. Spear, M.T. Esat, M.P. Fewell {\it et al.},     \newblock {\sc Nucl. Phys.} {\bf A 345}, 252 (1980).
\bibitem[1980Sp09]{1980Sp09} R.H. Spear, M.P. Fewell,     \newblock {\sc Aust. J. Phys.} {\bf 33}, 509 (1980); {\sc Corrigendum Aust. J. Phys.} {\bf 34}, 609 (1981).
\bibitem[1981Ah02]{1981Ah02} J. Ahlert, M. Schumacher,     \newblock {\sc Z. Phys.} {\bf A301}, 75 (1981).
\bibitem[1981Bo32]{1981Bo32} H.H. Bolotin, A.E. Stuchbery, I. Morrison {\it et al.},     \newblock {\sc Nucl. Phys.} {\bf A 370}, 146 (1981).
\bibitem[1981Ca01]{1981Ca01} C.M. Cartwright, P.D. Forsyth, I. Hall {\it et al.},     \newblock {\sc J. Phys. (London)} {\bf G 7}, 65 (1981).
\bibitem[1981Ca10]{1981Ca10} Y. Cauchois, H. Ben Abdelaziz, R. Kherouf, C. Schloesing-Moller,     \newblock {\sc J. Phys. (London)} {\bf G 7}, 1539 (1981).
\bibitem[1981De03]{1981De03} A.P.de Lima, A.V. Ramayya, J.H. Hamilton {\it et al.},     \newblock {\sc Phys. Rev.} {\bf C 23}, 213 (1981); Erratum Phys.Rev. {\bf C 23}, 2380 (1981).
\bibitem[1981DeYW]{1981DeYW} A. Dewald, U. Kaup, W. Gast {\it et al.},     \newblock {\sc Proc. Int. Conf. Nuclei Far from Stability}, Helsingor, Denmark, {\bf 2}, 418 (1981); {\sc CERN-81-09} (1981).
\bibitem[1981Dy01]{1981Dy01} K. Dybdal, J.S. Forster, P. Hornshoj {\it et al.},    \newblock {\sc Nucl. Phys.} {\bf A 359}, 431 (1981).
\bibitem[1981Es03]{1981Es03} M.T. Esat, M.P. Fewell, R.H. Spear {\it et al.},     \newblock {\sc Nucl. Phys.} {\bf A 362}, 227 (1981).
\bibitem[1981Fu03]{1981Fu03} L. Funke, J. Doring, F. Dubbers {\it et al.},     \newblock {\sc Nucl. Phys.} {\bf A 355}, 228 (1981).
\bibitem[1981Ho22]{1981Ho22} M.V. Hoehn, E.B. Shera, H.D. Wohlfahrt {\it et al.},     \newblock {\sc  Phys. Rev.} {\bf C 24}, 1667 (1981).
\bibitem[1981Is14]{1981Is14} H.A. Ismail, A. El-Naem, S.A. El-Malalk {\it et al.},     \newblock {\sc Rev. Roum. Phys.} {\bf 26}, 461 (1981).
\bibitem[1981Ji03]{1981Ji03} Jiang Cheng-lie, S. Pontoppidan,     \newblock {\sc Phys. Rev.} {\bf C 24}, 1350 (1981).
\bibitem[1981Jo03]{1981Jo03} N.-G. Jonsson, A. Backlin, J. Kantele {\it et al.},     \newblock {\sc  Nucl. Phys.} {\bf A 371}, 333 (1981).
\bibitem[1981Le02]{1981Le02} M.J. LeVine, E.K. Warburton, D. Schwalm,     \newblock {\sc Phys. Rev.} {\bf C23}, 244 (1981).
\bibitem[1981Wa09]{1981Wa09} N.J. Ward, L.P. Ekstrom, G.D. Jones {\it et al.},     \newblock {\sc J. Phys. (London)} {\bf G7}, 815 (1981).
\bibitem[1981Yo07]{1981Yo07} Y. Yoshizawa, B. Herskind, M. Hoshi,     \newblock {\sc J. Phys. Soc. Jpn.} {\bf 50}, 2151 (1981).
\bibitem[1981Zh07]{1981Zh07} U.Yu. Zhovliev, M.F. Kudoyarov, I.Kh. Lemberg, A.A. Pasternak,     \newblock {\sc Izv. Akad. Nauk SSSR}, Ser.Fiz. {\bf 45}, 1879 (1981).
\bibitem[1982Al15]{1982Al15} T.K. Alexander, G.C. Ball, W.G. Davies {\it et al.},    \newblock {\sc Phys. Lett.} {\bf B 113}, 132 (1982).
\bibitem[1982Al22]{1982Al22} T.K. Alexander, G.C. Ball, J.S. Forster {\it et al.},    \newblock {\sc Phys. Rev. Lett.} {\bf 49}, 438 (1982).
\bibitem[1982An06]{1982An06} D.S. Andreev, K.I. Erokhina, I.Kh. Lemberg {\it et al.},    \newblock {\sc  Izv. Akad. Nauk SSSR, Ser. Fiz.} {\bf 46}, 30 (1982).
\bibitem[1982Ba06]{1982Ba06} G.C. Ball, T.K. Alexander, W.G. Davies {\it et al.},    \newblock {\sc Nucl. Phys.} {\bf A 377}, 268 (1982).
\bibitem[1982Be38]{1982Be38} J.A. Becker, J.B. Carlson, R.G. Lanier {\it et al.},     \newblock {\sc Phys. Rev.} {\bf C 26}, 914 (1982).
\bibitem[1982De05]{1982De05} A. Dewald, U. Kaup, W. Gast {\it et al.},     \newblock {\sc Phys. Rev.} {\bf C 25}, 226 (1982).
\bibitem[1982De36]{1982De36} S. Della Negra, H. Gauvin, D. Jacquet, Y. Le Beyec, \newblock {\sc  Z. Phys.} {\bf A 307}, 305 (1982).
\bibitem[1982GaZH]{1982GaZH} K.F.W. Gast,  \newblock {\sc  Thesis}, Univ.Koln (1982).
\bibitem[1982He03]{1982He03} J. Heisenberg, J. Lichtenstadt, C.N. Papanicolas, J.S. McCarthy,     \newblock {\sc Phys. Rev.} {\bf C 25}, 2292 (1982).
\bibitem[1982HiZT]{1982HiZT} T. Higo, S. Matsuki, T. Oshawa {\it et al.},     \newblock {\sc Contrib. Intern. Symp. Dynamics of Nucl. Collective Motion}, Yamanishi, Japan, 27 (1982).
\bibitem[1982Jo04]{1982Jo04} N.R. Johnson, I.Y. Lee, F.K. McGowan {\it et al.},     \newblock {\sc Phys. Rev.} {\bf C 26}, 1004 (1982).
\bibitem[1982Ke01]{1982Ke01} J. Keinonen, K.P. Lieb, H.P. Hellmeister {\it et al.},     \newblock {\sc Nucl. Phys.} {\bf A 376}, 246 (1982).
\bibitem[1982Li08]{1982Li08} C.J. Lister, B.J. Varley, H.G. Price, J.W. Olness,     \newblock {\sc Phys. Rev. Lett.} {\bf 49}, 308 (1982).
\bibitem[1982No04]{1982No04} B.E. Norum, M.V. Hynes, H. Miska {\it et al.},    \newblock {\sc Phys. Rev.} {\bf C 25}, 1778 (1982).
\bibitem[1982Pa03]{1982Pa03} T. Paradellis, C.A. Kalfas,     \newblock {\sc Phys. Rev.} {\bf C 25}, 350 (1982).
\bibitem[1982Pa10]{1982Pa10} A. Pakkanen, Y.H. Chung, P.J. Daly {\it et al.},     \newblock {\sc Phys. Rev. Lett.} {\bf 48}, 1530 (1982).
\bibitem[1982Ro05]{1982Ro05} S. Rozak, E.G. Funk, J.W. Mihelich,     \newblock {\sc Phys. Rev.} {\bf C 25}, 3000 (1982).
\bibitem[1982Sp02]{1982Sp02} K.-H. Speidel, P.N. Tandon, V. Mertens {\it et al.},    \newblock {\sc Nucl. Phys.} {\bf A 378}, 130 (1982).
\bibitem[1982Sp05]{1982Sp05} R.H. Spear, T.H. Zabel, M.T. Esat {\it et al.},    \newblock {\sc Nucl. Phys.} {\bf A 378}, 559 (1982).
\bibitem[1983Bi08]{1983Bi08} E. Bitterwolf, A. Burkard, P. Betz {\it et al.},    \newblock {\sc Z. Phys.} {\bf A 313}, 123 (1983).
\bibitem[1983El02]{1983El02} M.S.S. El-Daghmah, N.M. Stewart,     \newblock {\sc Z. Phys.} {\bf A 309}, 219 (1983).
\bibitem[1983El03]{1983El03} D.V. Elenkov, D.P. Lefterov, G.Kh. Tumbev, \newblock {\sc  Izv.Akad.Nauk SSSR}, Ser.Fiz. {\bf 47}, 56 (1983).
\bibitem[1983Ga11]{1983Ga11} M. Gai, J.F. Ennis, M. Ruscev {\it et al.},     \newblock {\sc Phys.Rev.Lett.} {\bf 51}, 646 (1983).
\bibitem[1983He21]{1983He21} F.W. Hersman, W. Bertozzi, T.N. Buti {\it et al.},     \newblock {\sc  Phys. Lett.} {\bf B 132}, 47 (1983).
\bibitem[1983Kl09]{1983Kl09} R. Klein, P. Grabmayr, Y. Kawazoe {\it et al.},     \newblock {\sc Nuovo Cim.} {\bf 76 A}, 369 (1983).
\bibitem[1983Ko01]{1983Ko01} R.L. Kozub, J. Lin, J.F. Mateja {\it et al.},    \newblock {\sc Phys. Rev.} {\bf C 27}, 158 (1983).
\bibitem[1983La08]{1983La08} D.B. Laubacher, Y. Tanaka, R.M. Steffen {\it et al.},     \newblock {\sc Phys. Rev.} {\bf C 27}, 1772 (1983).
\bibitem[1983Li02]{1983Li02} J.W. Lightbody, Jr., J.W. Lightbody, J.B. Bellicard {\it et al.},     \newblock {\sc Phys. Rev.} {\bf C27}, 113 (1983).
\bibitem[1983MaYT]{1983MaYT} G. Mamane,     \newblock {\sc Thesis}, Weizmann Inst.Science, Rehovot (1983).
\bibitem[1983Pr08]{1983Pr08} H.G. Price, C.J. Lister, B.J. Varley {\it et al.},     \newblock {\sc Phys. Rev. Lett.} {\bf 51}, 1842 (1983).
\bibitem[1983Ra17]{1983Ra17} M. Rahman, H.P. Nottrodt, F. Rauch,    \newblock {\sc Nucl. Phys.} {\bf A 401}, 253 (1983).
\bibitem[1984Al06]{1984Al06} A. Alzner, D. Best, E. Bodenstedt {\it et al.},     \newblock {\sc Z. Phys.} {\bf A 316}, 87 (1984).
\bibitem[1984Be20]{1984Be20} M.J. Bechara, O. Dietzsch, J.H. Hirata,     \newblock {\sc Phys. Rev.} {\bf C 29}, 1672 (1984).
\bibitem[1984Bh03]{1984Bh03} R.K. Bhalla, A.R. Poletti,    \newblock {\sc Nucl. Phys.} {\bf A 420}, 96 (1984).
\bibitem[1984Dr02]{1984Dr02} G.D. Dracoulis, G.D. Sprouse, O.C. Kistner, M.H. Rafailovich,     \newblock {\sc Phys. Rev.} {\bf C 29}, 1576 (1984).
\bibitem[1984Ef01]{1984Ef01} A.D. Efimov, K.I. Erokhina, V.G. Kiptilyi {\it et al.},     \newblock {\sc Izv. Akad. Nauk SSSR}, Ser.Fiz. {\bf 48}, 10 (1984).
\bibitem[1984El12]{1984El12} D. Elenkov, D. Lefterov, G. Toumbev,    \newblock {\sc Nucl. Instrum. Methods} {\bf 228}, 62 (1984).
\bibitem[1984EnZY]{1984EnZY} J.F. Ennis, M. Gai, D.A. Bromley {\it et al.},     \newblock {\sc Bull. Am. Phys. Soc.} {\bf 29}, No.7, 1050, DC10 (1984).
\bibitem[1984Fe08]{1984Fe08} M.P. Fewell,     \newblock {\sc Nucl. Phys.} {\bf A 425}, 373 (1984).
\bibitem[1984He02]{1984He02} J. Heisenberg, J. Dawson, T. Milliman {\it et al.},    \newblock {\sc  Phys. Rev.} {\bf C 29}, 97 (1984).
\bibitem[1984Ke10]{1984Ke10} P. Kemnitz, P. Ojeda, J. Doring {\it et al.},     \newblock {\sc Nucl.Phys.} {\bf A 425}, 493 (1984).
\bibitem[1984Mu19]{1984Mu19} S.J. Mundy, J. Lukasiak, W.R. Phillips,     \newblock {\sc Nucl. Phys.} {\bf A 426}, 144 (1984).
\bibitem[1984Pa02]{1984Pa02} C.N. Papanicolas, J. Heisenberg, J. Lichtenstadt {\it et al.},     \newblock {\sc Phys. Rev. Lett.} {\bf 52}, 247 (1984).
\bibitem[1984Re10]{1984Re10} W. Reuter, E.B. Shera, M.V. Hoehn {\it et al.},     \newblock {\sc Phys. Rev.} {\bf C 30}, 1465 (1984).
\bibitem[1984Ro01]{1984Ro01} J. Roth, L. Cleemann, J. Eberth {\it et al.},     \newblock {\sc J. Phys. (London)} {\bf G 10}, L25 (1984).
\bibitem[1984Ta10]{1984Ta10} T. Tanaka, R.M. Steffen, E.B. Shera {\it et al.},     \newblock {\sc Phys. Rev.} {\bf C 30}, 350 (1984).
\bibitem[1984To10]{1984To10} D.M. Todd, R. Aryaeinejad, D.J.G. Love {\it et al.},     \newblock {\sc J. Phys. (London)} {\bf G 10}, 1407 (1984).
\bibitem[1984Ve07]{1984Ve07} W.J. Vermeer, M.T. Esat, J.A. Kuehne {\it et al.},     \newblock {\sc Aust. J. Phys.} {\bf 37}, 123 (1984).
\bibitem[1984We17]{1984We17} J.C. Wells, N.R. Johnson, J. Hattula {\it et al.},     \newblock {\sc Phys. Rev.} {\bf C 30}, 1532 (1984).
\bibitem[1984Wo10]{1984Wo10} B. Wormann, K.P. Lieb, R. Diller {\it et al.},     \newblock {\sc Nucl. Phys.} {\bf A 431}, 170 (1984).
\bibitem[1984ZoZZ]{1984ZoZZ} A.E. Zobov, V.G. Kiptily, I.Kh. Lemberg {\it et al.},     \newblock {\sc Program and Theses, Proc. 34th Ann. Conf. Nucl. Spectrosc. Struct. At. Nuclei}, Alma-Ata, 73 (1984).
\bibitem[1985Al18]{1985Al18} T.K. Alexander, G.C. Ball, D. Horn {\it et al.},    \newblock {\sc Nucl. Phys.} {\bf A 444}, 285 (1985).
\bibitem[1985Az02]{1985Az02} F. Azgui, H. Emling, E. Grosse {\it et al.},     \newblock {\sc Nucl. Phys.} {\bf A 439}, 573 (1985).
\bibitem[1985AzZY]{1985AzZY} F. Azgui,     \newblock {\sc Thesis}, Univ. Louis Pasteur de Strasbourg (1985); {\sc CRN/PN} 85 31 (1985).
\bibitem[1985Bo31]{1985Bo31} D. Bohle, A. Richter, K. Heyde {\it et al.},     \newblock {\sc Phys. Rev. Lett.} {\bf 55}, 1661 (1985).
\bibitem[1985Bo32]{1985Bo32} W. Bonin, H. Backe, M. Dahlinger {\it et al.},     \newblock {\sc Z. Phys.} {\bf A 322}, 59 (1985).
\bibitem[1985Bu01]{1985Bu01} S.M. Burnett, A.M. Baxter, S. Hinds {\it et al.},     \newblock {\sc Nucl. Phys.} {\bf A 432}, 514 (1985).
\bibitem[1985ChZY]{1985ChZY} A. Chaudhury, M.W. Drigert, E.G. Funk {\it et al.},     \newblock {\sc  Bull. Am. Phys. Soc.} {\bf 30}, No.4, 742, DE9 (1985).
\bibitem[1985Fe03]{1985Fe03} M.P. Fewell, G.J. Gyapong, R.H. Spear {\it et al.},     \newblock {\sc Phys. Lett.} {\bf B 157}, 353 (1985).
\bibitem[1985Lu06]{1985Lu06} S. Lunardi, F. Scarlassara, F. Soramel {\it et al.},     \newblock {\sc Z. Phys.} {\bf A 321}, 177 (1985) .
\bibitem[1985Si01]{1985Si01} K.P. Singh, D.C. Tayal, G. Singh, H.S. Hans,     \newblock {\sc Phys. Rev.} {\bf C 31}, 79 (1985).
\bibitem[1985VoZY]{1985VoZY} P. von Brentano,     \newblock {\sc Priv.Comm.} (June 1985).
\bibitem[1985Wi01]{1985Wi01} G. Winter, F. Dubbers, J. Doring {\it et al.},     \newblock {\sc J. Phys. (London)} {\bf G 11}, 277 (1985).
\bibitem[1985Wi06]{1985Wi06} J.E. Wise, J.S. McCarthy, R. Altemus {\it et al.},    \newblock {\sc Phys. Rev.} {\bf C 31}, 1699 (1985).
\bibitem[1985Wo06]{1985Wo06} U. Worsdorfer, H.J. Emrich, H. Miska  {\it et al.},    \newblock {\sc Nucl. Phys.} {\bf A 438}, 711 (1985).
\bibitem[1986Bi13]{1986Bi13} J. Billowes,     \newblock {\sc Hyperfine Interactions} {\bf 30}, 265 (1986).
\bibitem[1986Cz02]{1986Cz02} T. Czosnyka, D. Cline, L. Hasselgren {\it et al.},     \newblock {\sc Nucl. Phys.} {\bf A 458}, 123 (1986).
\bibitem[1986Dr05]{1986Dr05} G.D. Dracoulis, A.E. Stuchbery, A.P. Byrne {\it et al.},     \newblock {\sc J. Phys. (London)} {\bf G 12}, L97 (1986).
\bibitem[1986Ga21]{1986Ga21} U. Garg, A. Chaudhury, M.W. Drigert {\it et al.},     \newblock {\sc Phys. Lett.} {\bf B 180}, 319 (1986).
\bibitem[1986Gy04]{1986Gy04} G.J. Gyapong, R.H. Spear, M.T. Esat {\it et al.},     \newblock {\sc Nucl. Phys.} {\bf A 458}, 165 (1986).
\bibitem[1986He09]{1986He09} F.W. Hersman, W. Bertozzi, T.N. Buti  {\it et al.},     \newblock {\sc  Phys. Rev.} {\bf C 33}, 1905 (1986).
\bibitem[1986He17]{1986He17} J. Heese, K.P. Lieb, L. Luhmann {\it et al.},     \newblock {\sc Z. Phys.} {\bf A 325}, 45 (1986).
\bibitem[1986Ma22]{1986Ma22} G. Mamane, E. Cheifetz, E. Dafni {\it et al.},     \newblock {\sc Nucl. Phys.} {\bf A 454}, 213 (1986).
\bibitem[1986Ma39]{1986Ma39} A. Makishima, M. Adachi, H. Taketani, M. Ishii,     \newblock {\sc Phys. Rev.} {\bf C 34}, 576 (1986).
\bibitem[1986Os02]{1986Os02} M. Oshima, N.R. Johnson, F.K. McGowan {\it et al.},     \newblock {\sc Phys. Rev.} {\bf C 33}, 1988 (1986).
\bibitem[1986Ra07]{1986Ra07} M.N. Rao, N.R. Johnson, F.K. McGowan {\it et al.},     \newblock {\sc Phys. Rev. Lett.} {\bf 57}, 667 (1986).
\bibitem[1986Ro15]{1986Ro15} P.J. Rothschild, A.M. Baxter, S.M. Burnett {\it et al.},     \newblock {\sc Phys. Rev.} {\bf C 34}, 732 (1986).
\bibitem[1986Sc18]{1986Sc18} P. Schuler, Ch. Lauterbach, Y.K. Agarwal {\it et al.},     \newblock {\sc Phys. Lett.} {\bf B 174}, 241 (1986).
\bibitem[1987Bi13]{1987Bi13} J. Billowes, K.P. Lieb, J.W. Noe {\it et al.},     \newblock {\sc Phys. Rev.} {\bf C 36}, 974 (1987).
\bibitem[1987Dr08]{1987Dr08} G.D. Dracoulis, A.P. Byrne, A.E. Stuchbery {\it et al.},     \newblock {\sc Nucl. Phys.} {\bf A 467}, 305 (1987).
\bibitem[1987Ga12]{1987Ga12} J. Gascon, F. Banville, P. Taras {\it et al.},     \newblock {\sc Nucl. Phys.} {\bf A 470}, 230 (1987).
\bibitem[1987Ga14]{1987Ga14} J. Gascon, P. Taras, P. van Esbroek {\it et al.},     \newblock {\sc Nucl. Phys.} {\bf A 472}, 558 (1987).
\bibitem[1987Gy01]{1987Gy01} G.J. Gyapong, R.H. Spear, M.P. Fewell {\it et al.},     \newblock {\sc Nucl. Phys.} {\bf A 470}, 415 (1987).
\bibitem[1987IsZX]{1987IsZX} T. Ishii, M. Ishii, K. Yanagida, M. Ogawa,  \newblock {\sc  Japan Atomic Energy Res. Inst. Tandem Linac VDG, Ann. Rept., 1986}, 167 (1987).
\bibitem[1987MiZL]{1987MiZL} T.E. Milliman,     \newblock {\sc Thesis, Univ. New Hampshire} (1987).
\bibitem[1987Oh05]{1987Oh05} H. Ohm, G. Lhersonneau, K. Sistemich {\it et al.},     \newblock {\sc Z. Phys.} {\bf A 327}, 483 (1987).
\bibitem[1987Sc07]{1987Sc07} R. Schwengner, G. Winter, J. Doring {\it et al.},     \newblock {\sc Z. Phys.} {\bf A 326}, 287 (1987).
\bibitem[1987Wa02]{1987Wa02} R. Wadsworth, J.M. O'Donnell, D.L. Watson {\it et al.},     \newblock {\sc J. Phys. (London)} {\bf G 13}, 205 (1987).
\bibitem[1988Ah01]{1988Ah01} A. Ahmad, G. Bomar, H. Crowell {\it et al.},     \newblock {\sc Phys. Rev.} {\bf C 37}, 1836 (1988).
\bibitem[1988Bi03]{1988Bi03} P.J. Bishop, M.J. Godfrey, A.J. Kirwan {\it et al.},     \newblock {\sc J. Phys. (London)} {\bf G 14}, 995 (1988).
\bibitem[1988Bo08]{1988Bo08} W. Boeglin, P. Egelhof, I. Sick {\it et al.},     \newblock {\sc Nucl. Phys.} {\bf A 477}, 399 (1988).
\bibitem[1988Br10]{1988Br10} M.R. Braunstein, J.J. Kraushaar, R.P. Michel {\it et al.},     \newblock {\sc Phys. Rev.} {\bf C 37}, 1870 (1988).
\bibitem[1988DoZU]{1988DoZU} G.A. Dostemesova, D.K. Kaipov, Yu.G. Kosyak,        \newblock {\sc Program and Theses, Proc. 38th Ann. Conf. Nucl. Spectrosc. Struct. At. Nuclei}, Baku, 59 (1988).
\bibitem[1988Fa07]{1988Fa07} C. Fahlander, A. Backlin, L. Hasselgren {\it et al.},     \newblock {\sc Nucl. Phys.} {\bf A 485}, 327 (1988).
\bibitem[1988Fe01]{1988Fe01} M.P. Fewell, N.R. Johnson, F.K. McGowan {\it et al.},     \newblock {\sc Phys. Rev.} {\bf C 37}, 101 (1988).
\bibitem[1988Ga33]{1988Ga33} M. Gai, J.F. Ennis, D.A. Bromley {\it et al.},     \newblock {\sc Phys. Lett.} {\bf B 215}, 242 (1988).
\bibitem[1988Ka21]{1988Ka21} W. Karle, M. Knopp, K.-H. Speidel,     \newblock {\sc Nucl. Instrum. Methods Phys. Res.} {\bf A 271}, 507 (1988).
\bibitem[1988Ku01]{1988Ku01} A.I. Kucharska, J. Billowes, M.A. Grace,     \newblock {\sc J. Phys. (London)} {\bf G 14}, 65 (1988).
\bibitem[1988Ku33]{1988Ku33} A. Kuronen, J. Raisanen, J. Keinonen {\it et al.},    \newblock {\sc Nucl. Instrum. Methods Phys. Res.} {\bf B35}, 1 (1988).
\bibitem[1988Li22]{1988Li22} C.S. Lim, R.H. Spear, W.J. Vermeer {\it et al.},     \newblock {\sc Nucl. Phys.} {\bf A 485}, 399 (1988).
\bibitem[1988Lu04]{1988Lu04} Lu Xiting, Ban Yong, Liu Hongtao {\it et al.},    \newblock {\sc Nucl. Instrum. Methods Phys. Res.} {\bf A272}, 909 (1988).
\bibitem[1988Mo08]{1988Mo08} R. Moscrop, M. Campbell, W. Gelletly {\it et al.},     \newblock {\sc Nucl. Phys.} {\bf A 481}, 559 (1988).
\bibitem[1988MyZY]{1988MyZY} T. Mylaeus, J. Busch, J. Eberth {\it et al.},     \newblock {\sc HMI-466}, 210 (1988).
\bibitem[1988PeZW]{1988PeZW} B.A. Peterson, G.A. Rebka, S. Kowalski {\it et al.},     \newblock {\sc Bull. Am. Phys. Soc.} {\bf 33}, No.4, 1097, KI2 (1988).
\bibitem[1988Sa32]{1988Sa32} S. Salem-Vasconcelos, M.J. Bechara, J.H. Hirata, O. Dietzsch,     \newblock {\sc Phys. Rev.} {\bf C38}, 2439 (1988).
\bibitem[1988So06]{1988So06} F. Soramel, S. Lunardi, S. Beghini {\it et al.},     \newblock {\sc Phys. Rev.}  {\bf C 38}, 537 (1988).
\bibitem[1988Ve08]{1988Ve08} W.J. Vermeer, C.S. Lim, R.H. Spear,     \newblock {\sc Phys. Rev.} {\bf C 38}, 2982 (1988).
\bibitem[1989Ad01]{1989Ad01} J. Adam, M. Honusek, A. Spalek {\it et al.},     \newblock {\sc  Z. Phys.} {\bf A 332}, 143 (1989).
\bibitem[1989Bu07]{1989Bu07} S.M. Burnett, A.M. Baxter, G.J. Gyapong {\it et al.},     \newblock {\sc Nucl. Phys.} {\bf A 494}, 102 (1989).
\bibitem[1989BuZP]{1989BuZP} A.J.C. Burghardt,     \newblock {\sc Thesis}, Univ.Amsterdam (1989); {\sc INIS-mf-11407} (1989).
\bibitem[1989Ga24]{1989Ga24} Yu.P. Gangrsky, S.G. Zemlyanoi, N.N. Kolesnikov {\it et al.},     \newblock {\sc  Yad. Fiz.} {\bf 50}, 1217 (1989).
\bibitem[1989Ge09]{1989Ge09} M.K. Georgieva, D.V. Elenkov, D.P. Lefterov, G.H. Toumbev,    \newblock {\sc Fiz. Elem. Chastits At. Yadra} {\bf 20}, 930 (1989); {\sc Sov. J. Part. Nucl.} {\bf 20}, 393 (1989).
\bibitem[1989It02]{1989It02} K. Itoh, Y.M. Shin, W.J. Gerace, Y. Torizuka,    \newblock {\sc Nucl. Phys.} {\bf A 492}, 426 (1989).
\bibitem[1989Ke04]{1989Ke04} J. Keinonen, P. Tikkanen, A. Kuronen {\it et al.},    \newblock {\sc Nucl. Phys.} {\bf A 493}, 124 (1989).
\bibitem[1989Ki01]{1989Ki01} A.J. Kirwan, P.J. Bishop, D.J.G. Love {\it et al.},     \newblock {\sc J. Phys. (London)} {\bf G 15}, 85 (1989).
\bibitem[1989Ko40]{1989Ko40} B. Kotlinski, D. Cline, A. Backlin, D. Clark,     \newblock {\sc Nucl. Phys.} {\bf A 503}, 575 (1989).
\bibitem[1989Ku04]{1989Ku04} R. Kulessa, R. Bengtsson, H. Bohn {\it et al.},     \newblock {\sc Phys. Lett.} {\bf B 218}, 421 (1989).
\bibitem[1989Lh01]{1989Lh01} G. Lhersonneau, H. Gabelmann, N. Kaffrell {\it et al.},     \newblock {\sc Z. Phys.} {\bf A 332}, 243 (1989).
\bibitem[1989Lo01]{1989Lo01} G. Lo Bianco, K.P. Schmittgen, K.O. Zell, P. v.Brentano,     \newblock {\sc Z. Phys.} {\bf A 332}, 103 (1989).
\bibitem[1989Lo08]{1989Lo08} M. Loiselet, O. Naviliat-Cuncic, J. Vervier,     \newblock {\sc Nucl. Phys.} {\bf A 496}, 559 (1989).
\bibitem[1989Ma33]{1989Ma33} H. Mach, M. Moszynski, R.F. Casten {\it et al.},     \newblock {\sc Phys. Rev. Lett.} {\bf 63}, 143 (1989).
\bibitem[1989Ma38]{1989Ma38} H. Mach, R.L. Gill, M. Moszynski,     \newblock {\sc Nucl. Instrum. Methods Phys. Res.} {\bf A 280}, 49 (1989).
\bibitem[1989Ma47]{1989Ma47} H. Mach, M. Moszynski, R.L. Gill {\it et al.},     \newblock {\sc Phys. Lett.} {\bf B 230}, 21 (1989).
\bibitem[1989Mo06]{1989Mo06} M. Moszynski, H. Mach,     \newblock {\sc Nucl. Instrum. Methods Phys. Res.} {\bf A 277}, 407 (1989).
\bibitem[1989Mo10]{1989Mo10} R. Moscrop, M. Campbell, W. Gelletly {\it et al.},     \newblock {\sc Nucl. Phys.} {\bf A 499}, 565 (1989).
\bibitem[1989Mu13]{1989Mu13} M. Murzel, S.C. Pancholi, U. Birkental {\it et al.},     \newblock {\sc Z. Phys.} {\bf A 334}, 125 (1989).
\bibitem[1989Oh06]{1989Oh06} H. Ohm, M. Liang, G. Molnar, K. Sistemich,     \newblock {\sc Z. Phys.} {\bf A 334}, 519 (1989).
\bibitem[1989Sc06]{1989Sc06} K. Schiffer, S. Harissopulos, A. Dewald {\it et al.},     \newblock {\sc J. Phys. (London)} {\bf G 15}, L85 (1989).
\bibitem[1989Sp03]{1989Sp03} R.H. Spear, A.M. Baxter, S.M. Burnett, C.L. Miller,  \newblock {\sc  Aust. J. Phys.} {\bf 42}, 41 (1989).
\bibitem[1989Sp07]{1989Sp07} R.H. Spear, W.J. Vermeer, S.M. Burnett {\it et al.},     \newblock {\sc Aust. J. Phys.} {\bf 42}, 345 (1989).
\bibitem[1989SvZZ]{1989SvZZ} L.E. Svensson,  Thesis, Uppsala Univ. (1989).
\bibitem[1989VoZT]{1989VoZT} D.A. Volkov, A.I. Kovalenko, A.I. Levon {\it et al.},     \newblock {\sc Program and Thesis, Proc. 39th Ann. Conf. Nucl. Spectrosc. Struct. At. Nuclei}, Tashkent, 74 (1989).
\bibitem[1989Wu04]{1989Wu04} C.Y. Wu, D. Cline, E.G. Vogt {\it et al.},     \newblock {\sc Phys. Rev.} {\bf C 40}, R3 (1989); Erratum Phys.Rev. {\bf C 40}, 2431 (1989).
\bibitem[1990De04]{1990De04} M.J.A. De Voigt, R. Kaczarowski, H.J. Riezebos {\it et al.},     \newblock {\sc Nucl. Phys.} {\bf A 507}, 472 (1990).
\bibitem[1990DeZN]{1990DeZN} A. Dewald, P. Petkov, R. Wrzal {\it et al.},     \newblock {\sc Proc. Intern. Conf. Nuclear Structure, 24th Zakopane School on Physics}, Zakopane, Poland, 28 April-12 May, 1990, 152 (1990).
\bibitem[1990Ga22]{1990Ga22} G. Garcia Bermudez, M.A. Cardona, A. Filevich,     \newblock {\sc Nucl. Instrum. Methods Phys. Res.} {\bf A 292}, 367 (1990).
\bibitem[1990He04]{1990He04} J. Heese, K.P. Lieb, S. Ulbig {\it et al.},     \newblock {\sc Phys. Rev.} {\bf C 41}, 603 (1990).
\bibitem[1990Ka11]{1990Ka11} R. Kaczarowski, U. Garg, A. Chaudhury {\it et al.},     \newblock {\sc Phys. Rev.} {\bf C 41}, 2069 (1990).
\bibitem[1990Ko30]{1990Ko30} B. Kotlinski, D. Cline, A. Backlin {\it et al.},     \newblock {\sc Nucl. Phys.} {\bf A 517}, 365 (1990).
\bibitem[1990Ko38]{1990Ko38} B. Kotlinski, T. Czosnyka, D. Cline {\it et al.},     \newblock {\sc Nucl. Phys.} {\bf A 519}, 646 (1990).
\bibitem[1990Lh01]{1990Lh01} G. Lhersonneau, H. Gabelmann, N. Kaffrell {\it et al.},     \newblock {\sc Z. Phys.} {\bf A 337}, 143 (1990).
\bibitem[1990Ma25]{1990Ma25} H. Mach, W. Nazarewicz, D. Kusnezov {\it et al.},     \newblock {\sc  Phys. Rev.} {\bf C 41}, R2469 (1990).
\bibitem[1990Pi04]{1990Pi04} H.H. Pitz, R.D. Heil, U. Kneissl {\it et al.},     \newblock {\sc   Nucl. Phys.} {\bf A 509}, 587 (1990); {\sc Errata Nucl. Phys.} {\bf A 514}, 749 (1990).
\bibitem[1990Ro10]{1990Ro10} H. Rotter, J. Doring, L. Funke {\it et al.},     \newblock {\sc  Nucl. Phys.} {\bf A 514}, 401 (1990).
\bibitem[1990Ta12]{1990Ta12} S.L. Tabor, P.D. Cottle, J.W. Holcomb {\it et al.},     \newblock {\sc Phys. Rev.} {\bf C 41}, 2658 (1990).
\bibitem[1990Wa13]{1990Wa13} Y. Wang, J. Rapaport,     \newblock {\sc Nucl. Phys.} {\bf A 517}, 301 (1990).
\bibitem[1990WeZZ]{1990WeZZ} J. Wei, K.B. Beard, J.C. Walpe {\it et al.},     \newblock {\sc Bull. Am. Phys. Soc.} {\bf 35}, No.4, 1016, H6 7 (1990).
\bibitem[1991Ba38]{1991Ba38} D. Bazzacco, F. Brandolini, K. Loewenich {\it et al.},     \newblock {\sc Nucl. Phys.} {\bf A 533}, 541 (1991).
\bibitem[1991He03]{1991He03} M. Hellstrom, H. Mach, B. Fogelberg {\it et al.},     \newblock {\sc Phys. Rev.} {\bf C 43}, 1462 (1991).
\bibitem[1991Ib01]{1991Ib01} R.Ibbotson, B.Kotlinski, D.Cline  {\it et al.},     \newblock {\sc Nucl. Phys.} {\bf A 530}, 199 (1991).
\bibitem[1991Ki13]{1991Ki13} W. Kim, J.R. Calarco, J.P. Connelly {\it et al.},     \newblock {\sc Phys. Rev.} {\bf C 44}, 2400 (1991).
\bibitem[1991Li39]{1991Li39} M. Liang, H. Ohm, B. De Sutter {\it et al.},     \newblock {\sc Z. Phys.} {\bf A 340}, 223 (1991).
\bibitem[1991Ma05]{1991Ma05} H. Mach, F.K. Wohn, G. Molnar {\it et al.},     \newblock {\sc Nucl. Phys.} {\bf A 523}, 197 (1991).
\bibitem[1991Mc04]{1991Mc04} F.K. McGowan, N.R. Johnson, I.Y. Lee {\it et al.},     \newblock {\sc Nucl. Phys.} {\bf A 530}, 490 (1991).
\bibitem[1991Pe07]{1991Pe07} R.J. Peterson, J.J. Kraushaar, M.R. Braunstein, J.H. Mitchell,     \newblock {\sc Phys. Rev.} {\bf C 44}, 136 (1991).
\bibitem[1991ViZW]{1991ViZW} I.N. Vishnevsky, M.F. Kudoyarov, E.V. Kuzmin {\it et al.},     \newblock {\sc Program and Thesis, Proc. 41st Ann. Conf. Nucl. Spectrosc. Struct. At. Nuclei, Minsk}, 71 (1991).
\bibitem[1991We15]{1991We15} J. Wesseling, C.W. de Jager, J.B. van der Laan {\it et al.},     \newblock {\sc Nucl. Phys.} {\bf A 535}, 285 (1991).
\bibitem[1991Wu05]{1991Wu05} C.Y. Wu, D. Cline, E.G. Vogt {\it et al.},     \newblock {\sc Nucl. Phys.} {\bf A 533}, 359 (1991).
\bibitem[1992De29]{1992De29} C.C. Dey, N.R. Das, B.K. Sinha, R. Bhattacharya,     \newblock {\sc Can. J. Phys.} {\bf 70}, 268 (1992).
\bibitem[1992De60]{1992De60} A. Dewald,     \newblock {\sc Prog. Part. Nucl. Phys.} {\bf 28}, 409 (1992).
\bibitem[1992Dr05]{1992Dr05} Ch. Droste, T. Morek, S.G. Rohozinski {\it et al.},     \newblock {\sc J. Phys. (London)} {\bf G 18}, 1763 (1992).
\bibitem[1992Fa01]{1992Fa01} C. Fahlander, I. Thorslund, B. Varnestig {\it et al.},     \newblock {\sc Nucl. Phys.} {\bf A 537}, 183 (1992).
\bibitem[1992Fa05]{1992Fa05} C. Fahlander, B. Varnestig, A. Backlin {\it et al.},     \newblock {\sc Nucl. Phys.} {\bf A 541}, 157 (1992).
\bibitem[1992Ki10]{1992Ki10} W. Kim, B.L. Miller, J.R. Calarco {\it et al.},     \newblock {\sc Phys. Rev.} {\bf C 45}, 2290 (1992).
\bibitem[1992Ki20]{1992Ki20} W. Kim, J.P. Connelly, J.H. Heisenberg {\it et al.},     \newblock {\sc Phys. Rev.} {\bf C 46}, 1656 (1992).
\bibitem[1992Li14]{1992Li14} C.S. Lim, R.H. Spear, M.P. Fewell, G.J. Gyapong,     \newblock {\sc Nucl. Phys.} {\bf A 548}, 308 (1992).
\bibitem[1992MaZR]{1992MaZR} P.F. Mantica, P.K. Joshi, S.J. Robinson {\it et al.},     \newblock {\sc Contrib. 6th Intern. Conf. on Nuclei Far from Stability + 9th Intern. Conf. on Atomic Masses and Fundamental Constant}, Bernkastel-Kues, Germany, PE27 (1992).
\bibitem[1992Mc02]{1992Mc02} F.K. McGowan, N.R. Johnson, C. Baktash {\it et al.},     \newblock {\sc Nucl. Phys.} {\bf A 539}, 276 (1992).
\bibitem[1992Mo13]{1992Mo13} T. Morikawa, M. Oshima, T. Sekine {\it et al.},     \newblock {\sc Phys. Rev.} {\bf C 46}, R6 (1992).
\bibitem[1992Pe06]{1992Pe06} P. Petkov, S. Harissopulos, A. Dewald {\it et al.},     \newblock {\sc Nucl. Phys.} {\bf A 543}, 589 (1992).
\bibitem[1992Po09]{1992Po09} V.Yu. Ponomarev, W.T.A. Borghols, S.A. Fayans {\it et al.},     \newblock {\sc Nucl. Phys.} {\bf A 549}, 180 (1992).
\bibitem[1992Va06]{1992Va06} J.R. Vanhoy, M.T. McEllistrem, S.F. Hicks {\it et al.},    \newblock {\sc Phys. Rev.} {\bf C 45}, 1628 (1992).
\bibitem[1992Wi06]{1992Wi06} J.E. Wise, J.P. Connelly, F.W. Hersman {\it et al.},     \newblock {\sc Phys. Rev.} {\bf C 45}, 2701 (1992).
\bibitem[1993Be03]{1993Be03} T. Belgya, D. Seckel, E.L. Johnson {\it et al.},     \newblock {\sc Phys. Rev.} {\bf C 47}, 392 (1993).
\bibitem[1993Ch05]{1993Ch05} W.-T. Chou, D.S. Brenner, R.F. Casten, R.L. Gill,     \newblock {\sc  Phys.Rev.} {\bf C 47}, 157 (1993).
\bibitem[1993Ch41]{1993Ch41} A.A. Chishti, P. Chowdhury, D.J. Blumenthal {\it et al.},     \newblock {\sc Phys. Rev.} {\bf C 48}, 2607 (1993).
\bibitem[1993Ga16]{1993Ga16} R.A. Gatenby, E.L. Johnson, E.M. Baum {\it et al.},     \newblock {\sc Nucl. Phys.} {\bf A 560}, 633 (1993).
\bibitem[1993Ho12]{1993Ho12} D.J. Horen, R.L. Auble, C.Y. Wu {\it et al.},     \newblock {\sc Phys. Rev.} {\bf C 48}, 433 (1993).
\bibitem[1993Pe10]{1993Pe10} R. Perrino, N. Blasi, R. De Leo {\it et al.},     \newblock {\sc Nucl. Phys.} {\bf A 561}, 343 (1993).
\bibitem[1993Pi16]{1993Pi16} M. Piiparinen, R. Julin, S. Juutinen {\it et al.},     \newblock {\sc Nucl. Phys.} {\bf A 565}, 671 (1993).
\bibitem[1993Sa06]{1993Sa06} R.K.J. Sandor, H.P. Blok, M. Girod  {\it et al.},     \newblock {\sc Nucl. Phys.} {\bf A 551}, 349 (1993).
\bibitem[1993Sa38]{1993Sa38} G.P.S. Sahota, V.K. Mittal, H.S. Sahota,     \newblock {\sc J. Phys. Soc. Jpn.} {\bf 62}, 2958 (1993).
\bibitem[1993SaZT]{1993SaZT} P. Sala da Milano,     \newblock {\sc Thesis}, Universitat Kohn (1993).
\bibitem[1993Se08]{1993Se08} T. Seo,     \newblock {\sc Nucl. Instrum. Methods Phys. Res.} {\bf A 325}, 176 (1993).
\bibitem[1993Sp01]{1993Sp01} K.-H. Speidel, H. Busch, S. Kremeyer {\it et al.},     \newblock {\sc Nucl. Phys.} {\bf A 552}, 140 (1993).
\bibitem[1993Sr01]{1993Sr01} J. Srebrny, T. Czosnyka, W. Karczmarczyk {\it et al.},     \newblock {\sc  Nucl. Phys.} {\bf A 557}, 663c (1993).
\bibitem[1993Su16]{1993Su16} M. Sugawara, H. Kusakari, T. Morikawa {\it et al.},     \newblock {\sc   Nucl. Phys.} {\bf A 557}, 653c (1993).
\bibitem[1993Wo05]{1993Wo05} H.J. Wollersheim, H. Emling, H. Grein {\it et al.},     \newblock {\sc Nucl. Phys.} {\bf A 556}, 261 (1993).
\bibitem[1994Ch28]{1994Ch28} S. Chattopadhyay, H.C. Jain, M.L. Jhingan, C.R. Praharaj, \newblock {\sc  Phys. Rev.} {\bf C 50}, 93 (1994).
\bibitem[1994Go25]{1994Go25} K. Govaert, L. Govor, E. Jacobs {\it et al.},     \newblock {\sc Phys. Lett.} {\bf B 335}, 113 (1994).
\bibitem[1994Jo13]{1994Jo13} P.K. Joshi, E.F. Zganjar, D. Rupnik {\it et al.},     \newblock {\sc Int. J. Mod. Phys.} {\bf E 3}, 757 (1994).
\bibitem[1994Mc06]{1994Mc06} F.K. McGowan, N.R. Johnson, M.N. Rao {\it et al.},     \newblock {\sc Nucl. Phys.} {\bf A 580}, 335 (1994).
\bibitem[1994Pe02]{1994Pe02} P. Petkov, R. Kruecken, A. Dewald {\it et al.},     \newblock {\sc Nucl. Phys.} {\bf A 568}, 572 (1994).
\bibitem[1994Th01]{1994Th01} I. Thorslund, C. Fahlander, J. Nyberg {\it et al.},     \newblock {\sc Nucl. Phys.} {\bf A 568}, 306 (1994).
\bibitem[1995An15]{1995An15} S.S. Anderssen, A.E. Stuchbery, S. Kuyucak,     \newblock {\sc Nucl. Phys.} {\bf A 593}, 212 (1995).
\bibitem[1995Ef01]{1995Ef01} A.D. Efimov, M.F. Kudoyarov, A.S. Li,     \newblock {\sc Yad. Fiz.} {\bf 58}, No 1, 3 (1995); {\sc Phys. Atomic Nuclei} {\bf 58}, 1 (1995).
\bibitem[1995He25]{1995He25} R.-D. Herzberg, I. Bauske, P. von Brentano {\it et al.},     \newblock {\sc Nucl. Phys.} {\bf A 592}, 211 (1995).
\bibitem[1995Ka29]{1995Ka29} A.E. Kavka, C. Fahlander, A. Backlin {\it et al.},     \newblock {\sc Nucl. Phys.} {\bf A 593}, 177 (1995).
\bibitem[1995Ma03]{1995Ma03} H. Mach, B. Fogelberg,     \newblock {\sc Phys. Rev.} {\bf C 51}, 509 (1995).
\bibitem[1995Ma96]{1995Ma96} A. Makishima, T. Ishii, M. Ogawa, M. Ishii,     \newblock {\sc Nucl. Instrum. Methods Phys. Res.} {\bf A 363}, 591 (1995).
\bibitem[1995Mo16]{1995Mo16} T. Motobayashi, Y. Ikeda, Y. Ando {\it et al.},     \newblock {\sc  Phys.Lett.} {\bf B 346}, 9 (1995).
\bibitem[1995Sc24]{1995Sc24} S. Schoedder, G. Lhersonneau, A. Wohr {\it et al.},     \newblock {\sc Z. Phys.} {\bf A 352}, 237 (1995).
\bibitem[1995Sv01]{1995Sv01} L.E. Svensson, C. Fahlander, L. Hasselgren {\it et al.},     \newblock {\sc Nucl. Phys.} {\bf A 584}, 547 (1995).
\bibitem[1995Va25]{1995Va25} J.R. Vanhoy, J.M. Anthony, B.M. Haas {\it et al.},     \newblock {\sc Phys. Rev.} {\bf C 52}, 2387 (1995).
\bibitem[1995Vi05]{1995Vi05} A. Virtanen, N.R. Johnson, F.K. McGowan {\it et al.},     \newblock {\sc Nucl. Phys.} {\bf A 591}, 145 (1995).
\bibitem[1995Wa25]{1995Wa25} J.C. Walpe, B.F. Davis, S. Naguleswaran {\it et al.},     \newblock {\sc Phys. Rev.} {\bf C 52}, 1792 (1995).
\bibitem[1995Za01]{1995Za01} N.V. Zamfir, R.L. Gill, D.S. Brenner {\it et al.},     \newblock {\sc Phys. Rev.} {\bf C 51}, 98 (1995).
\bibitem[1996Al20]{1996Al20} I. Alfter, E. Bodenstedt, W. Knichel {\it et al.},     \newblock {\sc Z. Phys.} {\bf A 355}, 363 (1996).
\bibitem[1996Ch02]{1996Ch02} S. Chattopadhyay, H.C. Jain, J.A. Sheikh,     \newblock {\sc  Phys. Rev.} {\bf C 53}, 1001 (1996).
\bibitem[1996De50]{1996De50} A. Dewald, D. Weil, R. Krucken {\it et al.},     \newblock {\sc Phys. Rev.} {\bf C 54}, R2119 (1996).
\bibitem[1996Go36]{1996Go36} L.C. Gomes, L.B. Horodynski-Matsushigue, T. Borello-Lewin {\it et al.},     \newblock {\sc Phys. Rev.} {\bf C 54}, 2296 (1996).
\bibitem[1996Jo05]{1996Jo05} G.D. Johns, K.A. Christian, R.A. Kaye {\it et al.},     \newblock {\sc Phys. Rev.} {\bf C 53}, 1541 (1996).
\bibitem[1996Ma16]{1996Ma16} P.F. Mantica, W.B. Walters,     \newblock {\sc Phys. Rev.} {\bf C 53}, R2586 (1996).
\bibitem[1996Pe25]{1996Pe25} H. Penttila, P. Dendooven, A. Honkanen {\it et al.},     \newblock {\sc Phys. Rev.} {\bf C 54}, 2760 (1996).
\bibitem[1996Sc31]{1996Sc31} H. Scheit, T. Glasmacher, B.A. Brown {\it et al.},    \newblock {\sc Phys. Rev. Lett.} {\bf 77}, 3967 (1996).
\bibitem[1996Wu07]{1996Wu07} C.Y. Wu, D. Cline, T. Czosnyka {\it et al.},     \newblock {\sc Nucl. Phys.} {\bf A 607}, 178 (1996).
\bibitem[1997Bb08]{1997Bb08} C.J. Barton, R.L. Gill, R.F. Casten {\it et al.},     \newblock {\sc Nucl. Instrum. Methods Phys. Res.} {\bf A 391}, 289 (1997).
\bibitem[1997Gl02]{1997Gl02} T. Glasmacher, B.A. Brown, M.J. Chromik {\it et al.},    \newblock {\sc Phys. Lett.} {\bf B 395}, 163 (1997).
\bibitem[1997Ib01]{1997Ib01} R.W. Ibbotson, C.A. White, T. Czosnyka  {\it et al.},     \newblock {\sc Nucl. Phys.} {\bf A 619}, 213 (1997).
\bibitem[1997Ke07]{1997Ke07} J.H. Kelley, T. Suomijarvi, S.E. Hirzebruch {\it et al.},    \newblock {\sc Phys. Rev.} {\bf C 56}, R1206 (1997).
\bibitem[1997Pa07]{1997Pa07} S.D. Paul, H.C. Jain, J.A. Sheikh,     \newblock {\sc Phys. Rev.} {\bf C 55}, 1563 (1997).
\bibitem[1998De29]{1998De29} J. DeGraaf, M. Cromaz, T.E. Drake {\it et al.},     \newblock {\sc Phys. Rev.} {\bf C 58}, 164 (1998); Erratum Phys.Rev. {\bf C 59}, 1818 (1999).
\bibitem[1998Go03]{1998Go03} I.M. Govil, A. Kumar, H. Iyer {\it et al.},     \newblock {\sc Phys. Rev.} {\bf C 57}, 632 (1998).
\bibitem[1998Gu09]{1998Gu09} K. Gulda, and the ISOLDE Collaboration,     \newblock {\sc Nucl. Phys.} {\bf A 636}, 28 (1998).
\bibitem[1998Hi01]{1998Hi01} J.H. Hirata, S. Salem-Vasconcelos, M.J. Bechara {\it et al.},     \newblock {\sc Phys. Rev.} {\bf C 57}, 76 (1998).
\bibitem[1998Ib01]{1998Ib01} R.W. Ibbotson, T. Glasmacher, B.A. Brown {\it et al.},    \newblock {\sc Phys. Rev. Lett.} {\bf 80}, 2081 (1998).
\bibitem[1998Ka19]{1998Ka19} R.A. Kaye, J.B. Adams, A. Hale {\it et al.},     \newblock {\sc Phys. Rev.} {\bf C 57}, 2189 (1998).
\bibitem[1998Ka31]{1998Ka31} A. Kangasmaki, P. Tikkanen, J. Keinonen {\it et al.},    \newblock {\sc Phys. Rev.} {\bf C 58}, 699 (1998).
\bibitem[1998LhZZ]{1998LhZZ} G. Lhersonneau, P. Dendooven, A. Honkanen {\it et al.},     \newblock {\sc JYFL Ann. Rept.}, 1997, 21 (1998).
\bibitem[1998Si25]{1998Si25} K.P. Singh, D.C. Tayal, H.S. Hans,     \newblock {\sc Phys. Rev.} {\bf C58}, 1980 (1998).
\bibitem[1998Sk01]{1998Sk01} S. Skoda, F. Becker, T. Burkardt {\it et al.},     \newblock {\sc Nucl. Phys.} {\bf A 633}, 565 (1998).
\bibitem[1998StZX]{1998StZX} O. Stuch, K. Jessen, A. Dewald {\it et al.},     \newblock {\sc Contrib. Nuclear Structure '98}, Gatlinburg, 128 (1998).
\bibitem[1998Uc01]{1998Uc01} K. Uchiyama, K. Furuno, T. Shizuma {\it et al.},     \newblock {\sc Eur. Phys. J.} {\bf A 2}, 13 (1998).
\bibitem[1998We02]{1998We02} L. Weissman, M. Hass, C. Broude,     \newblock {\sc Phys. Rev.} {\bf C 57}, 621 (1998).
\bibitem[1998YaZR]{1998YaZR} Y. Yanagisawa, T. Motobayashi, S. Shimoura {\it et al.},     \newblock  {\sc Proc.Conf on Exotic Nuclei and Atomic Masses}, Bellaire, Michigan, June 23-27, 1998, 610 (1998); {\sc AIP Conf. Proc.} 455 (1998).
\bibitem[1999Au01]{1999Au01} T. Aumann, D. Aleksandrov, L. Axelsson {\it et al.},    \newblock {\sc Phys. Rev.} {\bf C 59}, 1252 (1999).
\bibitem[1999Co23]{1999Co23} P.D. Cottle, M. Fauerbach, T. Glasmacher {\it et al.},    \newblock {\sc Phys. Rev.} {\bf C 60}, 031301 (1999).
\bibitem[1999Kl11]{1999Kl11} T. Klemme, A. Fitzler, A. Dewald {\it et al.},     \newblock {\sc Phys. Rev.} {\bf C 60}, 034301 (1999).
\bibitem[1999Li18]{1999Li18} A. Lindroth, B. Fogelberg, H. Mach {\it et al.},     \newblock {\sc Phys. Rev. Lett.} {\bf 82}, 4783 (1999).
\bibitem[1999Ma63]{1999Ma63} F. Marechal, T. Suomijarvi, Y. Blumenfeld {\it et al.},    \newblock {\sc Phys. Rev.} {\bf C 60}, 034615 (1999).
\bibitem[1999Pr09]{1999Pr09} B.V. Pritychenko, T. Glasmacher, P.D. Cottle {\it et al.},    \newblock {\sc Phys. Lett.} {\bf B 461}, 322 (1999); {\sc Erratum Phys. Lett.} {\bf B 467}, 309 (1999).
\bibitem[1999To04]{1999To04} Y. Toh, S. Yamada, A. Taniguchi, Y. Kawase,     \newblock {\sc Eur. Phys. J.} {\bf A 4}, 233 (1999).
\bibitem[2000Br05]{2000Br05} J. Bryssinck, L. Govor, V.Yu. Ponomarev {\it et al.},     \newblock {\sc Phys. Rev.} {\bf C 61}, 024309 (2000).
\bibitem[2000En08]{2000En08} J. Enders, P. von Brentano, J. Eberth {\it et al.},     \newblock {\sc Nucl. Phys.} {\bf A 674}, 3 (2000).
\bibitem[2000Er06]{2000Er06} R. Ernst, K.-H. Speidel, O. Kenn {\it et al.},    \newblock {\sc Phys. Rev.} {\bf C 62}, 024305 (2000); {\sc Comment Phys. Rev.} {\bf C 64}, 069801 (2001).
\bibitem[2000Er01]{2000Er01} R. Ernst, K.-H. Speidel, O. Kenn {\it et al.},     \newblock {\sc Phys. Rev. Lett.} {\bf 84}, 416 (2000).
\bibitem[2000Ga08]{2000Ga08} A. Gade, I. Wiedenhover, J. Gableske {\it et al.},     \newblock {\sc Nucl. Phys.} {\bf A 665}, 268 (2000).
\bibitem[2000Iw02]{2000Iw02} H. Iwasaki, T. Motobayashi, H. Akiyoshi {\it et al.},    \newblock {\sc Phys. Lett.} {\bf 481B}, 7 (2000).
\bibitem[2000Kh02]{2000Kh02} B. Kharraja, U. Garg, S.S. Ghugre {\it et al.},     \newblock {\sc Phys. Rev.} {\bf C 61}, 024301 (2000).
\bibitem[2000Pe20]{2000Pe20} P. Petkov, A. Dewald, R. Kuhn {\it et al.},     \newblock {\sc Phys. Rev.} {\bf C 62}, 014314 (2000).
\bibitem[2000Ri15]{2000Ri15} L.A. Riley, P.D. Cottle, M. Fauerbach {\it et al.},    \newblock {\sc Phys. Rev.} {\bf C 62}, 034306 (2000).
\bibitem[2000Sp08]{2000Sp08} K.-H. Speidel, R. Ernst, O. Kenn {\it et al.},    \newblock {\sc Phys. Rev.} {\bf C 62}, 031301 (2000).
\bibitem[2000St07]{2000St07} O. Stuch, K. Jessen, R.S. Chakrawarthy {\it et al.},     \newblock {\sc Phys. Rev.} {\bf C 61}, 044325 (2000).
\bibitem[2000Th11]{2000Th11} P.G. Thirolf, B.V. Pritychenko, B.A. Brown {\it et al.},    \newblock {\sc Phys. Lett.} {\bf B 485}, 16 (2000).
\bibitem[2000To12]{2000To12} Y. Toh, T. Czosnyka, M. Oshima {\it et al.},     \newblock {\sc Eur. Phys. J.} {\bf A 9}, 353 (2000).
\bibitem[2001Ch56]{2001Ch56} V. Chiste, A. Gillibert, A. Lepine-Szily {\it et al.},    \newblock {\sc Phys. Lett.} {\bf B 514}, 233 (2001).
\bibitem[2001Co20]{2001Co20} P.D. Cottle, B.V. Pritychenko, J.A. Church {\it et al.},    \newblock {\sc Phys. Rev.} {\bf C 64}, 057304 (2001).
\bibitem[2001Ga52]{2001Ga52} J. Gableske, A. Dewald, H. Tiesler {\it et al.},     \newblock {\sc Nucl. Phys.} {\bf A 691}, 551 (2001).
\bibitem[2001Ge07]{2001Ge07} J. Genevey, J.A. Pinston, C. Foin {\it et al.},     \newblock {\sc Phys. Rev.} {\bf C 63}, 054315 (2001).
\bibitem[2001Ha09]{2001Ha09} S. Harissopulos, A. Dewald, A. Gelberg {\it et al.},     \newblock {\sc Nucl. Phys.} {\bf A 683}, 157 (2001).
\bibitem[2001Iw07]{2001Iw07} H. Iwasaki, T. Motobayashi, H. Sakurai {\it et al.},    \newblock {\sc Phys. Lett.} {\bf B 522}, 227 (2001).
\bibitem[2001Ke02]{2001Ke02} O. Kenn, K.-H. Speidel, R. Ernst {\it et al.},     \newblock {\sc Phys. Rev.} {\bf C 63}, 021302 (2001).
\bibitem[2001Ke08]{2001Ke08} O. Kenn, K.-H. Speidel, R. Ernst {\it et al.},     \newblock {\sc Phys. Rev.} {\bf C 63}, 064306 (2001).
\bibitem[2001Li24]{2001Li24} K.P. Lieb, D. Kast, A. Jungclaus {\it et al.},     \newblock {\sc Phys. Rev.} {\bf C 63}, 054304 (2001).
\bibitem[2001Me20]{2001Me20} T.J. Mertzimekis, N. Benczer-Koller, J. Holden {\it et al.},     \newblock {\sc Phys. Rev.} {\bf C 64}, 024314 (2001).
\bibitem[2001Mu19]{2001Mu19} G.A. Muller, A. Jungclaus, O. Yordanov {\it et al.},     \newblock {\sc Phys. Rev.} {\bf C 64}, 014305 (2001).
\bibitem[2001Mu25]{2001Mu25} G. Mukherjee, H.C. Jain, R. Palit {\it et al.},     \newblock {\sc Phys. Rev.} {\bf C 64}, 034316 (2001).
\bibitem[2001Pa03]{2001Pa03} R. Palit, H.C. Jain, P.K. Joshi {\it et al.},     \newblock {\sc Phys. Rev.} {\bf C 63}, 024313 (2001).
\bibitem[2001RyZZ]{2001RyZZ} N. Ryezayeva, \newblock {\sc  Thesis}, Technische Universitat Darmstadt, Germany (2001).
\bibitem[2001Wu03]{2001Wu03} C.Y. Wu, D. Cline, A.B. Hayes {\it et al.},     \newblock {\sc Phys. Rev.} {\bf C 64}, 014307 (2001); {\sc Comment Phys. Rev.} {\bf C 66}, 039801 (2002).
\bibitem[2002Ba28]{2002Ba28} M. Babilon, T. Hartmann, P. Mohr {\it et al.},    \newblock {\sc Phys. Rev.} {\bf C 65}, 037303 (2002).
\bibitem[2002Co09]{2002Co09} P.D. Cottle, Z. Hu, B.V. Pritychenko {\it et al.},    \newblock {\sc Phys. Rev. Lett.} {\bf 88}, 172502 (2002).
\bibitem[2002De26]{2002De26} G. de Angelis, A. Gadea, E. Farnea {\it et al.},     \newblock {\sc Phys. Lett.} {\bf B 535}, 93 (2002).
\bibitem[2002Go36]{2002Go36} I.M. Govil, A. Kumar, H. Iyer {\it et al.},     \newblock {\sc Phys. Rev.} {\bf C 66}, 064318 (2002).
\bibitem[2002Ha13]{2002Ha13} T. Hartmann, J. Enders, P. Mohr {\it et al.},    \newblock {\sc Phys. Rev.} {\bf C 65}, 034301 (2002).
\bibitem[2002Ja02]{2002Ja02} G. Jakob, N. Benczer-Koller, G. Kumbartzki {\it et al.},    \newblock {\sc  Phys. Rev.} {\bf C 65}, 024316 (2002).
\bibitem[2002Jo07]{2002Jo07} P.K. Joshi, H.C. Jain, R. Palit {\it et al.},     \newblock {\sc Nucl. Phys.} {\bf A 700}, 59 (2002).
\bibitem[2002Ka80]{2002Ka80} S. Kanno, T. Gomi, T. Motobayashi {\it et al.},    \newblock {\sc Prog. Theor. Phys. (Kyoto)}, Suppl. {\bf 146}, 575 (2002).
\bibitem[2002Ke02]{2002Ke02} O. Kenn, K.-H. Speidel, R. Ernst {\it et al.},     \newblock {\sc Phys. Rev.} {\bf C65}, 034308 (2002).
\bibitem[2002Kl07]{2002Kl07} H. Klein, A.F. Lisetskiy, N. Pietralla {\it et al.},     \newblock {\sc Phys. Rev.} {\bf C 65}, 044315 (2002).
\bibitem[2002Le17]{2002Le17} S. Leenhardt, O. Sorlin, M.G. Porquet {\it et al.},     \newblock {\sc Eur. Phys. J.} {\bf A 14}, 1 (2002).
\bibitem[2002Os07]{2002Os07} A. Osa, T. Czosnyka, Y. Utsuno {\it et al.},     \newblock {\sc Phys. Lett.} {\bf B 546}, 48 (2002).
\bibitem[2002Pa19]{2002Pa19} A.A. Pasternak, J. Srebrny, A.D. Efimov {\it et al.},     \newblock {\sc Eur. Phys. J.} {\bf A 13}, 435 (2002).
\bibitem[2002Ra21]{2002Ra21} D.C. Radford, C. Baktash, J.R. Beene {\it et al.},     \newblock {\sc Phys. Rev. Lett.} {\bf 88}, 222501 (2002).
\bibitem[2002Sh09]{2002Sh09} S.L. Shepherd, J. Simpson, A. Dewald {\it et al.},     \newblock {\sc Phys. Rev.} {\bf C 65}, 034320 (2002).
\bibitem[2002Sm10]{2002Sm10} A.G. Smith, R.M. Wall, D. Patel {\it et al.},     \newblock {\sc J.Phys.(London)} {\bf G 28}, 2307 (2002).
\bibitem[2002So03]{2002So03} O. Sorlin, S. Leenhardt, C. Donzaud {\it et al.},     \newblock {\sc Phys. Rev. Lett.} {\bf 88}, 092501 (2002).
\bibitem[2002We15]{2002We15} V. Werner, D. Belic, P. von Brentano {\it et al.},     \newblock {\sc Phys. Lett.} {\bf B 550}, 140 (2002).
\bibitem[2002Zi06]{2002Zi06} M. Zielinska, T. Czosnyka, J. Choinski {\it et al.},     \newblock {\sc  Nucl. Phys.} {\bf A 712}, 3 (2002).
\bibitem[2003Ba01]{2003Ba01} C.J. Barton, M.A. Caprio, D. Shapira {\it et al.},     \newblock {\sc Phys. Lett.} {\bf B 551}, 269 (2003).
\bibitem[2003Ca03]{2003Ca03} M.A. Caprio, N.V. Zamfir, E.A. McCutchan {\it et al.},     \newblock {\  Eur. Phys. J.} {\bf A 16}, 177 (2003).
\bibitem[2003De24]{2003De24} A. Dewald, R. Peusquens, B. Saha {\it et al.},     \newblock {\sc Phys. Rev.} {\bf C 68}, 034314 (2003).
\bibitem[2003En07]{2003En07} J. Enders, P. von Brentano, J. Eberth {\it et al.},     \newblock {\sc Nucl. Phys.} {\bf A 724}, 243 (2003).
\bibitem[2003Ga20]{2003Ga20} A. Gade, D. Bazin, C.M. Campbell {\it et al.},    \newblock {\sc Phys. Rev.} {\bf C 68}, 014302 (2003).
\bibitem[2003Ha15]{2003Ha15} T. Hayakawa, Y. Toh, M. Oshima {\it et al.},     \newblock {\sc Phys. Rev.} {\bf C 67}, 064310 (2003).
\bibitem[2003Ko51]{2003Ko51} M.Koizumi, A.Seki, Y.Toh {\it et al.},     \newblock {\sc Eur. Phys. J.} {\bf A 18}, 87 (2003).
\bibitem[2003Ku11]{2003Ku11} G. Kumbartzki, N. Benczer-Koller, J. Holden {\it et al.},     \newblock {\sc Phys. Lett.} {\bf B 562}, 193 (2003).
\bibitem[2003Mo02]{2003Mo02} O. Moller, K. Jessen, A. Dewald {\it et al.},    \newblock {\sc Phys. Rev.} {\bf C 67}, 011301 (2003).
\bibitem[2003Po02]{2003Po02} Zs. Podolyak, P.G. Bizzeti, A.M. Bizzeti-Sona {\it et al.},     \newblock {\sc Eur. Phys. J.} {\bf A 17}, 29 (2003).
\bibitem[2003Ri08]{2003Ri08} L.A. Riley, P.D. Cottle, M. Brown-Hayes {\it et al.},    \newblock {\sc Phys. Rev.} {\bf C 68}, 044309 (2003).
\bibitem[2003Sc19]{2003Sc19} S. Schielke, K.-H. Speidel, O. Kenn {\it et al.},    \newblock {\sc Phys.Lett.} {\bf B 567}, 153 (2003).
\bibitem[2003Sc21]{2003Sc21} S.Schielke, D.Hohn, K.-H.Speidel {\it et al.},     \newblock {\sc Phys. Lett.} {\bf B 571}, 29 (2003).
\bibitem[2003Sp04]{2003Sp04} K.-H. Speidel, S. Schielke, O. Kenn {\it et al.},     \newblock {\sc Phys. Rev.} {\bf C 68}, 061302 (2003).
\bibitem[2003Ya05]{2003Ya05} Y. Yanagisawa, M. Notani, H. Sakurai {\it et al.},    \newblock {\sc Phys. Lett.} {\bf B 566}, 84 (2003).
\bibitem[2004Im01]{2004Im01} N. Imai, H.J. Ong, N. Aoi {\it et al.},    \newblock {\sc Phys. Rev. Lett.} {\bf 92}, 062501 (2004); Comment Phys.Rev.Lett. {\bf 94}, 199201 (2005).
\bibitem[2004Ko03]{2004Ko03} M. Koizumi, A. Seki, Y. Toh {\it et al.},     \newblock {\sc Nucl. Phys.} {\bf A730}, 46 (2004).
\bibitem[2004Ra27]{2004Ra27} D.C. Radford, C. Baktash, J.R. Beene {\it et al.},     \newblock {\sc Nucl. Phys.} {\bf A 746}, 83c (2004).
\bibitem[2004Sa47]{2004Sa47} B. Saha, A. Dewald, O. Moller {\it et al.},     \newblock {\sc Phys. Rev.} {\bf C 70}, 034313 (2004); {\sc Erratum Phys. Rev.} {\bf C 71}, 039902 (2005); {\sc Comment Phys. Rev.} {\bf C 72}, 029801 (2005).
\bibitem[2004Yu07]{2004Yu07} K.L. Yurkewicz, D. Bazin, B.A. Brown {\it et al.},     \newblock {\sc Phys. Rev.} {\bf C 70}, 034301 (2004).
\bibitem[2004Yu10]{2004Yu10} K.L. Yurkewicz, D. Bazin, B.A. Brown {\it et al.},     \newblock {\sc Phys. Rev.} {\bf C 70}, 054319 (2004).
\bibitem[2005Bb09]{2005Bb09} A. Banu, J. Gerl, C. Fahlander {\it et al.},     \newblock {\sc Phys. Rev.} {\bf C 72}, 061305 (2005).
\bibitem[2005Bi02]{2005Bi02} D.C. Biswas, A.G. Smith, R.M. Wall {\it et al.},     \newblock {\sc  Phys. Rev.} {\bf C 71}, 011301 (2005); {\sc Erratum Phys. Rev.} {\bf C 71}, 019901 (2005).
\bibitem[2005Bu29]{2005Bu29} A. Burger, T.R. Saito, H. Grawe {\it et al.},     \newblock {\sc Phys. Lett.} {\bf B 622}, 29 (2005).
\bibitem[2005Ch66]{2005Ch66} J.A. Church, C.M. Campbell, D.-C. Dinca {\it et al.},    \newblock {\sc Phys. Rev.} {\bf C 72}, 054320 (2005).
\bibitem[2005Di05]{2005Di05} D.-C. Dinca, R.V.F. Janssens, A. Gade {\it et al.},    \newblock {\sc Phys. Rev.} {\bf C 71}, 041302 (2005).
\bibitem[2005Fo17]{2005Fo17} D. Fong, J.K. Hwang, A.V. Ramayya {\it et al.},     \newblock {\sc Eur. Phys. J.} {\bf A 25}, Supplement 1, 465 (2005).
\bibitem[2005Ga22]{2005Ga22} A. Gade, D. Bazin, A. Becerril {\it et al.},     \newblock {\sc Phys. Rev. Lett.} {\bf 95}, 022502 (2005); {\sc Erratum Phys. Rev. Lett.} {\bf 96}, 189901 (2006).
\bibitem[2005Go43]{2005Go43} A. Gorgen, E. Clement, A. Chatillon {\it et al.},     \newblock {\sc Eur. Phys. J.} {\bf A 26}, 153 (2005).
\bibitem[2005Hi04]{2005Hi04} S.F. Hicks, G.K. Alexander, C.A. Aubin {\it et al.},     \newblock {\sc Phys. Rev.} {\bf C 71}, 034307 (2005).
\bibitem[2005Iw02]{2005Iw02} H. Iwasaki, T. Motobayashi, H. Sakurai {\it et al.},    \newblock {\sc Phys. Lett.} {\bf B 620}, 118 (2005).
\bibitem[2005Iw03]{2005Iw03} H. Iwasaki, N. Aoi, S. Takeuchi {\it et al.},     \newblock {\sc Eur. Phys. J.} {\bf A 25}, Supplement 1, 415 (2005).
\bibitem[2005Le12]{2005Le12} J. Leske, K.-H. Speidel, S. Schielke {\it et al.},     \newblock {\sc Phys. Rev.} {\bf C 71}, 034303 (2005).
\bibitem[2005Le19]{2005Le19} J. Leske, K.-H. Speidel, S. Schielke {\it et al.},     \newblock {\sc Phys. Rev.} {\bf C 71}, 044316 (2005).
\bibitem[2005Le38]{2005Le38} J. Leske, K.-H. Speidel, S. Schielke {\it et al.},     \newblock {\sc  Phys. Rev.} {\bf C 72}, 044301 (2005).
\bibitem[2005Ma81]{2005Ma81} H. Mach, P.M. Walker, R. Julin {\it et al.},    \newblock {\sc J. Phys. (London)} {\bf G 31}, S1421 (2005).
\bibitem[2005Mo20]{2005Mo20} O. Moller, N. Warr, J. Jolie {\it et al.},     \newblock {\sc Phys. Rev.} {\bf C 71}, 064324 (2005).
\bibitem[2005Mo33]{2005Mo33} O. Moller, P. Petkov, B. Melon {\it et al.},     \newblock {\sc Phys. Rev.} {\bf C 72}, 034306 (2005).
\bibitem[2005Ni11]{2005Ni11} O. Niedermaier, H. Scheit, V. Bildstein {\it et al.},    \newblock {\sc Phys. Rev. Lett.} {\bf 94}, 172501 (2005).
\bibitem[2005NiZS]{2005NiZS} O.T. Niedermaier,    \newblock {\sc Thesis}, Univ. Heidelberg, Germany (2005).
\bibitem[2005Pa23]{2005Pa23} E. Padilla-Rodal, A. Galindo-Uribarri, C. Baktash {\it et al.},     \newblock {\sc Phys. Rev. Lett.} {\bf 94}, 122501 (2005).
\bibitem[2005Va31]{2005Va31} R.L. Varner, J.R. Beene, C. Baktash {\it et al.},     \newblock {\sc Eur. Phys. J.} {\bf A 25}, Supplement 1, 391 (2005).
\bibitem[2005Ya26]{2005Ya26} K. Yamada, T. Motobayashi, N. Aoi {\it et al.},     \newblock {\sc Eur. Phys. J.} {\bf A 25}, Supplement 1, 409 (2005).
\bibitem[2006Be04]{2006Be04} E. Becheva, Y. Blumenfeld, E. Khan {\it et al.},    \newblock {\sc Phys. Rev. Lett.} {\bf 96}, 012501 (2006).
\bibitem[2006Be18]{2006Be18} F. Becker, A. Petrovici, J. Iwanicki {\it et al.},     \newblock {\sc Nucl. Phys.} {\bf A 770}, 107 (2006).
\bibitem[2006Ch26]{2006Ch26} A. Chester, P. Adrich, A. Becerril {\it et al.},     \newblock {\sc Nucl. Instrum. Methods Phys. Res.} {\bf A 562}, 230 (2006).
\bibitem[2006Co20]{2006Co20} A. Costin, T. Ahn, B. Bochev {\it et al.},     \newblock {\sc Phys. Rev.} {\bf C 74}, 067301 (2006).
\bibitem[2006Ek01]{2006Ek01} A. Ekstrom, J. Cederkall, A. Hurst {\it et al.},     \newblock {\sc Phys. Scr.} {\bf T 125}, 190 (2006).
\bibitem[2006El03]{2006El03} Z. Elekes, Zs. Dombradi, A. Saito {\it et al.},    \newblock {\sc Phys. Rev.} {\bf C 73}, 044314 (2006).
\bibitem[2006El05]{2006El05} Z. Elekes, Zs. Dombradi, N. Aoi {\it et al.},    \newblock {\sc Phys. Rev.} {\bf C 74}, 017306 (2006).
\bibitem[2006Gr16]{2006Gr16} T. Grahn, A. Dewald, O. Moller {\it et al.},     \newblock {\sc Phys. Rev. Lett.} {\bf 97}, 062501 (2006).
\bibitem[2006Hw01]{2006Hw01} J.K. Hwang, A.V. Ramayya, J.H. Hamilton {\it et al.},     \newblock {\sc Phys. Rev.} {\bf C 73}, 044316 (2006).
\bibitem[2006Je04]{2006Je04} K. Jessen, O. Moller, A. Dewald {\it et al.},    \newblock {\sc Phys. Rev.} {\bf C 74}, 021304 (2006).
\bibitem[2006KrZV]{2006KrZV} Th.Kroll, and the REX-ISOLDE and MINIBALL Collaborations,     \newblock {\sc Proc. Frontiers in Nuclear Structure, Astrophysics, and Reactions}, Isle of Kos, Greece, 12-17 Sept. 2005, S.V Harissopulos, P.Demetriou, R.Julin, Eds., 119 (2006); {\sc AIP Conf. Proc.} 831 (2006).
\bibitem[2006Le24]{2006Le24} J. Leske, K.-H. Speidel, S. Schielke {\it et al.},     \newblock {\sc Phys. Rev.} {\bf C 73}, 064305 (2006).
\bibitem[2006Le31]{2006Le31} J. Leske, K.-H. Speidel, S. Schielke {\it et al.},     \newblock {\sc Phys. Rev.} {\bf C 74}, 024315 (2006).
\bibitem[2006Mo22]{2006Mo22} O. Moller, A. Dewald, P. Petkov {\it et al.},     \newblock {\sc Phys. Rev.} {\bf C 74}, 024313 (2006).
\bibitem[2006Mu04]{2006Mu04} W.F. Mueller, M.P. Carpenter, J.A. Church {\it et al.},     \newblock {\sc Phys. Rev.} {\bf C 73}, 014316 (2006).
\bibitem[2006Pe13]{2006Pe13} O. Perru, O. Sorlin, S. Franchoo {\it et al.},     \newblock {\sc Phys. Rev. Lett.} {\bf 96}, 232501 (2006).
\bibitem[2006Sp01]{2006Sp01} K.-H. Speidel, S. Schielke, J. Leske {\it et al.},    \newblock {\sc Phys. Lett.} {\bf B 632}, 207 (2006).
\bibitem[2006Sp02]{2006Sp02} K.-H. Speidel, J. Leske, S. Schielke {\it et al.},    \newblock {\sc Phys. Lett.} {\bf B 633}, 219 (2006).
\bibitem[2006Sr01]{2006Sr01} J. Srebrny, T. Czosnyka, Ch. Droste {\it et al.},     \newblock {\sc Nucl. Phys.} {\bf A 766}, 25 (2006).
\bibitem[2006YaZV]{2006YaZV} K. Yamada, N. Iwasa, S. Bishop {\it et al.},    \newblock {\sc RIKEN Accelerator Progress Report 2005}, 55 (2006).
\bibitem[2007Bo17]{2007Bo17} N. Boelaert, A. Dewald, C. Fransen {\it et al.},     \newblock {\sc Phys. Rev.} {\bf C 75}, 054311 (2007); {\sc Erratum Phys. Rev.} {\bf C 77}, 019901 (2008).
\bibitem[2007Ca35]{2007Ca35} C.M. Campbell, N. Aoi, D. Bazin {\it et al.},    \newblock {\sc Phys. Lett.} {\bf B 652}, 169 (2007); Addendum Phys.Lett. {\bf B 656}, 272 (2007).
\bibitem[2007Ce02]{2007Ce02} J. Cederkall, A. Ekstrom, C. Fahlander {\it et al.},     \newblock {\sc Phys. Rev. Lett.} {\bf 98}, 172501 (2007).
\bibitem[2007Cl02]{2007Cl02} E. Clement, A. Gorgen, W. Korten {\it et al.},     \newblock {\sc Phys. Rev.} {\bf C 75}, 054313 (2007).
\bibitem[2007Gi06]{2007Gi06} J. Gibelin, D. Beaumel, T. Motobayashi {\it et al.},    \newblock {\sc Phys. Rev.} {\bf C 75}, 057306 (2007).
\bibitem[2007Hu03]{2007Hu03} A.M. Hurst, P.A. Butler, D.G. Jenkins {\it et al.},     \newblock {\sc Phys. Rev. Lett.} {\bf 98}, 072501 (2007).
\bibitem[2007Ko23]{2007Ko23} J.J. Kolata, H. Amro, F.D. Becchetti {\it et al.},    \newblock {\sc Phys. Rev.} {\bf C 75}, 031302 (2007).
\bibitem[2007Kr12]{2007Kr12} Th.Kroll, on behalf of the MINIBALL and REX-ISOLDE Collaborations,     \newblock {\sc Phys. Atomic Nuclei} {\bf 70}, 1369 (2007).
\bibitem[2007Kr19]{2007Kr19} Th. Kroll, T. Behrens, R. Krucken {\it et al.},     \newblock {\sc Eur. Phys. J. Special Topics} {\bf 150}, 127 (2007).
\bibitem[2007Or04]{2007Or04} J.N. Orce, S.N. Choudry, B. Crider {\it et al.},     \newblock {\sc  Phys. Rev.} {\bf C 76}, 021302 (2007); {\sc Erratum Phys. Rev.} {\bf C 77}, 029902 (2008).
\bibitem[2007St16]{2007St16} K. Starosta, A. Dewald, A. Dunomes {\it et al.},     \newblock {\sc Phys. Rev. Lett.} {\bf 99}, 042503 (2007).
\bibitem[2007Su20]{2007Su20} T. Sugimoto, T. Nakamura, Y. Kondo {\it et al.},    \newblock {\sc Phys. Lett.} {\bf B 654}, 160 (2007).
\bibitem[2007Va20]{2007Va20} J. Van de Walle, F. Aksouh, F. Ames {\it et al.},     \newblock {\sc Phys. Rev. Lett.} {\bf 99}, 142501 (2007).
\bibitem[2007Va22]{2007Va22} C. Vaman, C. Andreoiu, D. Bazin {\it et al.},     \newblock {\sc Phys. Rev. Lett.} {\bf 99}, 162501 (2007).
\bibitem[2008Br18]{2008Br18} N. Bree, I. Stefanescu, P.A. Butler {\it et al.},     \newblock {\sc Phys. Rev.} {\bf C 78}, 047301 (2008).
\bibitem[2008De30]{2008De30} A. Dewald, K. Starosta, P. Petkov {\it et al.},     \newblock {\sc Phys. Rev.} {\bf C 78}, 051302 (2008).
\bibitem[2008Do19]{2008Do19} P. Doornenbal, P. Reiter, H. Grawe {\it et al.},     \newblock {\sc Phys. Rev.} {\bf C 78}, 031303 (2008).
\bibitem[2008Ek01]{2008Ek01} A. Ekstrom, J. Cederkall, C. Fahlander {\it et al.},     \newblock {\sc Phys. Rev. Lett.} {\bf 101}, 012502 (2008).
\bibitem[2008EkZZ]{2008EkZZ} A. Ekstrom, J. Cederkall, C. Fahlander {\it et al.},     \newblock {\sc Proc. Frontiers in Nuclear Structure, and Reactions (FINUSTAR 2)}, Crete, Greece, 10-14 Sept. 2007, P.Demetriou, R.Julin, S.V.Harissopulos, Eds. 296 (2008); {\sc AIP Conf. Proc} 1012 (2008).
\bibitem[2008Go25]{2008Go25} L.I. Govor, A.M. Demidov, V.A. Kurkin, I.V. Mikhailov,     \newblock {\sc Phys. Atomic Nuclei} {\bf 71}, 1339 (2008); {\sc Yad. Fiz.} {\bf 71}, 1367 (2008).
\bibitem[2008Gr04]{2008Gr04} T. Grahn, A. Dewald, O. Moller {\it et al.},     \newblock {\sc Nucl. Phys.} {\bf A 801}, 83 (2008).
\bibitem[2008Iw04]{2008Iw04} N. Iwasa, T. Motobayashi, S. Bishop {\it et al.},    \newblock {\sc Phys. Rev.} {\bf C 78}, 024306 (2008); {\sc Publishers's Note Phys. Rev.} {\bf C 78}, 029902 (2008).
\bibitem[2008KrZZ]{2008KrZZ} Th.Kroll, for the REX-ISOLDE and MINIBALL collaborations,     \newblock {\sc Proc. Frontiers in Nuclear Structure, and Reactions (FINUSTAR 2)}, Crete, Greece, 10-14 Sept. 2007, P.Demetriou, R.Julin, S.V.Harissopulos, Eds. 84 (2008); {\sc AIP Conf. Proc.} 1012 (2008).
\bibitem[2008Lj01]{2008Lj01} J. Ljungvall, A. Gorgen, M. Girod {\it et al.},     \newblock {\sc Phys. Rev. Lett.} {\bf 100}, 102502 (2008).
\bibitem[2008On02]{2008On02} H.J. Ong, N. Imai, D. Suzuki {\it et al.},    \newblock {\sc Phys. Rev.} {\bf C 78}, 014308 (2008).
\bibitem[2008Or02]{2008Or02} J.N. Orce, B. Crider, S. Mukhopadhyay {\it et al.},     \newblock {\sc Phys. Rev.} {\bf C 77}, 064301 (2008).
\bibitem[2008Sa05]{2008Sa05} M. Sanchez-Vega, H. Mach, R.B.E. Taylor {\it et al.},     \newblock {\sc Eur. Phys. J.} {\bf A 35}, 159 (2008).
\bibitem[2008Sa35]{2008Sa35} T.R. Saito, N. Saito, K. Starosta {\it et al.},     \newblock {\sc Phys. Lett.} {\bf B 669}, 19 (2008).
\bibitem[2008Sh23]{2008Sh23} T. Shizuma, T. Hayakawa, H. Ohgaki {\it et al.},     \newblock {\sc Phys. Rev.} {\bf C 78}, 061303 (2008).
\bibitem[2008Sp01]{2008Sp01} K.-H.Speidel, S.Schielke, J.Leske {\it et al.},    \newblock {\sc Phys. Lett.} {\bf B 659}, 101 (2008).
\bibitem[2008Sp04]{2008Sp04} K.-H. Speidel, S. Schielke, J. Leske {\it et al.},    \newblock {\sc Phys. Rev.} {\bf C 78}, 017304 (2008).
\bibitem[2008Wi04]{2008Wi04} M. Wiedeking, P. Fallon, A.O. Macchiavelli {\it et al.},    \newblock {\sc Phys. Rev. Lett.} {\bf 100}, 152501 (2008).
\bibitem[2009Ao01]{2009Ao01} N. Aoi, E. Takeshita, H. Suzuki {\it et al.},     \newblock {\sc Phys. Rev. Lett.} {\bf 102}, 012502 (2009).
\bibitem[2009Ek01]{2009Ek01} A. Ekstrom, J. Cederkall, D.D. DiJulio {\it et al.},     \newblock {\sc Phys. Rev.} {\bf C 80}, 054302 (2009).
\bibitem[2009El03]{2009El03} Z. Elekes, Zs. Dombradi, T. Aiba {\it et al.},    \newblock {\sc Phys. Rev.} {\bf C 79}, 011302 (2009).
\bibitem[2009FrZZ]{2009FrZZ} C. Fransen, A. Blazhev, A. Dewald {\it et al.},     \newblock {\sc Proc. 13th Intern. Symposium on Capture Gamma-Ray Spectroscopy and Related Topics}, Cologne, Germany, 25-29 Aug.2008, J.Jolie, A.Zilges, N.Warr, A.Blazhev, Eds., 529 (2009); {\sc AIP Conf. Proc.} {\bf 1090} (2009).
\bibitem[2009Gr08]{2009Gr08} T. Grahn, A. Dewald, P.T. Greenlees {\it et al.},     \newblock {\sc Phys. Rev.} {\bf C 80}, 014323 (2009).
\bibitem[2009Gr09]{2009Gr09} T. Grahn, A. Petts, M. Scheck {\it et al.},     \newblock {\sc Phys. Rev.} {\bf C 80}, 014324 (2009).
\bibitem[2009Im01]{2009Im01} N. Imai, N. Aoi, H.J. Ong {\it et al.},    \newblock {\sc Phys. Lett.} {\bf B 673}, 179 (2009).
\bibitem[2009Mc02]{2009Mc02} E.A. McCutchan, C.J. Lister, R.B. Wiringa {\it et al.},    \newblock {\sc Phys. Rev. Lett.} {\bf 103}, 192501 (2009).
\bibitem[2009Me23]{2009Me23} D. Mengoni, J.J. Valiente-Dobon, E. Farnea {\it et al.},     \newblock {\sc Eur. Phys. J.} {\bf A 42}, 387 (2009).
\bibitem[2009MuZU]{2009MuZU} D. Mucher, \newblock {\sc Thesis  Koln Universitat} (2009).
\bibitem[2009MuZW]{2009MuZW} D. Mucher, J. Iwanicki, J. Jolie {\it et al.},     \newblock {\sc Proc. 13th Intern.Symposium on Capture Gamma-Ray Spectroscopy and Related Topics}, Cologne, Germany, 25-29 Aug.2008, J.Jolie, A.Zilges, N.Warr, A.Blazhev, Eds., 587 (2009); {\sc AIP Conf.Proc.} 1090 (2009).
\bibitem[2009Ob02]{2009Ob02} A. Obertelli, T. Baugher, D. Bazin {\it et al.},     \newblock {\sc Phys. Rev.} {\bf C 80}, 031304 (2009).
\bibitem[2009Ra28]{2009Ra28} D. Radeck, A. Blazhev, M. Albers {\it et al.},     \newblock {\sc Phys. Rev.} {\bf C 80}, 044331 (2009).
\bibitem[2009Re20]{2009Re20} J.-M. Regis, Th. Materna, S. Christen {\it et al.},     \newblock {\sc Nucl. Instrum. Methods Phys. Res.} {\bf A 606}, 466 (2009).
\bibitem[2009Va01]{2009Va01} J. Van de Walle, F. Aksouh, T. Behrens {\it et al.},     \newblock {\sc Phys. Rev.} {\bf C 79}, 014309 (2009).
\bibitem[2009Va06]{2009Va06} J.J. Valiente-Dobon, D. Mengoni, A. Gadea {\it et al.},    \newblock {\sc Phys. Rev. Lett.} {\bf 102}, 242502 (2009).
\bibitem[2009Zi01]{2009Zi01} M. Zielinska, A. Gorgen, E. Clement {\it et al.},    \newblock {\sc Phys. Rev.} {\bf C 80}, 014317 (2009).
\bibitem[2010Ao01]{2010Ao01} N. Aoi, S. Kanno, S. Takeuchi {\it et al.},     \newblock {\sc Phys. Lett.} {\bf B 692} 302 (2010).
\bibitem[2010Be30]{2010Be30} L. Bettermann, J.-M. Regis, T. Materna {\it et al.},     \newblock {\sc Phys. Rev.} {\bf C 82}, 044310 (2010).
\bibitem[2010Bi11]{2010Bi11} P.G. Bizzeti, A.M. Bizzeti-Sona, D. Tonev {\it et al.},     \newblock {\sc Phys. Rev.} {\bf C 82}, 054311 (2010).
\bibitem[2010Ga14]{2010Ga14} A. Gade, T. Baugher, D. Bazin {\it et al.},     \newblock {\sc Phys. Rev.} {\bf C 81}, 064326 (2010).
\bibitem[2010Ku07]{2010Ku07} R. Kumar, P. Doornenbal, A. Jhingan {\it et al.},     \newblock {\sc Phys. Rev.} {\bf C 81}, 024306 (2010).
\bibitem[2010Lj01]{2010Lj01} J. Ljungvall, A. Gorgen, A. Obertelli {\it et al.},     \newblock {\sc Phys. Rev.} {\bf C 81}, 061301 (2010).
\bibitem[2010Me07]{2010Me07} D. Mengoni, J.J. Valiente-Dobon, A. Gadea {\it et al.},    \newblock {\sc Phys. Rev.} {\bf C 82}, 024308 (2010).
\bibitem[2010NaZY]{2010NaZY} D. Nagae, T. Ishii, R. Takahashi {\it et al.},     \newblock {\sc Proc. Intern. Symposium Exotic Nuclei}, Sochi, (Russia), 28 Sept.-2 Oct. 2009, Yu.E.Penionzhkevich, S.M.Lukyanov, Eds., 156 (2010); {\sc AIP Conf. Proc.} {\bf 1224} (2010).
\bibitem[2010Ru12]{2010Ru12} M. Rudigier, J.-M. Regis, J. Jolie {\it et al.},     \newblock {\sc Nucl. Phys.} {\bf A 847}, 89 (2010).
\bibitem[2010Sc03]{2010Sc03} M. Scheck, T. Grahn, A. Petts {\it et al.},     \newblock {\sc  Phys. Rev.} {\bf C 81}, 014310 (2010).
\bibitem[2010We12]{2010We12} V. Werner, J.R. Terry, M. Bunce, Z. Berant,     \newblock {\sc J. Phys.: Conf. Ser.} {\bf 205}, 012025 (2010).
\bibitem[2011Al25]{2011Al25} J.M. Allmond, D.C. Radford, C. Baktash {\it et al.},     \newblock {\sc Phys. Rev.} {\bf C 84}, 061303 (2011).
\bibitem[2011An04]{2011An04} V. Anagnostatou, P.H. Regan, M.R. Bunce {\it et al.},     \newblock {\sc Acta Phys. Pol.} {\bf B 42}, 807 (2011).
\bibitem[2011Ba37]{2011Ba37} T. Back, C. Qi, F. Ghazi Moradi {\it et al.},     \newblock {\sc Phys. Rev.} {\bf C 84}, 041306 (2011).
\bibitem[2011Ch05]{2011Ch05} A. Chakraborty, J.N. Orce, S.F. Ashley  {\it et al.},     \newblock {\sc Phys. Rev.} {\bf C 83}, 034316 (2011).
\bibitem[2011Cl03]{2011Cl03} E. Clement, G. De France, J.M. Casandjian {\it et al.},     \newblock {\sc Int. J. Mod. Phys.} {\bf E 20}, 415 (2011).
\bibitem[2011Da21]{2011Da21} M. Danchev, G. Rainovski, N. Pietralla {\it et al.},     \newblock {\sc Phys. Rev.} {\bf C 84}, 061306 (2011).
\bibitem[2011Di07]{2011Di07} D.D. DiJulio, J. Cederkall, C. Fahlander {\it et al.},     \newblock {\sc Eur. Phys. J.} {\bf A 47}, 25 (2011).
\bibitem[2011Ju01]{2011Ju01} A. Jungclaus, J. Walker, J. Leske {\it et al.},     \newblock {\sc Phys. Lett.} {\bf B 695}, 110 (2011).
\bibitem[2011Ku05]{2011Ku05} R.Kumar, P.Doornenbal, A.Jhingan {\it et al.},     \newblock {\sc Acta Phys. Pol.} {\bf B 42}, 813 (2011).
\bibitem[2011Mc01]{2011Mc01} E.A. McCutchan, C.J. Lister, T. Ahn {\it et al.},     \newblock {\sc Phys. Rev.} {\bf C 83}, 024310 (2011).
\bibitem[2011Ni03]{2011Ni03} M. Niikura, B. Mouginot, F. Azaiez  {\it et al.},     \newblock {\sc Acta Phys. Pol.} {\bf B 42}, 537 (2011).
\bibitem[2011Pe21]{2011Pe21} M. Petri, P. Fallon, A.O. Macchiavelli {\it et al.},    \newblock {\sc Phys. Rev. Lett.} {\bf 107}, 102501 (2011).
\bibitem[2011Pr10]{2011Pr10} M.G. Procter, D.M. Cullen, P. Ruotsalainen {\it et al.},     \newblock {\sc Phys. Rev.} {\bf C 84}, 024314 (2011).
\bibitem[2011ReZZ]{2011ReZZ} J.-M.Regis,     \newblock {\sc Thesis  Univ. Cologne} (2011).
\bibitem[2011Ro02]{2011Ro02} W. Rother, A. Dewald, H. Iwasaki {\it et al.},     \newblock {\sc Phys. Rev. Lett.} {\bf 106}, 022502 (2011).
\bibitem[2011Ro53]{2011Ro53} W. Rother, A. Dewald, G. Pascovici {\it et al.},     \newblock {\sc Nucl. Instrum. Methods Phys. Res.} {\bf A 654}, 196 (2011).
\bibitem[2011We08]{2011We08} V. Werner, N. Cooper, M. Bonett-Matiz {\it et al.},     \newblock {\sc J. Phys.: Conf. Ser.} {\bf 312}, 092062 (2011).
\bibitem[2012Al03]{2012Al03} M. Albers, N. Warr, K. Nomura {\it et al.},     \newblock {\sc Phys. Rev. Lett.} {\bf 108}, 062701 (2012); {\sc Erratum Phys. Rev. Lett.} {\bf 109}, 209904 (2012).
\bibitem[2012An17]{2012An17} V. Anagnostatou, P.H. Regan, V. Werner {\it et al.},     \newblock {\sc Appl. Radiat. Isot.} {\bf 70}, 1321 (2012).
\bibitem[2012Ba31]{2012Ba31} T. Baugher, A. Gade, R.V.F. Janssens {\it et al.},     \newblock {\sc Phys. Rev.} {\bf C 86}, 011305 (2012), Erratum Phys.Rev. {\bf C 86}, 049902 (2012).
\bibitem[2012Ba40]{2012Ba40} C. Bauer, T. Behrens, V. Bildstein {\it et al.},     \newblock {\sc Phys. Rev.} {\bf C 86}, 034310 (2012).
\bibitem[2012Gl01]{2012Gl01} K.A. Gladnishki, P. Petkov, A. Dewald {\it et al.},     \newblock {\sc Nucl. Phys.} {\bf A 877}, 19 (2012).
\bibitem[2012Ku14]{2012Ku14} G.J. Kumbartzki, K.-H. Speidel, N. Benczer-Koller {\it et al.},     \newblock {\sc Phys. Rev.} {\bf C 85}, 044322 (2012).
\bibitem[2012Ku24]{2012Ku24} G.J. Kumbartzki, N. Benczer-Koller, D.A. Torres {\it et al.},     \newblock {\sc Phys. Rev.} {\bf C 86}, 034319 (2012).
\bibitem[2012Le05]{2012Le05} A. Lemasson, H. Iwasaki, C. Morse {\it et al.},     \newblock {\sc Phys. Rev.} {\bf C 85}, 041303 (2012).
\bibitem[2012Li45]{2012Li45} K.-A.Li, Y.-L.Ye, H.Scheit {\it et al.},    \newblock {\sc Chin. Phys. Lett.} {\bf 29}, 102301 (2012).
\bibitem[2012Li50]{2012Li50} C.B. Li, X.G. Wu, X.F. Li {\it et al.},     \newblock {\sc Phys. Rev.} {\bf C 86}, 057303 (2012).
\bibitem[2012Lu03]{2012Lu03} R. Luttke, E.A. McCutchan,  V. Werner {\it et al.},     \newblock {\sc Phys. Rev.} {\bf C 85}, 017301 (2012).
\bibitem[2012MaZP]{2012MaZP} P.J.R. Mason, Zs. Podolyak, N. Marginean {\it et al.},     \newblock {\sc AIP Conf. Proc.} {\bf 1491}, 93 (2012).
\bibitem[2012Mc03]{2012Mc03} E.A. McCutchan, C.J. Lister, Steven C. Pieper, {\it et al.},    \newblock {\sc Phys. Rev.} {\bf C 86}, 014312 (2012).
\bibitem[2012Mo11]{2012Mo11} D. Montanari, S. Leoni, D. Mengoni {\it et al.},     \newblock {\sc Phys. Rev.} {\bf C 85}, 044301 (2012).
\bibitem[2012Ni09]{2012Ni09} M. Niikura, B. Mouginot, S. Franchoo {\it et al.},     \newblock {\sc Phys. Rev.} {\bf C 85}, 054321 (2012).
\bibitem[2012Pe16]{2012Pe16} M. Petri, S. Paschalis, R.M. Clark {\it et al.},    \newblock {\sc Phys. Rev.} {\bf C 86}, 044329 (2012).
\bibitem[2012Ra03]{2012Ra03} D. Radeck, V. Werner, G. Ilie {\it et al.},     \newblock {\sc Phys. Rev.} {\bf C 85}, 014301 (2012).
\bibitem[2012St03]{2012St03} I. Strojek, W. Czarnacki, W. Gawlikowicz {\it et al.},    \newblock {\sc Acta Phys. Pol.} {\bf B 43}, 339 (2012).
\bibitem[2012To01]{2012To01} D.A. Torres, G.J. Kumbartzki, Y.Y. Sharon {\it et al.},     \newblock {\sc Phys. Rev.} {\bf C 85}, 017305 (2012).
\bibitem[2012To06]{2012To06} Y. Togano, Y. Yamada, N. Iwasa {\it et al.},    \newblock {\sc Phys. Rev. Lett.} {\bf 108}, 222501 (2012).
\bibitem[2012Ts03]{2012Ts03} K. Tshoo, Y. Satou, H. Bhang {\it et al.},    \newblock {\sc Phys. Rev. Lett.} {\bf 109}, 022501 (2012).
\bibitem[2012Vo05]{2012Vo05} P. Voss, T. Baugher, D. Bazin {\it et al.},    \newblock {\sc Phys. Rev.} {\bf C 86}, 011303 (2012).
\bibitem[2012Wa16]{2012Wa16} J.C. Walpe, U. Garg, S. Naguleswaran {\it et al.},     \newblock {\sc Phys. Rev.} {\bf C 85}, 057302 (2012).
\bibitem[2012Wi05]{2012Wi05} R. Winkler, A. Gade, T. Baugher {\it et al.},    \newblock {\sc Phys. Rev. Lett.} {\bf 108}, 182501 (2012).
\bibitem[2013Al05]{2013Al05} M.Albers, K.Nomura, N.Warr {\it et al.},  \newblock {\sc  Nucl. Phys.} {\bf A 899}, 1 (2013).
\bibitem[2013Ba38]{2013Ba38} C. Bauer, G. Rainovski, N. Pietralla {\it et al.},     \newblock {\sc Phys. Rev.} {\bf C 88}, 021302 (2013).
\bibitem[2013Ba57]{2013Ba57} V.M. Bader, A. Gade, D. Weisshaar {\it et al.},     \newblock {\sc Phys. Rev.} {\bf C 88}, 051301(R) (2013).
\bibitem[2013Ce01]{2013Ce01} I. Celikovic, A. Dijon, E. Clement {\it et al.},     \newblock {\sc Acta Phys. Pol.} {\bf B 44}, 375 (2013).
\bibitem[2013Co23]{2013Co23} A. Corsi, J.-P. Delaroche, A. Obertelli {\it et al.},     \newblock {\sc Phys. Rev.} {\bf C 85}, 044311 (2013).
\bibitem[2013Cr02]{2013Cr02} H.L. Crawford, R.M. Clark, P. Fallon  {\it et al.},     \newblock {\sc Phys. Rev. Lett.} {\bf 110}, 242701 (2013).
\bibitem[2013DoZY]{2013DoZY} P. Doornenbal, S. Takeuchi, N. Aoi {\it et al.},     \newblock {\sc arXiv 1305.2877} (2013).
\bibitem[2013Ga23]{2013Ga23} L.P. Gaffney, P.A. Butler, M. Scheck {\it et al.},     \newblock {\sc Nature} {\bf 497}, 199 (2013).
\bibitem[2013Gu13]{2013Gu13} G. Guastalla, D.D. DiJulio, M. Gorska {\it et al.},     \newblock {\sc Phys. Rev. Lett.} {\bf 110}, 172501 (2013).
\bibitem[2013Lo04]{2013Lo04} C. Louchart, A. Obertelli, A. Gorgen {\it et al.},     \newblock {\sc Phys. Rev.} {\bf C 87}, 054302 (2013).
\bibitem[2013Ma66]{2013Ma66} P.J.R. Mason, Zs. Podolyak, N. Marginean {\it et al.},     \newblock {\sc Phys. Rev.} {\bf C 88}, 044391 (2013).
\bibitem[2013Pe16]{2013Pe16} E.E. Peters, A. Chakraborty, B.P. Crider {\it et al.},     \newblock {\sc Phys. Rev.} {\bf C 88}, 024317 (2013).
\bibitem[2013Sc01]{2013Sc01} A. Scheikh-Obeid, O. Burda, M. Chernykh {\it et al.},     \newblock {\sc Phys. Rev.} {\bf C 87}, 014337 (2013).
\bibitem[2013Sc20]{2013Sc20} M. Scheck, V.Yu. Ponomarev, M. Fritzsche {\it et al.},     \newblock {\sc Phys. Rev.} {\bf C 88}, 044304 (2013).
\bibitem[2013St24]{2013St24} A.E. Stuchbery, J.M. Allmond, A. Galindo-Uribarri {\it et al.},     \newblock {\sc Phys. Rev.} {\bf C 88}, 051304(R) (2013).
\bibitem[2013Su20]{2013Su20} H.Suzuki, N.Aoi, E.Takeshita {\it et al.},    \newblock {\sc Phys. Rev.} {\bf C 88}, 024326 (2013).


\bibitem[2014Al20]{2014Al20} J.M. Allmond, B.A. Brown, A.E. Stuchbery {\it et al.},  \newblock {\sc  Phys.Rev.} {\bf C 90}, 034309 (2014).
\bibitem[2014Br05]{2014Br05} N. Bree, K. Wrzosek-Lipska, A. Petts {\it et al.},  \newblock {\sc  Phys.Rev.Lett.} {\bf 112}, 162701 (2014).
\bibitem[2014Ca10]{2014Ca10} S.Calinescu, L.Caceres, S.Grevy {\it et al.},  \newblock {\sc   Acta Phys.Pol.} {\bf B 45}, 199 (2014).
\bibitem[2014Do19]{2014Do19} P. Doornenbal, S. Takeuchi, N. Aoi {\it et al.},  \newblock {\sc Phys. Rev.} {\bf C 90}, 061302(R) (2014).
\bibitem[2014Ga04]{2014Ga04} L.P. Gaffney, M. Hackstein, R.D. Page {\it et al.},  \newblock {\sc  Phys.Rev.} {\bf C 89}, 024307 (2014).
\bibitem[2014Il01]{2014Il01} S. Ilieva, M. Thurauf, Th. Kroll {\it et al.},  \newblock {\sc  Phys.Rev.} {\bf C 89}, 014313 (2014).
\bibitem[2014Iw01]{2014Iw01} H. Iwasaki, A. Lemasson, C. Morse {\it et al.},  \newblock {\sc  Phys.Rev.Lett.} {\bf 112}, 142502 (2014).
\bibitem[2014Li45]{2014Li45} C.B. Li, F.Q. Chen, X.G. Wu  {\it et al.},  \newblock {\sc  Phys.Rev.} {\bf C 90}, 047302 (2014).
\bibitem[2014Ma85]{2014Ma85} T. Marchi, G. de Angelis, J.J. Valiente-Dobon {\it et al.},  \newblock {\sc  Phys.Rev.Lett.} {\bf 113}, 182501 (2014).
\bibitem[2014Mi09]{2014Mi09} S. Michimasa, Y. Yanagisawa, K. Inafuku {\it et al.},  \newblock {\sc Phys. Rev.} {\bf C 89}, 054307 (2014).
\bibitem[2014Na15]{2014Na15} F. Naqvi, V. Werner, P. Petkov {\it et al.},  \newblock {\sc Phys. Lett.} {B 728}, 303 (2014).
\bibitem[2014Ni09]{2014Ni09} A.J. Nichols, R. Wadsworth, H. Iwasaki {\it et al.},  \newblock {\sc  Phys.Lett.} {\bf B 733}, 52 (2014).
\bibitem[2014Pl01]{2014Pl01} C. Plaisir, L. Gaudefroy, V. Meot {\it et al.},  \newblock {\sc  Phys.Rev.} {\bf C 89}, 021302 (2014).
\bibitem[2014Re15]{2014Re15}  J.-M. Regis, J. Jolie, N. Saed-Samii {\it et al.},  \newblock {\sc Phys. Rev.} {\bf C 90}, 067301 (2014).
\bibitem[2014Ri04]{2014Ri04} L.A. Riley, M.L. Agiorgousis, T.R. Baugher {\it et al.},  \newblock {\sc  Phys. Rev.} {\bf C 90}, 011305 (2014).
\bibitem[2014Sa49]{2014Sa49} M. Saxena, R. Kumar, A. Jhingan {\it et al.},  \newblock {\sc   Phys.Rev.} {\bf C 90}, 024316 (2014).

\bibitem[2015Be25]{2015Be25} F.L. Bello Garrote, A. Gorgen, J. Mierzejewski {\it et al.},  \newblock {\sc   Phys.Rev. C} {\bf 92}, 024317 (2015). 
\bibitem[2015Br03]{2015Br03} F. Browne, A.M. Bruce, T. Sumikama {\it et al.},  \newblock {\sc  Acta Phys.Pol. B} {\bf 46}, 721 (2015).
\bibitem[2015Br10]{2015Br10} T. Braunroth, A. Dewald, H. Iwasaki {\it et al.},  \newblock {\sc   Phys.Rev. C} {\bf 92}, 034306 (2015).
\bibitem[2015BrP]{2015BrP} A.M. Bruce, T. Sumikama, I. Nishizuka {\it et al.},  \newblock {\sc Phys. Lett. B} (2015), in Press. 
\bibitem[2015Do04]{2015Do04} M. Doncel, T. Back, D.M. Cullen {\it et al.} \newblock {\sc  Phys. Rev. C} {\bf 91}, 061304 (2015).
\bibitem[2015Ga19]{2015Ga19} L.P. Gaffney, A.P. Robinson, D.G. Jenkins {\it et al.},  \newblock {\sc  Phys. Rev. C} {\bf 91}, 064313 (2015).
\bibitem[2015Jo01]{2015Jo01} J. Jolie, J.-M. Regis, D. Wilmsen {\it et al.},  \newblock {\sc  Nucl. Phys. A} {\bf 934}, 1 (2015).
\bibitem[2015KeZZ]{2015KeZZ} N. Kesteloot,  \newblock {\sc Thesis}, Katholieke Universiteit Leuven Belgium (2015); N. Kesteloot,  B. Bastin, K. Auranen {\it et al.},  \newblock {\sc   Phys.Rev. C} (2015), in Press.
\bibitem[2015Li28]{2015Li28} K. Li, Y. Ye, T. Motobayashi {\it et al.},  \newblock {\sc  Phys. Rev. C} {\bf 92}, 014608 (2015)
\bibitem[2015Pa14]{2015Pa14} S. Pascu, D. Bucurescu, Gh. Cata-Danil {\it et al.},  \newblock {\sc  Phys. Rev. C} {\bf 91}, 034321 (2015).
\bibitem[2015Ru03]{2015Ru03} M. Rudigier, K. Nomura, M. Dannhoff {\it et al.},  \newblock {\sc  Phys. Rev. C} {\bf 91},  044301 (2015).
\bibitem[2015St08]{2015St08} R. Stegmann, C. Bauer, G. Rainovski   {\it et al.},  \newblock {\sc  Phys. Rev. C} {\bf 91},  054326 (2015).
\end{theDTbibliography}

\newpage

\begin{Dfigures}[ht!]
\begin{center}
\includegraphics[height=4in,width=\linewidth]{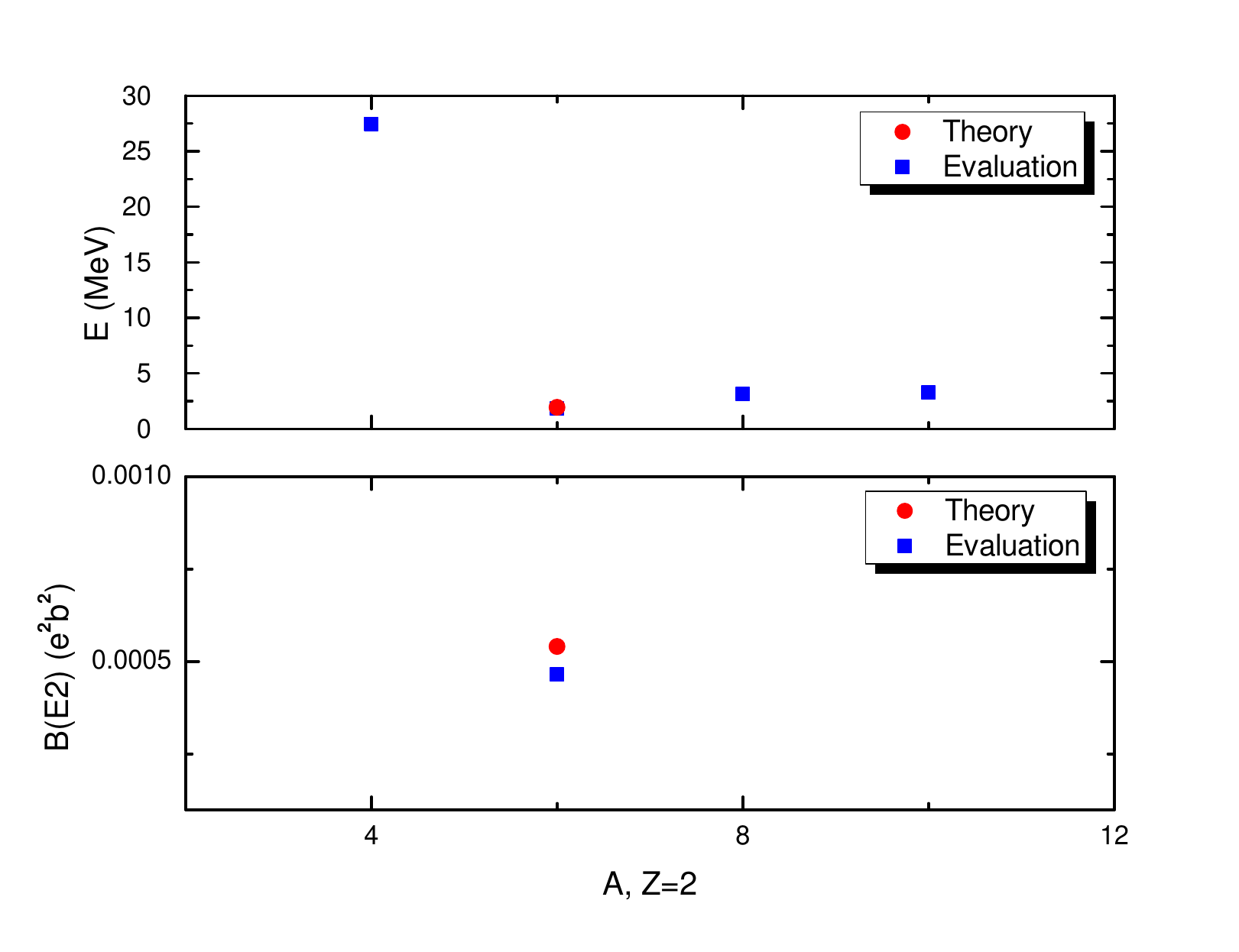}
\end{center}
\caption{Evaluated and shell model calculated energies, E($2^{+}_{1}$), and B(E; $0_{1}^{+} \rightarrow 2_{1}^{+}$) values for He nuclei.}\label{fig:graph2}
\end{Dfigures}

\begin{Dfigures}[ht!]
\begin{center}
\includegraphics[height=4in,width=\linewidth]{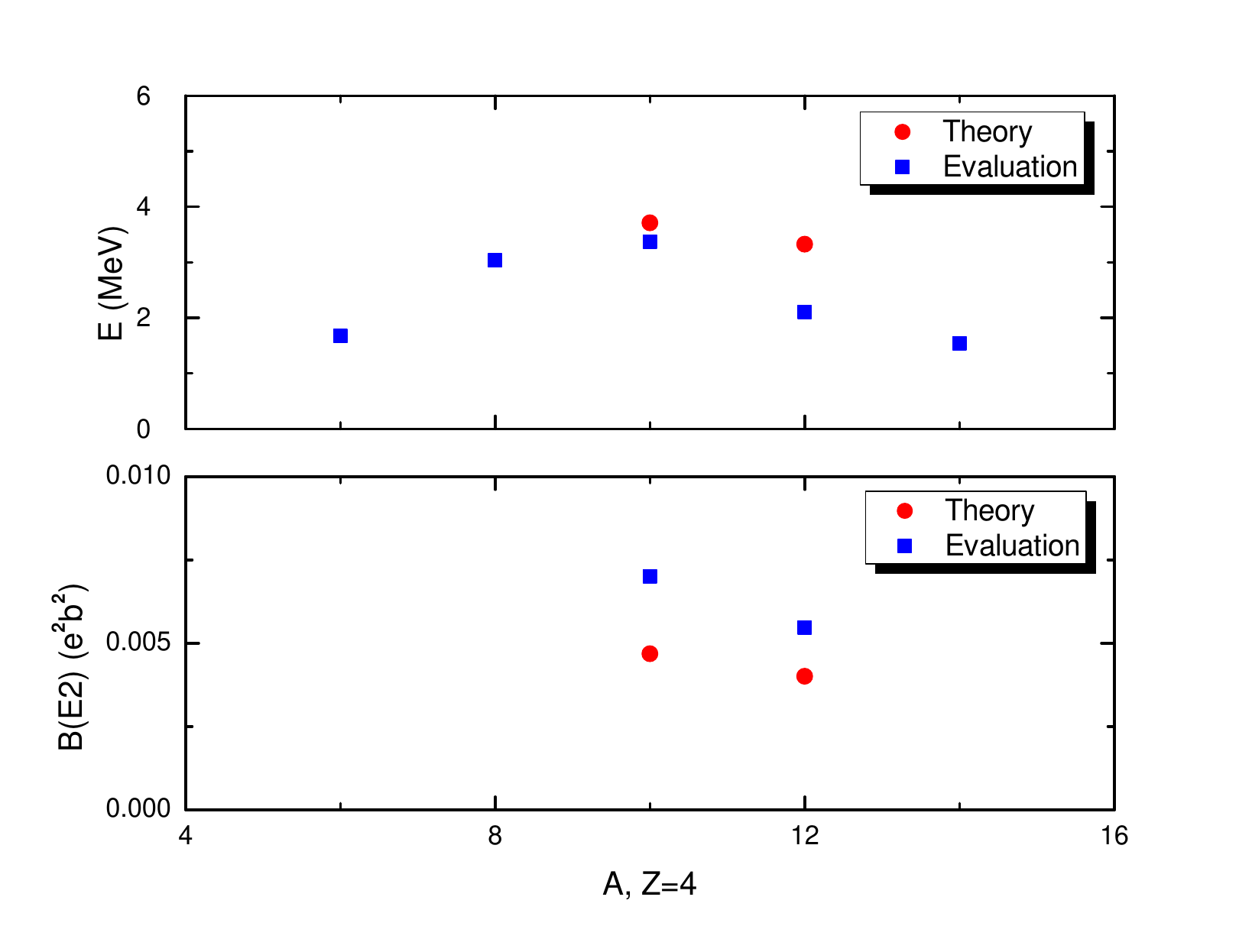}
\end{center}
\caption{Evaluated and shell model calculated energies, E($2^{+}_{1}$), and B(E; $0_{1}^{+} \rightarrow 2_{1}^{+}$) values for Be nuclei.}\label{fig:graph4}
\end{Dfigures}
\clearpage
\begin{Dfigures}[ht!]
\begin{center}
\includegraphics[height=4in,width=\linewidth]{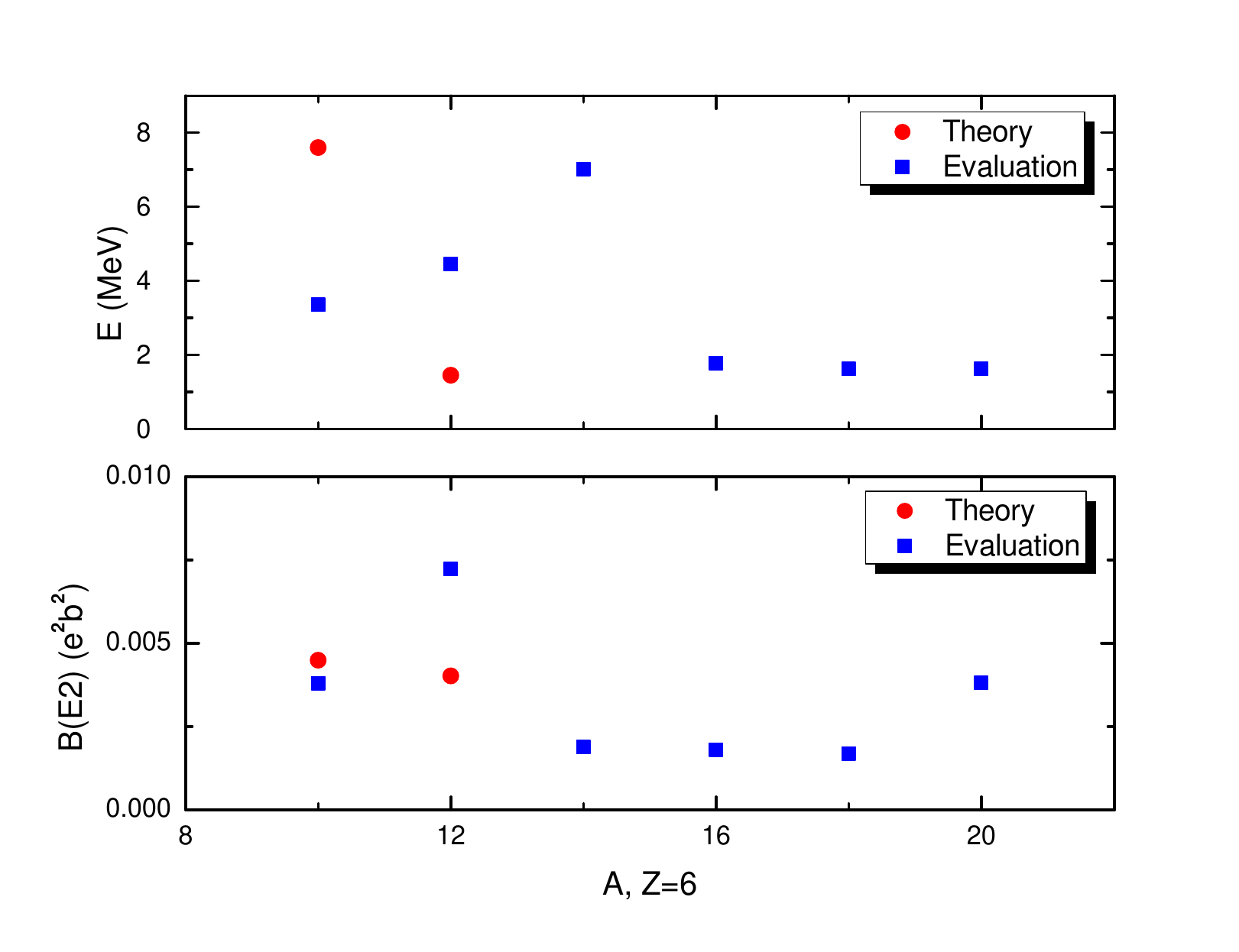}
\end{center}
\caption{Evaluated and shell model calculated energies, E($2^{+}_{1}$), and B(E; $0_{1}^{+} \rightarrow 2_{1}^{+}$) values for C nuclei.}\label{fig:graph6}
\end{Dfigures}

\begin{Dfigures}[ht!]
\begin{center}
\includegraphics[height=4in,width=\linewidth]{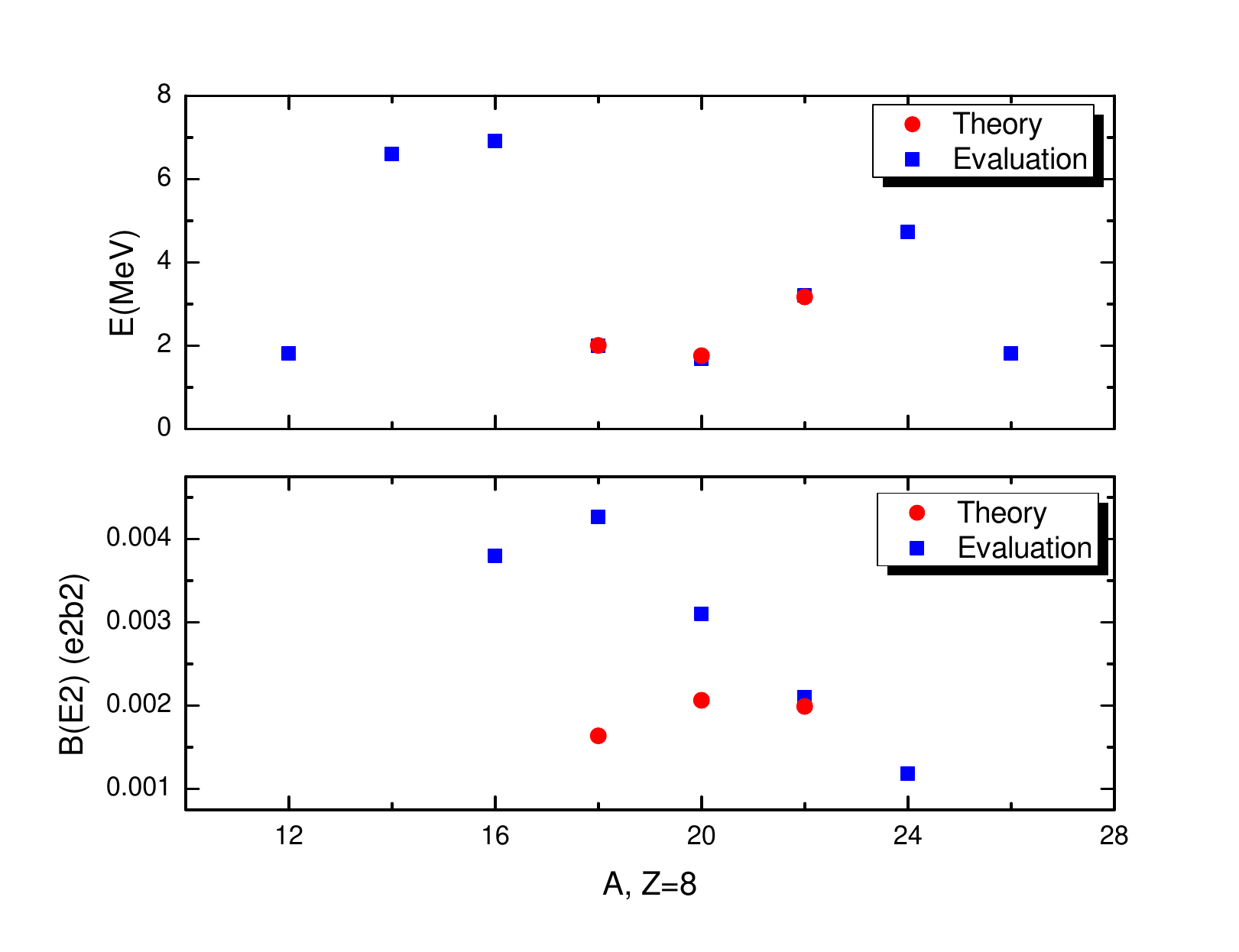}
\end{center}
\caption{Evaluated and shell model calculated energies, E($2^{+}_{1}$), and B(E; $0_{1}^{+} \rightarrow 2_{1}^{+}$) values for O nuclei.}\label{fig:graph8}
\end{Dfigures}
\clearpage
\begin{Dfigures}[ht!]
\begin{center}
\includegraphics[height=4in,width=\linewidth]{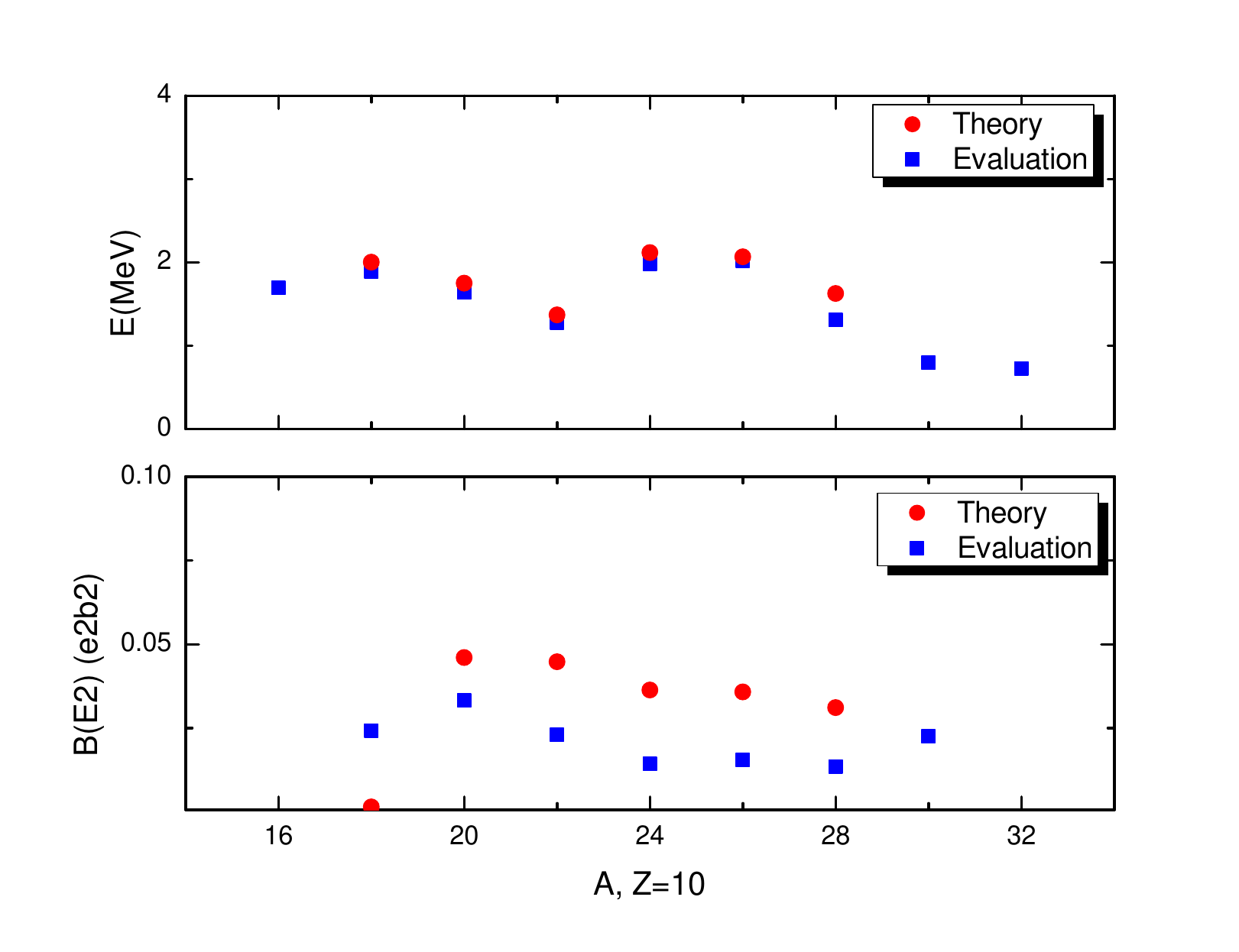}
\end{center}
\caption{Evaluated and shell model calculated energies, E($2^{+}_{1}$), and B(E; $0_{1}^{+} \rightarrow 2_{1}^{+}$) values for Ne nuclei.}\label{fig:graph10}
\end{Dfigures}

\begin{Dfigures}[ht!]
\begin{center}
\includegraphics[height=4in,width=\linewidth]{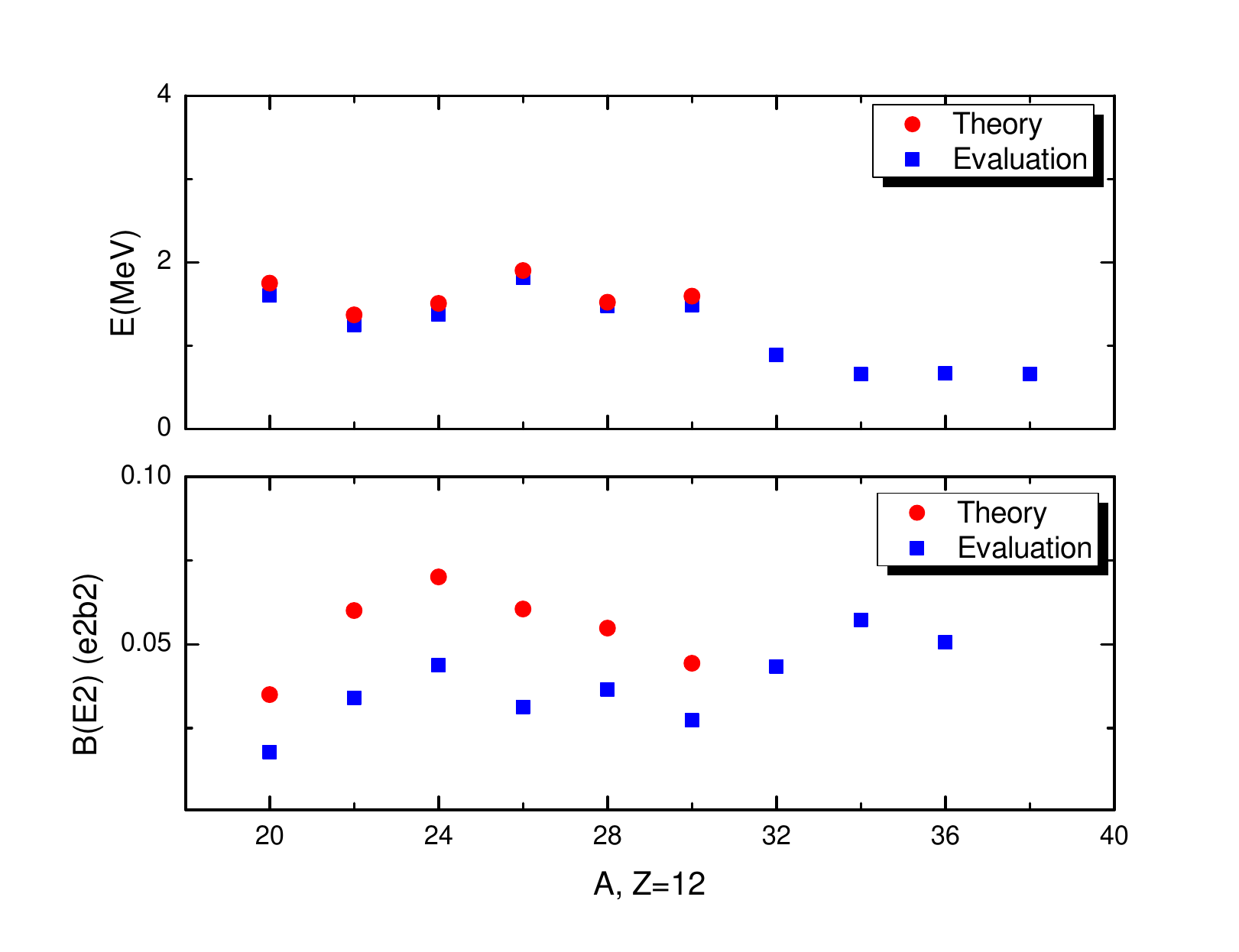}
\end{center}
\caption{Evaluated and shell model calculated energies, E($2^{+}_{1}$), and B(E; $0_{1}^{+} \rightarrow 2_{1}^{+}$) values for Mg nuclei.}\label{fig:graph12}
\end{Dfigures}
\clearpage
\begin{Dfigures}[ht!]
\begin{center}
\includegraphics[height=4in,width=\linewidth]{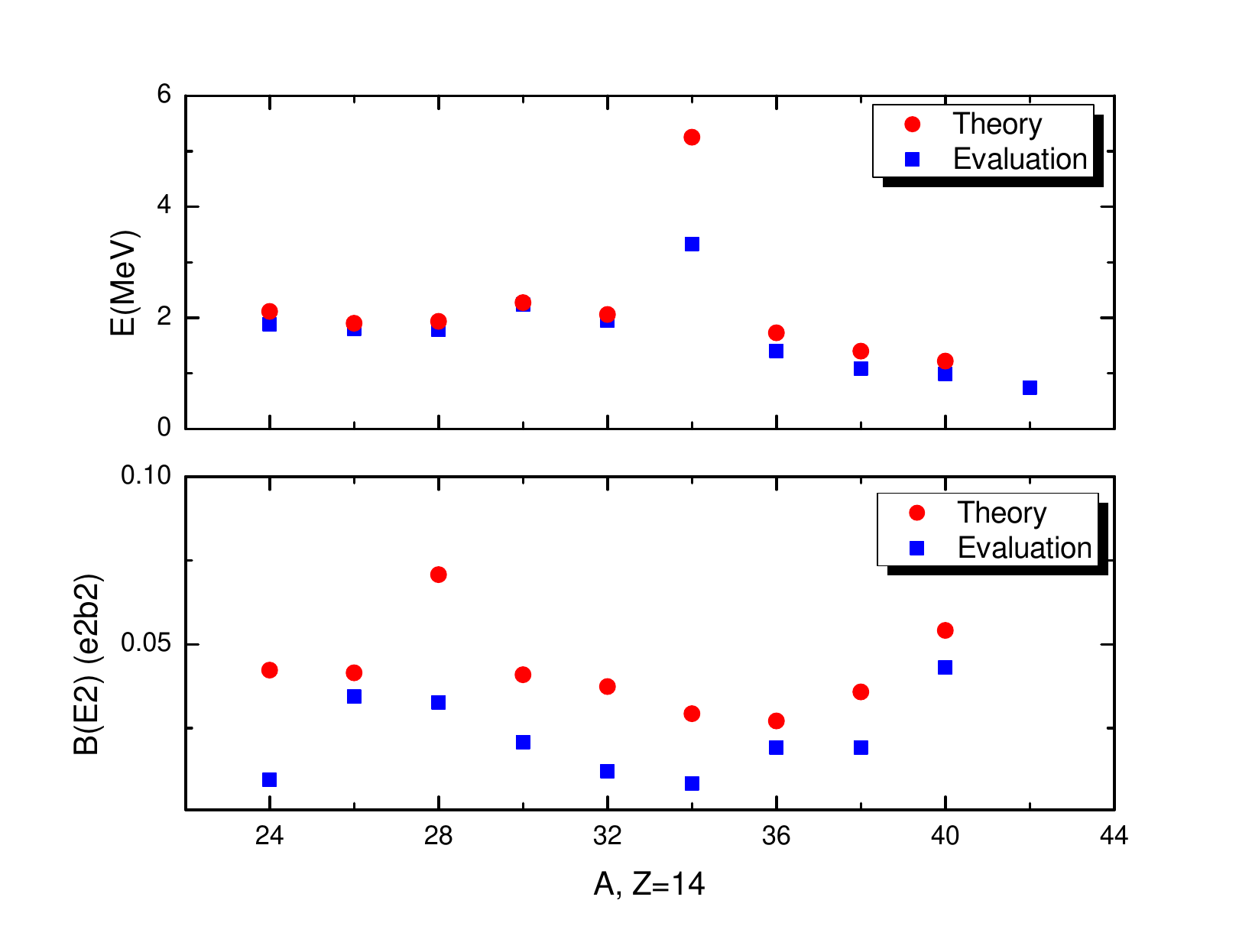}
\end{center}
\caption{Evaluated and shell model calculated energies, E($2^{+}_{1}$), and B(E; $0_{1}^{+} \rightarrow 2_{1}^{+}$) values for Si nuclei.}\label{fig:graph14}
\end{Dfigures}

\begin{Dfigures}[ht!]
\begin{center}
\includegraphics[height=4in,width=\linewidth]{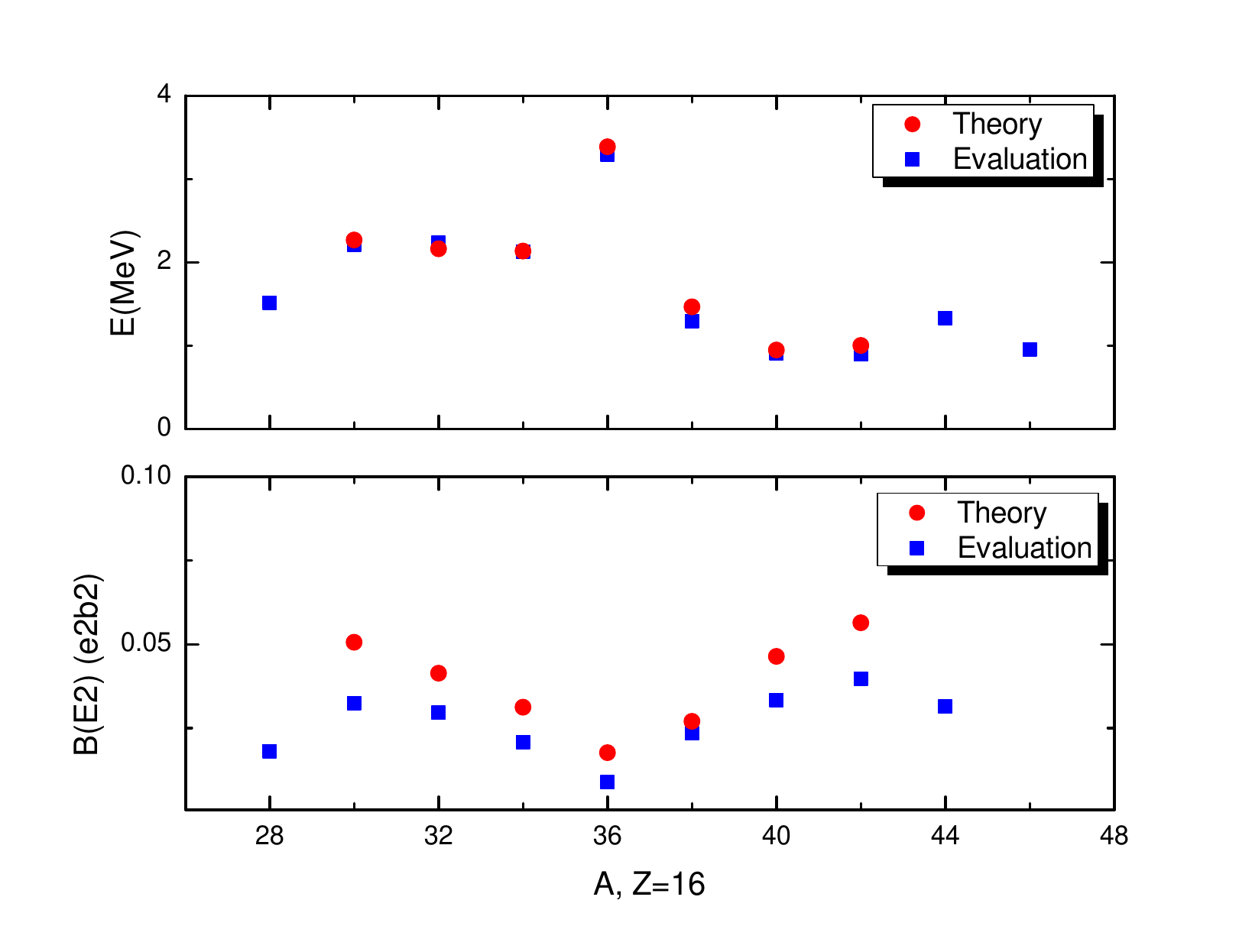}
\end{center}
\caption{Evaluated and shell model calculated energies, E($2^{+}_{1}$), and B(E; $0_{1}^{+} \rightarrow 2_{1}^{+}$) values for S nuclei.}\label{fig:graph16}
\end{Dfigures}
\clearpage
\begin{Dfigures}[ht!]
\begin{center}
\includegraphics[height=4in,width=\linewidth]{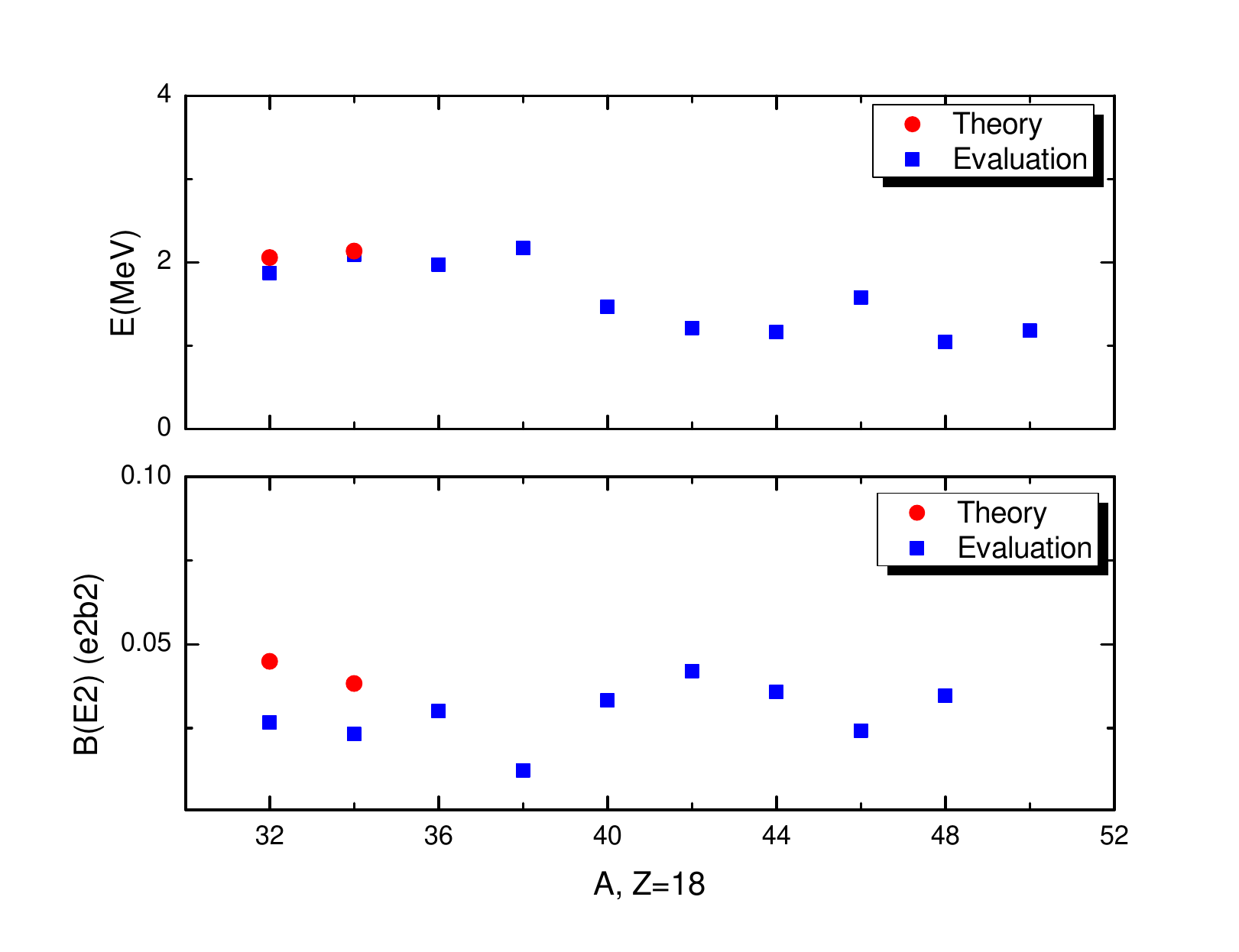}
\end{center}
\caption{Evaluated and shell model calculated energies, E($2^{+}_{1}$), and B(E; $0_{1}^{+} \rightarrow 2_{1}^{+}$) values for Ar nuclei.}\label{fig:graph18}
\end{Dfigures}

\begin{Dfigures}[ht!]
\begin{center}
\includegraphics[height=4in,width=\linewidth]{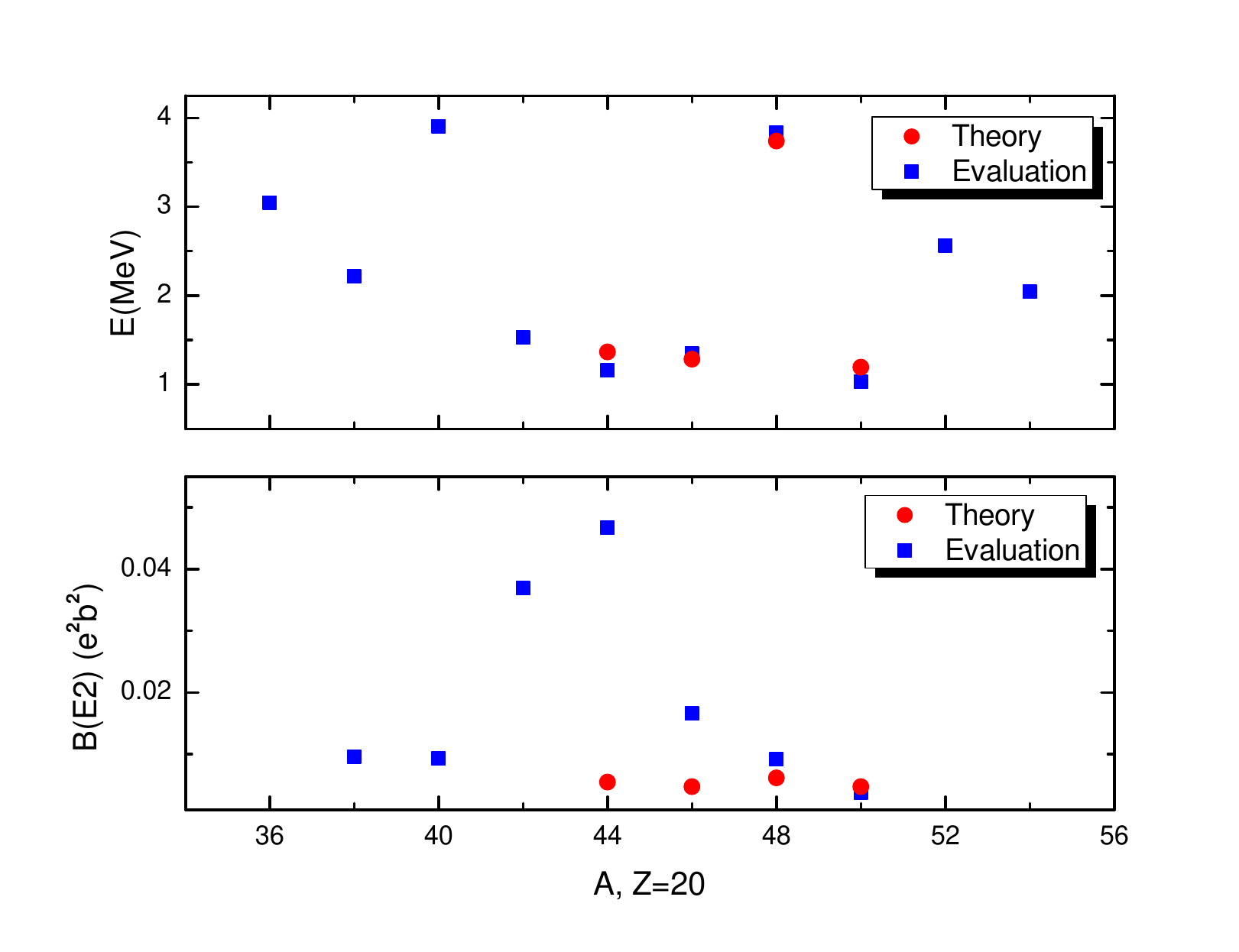}
\end{center}
\caption{Evaluated and shell model calculated energies, E($2^{+}_{1}$), and B(E; $0_{1}^{+} \rightarrow 2_{1}^{+}$) values for Ca nuclei.}\label{fig:graph20}
\end{Dfigures}
\clearpage
\begin{Dfigures}[ht!]
\begin{center}
\includegraphics[height=4in,width=\linewidth]{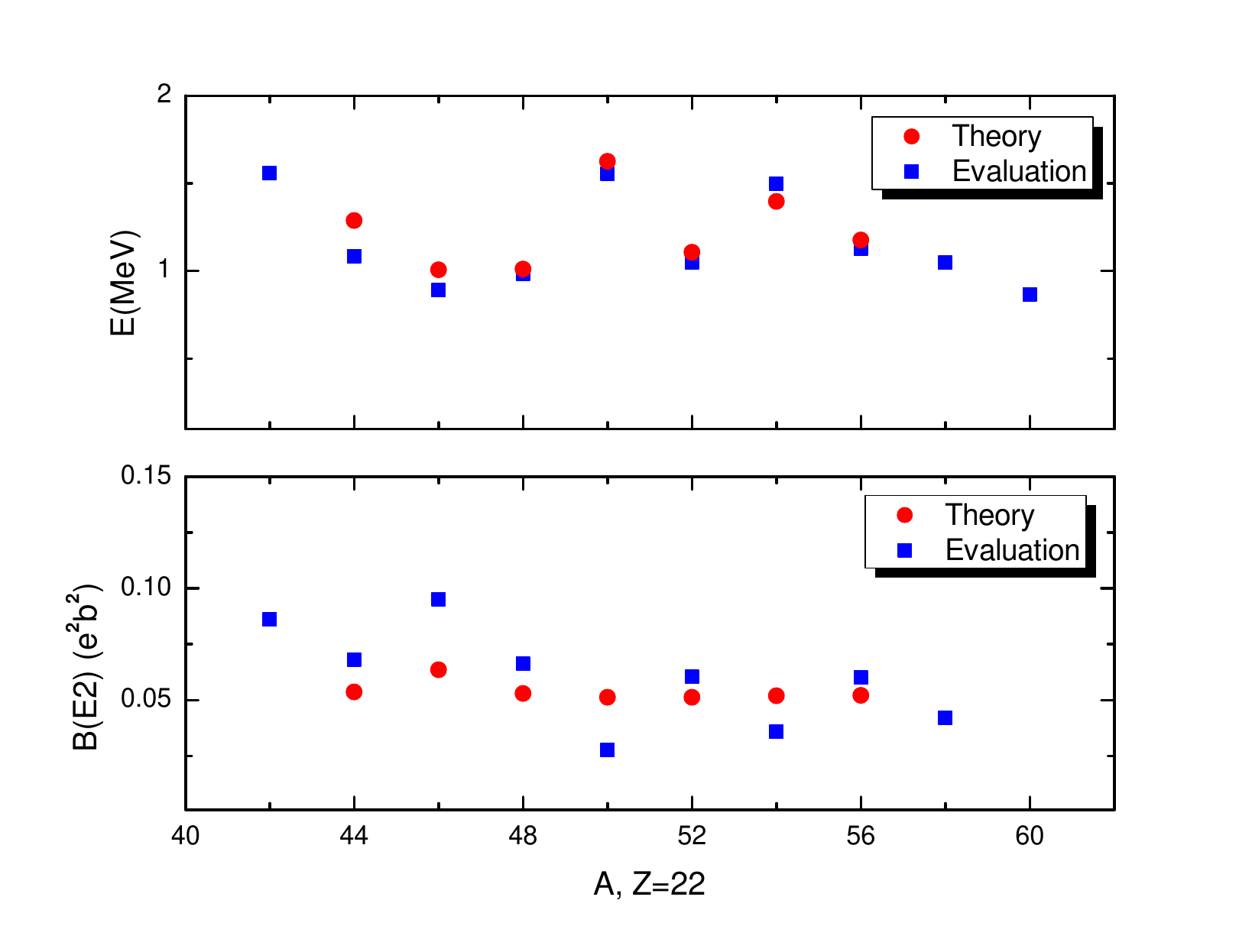}
\end{center}
\caption{Evaluated and shell model calculated energies, E($2^{+}_{1}$), and B(E; $0_{1}^{+} \rightarrow 2_{1}^{+}$) values for Ti nuclei.}\label{fig:graph22}
\end{Dfigures}

\begin{Dfigures}[ht!]
\begin{center}
\includegraphics[height=4in,width=\linewidth]{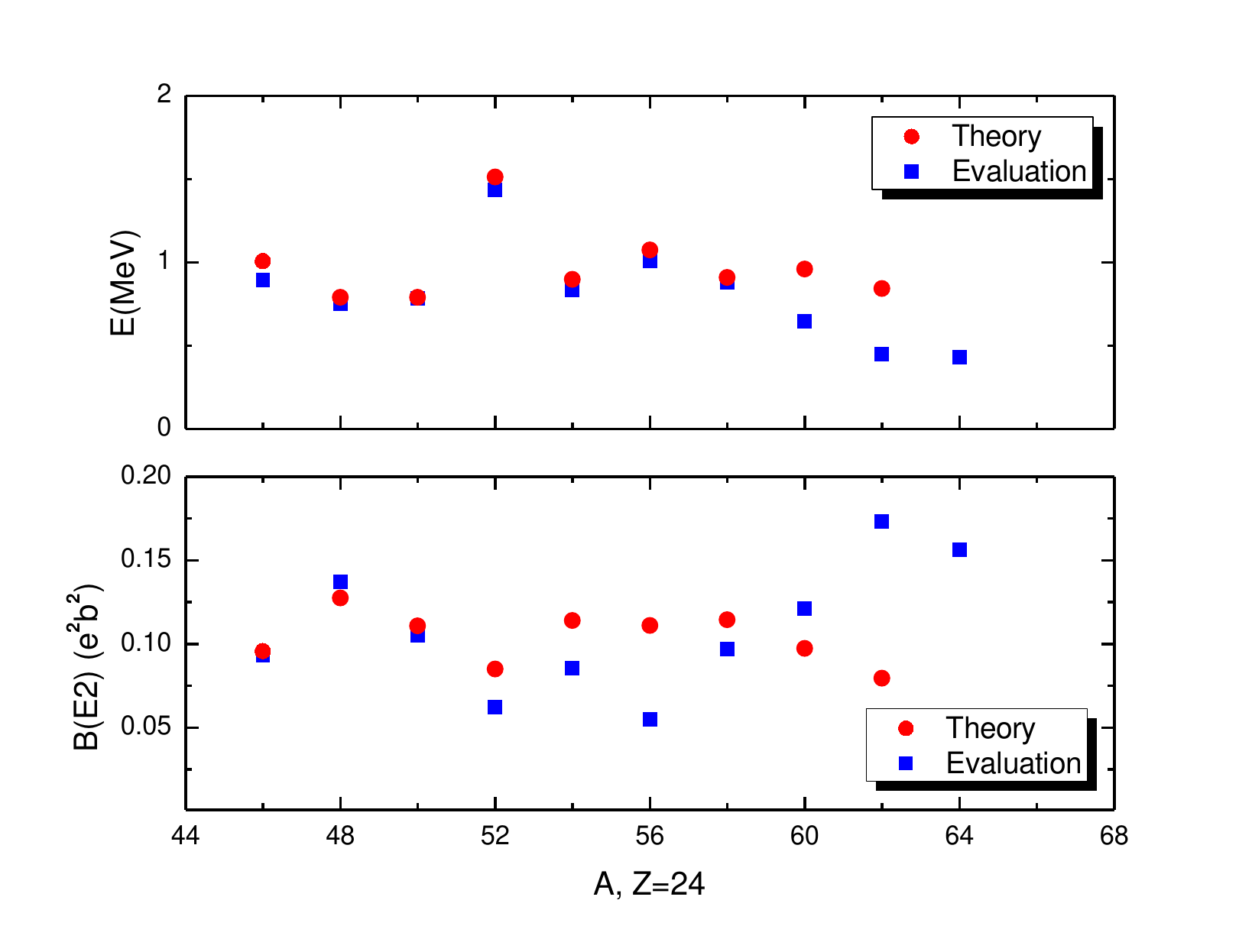}
\end{center}
\caption{Evaluated and shell model calculated energies, E($2^{+}_{1}$), and B(E; $0_{1}^{+} \rightarrow 2_{1}^{+}$) values for Cr nuclei.}\label{fig:graph24}
\end{Dfigures}
\clearpage
\begin{Dfigures}[ht!]
\begin{center}
\includegraphics[height=4in,width=\linewidth]{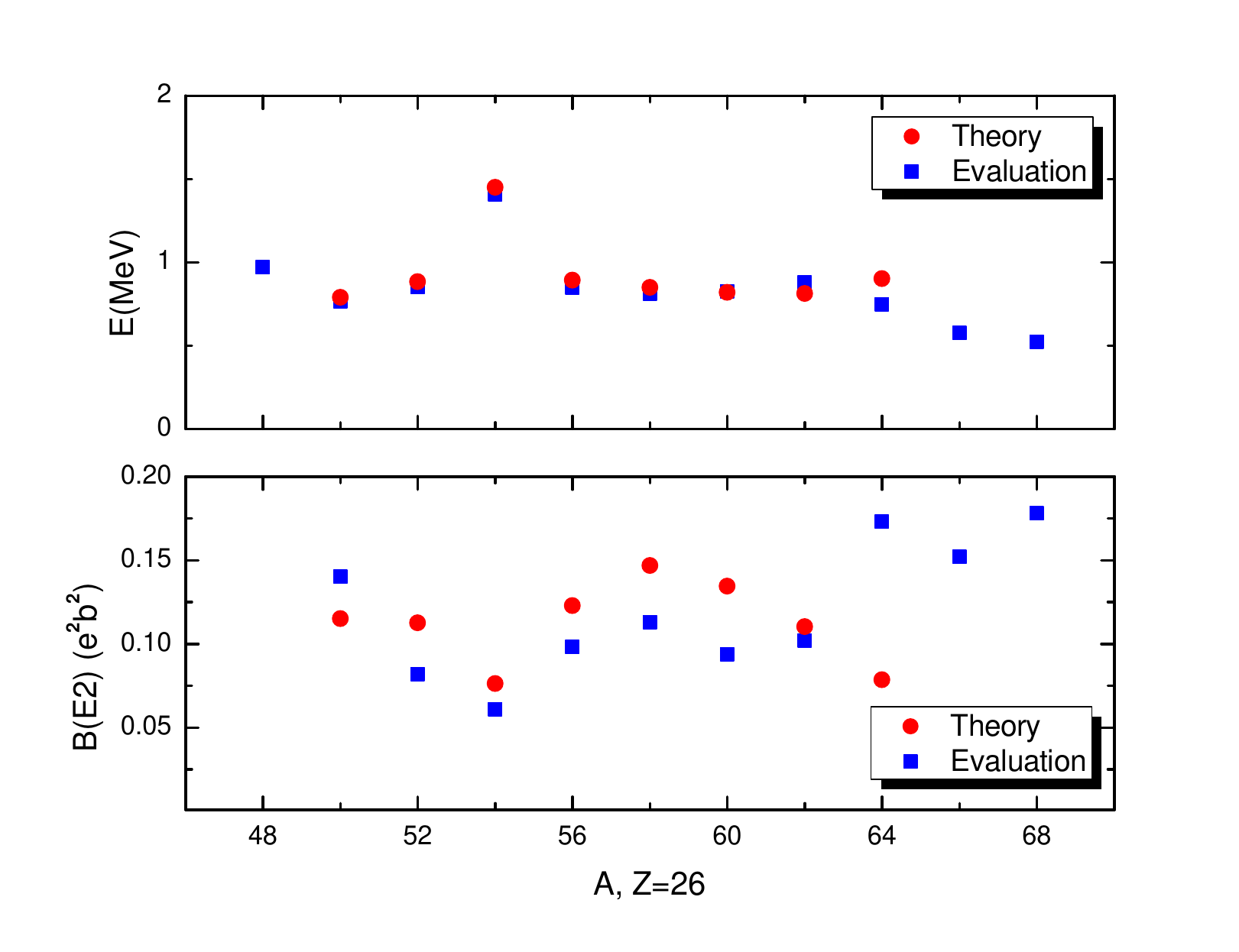}
\end{center}
\caption{Evaluated and shell model calculated energies, E($2^{+}_{1}$), and B(E; $0_{1}^{+} \rightarrow 2_{1}^{+}$) values for Fe nuclei.}\label{fig:graph26}
\end{Dfigures}

\begin{Dfigures}[ht!]
\begin{center}
\includegraphics[height=4in,width=\linewidth]{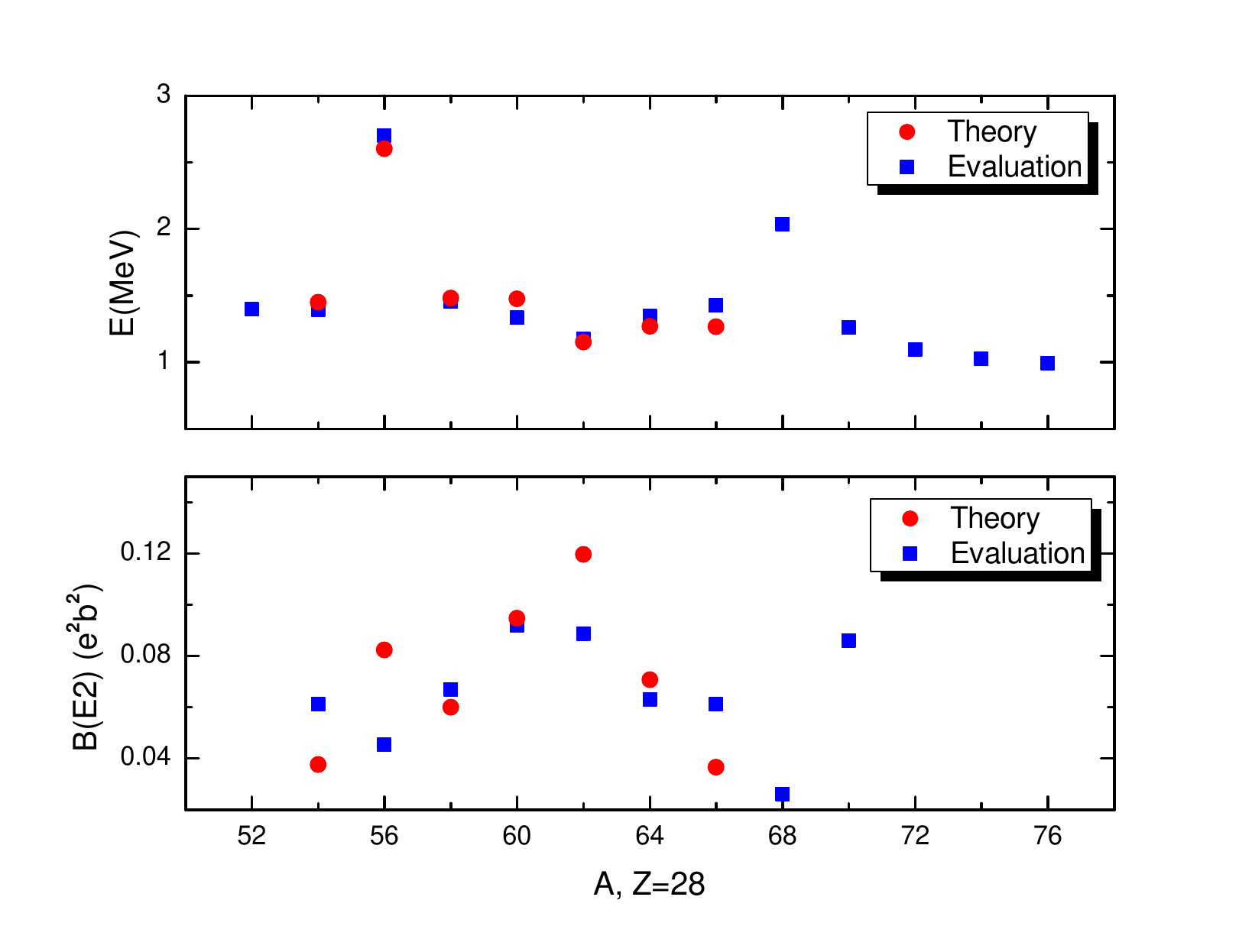}
\end{center}
\caption{Evaluated and shell model calculated energies, E($2^{+}_{1}$), and B(E; $0_{1}^{+} \rightarrow 2_{1}^{+}$) values for Ni nuclei.}\label{fig:graph28}
\end{Dfigures}
\clearpage
\begin{Dfigures}[ht!]
\begin{center}
\includegraphics[height=4in,width=\linewidth]{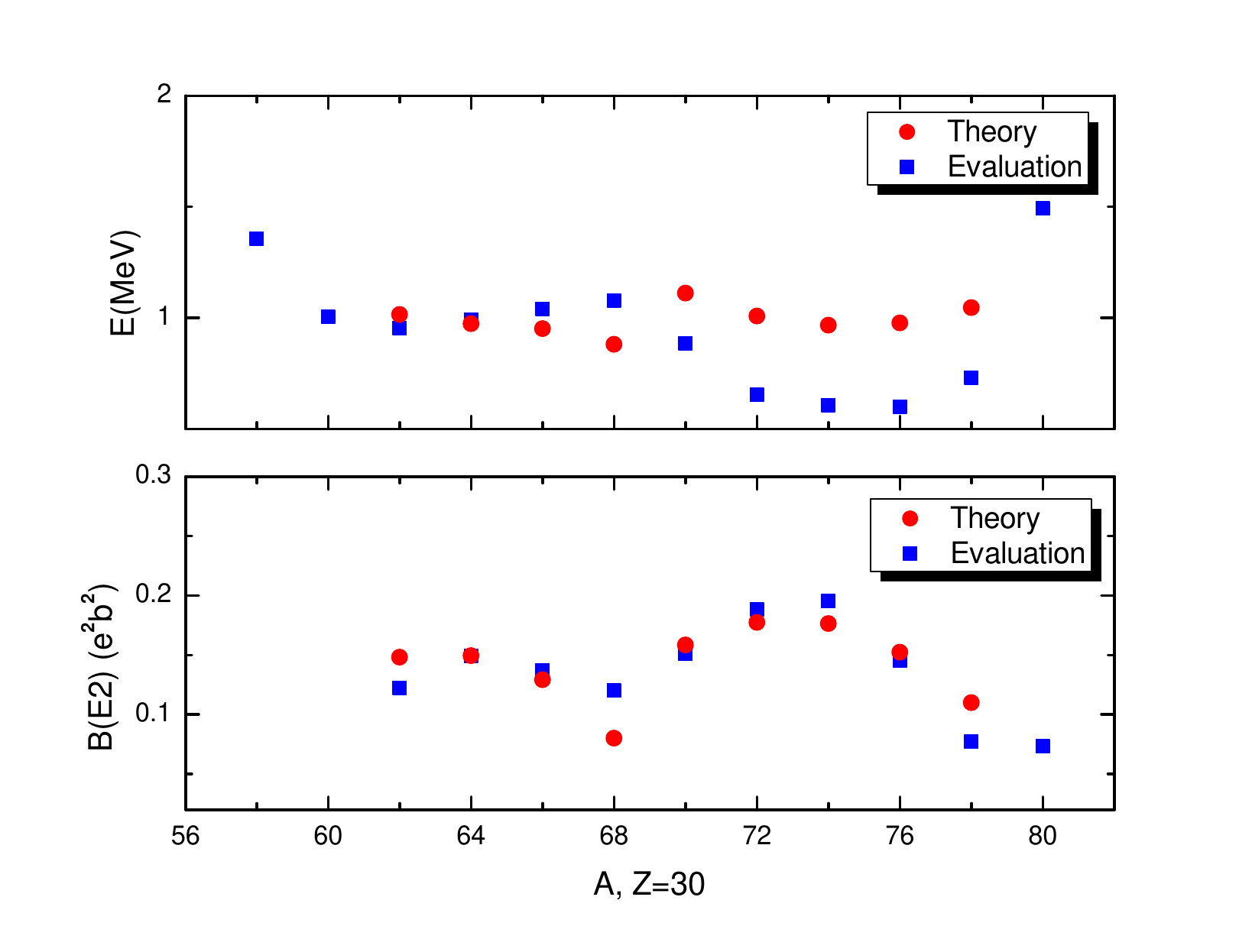}
\end{center}
\caption{Evaluated and shell model calculated energies, E($2^{+}_{1}$), and B(E; $0_{1}^{+} \rightarrow 2_{1}^{+}$) values for Zn nuclei.}\label{fig:graph30}
\end{Dfigures}

\begin{Dfigures}[ht!]
\begin{center}
\includegraphics[height=4in,width=\linewidth]{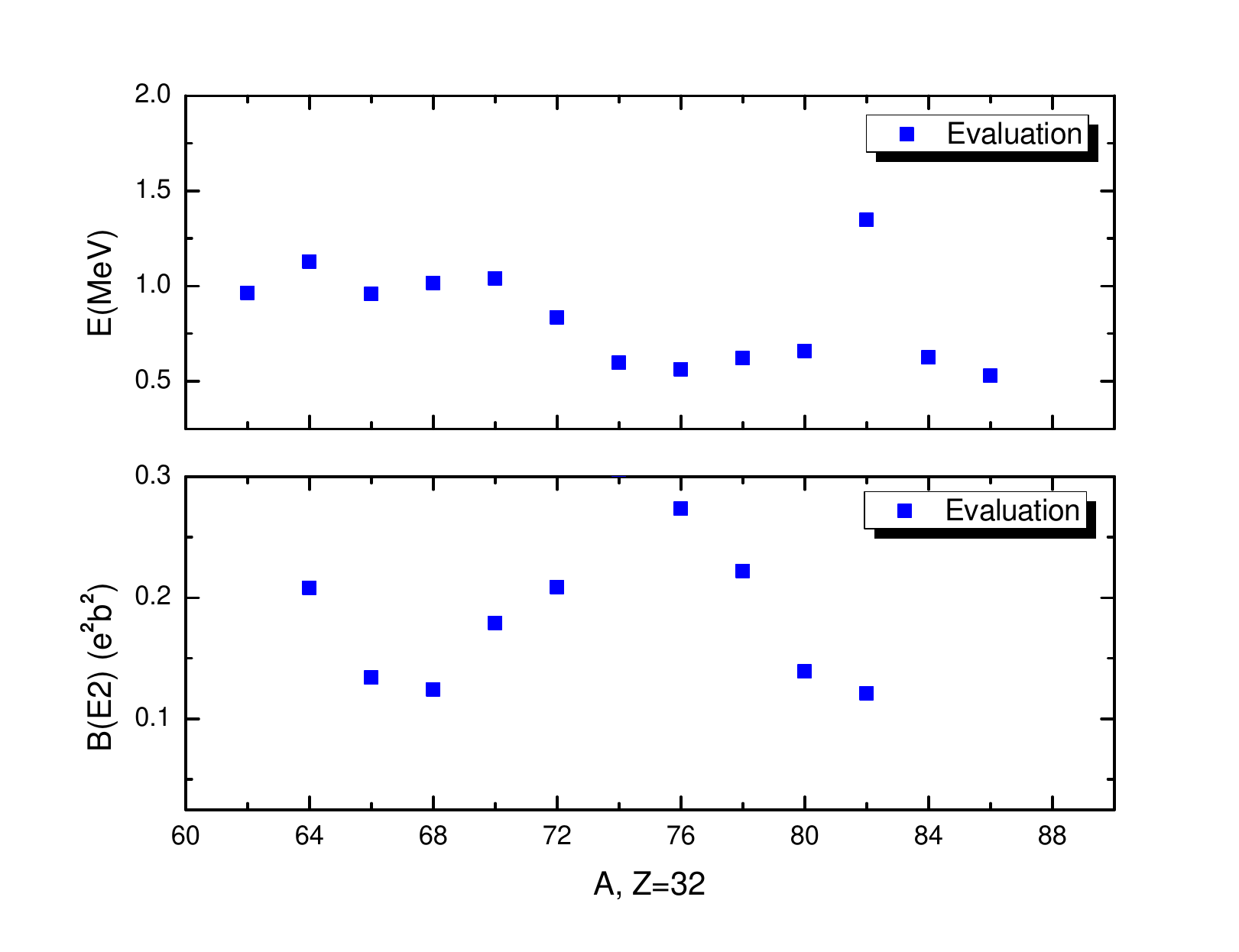}
\end{center}
\caption{Evaluated energies, E($2^{+}_{1}$), and B(E; $0_{1}^{+} \rightarrow 2_{1}^{+}$) values for Ge nuclei.}\label{fig:graph32}
\end{Dfigures}
\clearpage
\begin{Dfigures}[ht!]
\begin{center}
\includegraphics[height=4in,width=\linewidth]{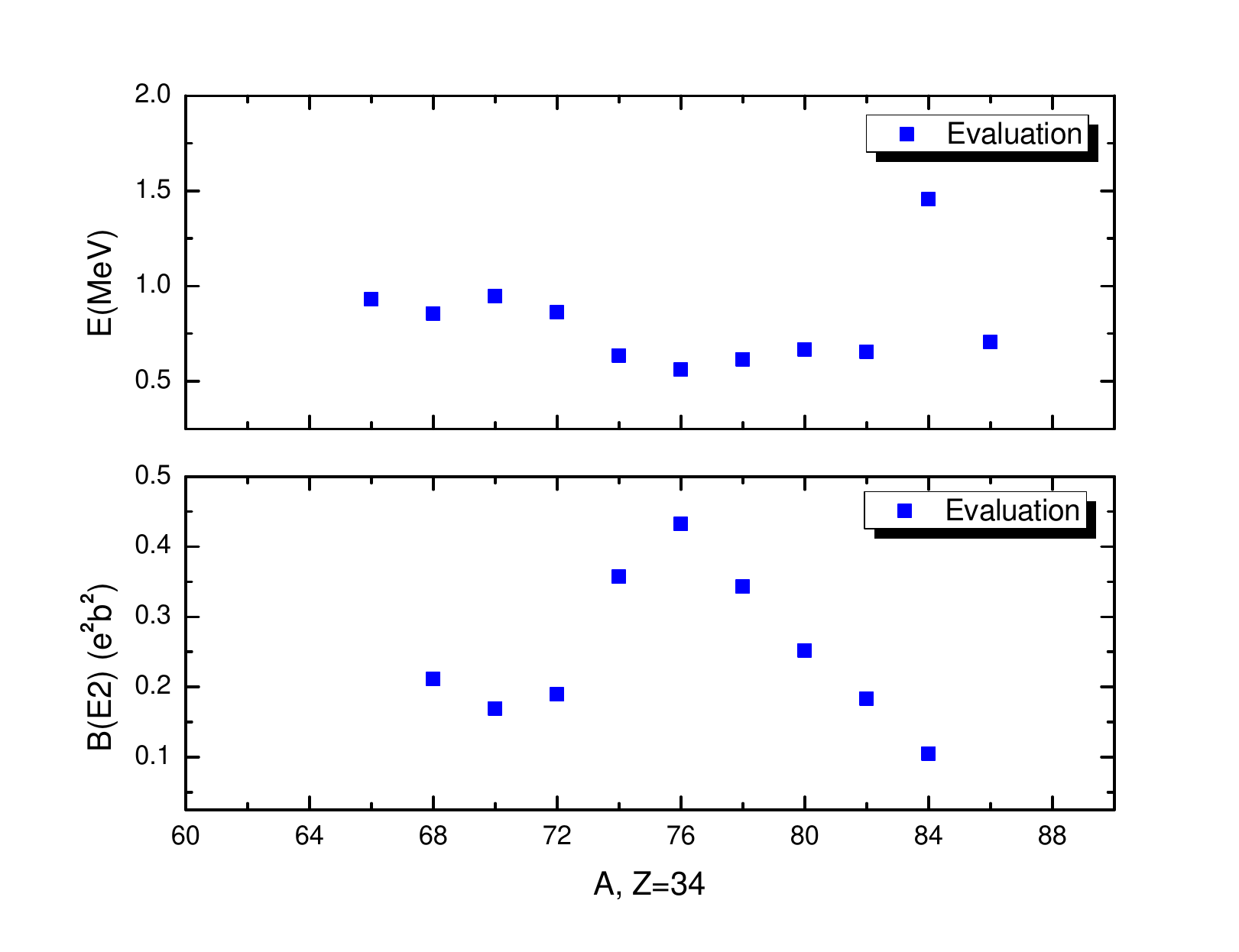}
\end{center}
\caption{Evaluated energies, E($2^{+}_{1}$), and B(E; $0_{1}^{+} \rightarrow 2_{1}^{+}$) values for Se nuclei.}\label{fig:graph34}
\end{Dfigures}

\begin{Dfigures}[ht!]
\begin{center}
\includegraphics[height=4in,width=\linewidth]{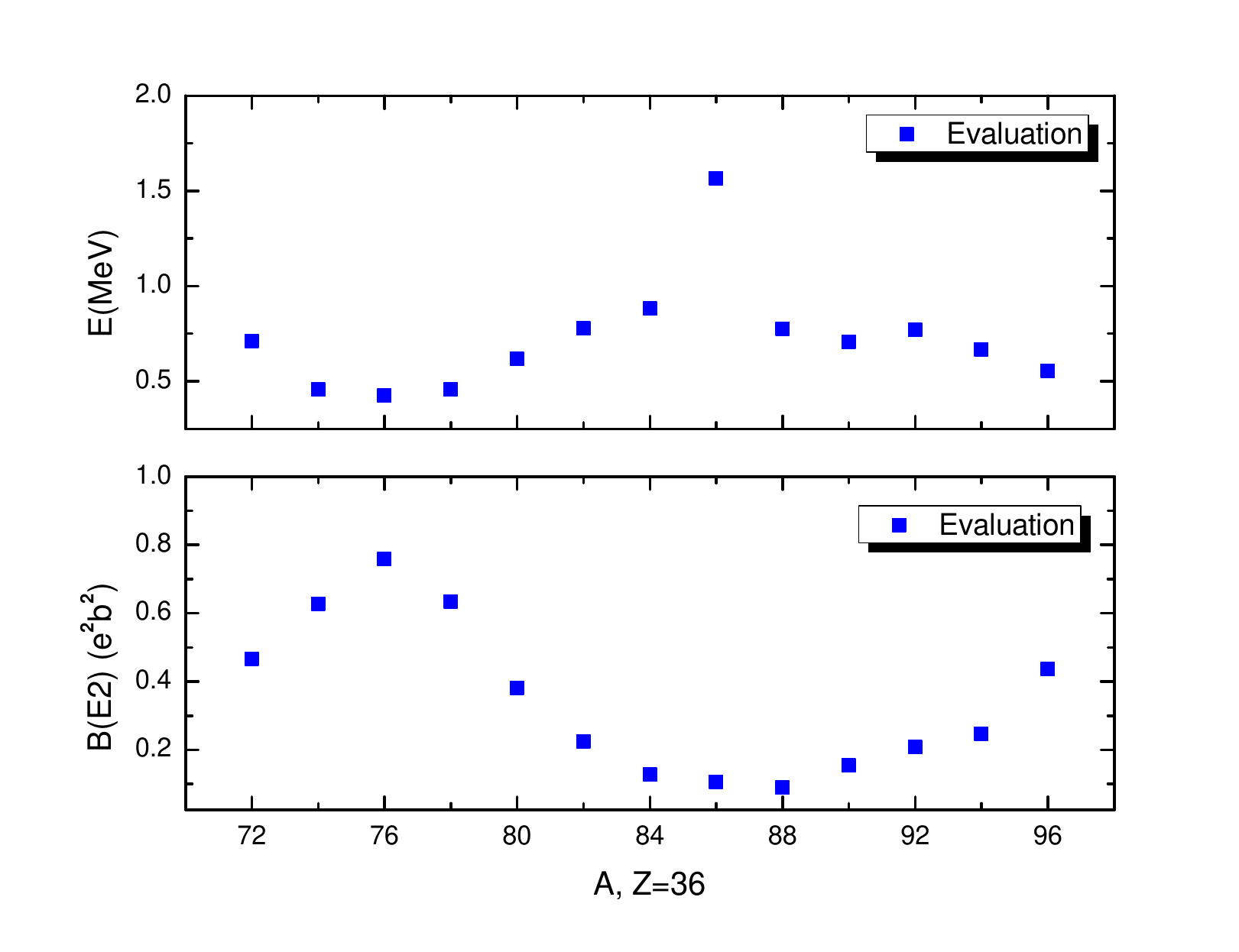}
\end{center}
\caption{Evaluated energies, E($2^{+}_{1}$), and B(E; $0_{1}^{+} \rightarrow 2_{1}^{+}$) values for Kr nuclei.}\label{fig:graph36}
\end{Dfigures}
\clearpage
\begin{Dfigures}[ht!]
\begin{center}
\includegraphics[height=4in,width=\linewidth]{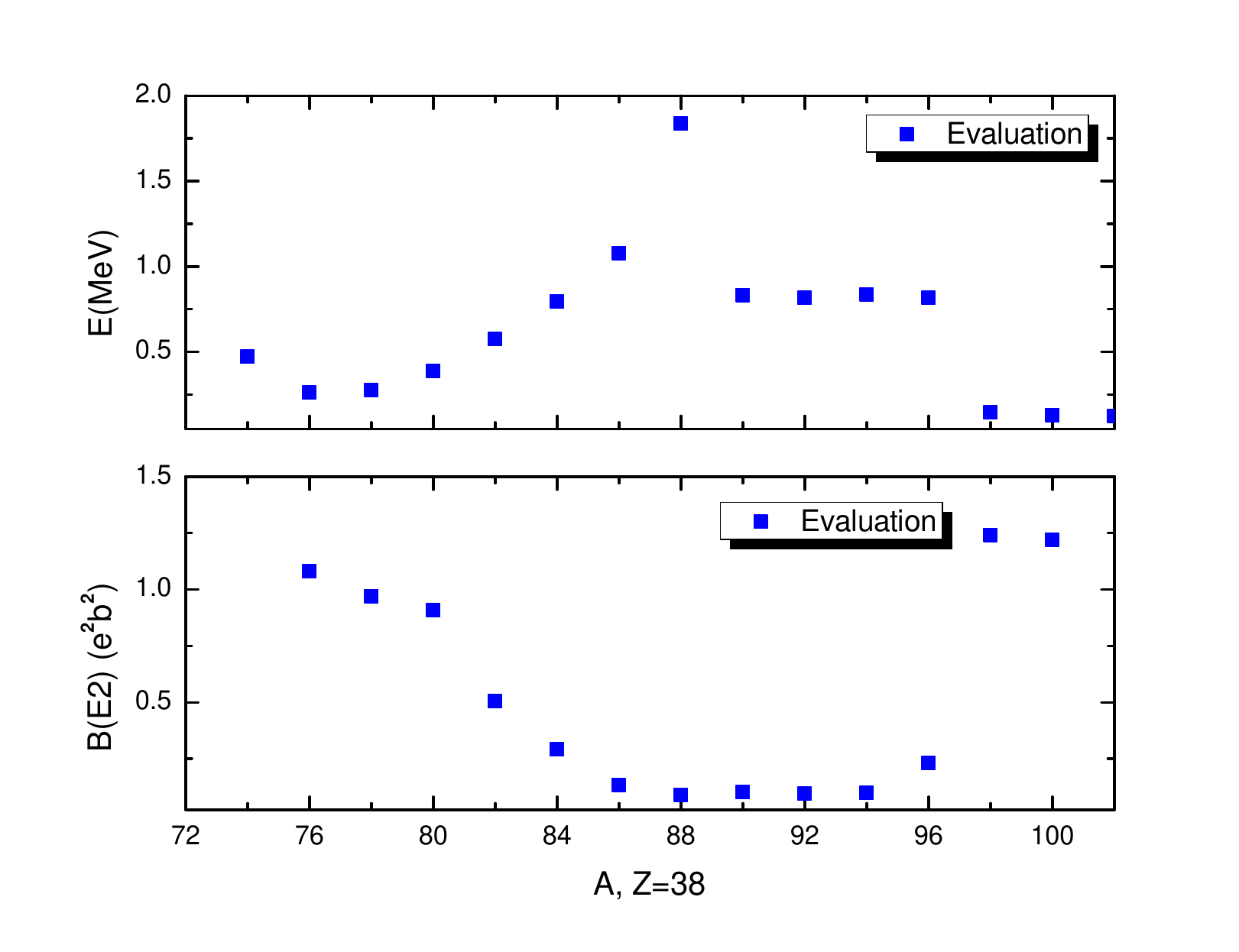}
\end{center}
\caption{Evaluated energies, E($2^{+}_{1}$), and B(E; $0_{1}^{+} \rightarrow 2_{1}^{+}$) values for Sr nuclei.}\label{fig:graph38}
\end{Dfigures}

\begin{Dfigures}[ht!]
\begin{center}
\includegraphics[height=4in,width=\linewidth]{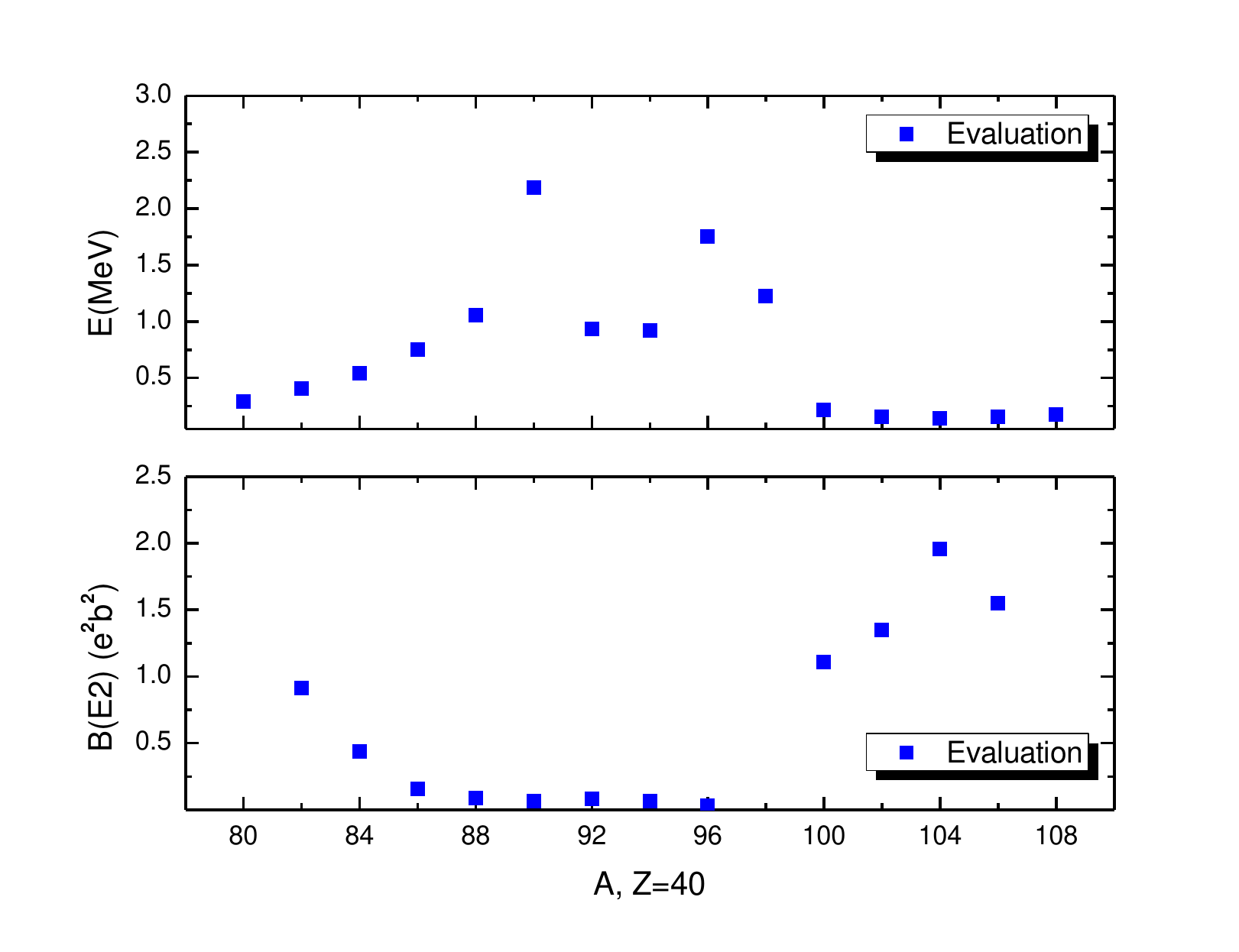}
\end{center}
\caption{Evaluated energies, E($2^{+}_{1}$), and B(E; $0_{1}^{+} \rightarrow 2_{1}^{+}$) values for Zr nuclei.}\label{fig:graph40}
\end{Dfigures}
\clearpage
\begin{Dfigures}[ht!]
\begin{center}
\includegraphics[height=4in,width=\linewidth]{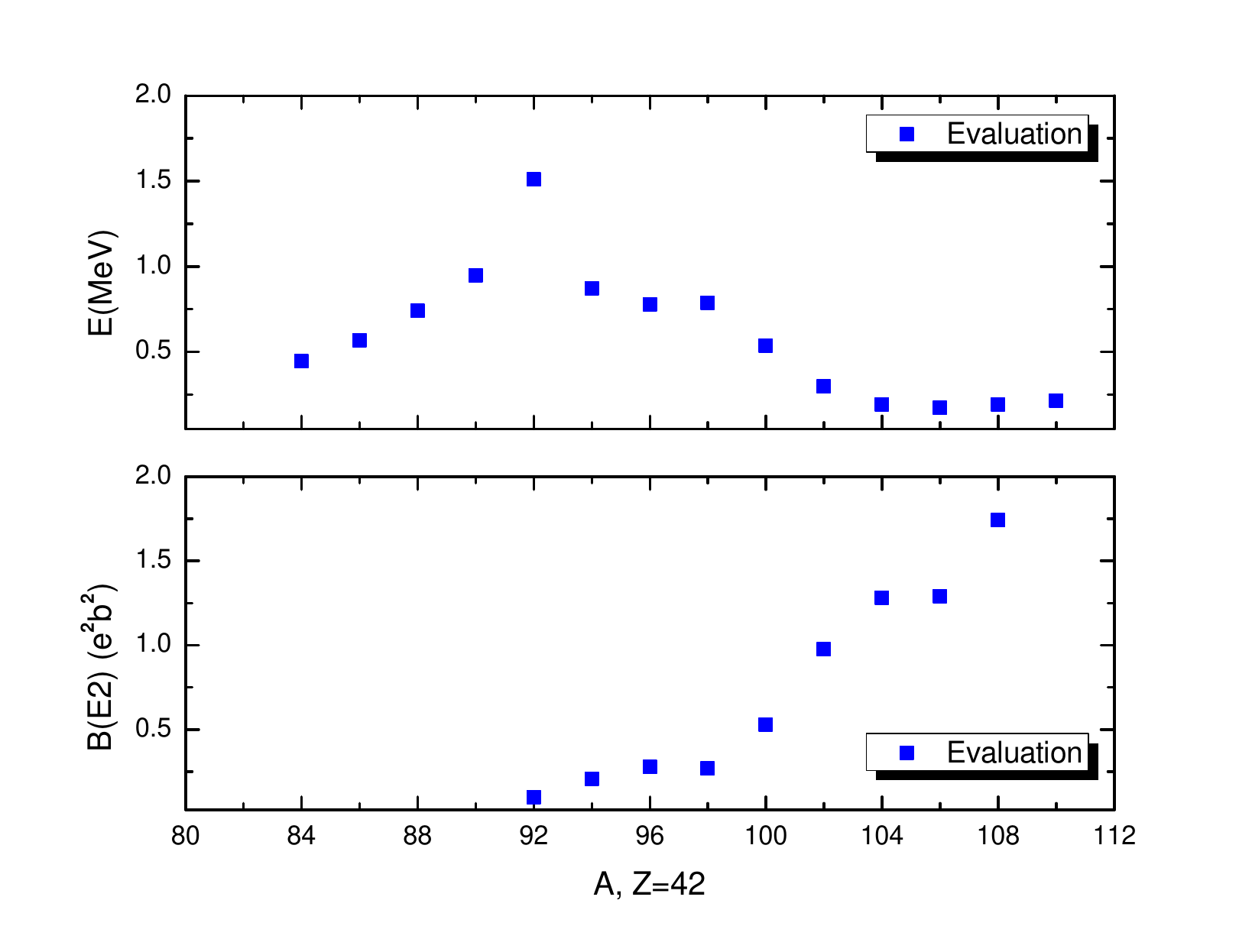}
\end{center}
\caption{Evaluated energies, E($2^{+}_{1}$), and B(E; $0_{1}^{+} \rightarrow 2_{1}^{+}$) values for Mo nuclei.}\label{fig:graph42}
\end{Dfigures}

\begin{Dfigures}[ht!]
\begin{center}
\includegraphics[height=4in,width=\linewidth]{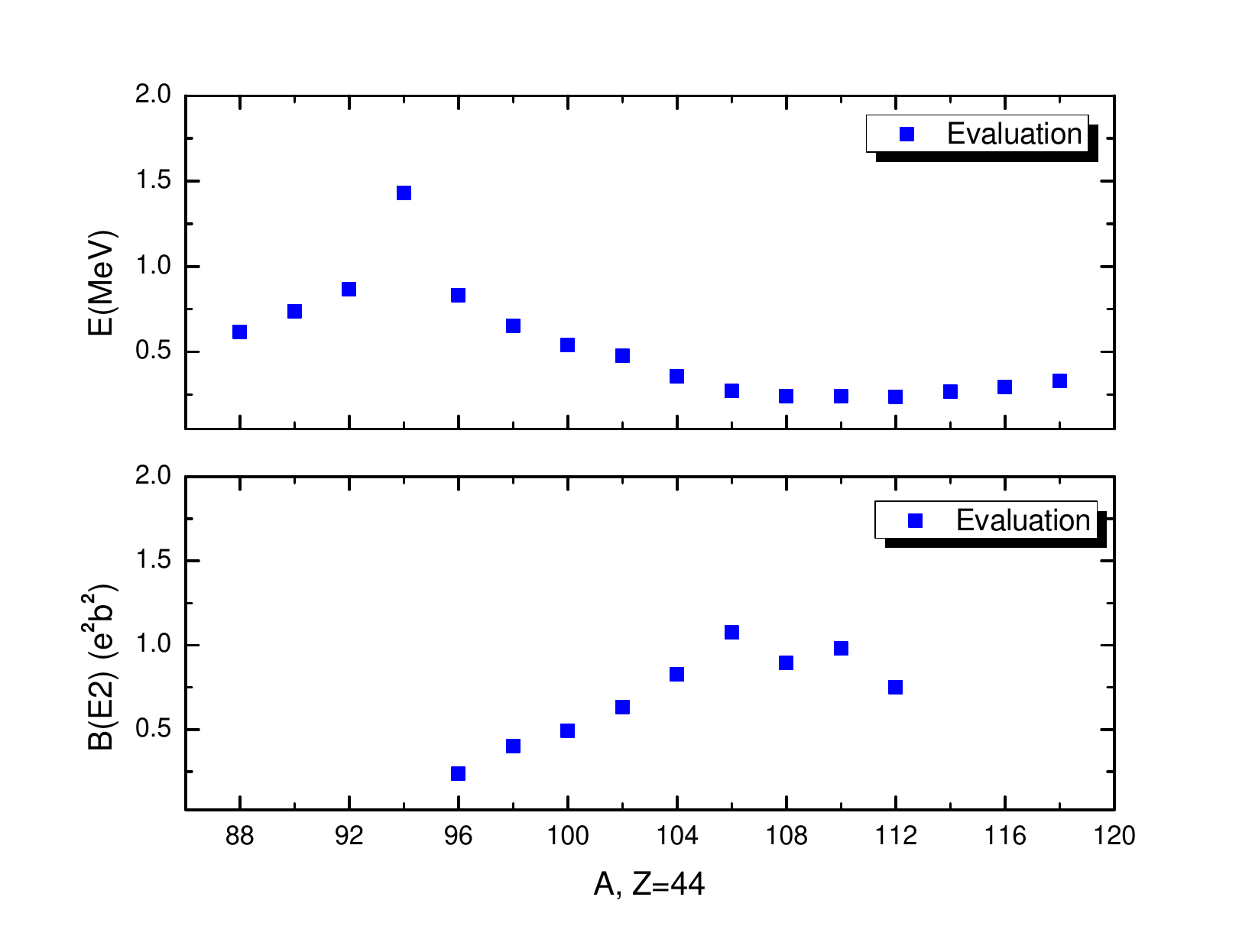}
\end{center}
\caption{Evaluated energies, E($2^{+}_{1}$), and B(E; $0_{1}^{+} \rightarrow 2_{1}^{+}$) values for Ru nuclei.}\label{fig:graph44}
\end{Dfigures}
\clearpage
\begin{Dfigures}[ht!]
\begin{center}
\includegraphics[height=4in,width=\linewidth]{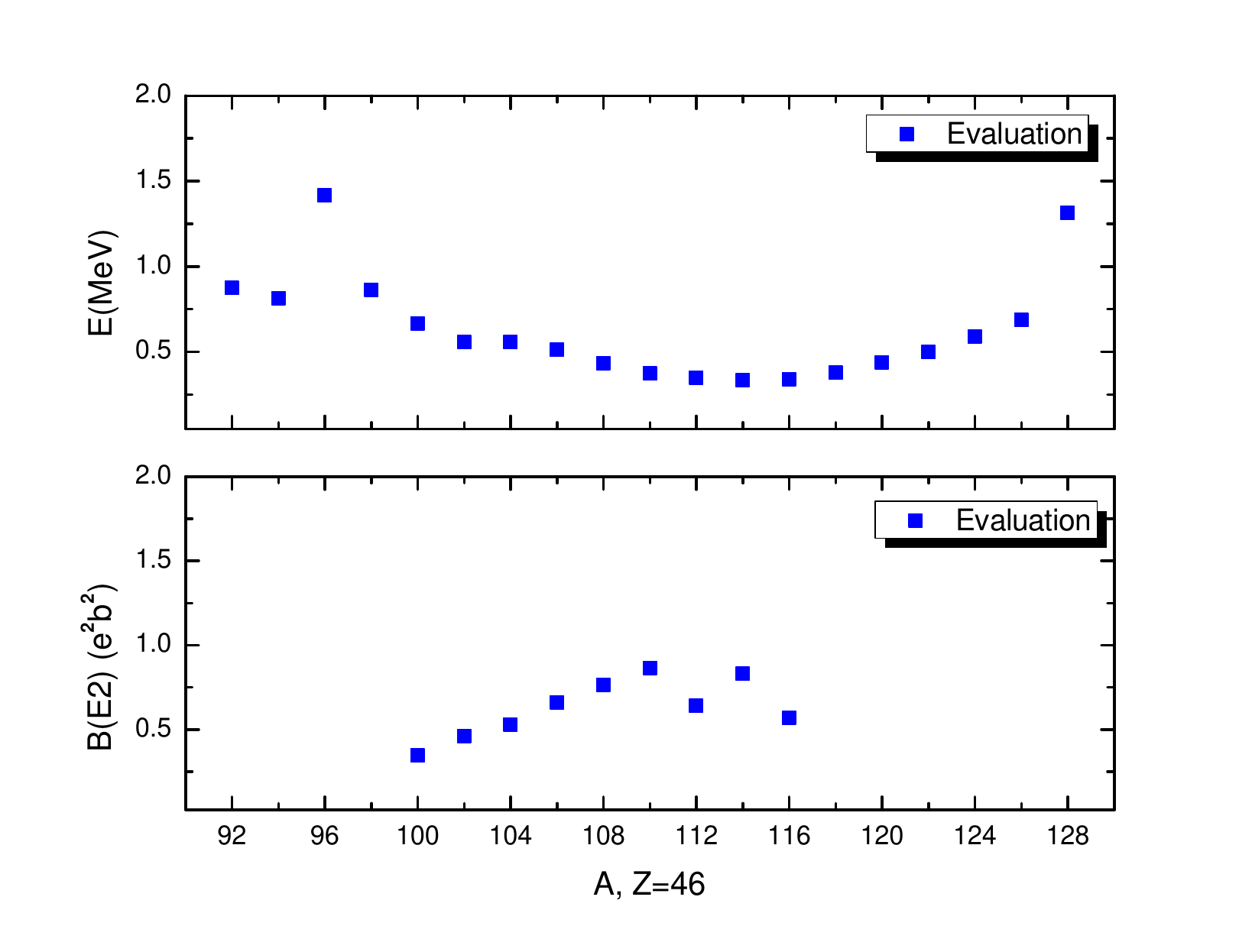}
\end{center}
\caption{Evaluated energies, E($2^{+}_{1}$), and B(E; $0_{1}^{+} \rightarrow 2_{1}^{+}$) values for Pd nuclei.}\label{fig:graph46}
\end{Dfigures}

\begin{Dfigures}[ht!]
\begin{center}
\includegraphics[height=4in,width=\linewidth]{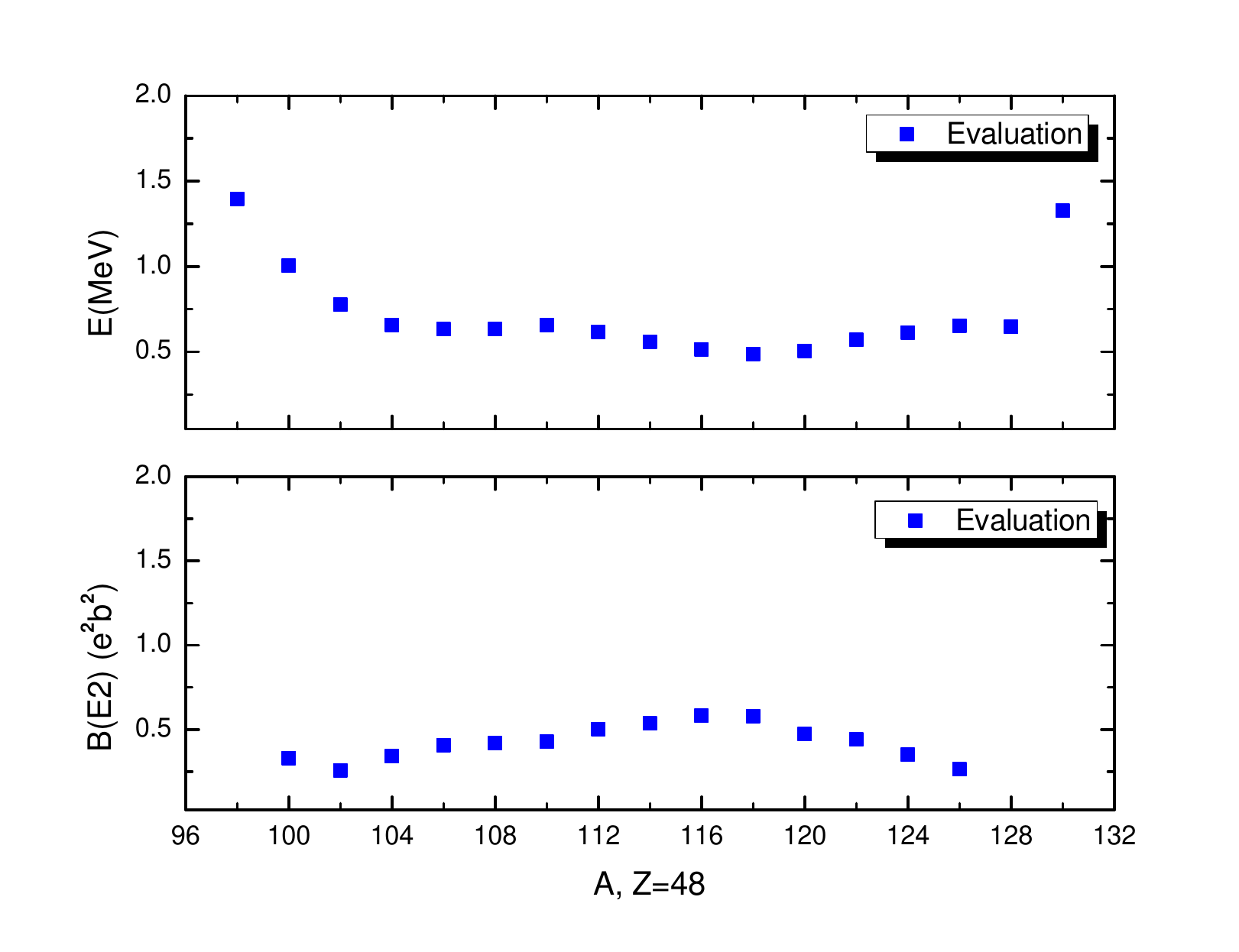}
\end{center}
\caption{Evaluated energies, E($2^{+}_{1}$), and B(E; $0_{1}^{+} \rightarrow 2_{1}^{+}$) values for Cd nuclei.}\label{fig:graph48}
\end{Dfigures}
\clearpage
\begin{Dfigures}[ht!]
\begin{center}
\includegraphics[height=4in,width=\linewidth]{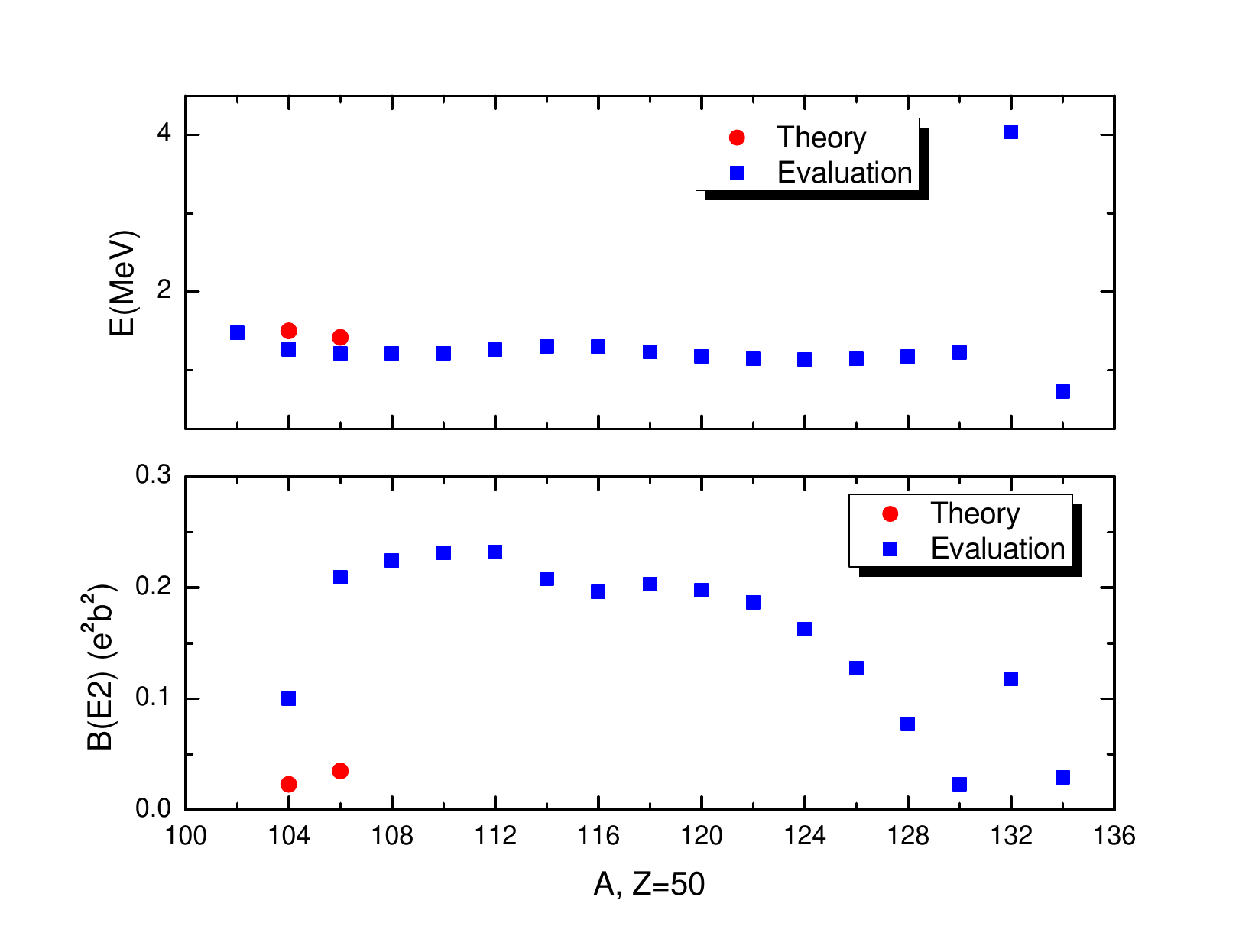}
\end{center}
\caption{Evaluated energies, E($2^{+}_{1}$), and B(E; $0_{1}^{+} \rightarrow 2_{1}^{+}$) values for Sn nuclei.}\label{fig:graph50}
\end{Dfigures}

\begin{Dfigures}[ht!]
\begin{center}
\includegraphics[height=4in,width=\linewidth]{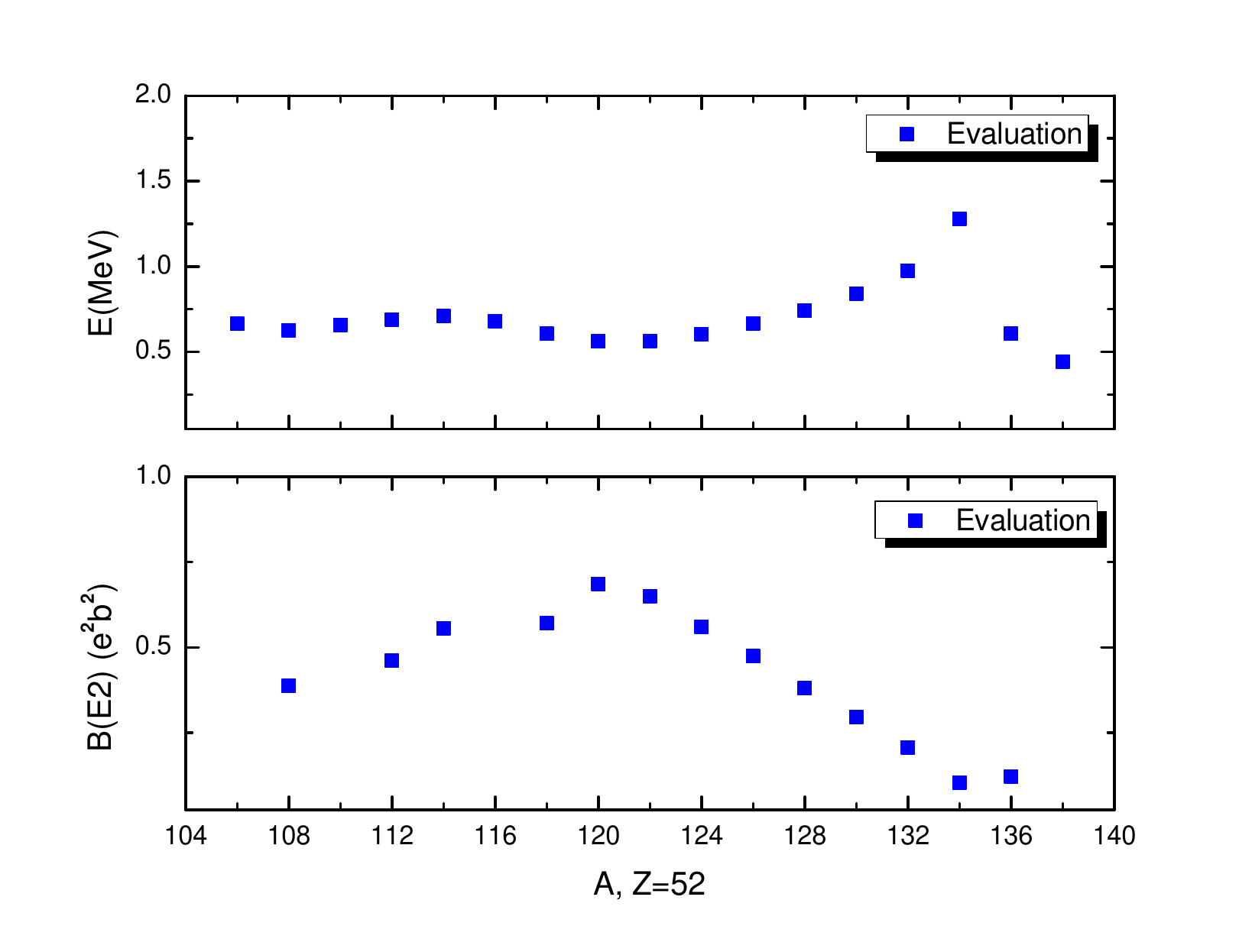}
\end{center}
\caption{Evaluated energies, E($2^{+}_{1}$), and B(E; $0_{1}^{+} \rightarrow 2_{1}^{+}$) values for Te nuclei.}\label{fig:graph52}
\end{Dfigures}
\clearpage
\begin{Dfigures}[ht!]
\begin{center}
\includegraphics[height=4in,width=\linewidth]{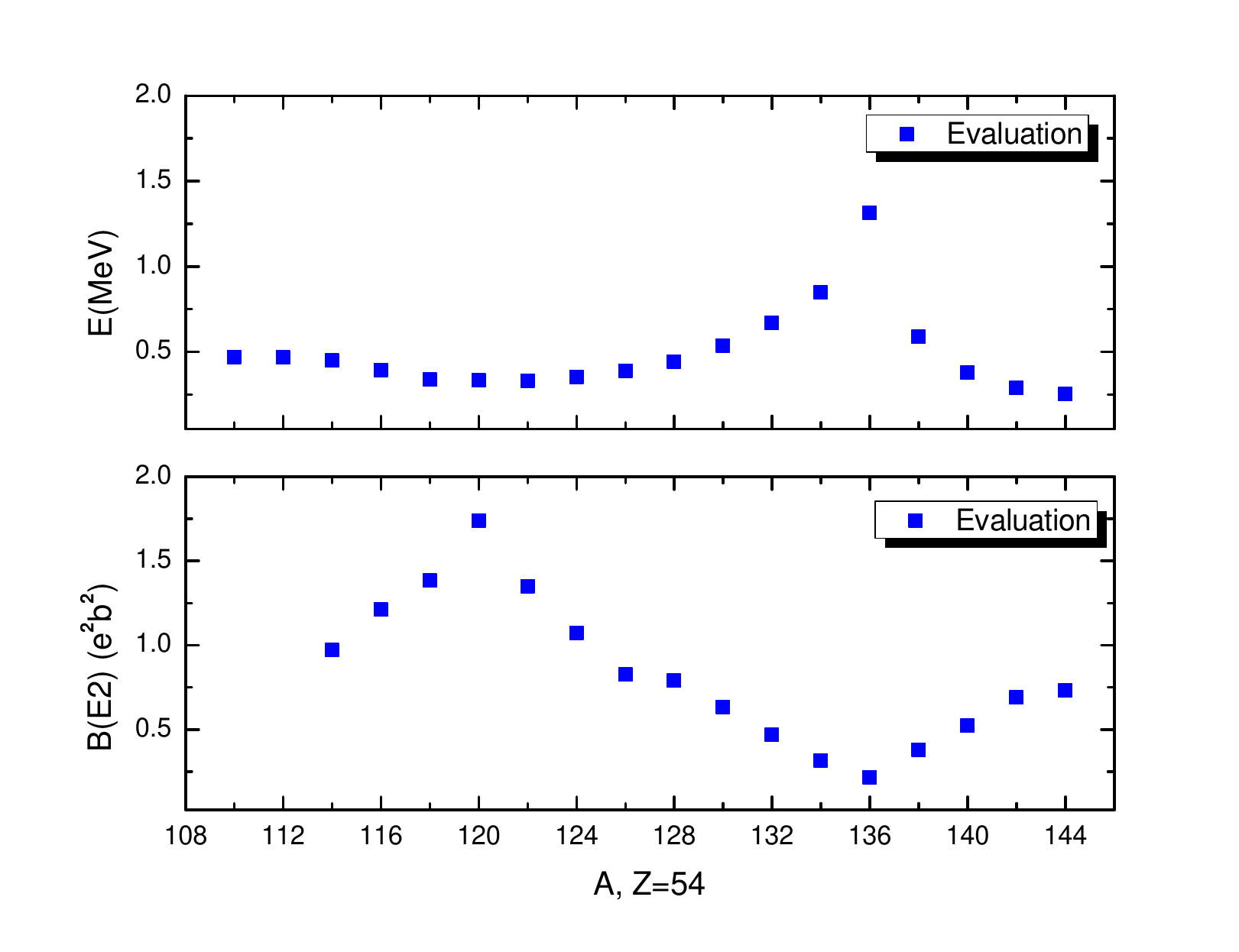}
\end{center}
\caption{Evaluated energies, E($2^{+}_{1}$), and B(E; $0_{1}^{+} \rightarrow 2_{1}^{+}$) values for Xe nuclei.}\label{fig:graph54}
\end{Dfigures}

\begin{Dfigures}[ht!]
\begin{center}
\includegraphics[height=4in,width=\linewidth]{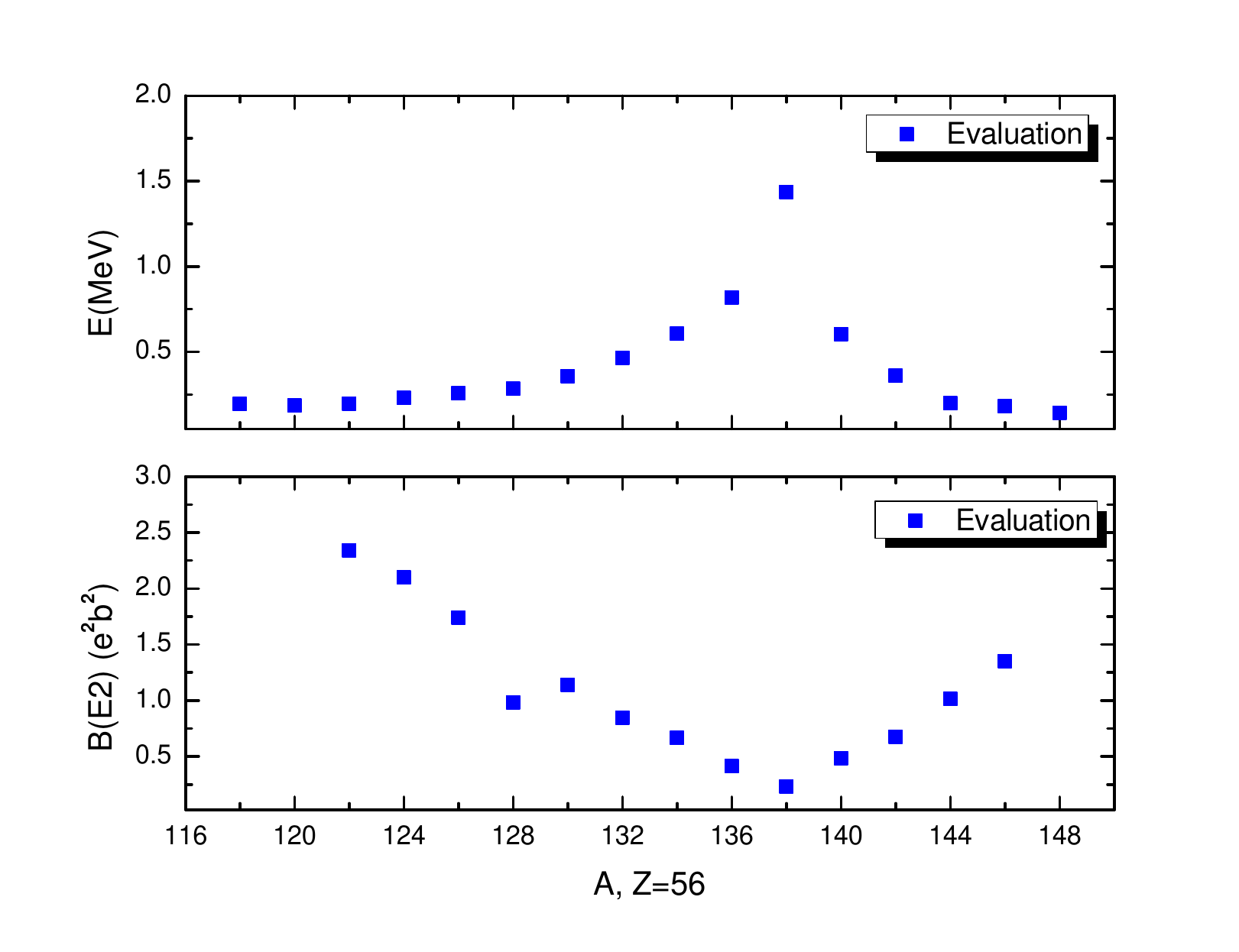}
\end{center}
\caption{Evaluated energies, E($2^{+}_{1}$), and B(E; $0_{1}^{+} \rightarrow 2_{1}^{+}$) values for Ba nuclei.}\label{fig:graph56}
\end{Dfigures}
\clearpage

\begin{Dfigures}[ht!]
\begin{center}
\includegraphics[height=4in,width=\linewidth]{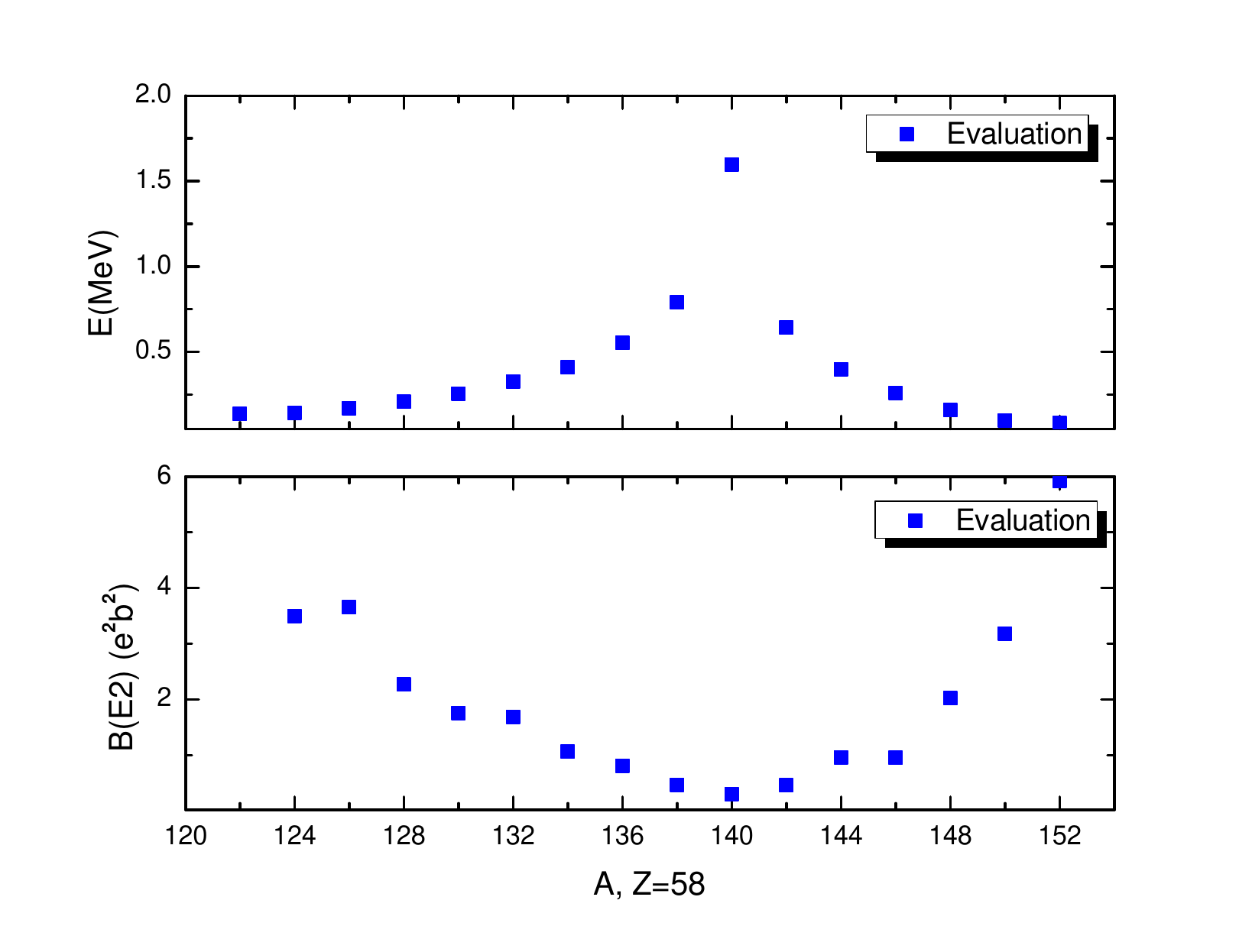}
\end{center}
\caption{Evaluated energies, E($2^{+}_{1}$), and B(E; $0_{1}^{+} \rightarrow 2_{1}^{+}$) values for Ce nuclei.}\label{fig:graph58}
\end{Dfigures}

\begin{Dfigures}[ht!]
\begin{center}
\includegraphics[height=4in,width=\linewidth]{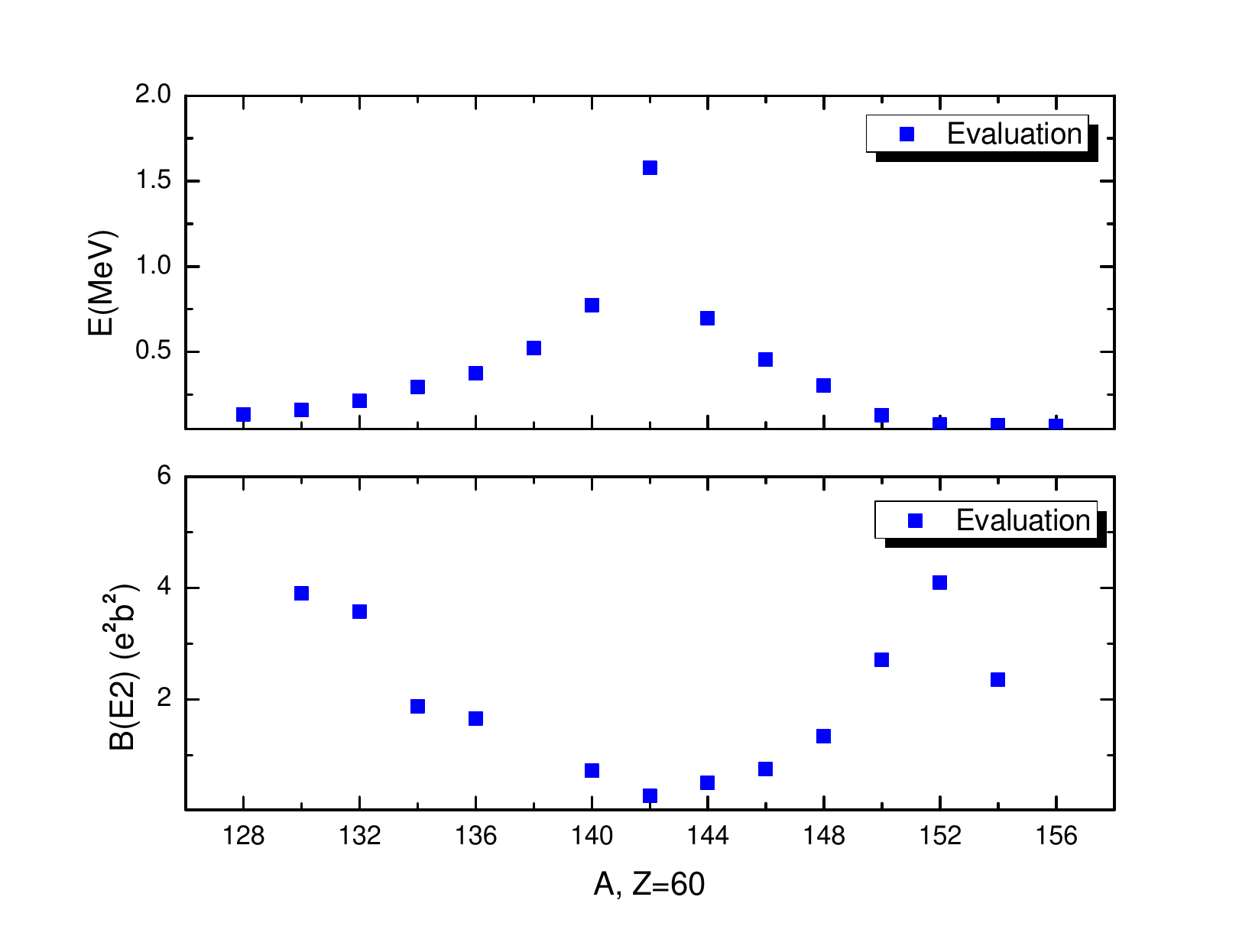}
\end{center}
\caption{Evaluated energies, E($2^{+}_{1}$), and B(E; $0_{1}^{+} \rightarrow 2_{1}^{+}$) values for Nd nuclei.}\label{fig:graph60}
\end{Dfigures}
\clearpage

\begin{Dfigures}[ht!]
\begin{center}
\includegraphics[height=4in,width=\linewidth]{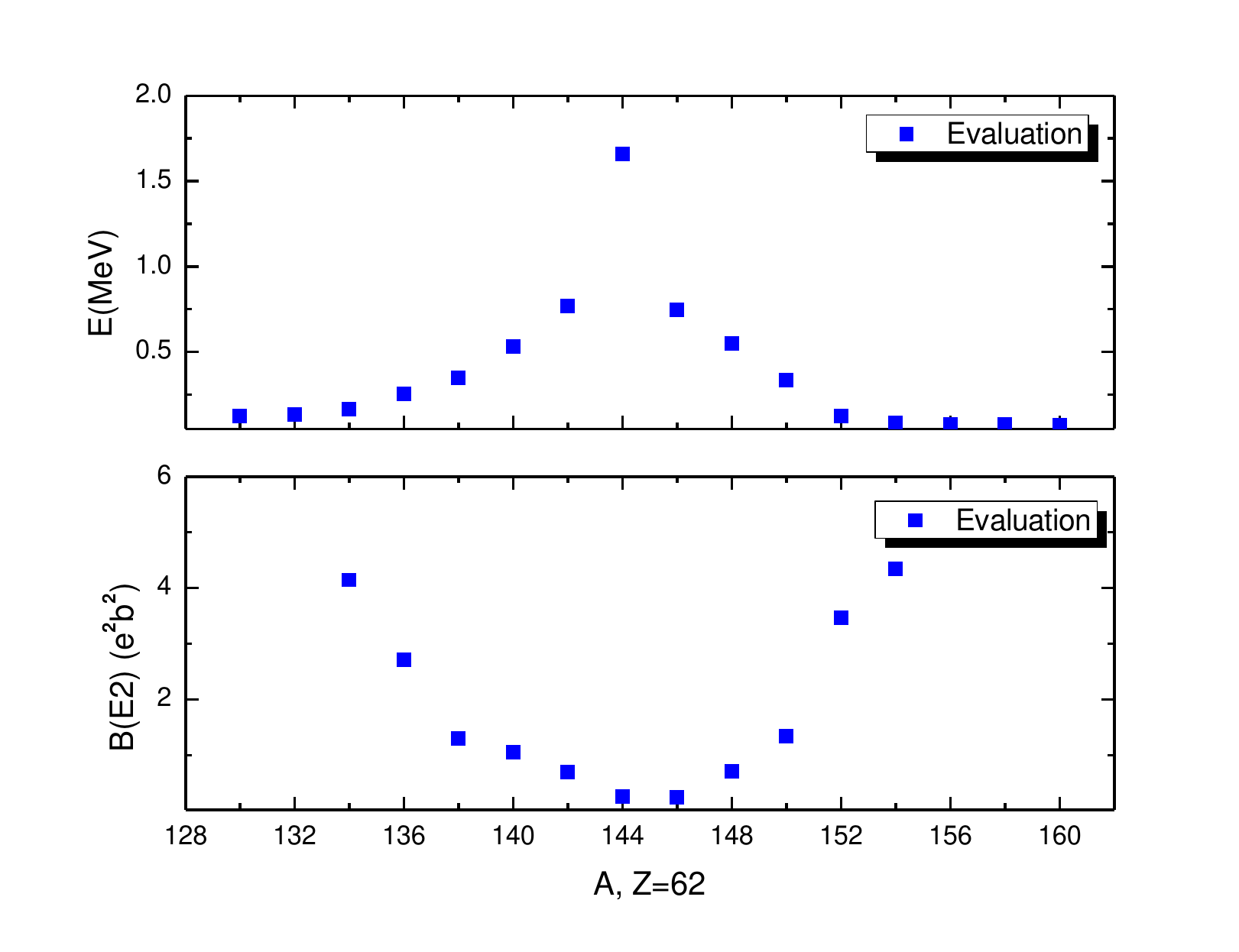}
\end{center}
\caption{Evaluated energies, E($2^{+}_{1}$), and B(E; $0_{1}^{+} \rightarrow 2_{1}^{+}$) values for Sm nuclei.}\label{fig:graph62}
\end{Dfigures}

\begin{Dfigures}[ht!]
\begin{center}
\includegraphics[height=4in,width=\linewidth]{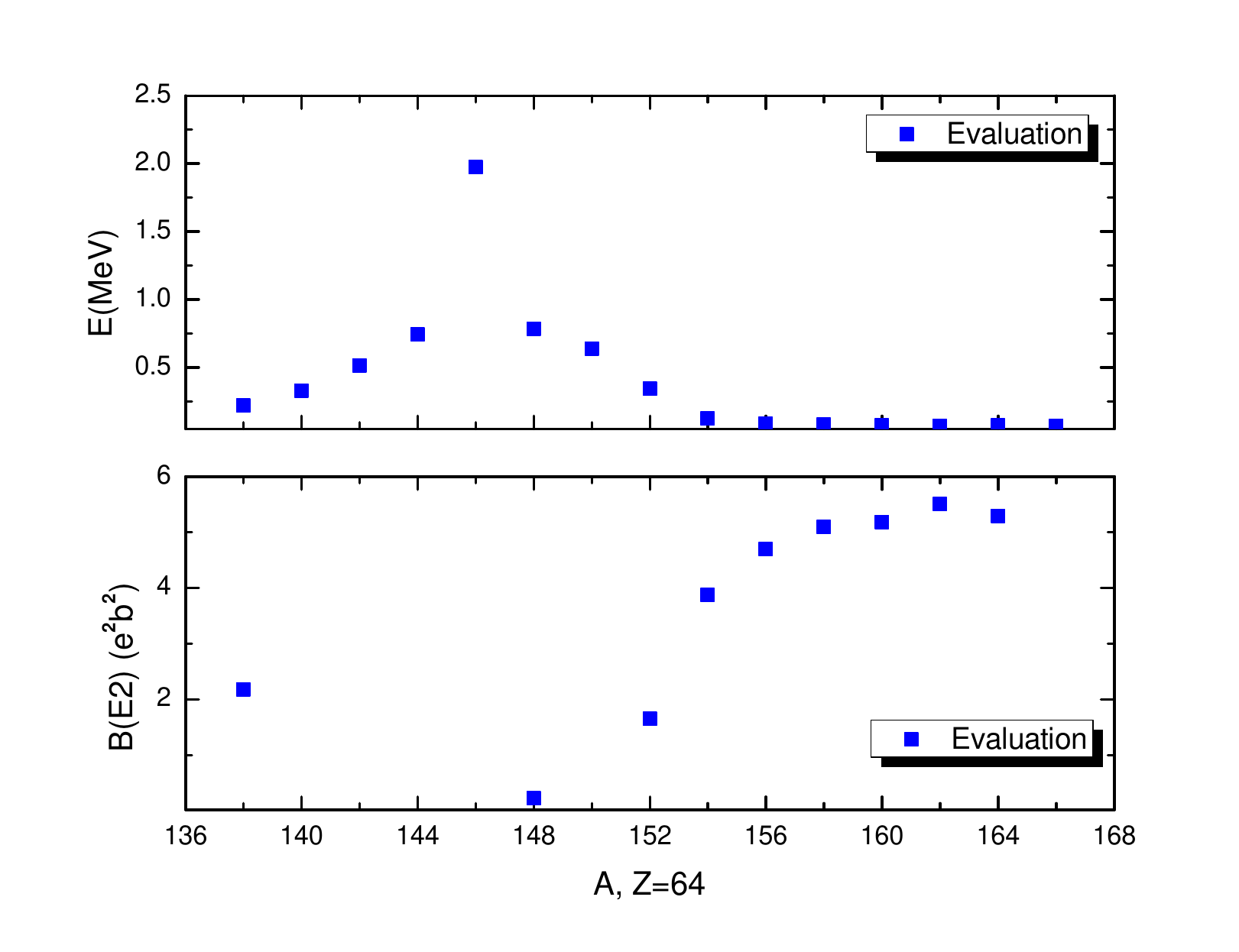}
\end{center}
\caption{Evaluated energies, E($2^{+}_{1}$), and B(E; $0_{1}^{+} \rightarrow 2_{1}^{+}$) values for Gd nuclei.}\label{fig:graph64}
\end{Dfigures}
\clearpage

\begin{Dfigures}[ht!]
\begin{center}
\includegraphics[height=4in,width=\linewidth]{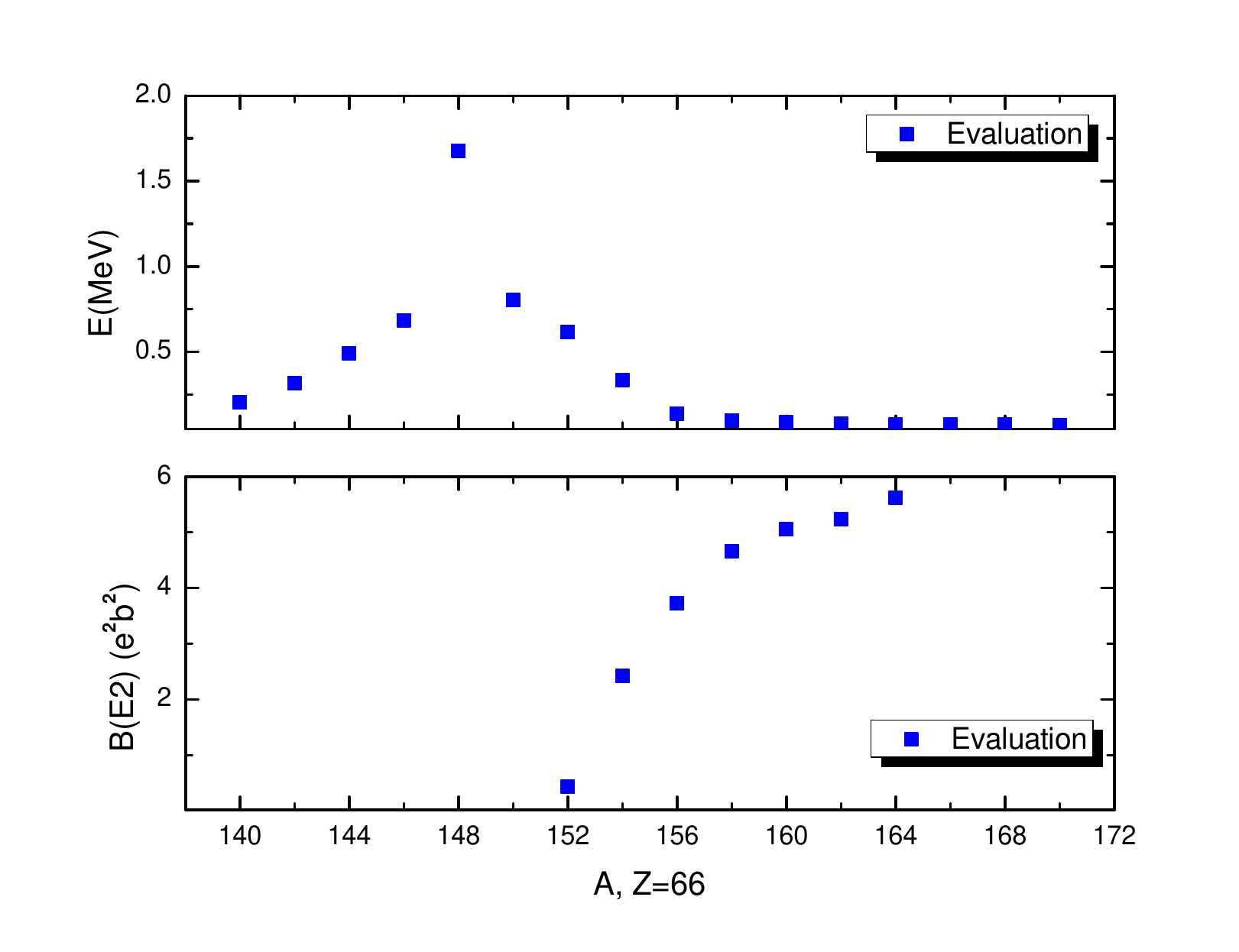}
\end{center}
\caption{Evaluated energies, E($2^{+}_{1}$), and B(E; $0_{1}^{+} \rightarrow 2_{1}^{+}$) values for Dy nuclei.}\label{fig:graph66}
\end{Dfigures}

\begin{Dfigures}[ht!]
\begin{center}
\includegraphics[height=4in,width=\linewidth]{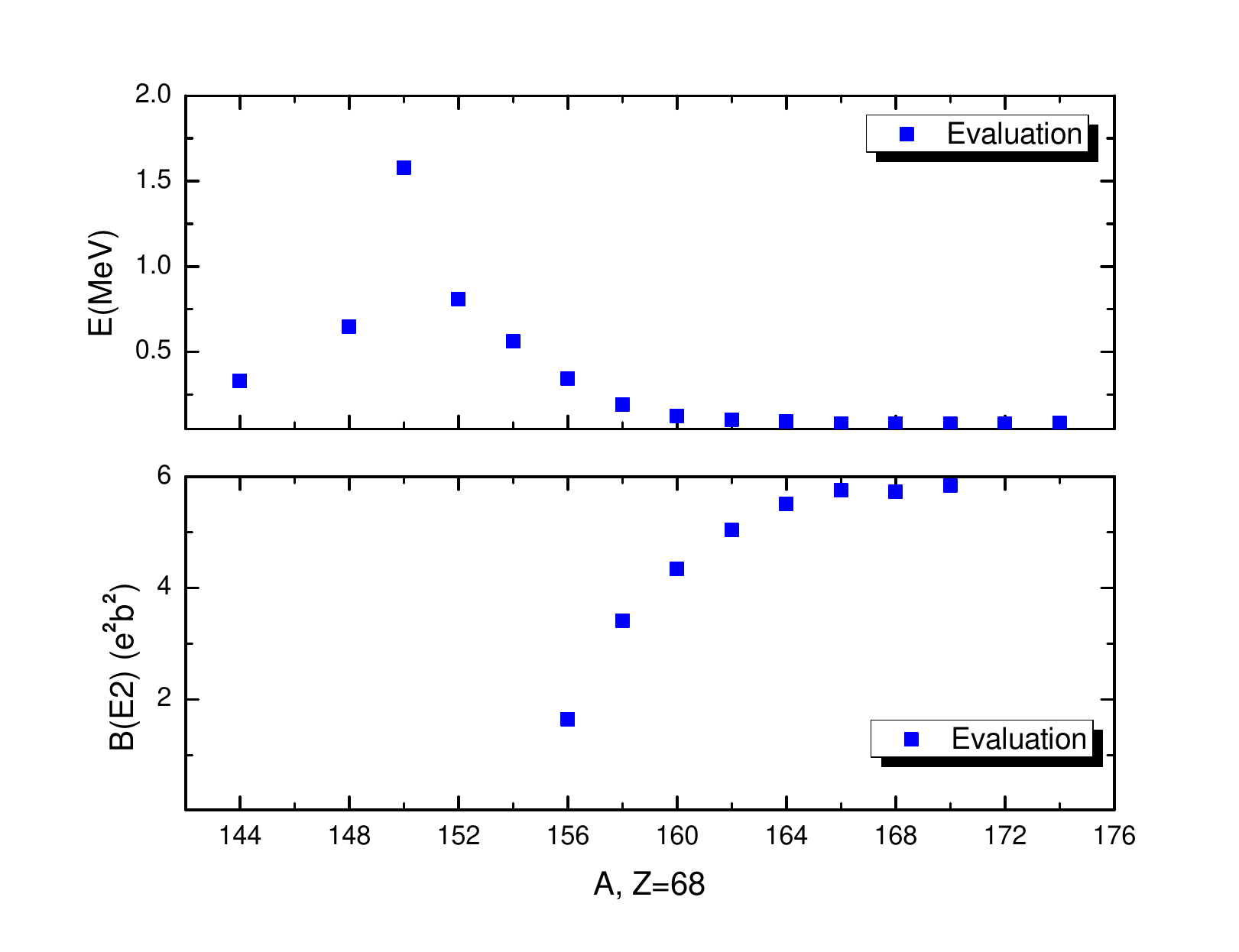}
\end{center}
\caption{Evaluated energies, E($2^{+}_{1}$), and B(E; $0_{1}^{+} \rightarrow 2_{1}^{+}$) values for Er nuclei.}\label{fig:graph68}
\end{Dfigures}
\clearpage

\begin{Dfigures}[ht!]
\begin{center}
\includegraphics[height=4in,width=\linewidth]{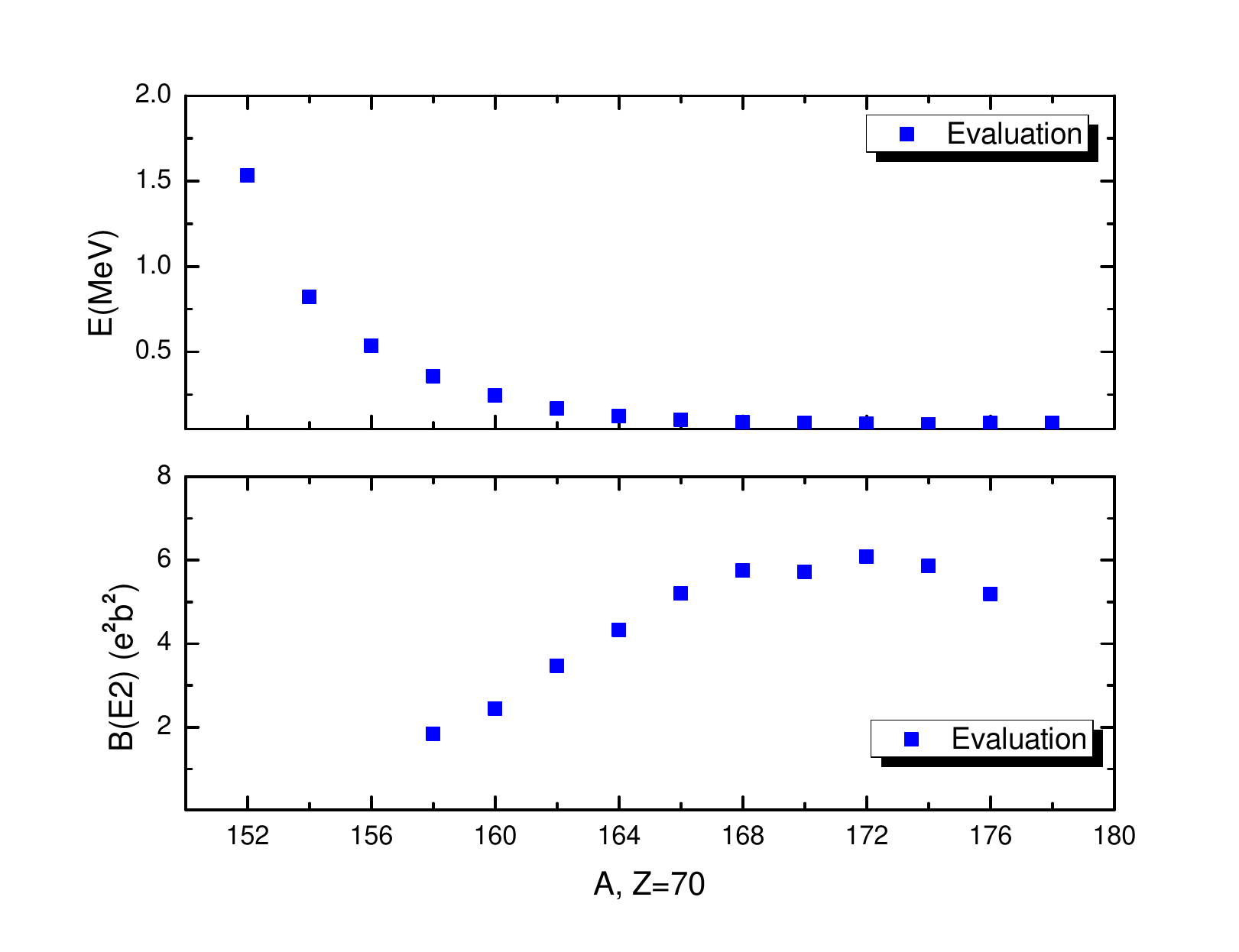}
\end{center}
\caption{Evaluated energies, E($2^{+}_{1}$), and B(E; $0_{1}^{+} \rightarrow 2_{1}^{+}$) values for Yb nuclei.}\label{fig:graph70}
\end{Dfigures}

\begin{Dfigures}[ht!]
\begin{center}
\includegraphics[height=4in,width=\linewidth]{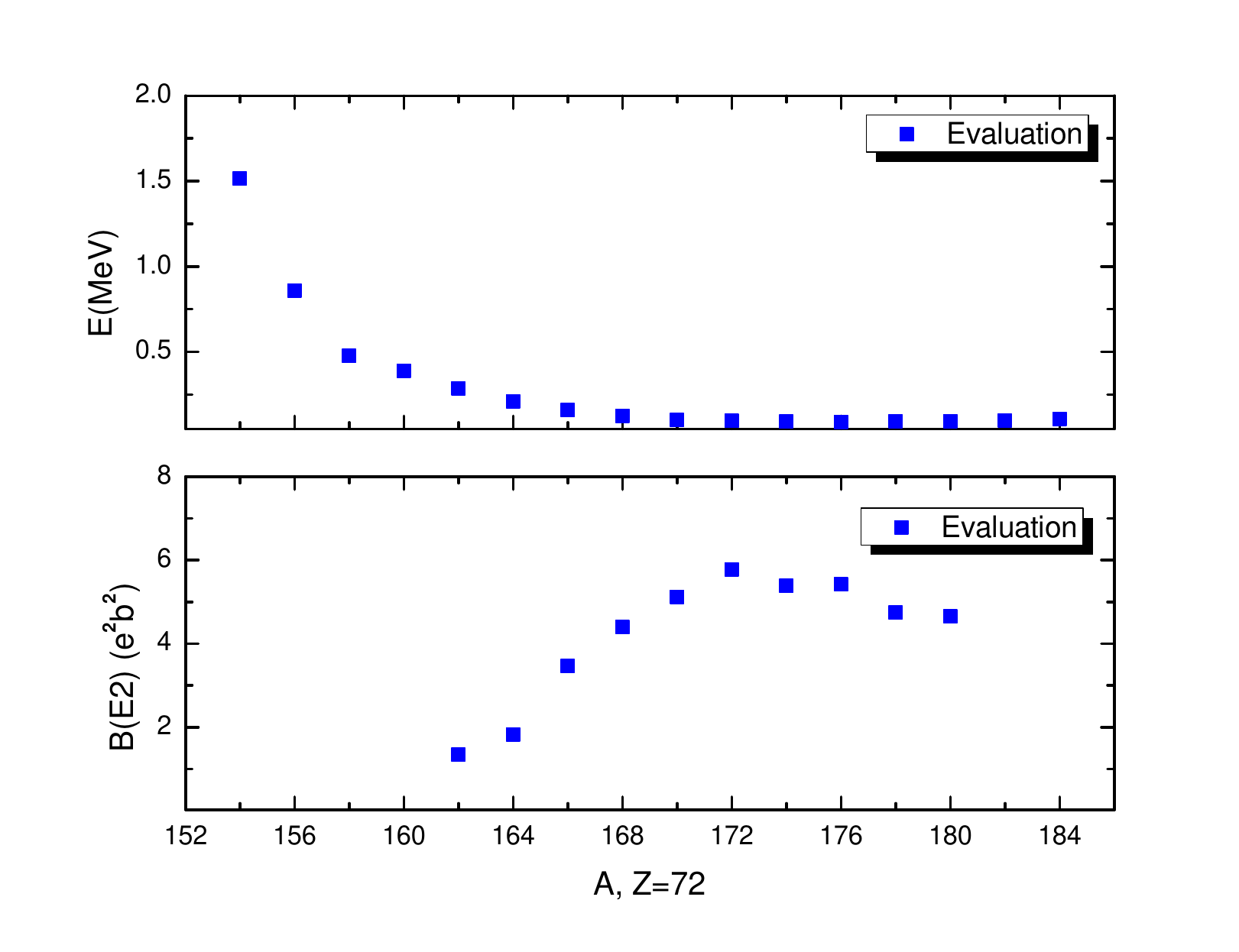}
\end{center}
\caption{Evaluated energies, E($2^{+}_{1}$), and B(E; $0_{1}^{+} \rightarrow 2_{1}^{+}$) values for Hf nuclei.}\label{fig:graph72}
\end{Dfigures}
\clearpage

\begin{Dfigures}[ht!]
\begin{center}
\includegraphics[height=4in,width=\linewidth]{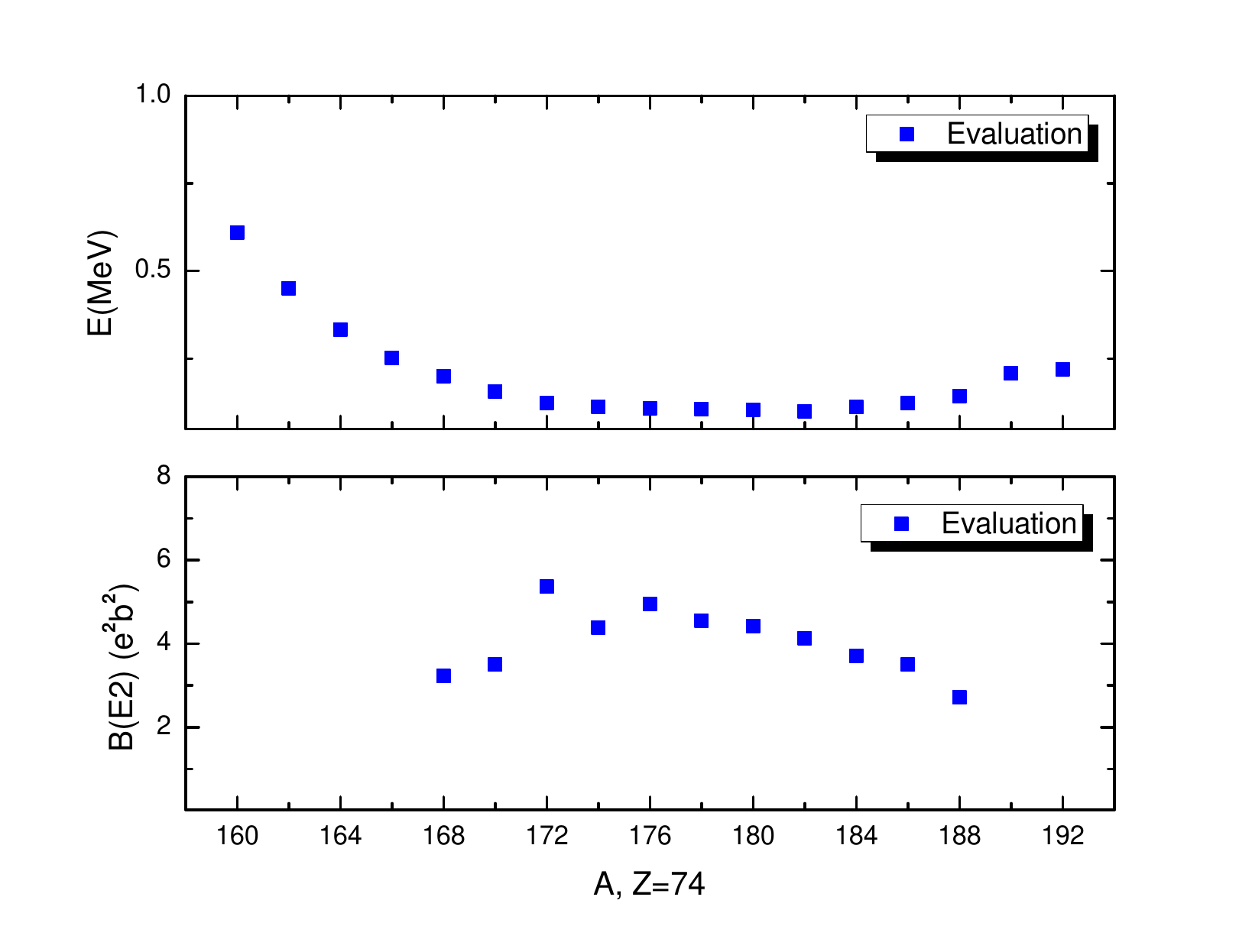}
\end{center}
\caption{Evaluated energies, E($2^{+}_{1}$), and B(E; $0_{1}^{+} \rightarrow 2_{1}^{+}$) values for W nuclei.}\label{fig:graph74}
\end{Dfigures}

\begin{Dfigures}[ht!]
\begin{center}
\includegraphics[height=4in,width=\linewidth]{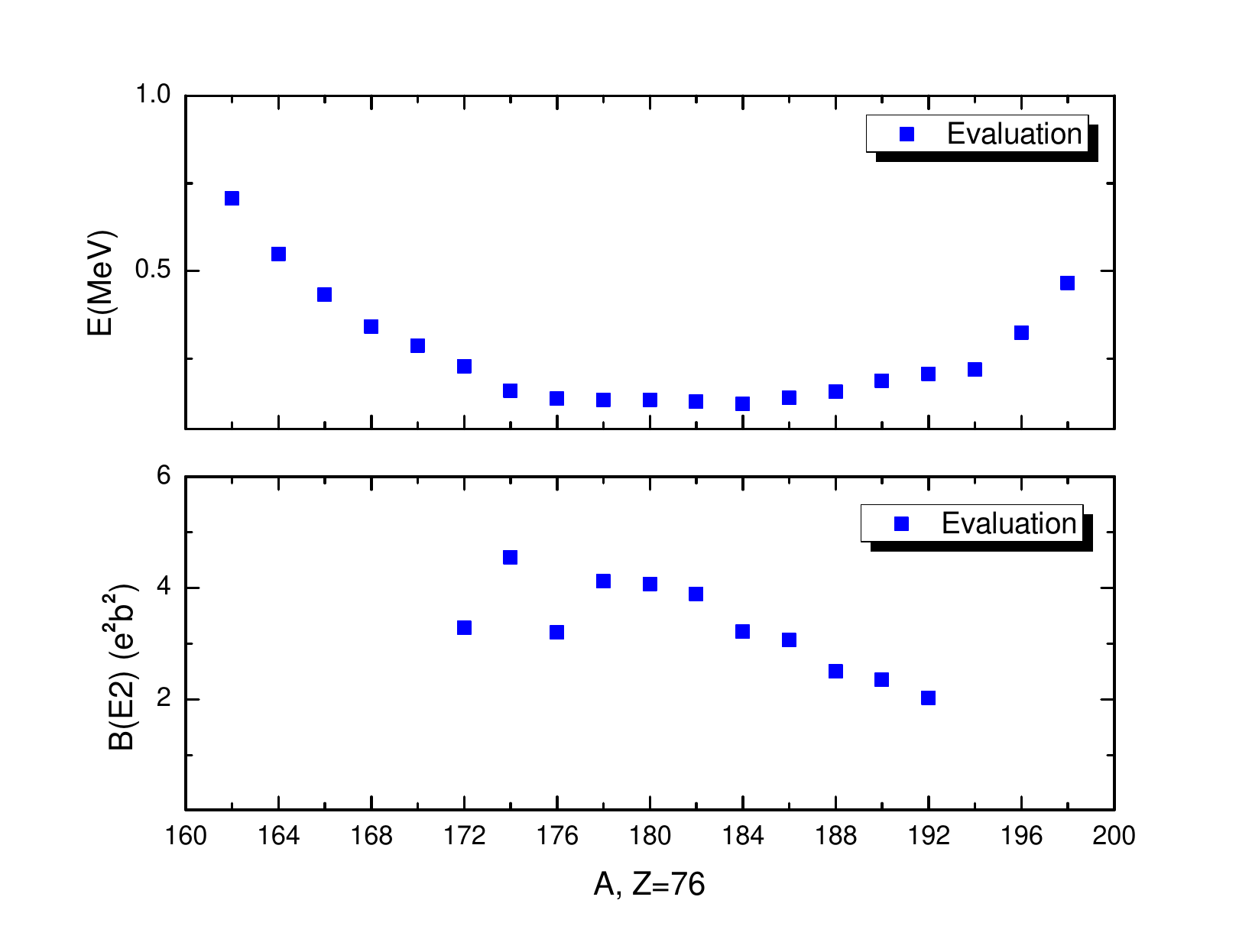}
\end{center}
\caption{Evaluated energies, E($2^{+}_{1}$), and B(E; $0_{1}^{+} \rightarrow 2_{1}^{+}$) values for Os nuclei.}\label{fig:graph76}
\end{Dfigures}
\clearpage

\begin{Dfigures}[ht!]
\begin{center}
\includegraphics[height=4in,width=\linewidth]{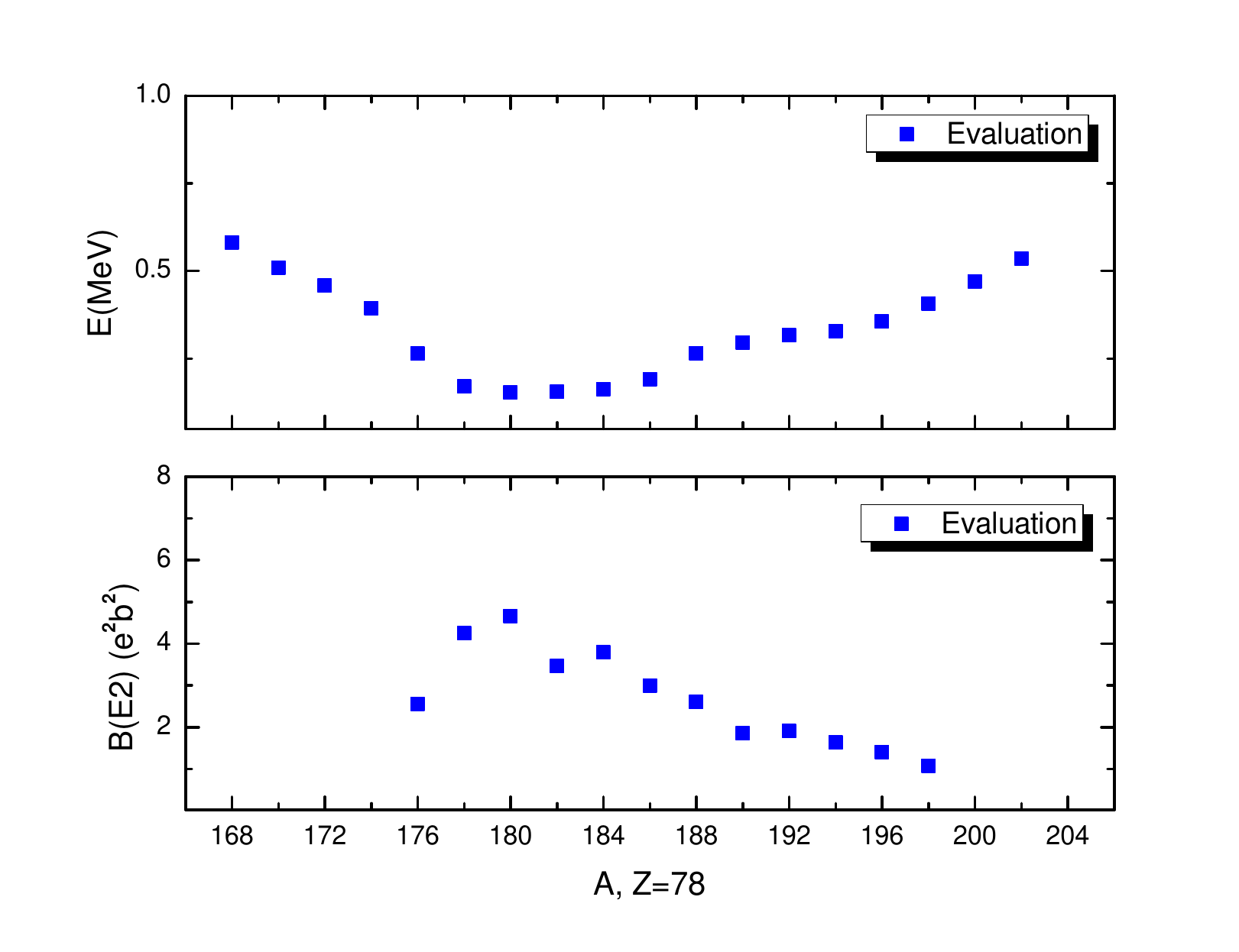}
\end{center}
\caption{Evaluated energies, E($2^{+}_{1}$), and B(E; $0_{1}^{+} \rightarrow 2_{1}^{+}$) values for Pt nuclei.}\label{fig:graph78}
\end{Dfigures}

\begin{Dfigures}[ht!]
\begin{center}
\includegraphics[height=4in,width=\linewidth]{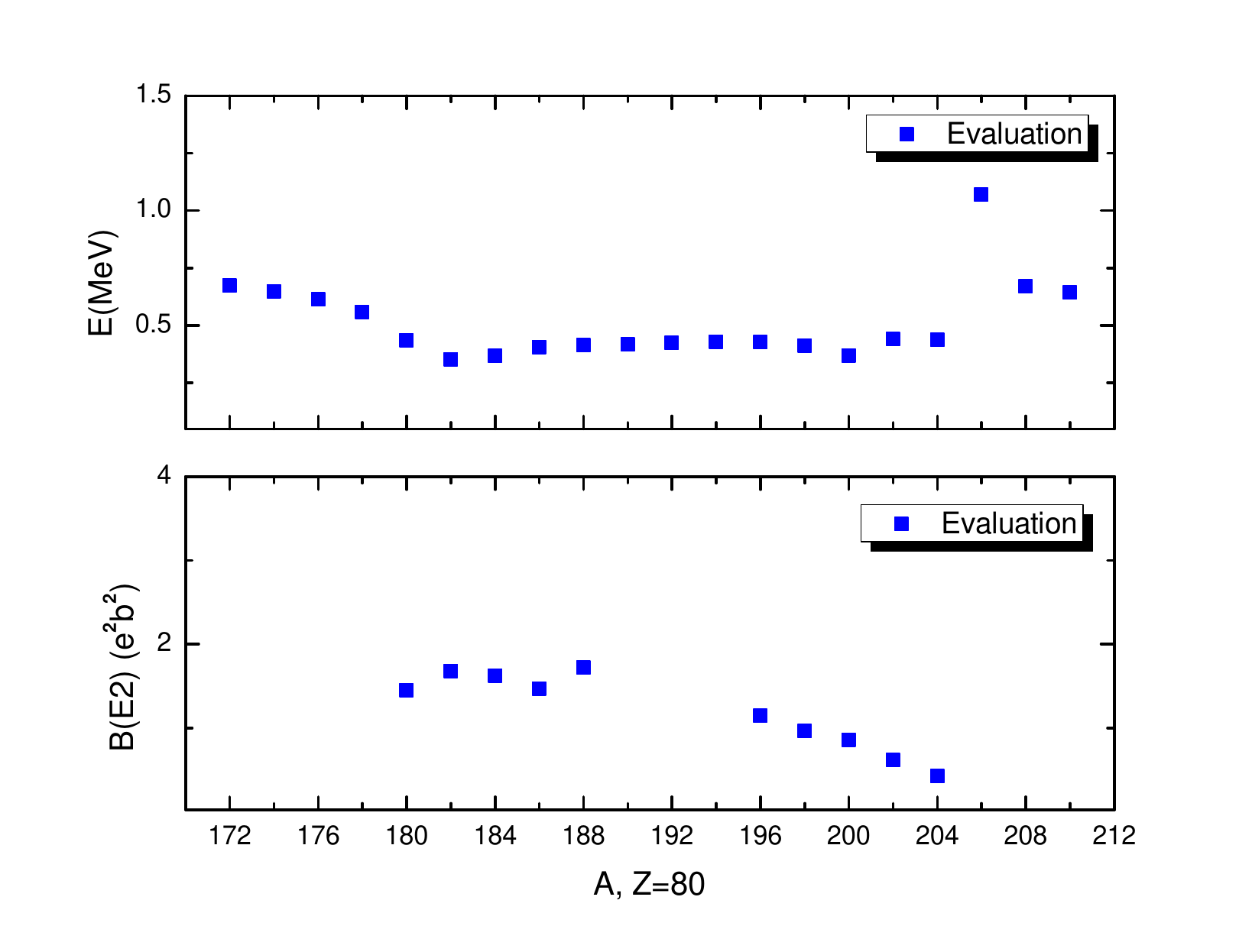}
\end{center}
\caption{Evaluated energies, E($2^{+}_{1}$), and B(E; $0_{1}^{+} \rightarrow 2_{1}^{+}$) values for Hg nuclei.}\label{fig:graph80}
\end{Dfigures}
\clearpage

\begin{Dfigures}[ht!]
\begin{center}
\includegraphics[height=4in,width=\linewidth]{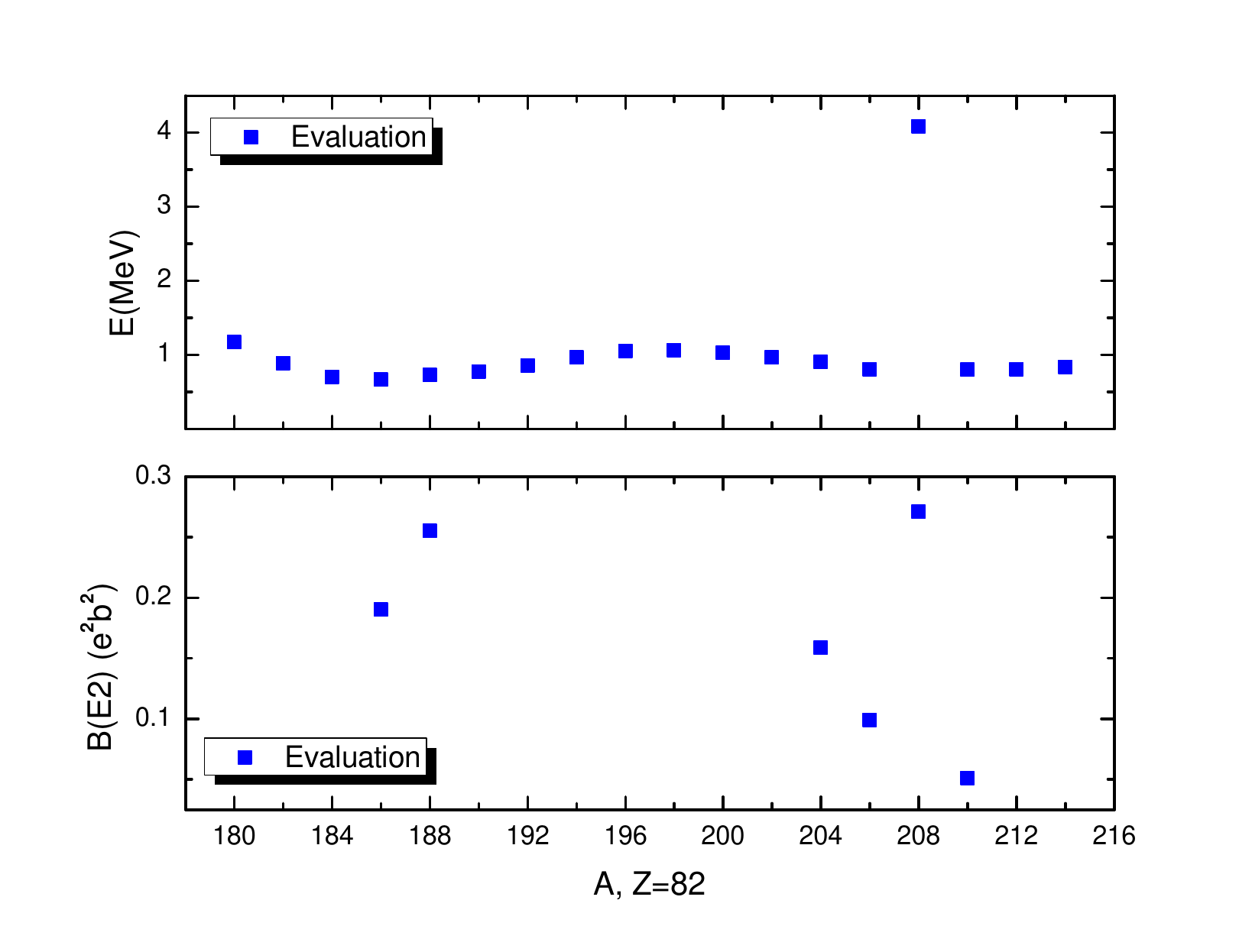}
\end{center}
\caption{Evaluated energies, E($2^{+}_{1}$), and B(E; $0_{1}^{+} \rightarrow 2_{1}^{+}$) values for Pb nuclei.}\label{fig:graph82}
\end{Dfigures}

\begin{Dfigures}[ht!]
\begin{center}
\includegraphics[height=4in,width=\linewidth]{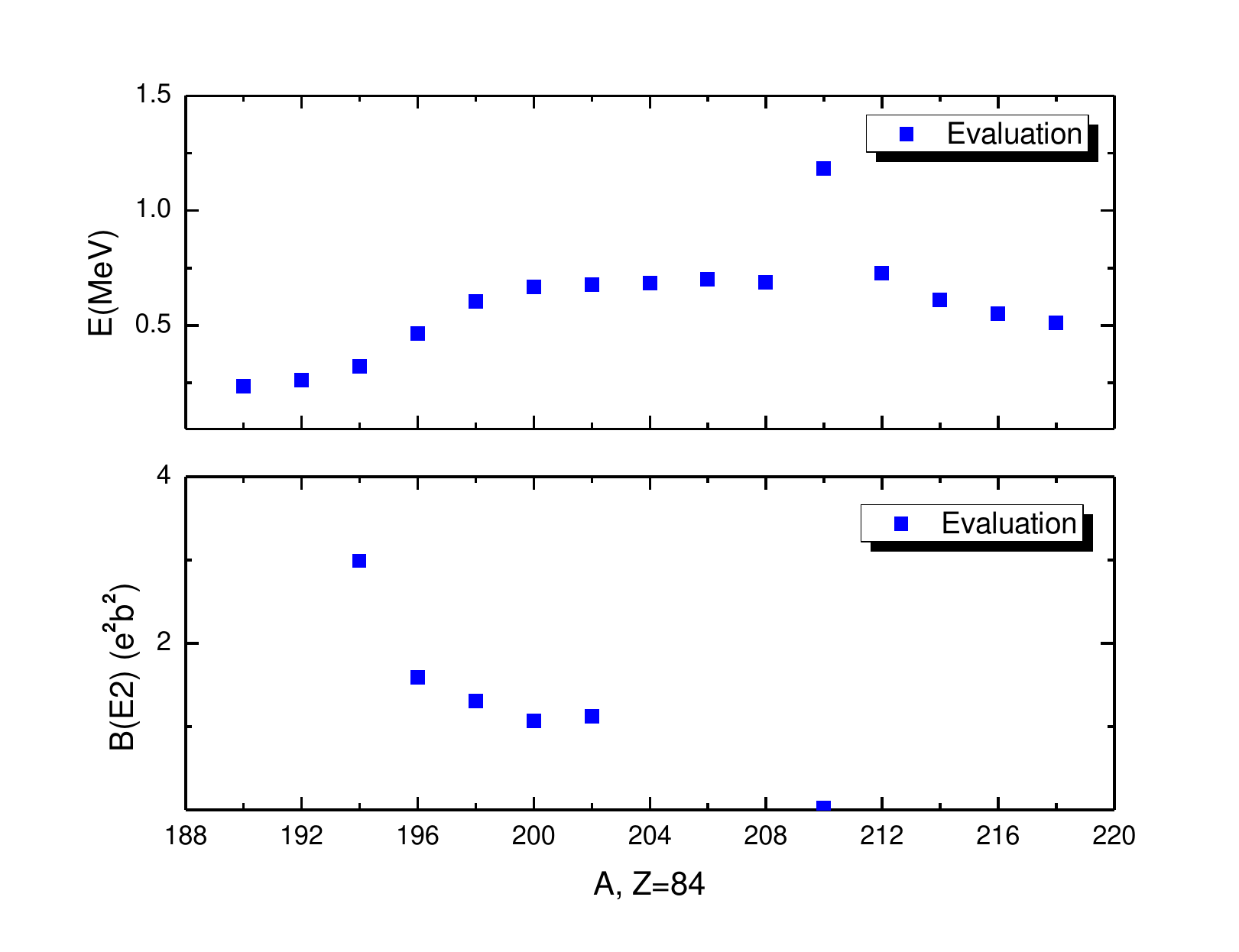}
\end{center}
\caption{Evaluated energies, E($2^{+}_{1}$), and B(E; $0_{1}^{+} \rightarrow 2_{1}^{+}$) values for Po nuclei.}\label{fig:graph84}
\end{Dfigures}
\clearpage

\begin{Dfigures}[ht!]
\begin{center}
\includegraphics[height=4in,width=\linewidth]{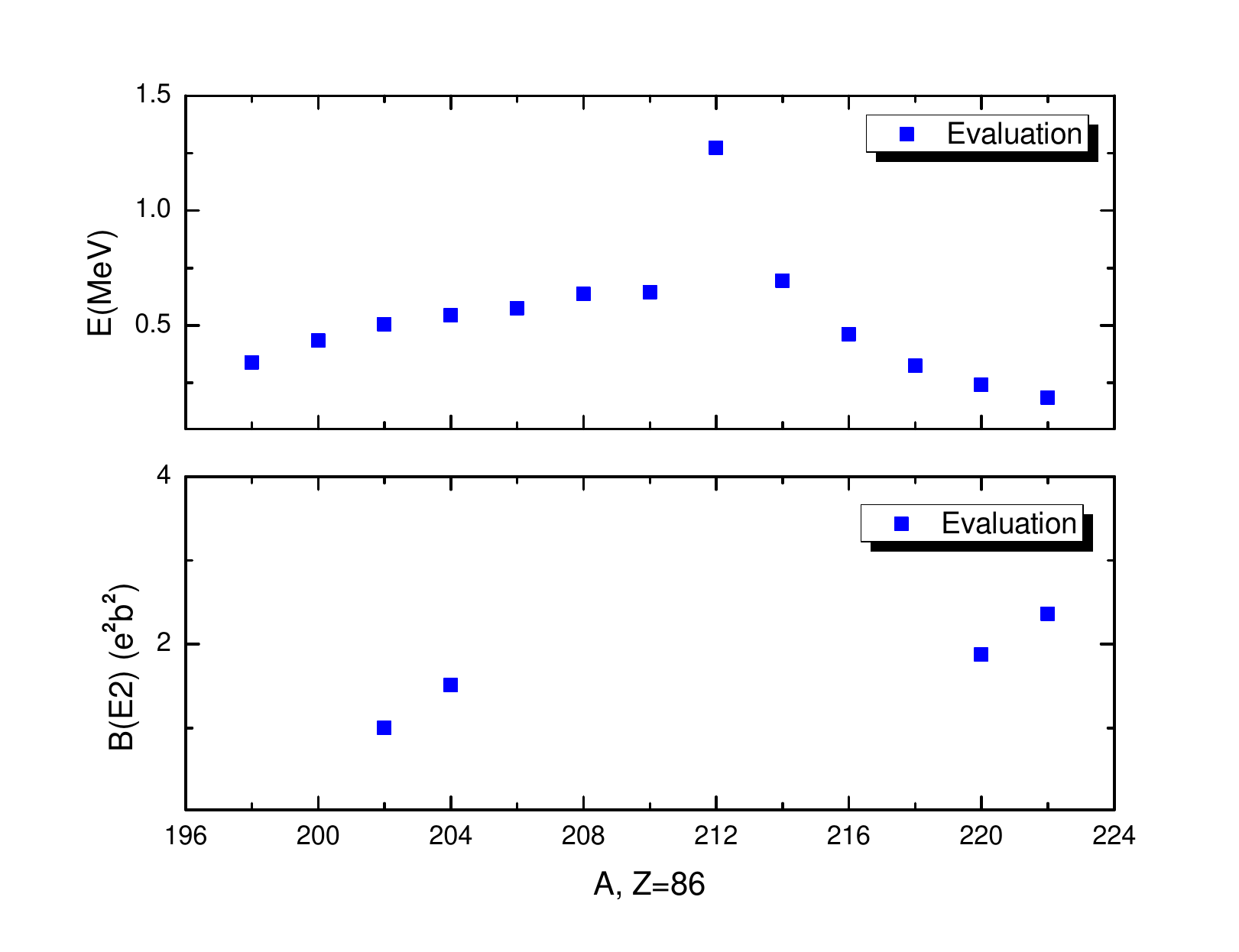}
\end{center}
\caption{Evaluated energies, E($2^{+}_{1}$), and B(E; $0_{1}^{+} \rightarrow 2_{1}^{+}$) values for Rn nuclei.}\label{fig:graph86}
\end{Dfigures}

\begin{Dfigures}[ht!]
\begin{center}
\includegraphics[height=4in,width=\linewidth]{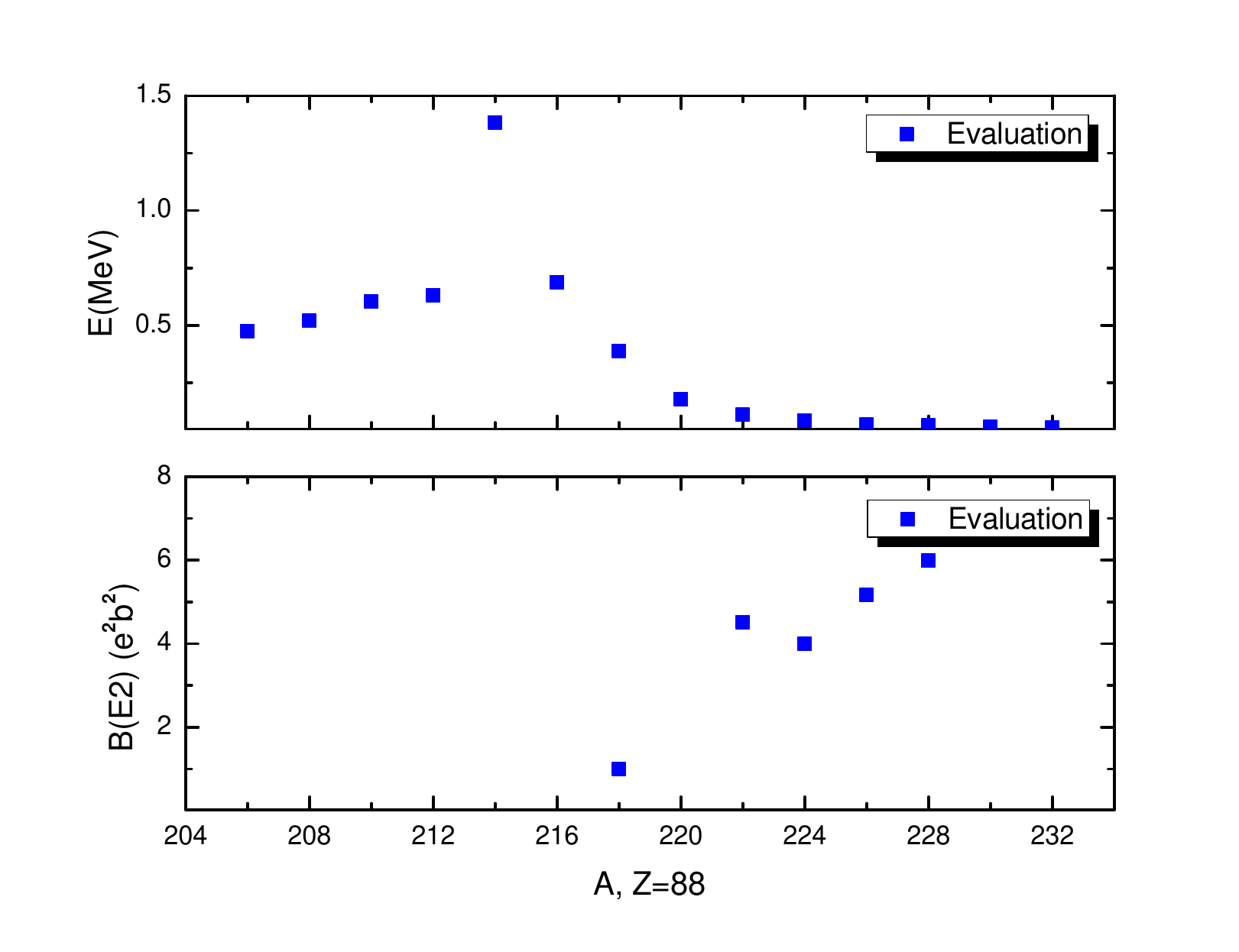}
\end{center}
\caption{Evaluated energies, E($2^{+}_{1}$), and B(E; $0_{1}^{+} \rightarrow 2_{1}^{+}$) values for Ra nuclei.}\label{fig:graph88}
\end{Dfigures}
\clearpage

\begin{Dfigures}[ht!]
\begin{center}
\includegraphics[height=4in,width=\linewidth]{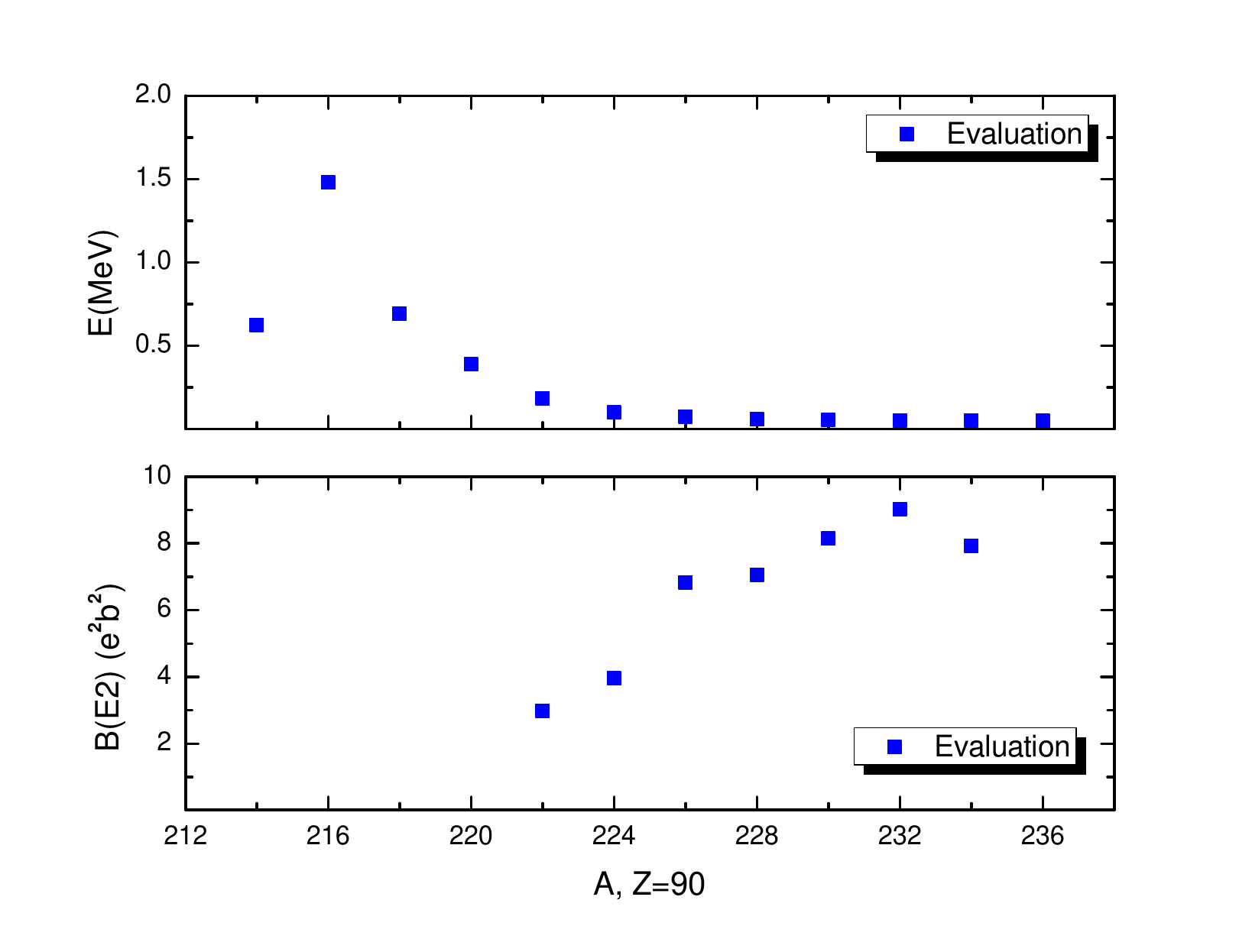}
\end{center}
\caption{Evaluated energies, E($2^{+}_{1}$), and B(E; $0_{1}^{+} \rightarrow 2_{1}^{+}$) values for Th nuclei.}\label{fig:graph90}
\end{Dfigures}

\begin{Dfigures}[ht!]
\begin{center}
\includegraphics[height=4in,width=\linewidth]{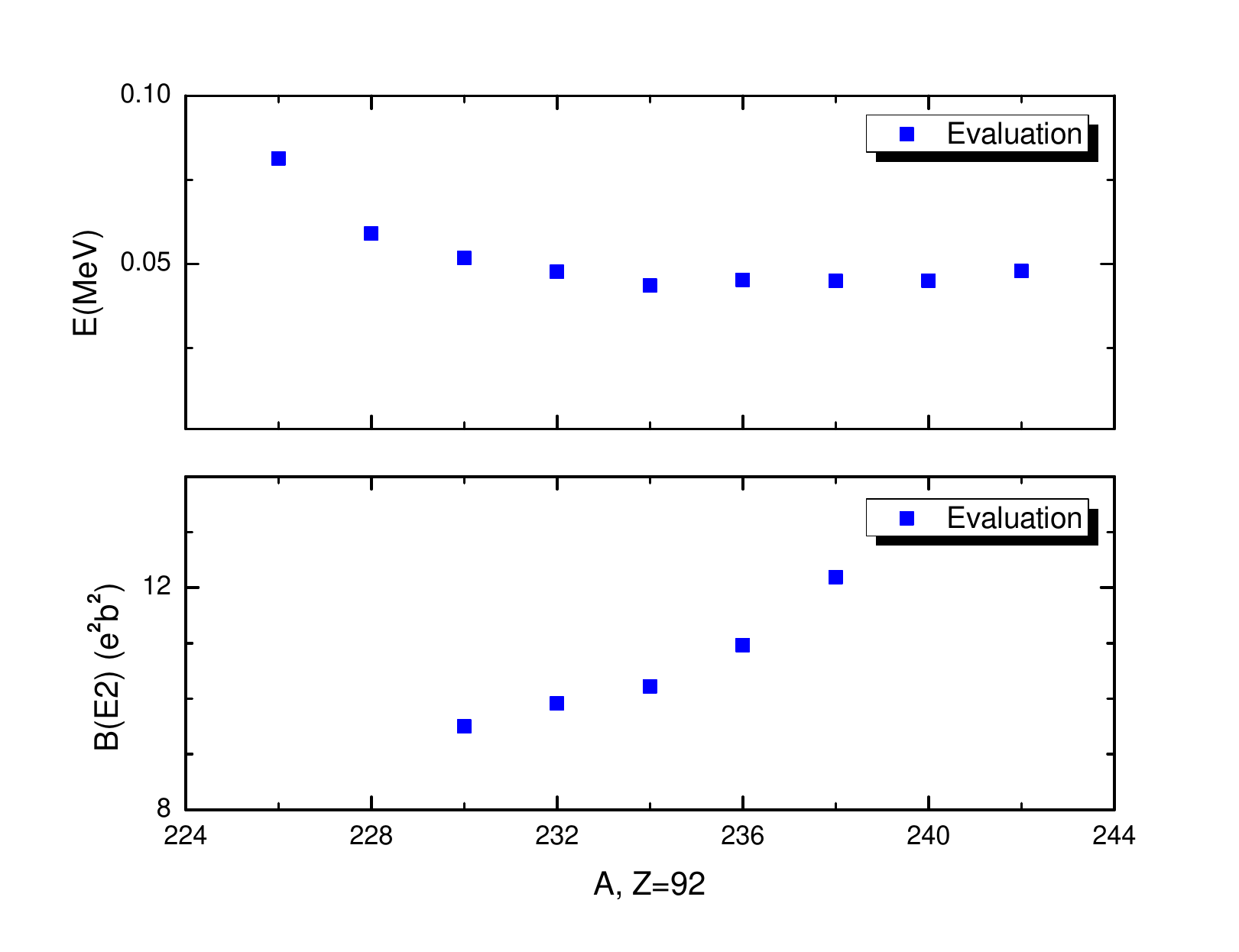}
\end{center}
\caption{Evaluated energies, E($2^{+}_{1}$), and B(E; $0_{1}^{+} \rightarrow 2_{1}^{+}$) values for U nuclei.}\label{fig:graph92}
\end{Dfigures}
\clearpage

\begin{Dfigures}[ht!]
\begin{center}
\includegraphics[height=4in,width=\linewidth]{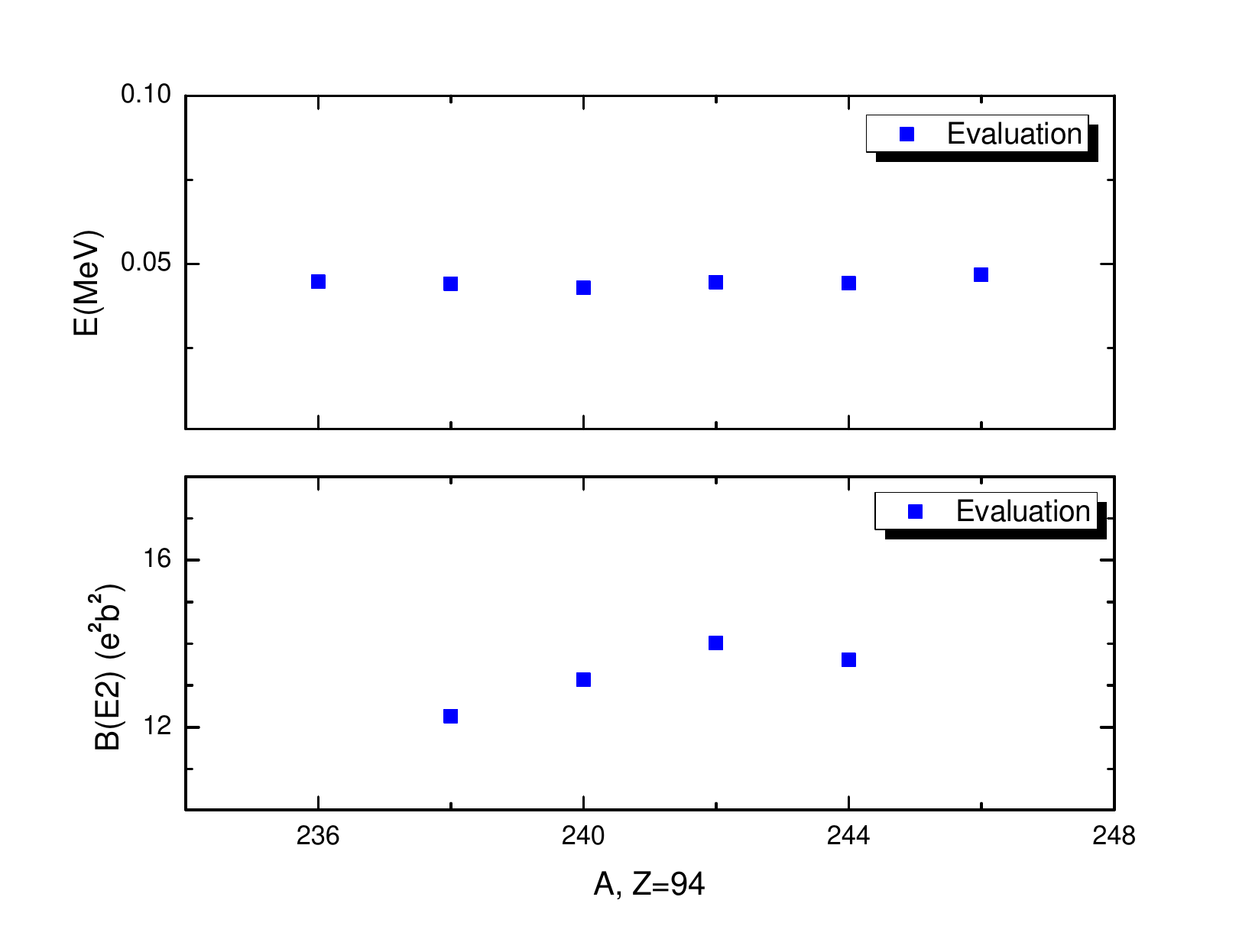}
\end{center}
\caption{Evaluated energies, E($2^{+}_{1}$), and B(E; $0_{1}^{+} \rightarrow 2_{1}^{+}$) values for Pu nuclei.}\label{fig:graph94}
\end{Dfigures}

\begin{Dfigures}[ht!]
\begin{center}
\includegraphics[height=4in,width=\linewidth]{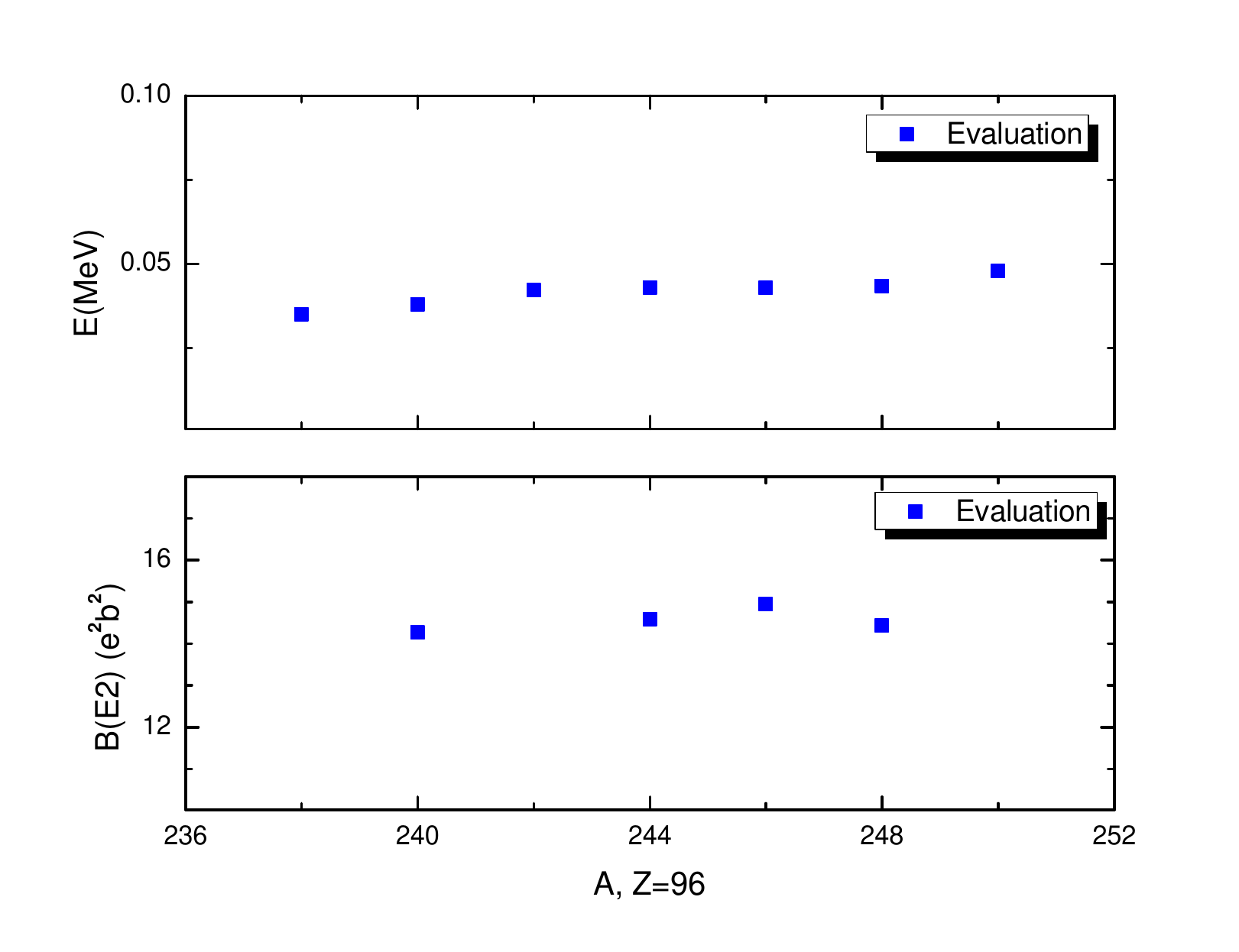}
\end{center}
\caption{Evaluated energies, E($2^{+}_{1}$), and B(E; $0_{1}^{+} \rightarrow 2_{1}^{+}$) values for Cm nuclei.}\label{fig:graph96}
\end{Dfigures}
\clearpage

\begin{Dfigures}[ht!]
\begin{center}
\includegraphics[height=4in,width=\linewidth]{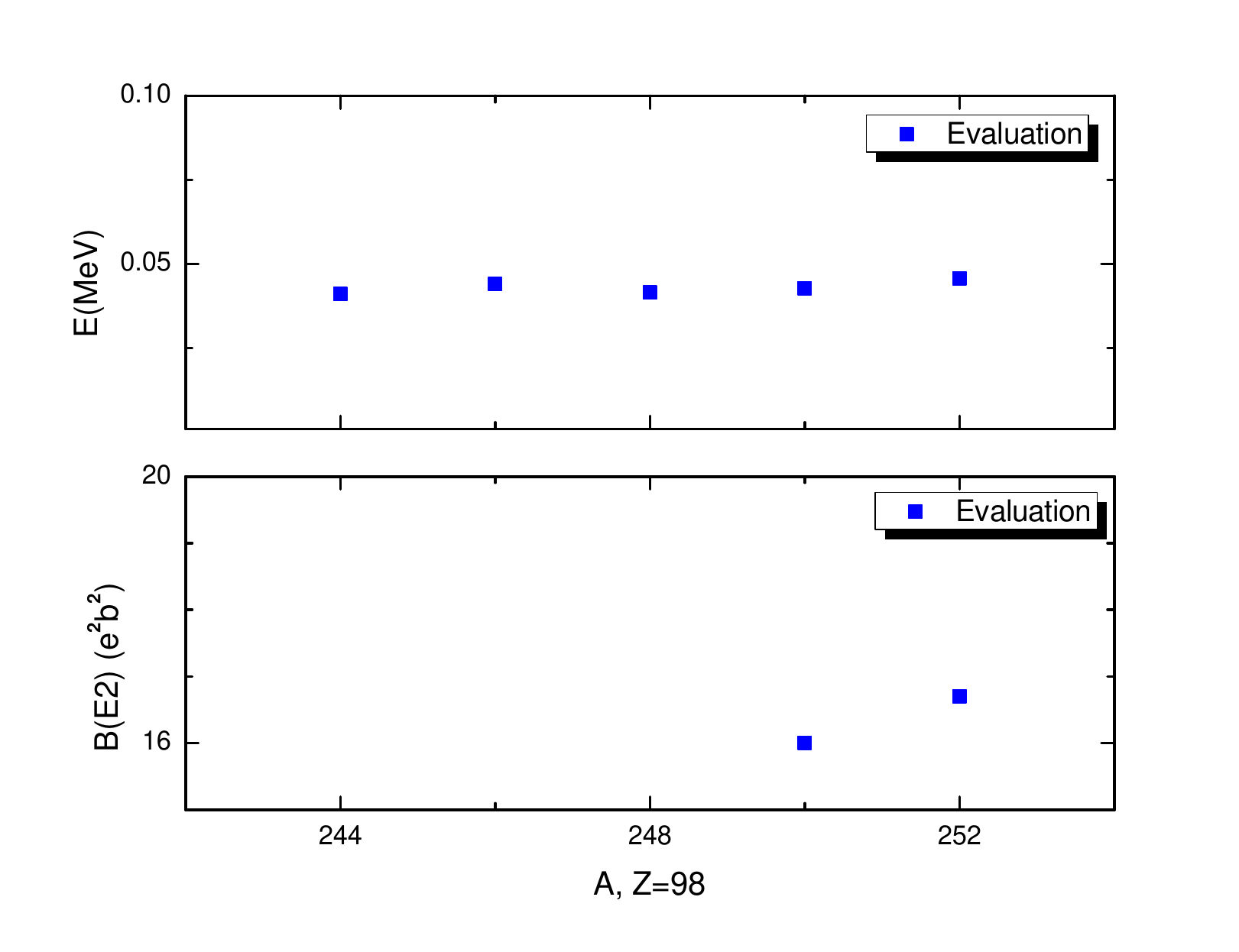}
\end{center}
\caption{Evaluated energies, E($2^{+}_{1}$), and B(E; $0_{1}^{+} \rightarrow 2_{1}^{+}$) values for Cf nuclei.}\label{fig:graph98}
\end{Dfigures}

\begin{Dfigures}[ht!]
\begin{center}
\includegraphics[height=2.5in,width=\linewidth]{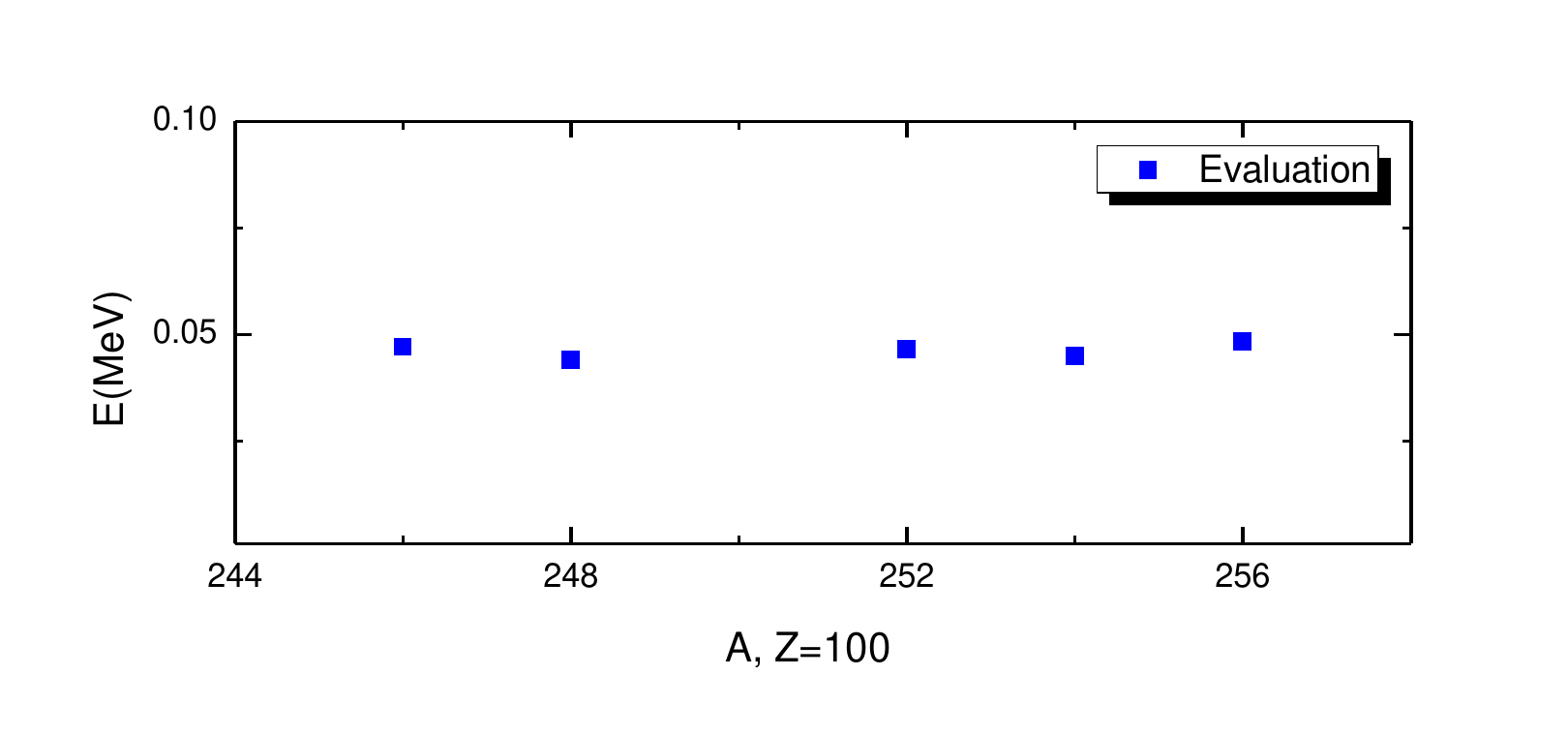}
\end{center}
\caption{Evaluated energies, E($2^{+}_{1}$)  for Fm nuclei.}\label{fig:graph100}
\end{Dfigures}
\clearpage

\begin{Dfigures}[ht!]
\begin{center}
\includegraphics[height=2.5in,width=\linewidth]{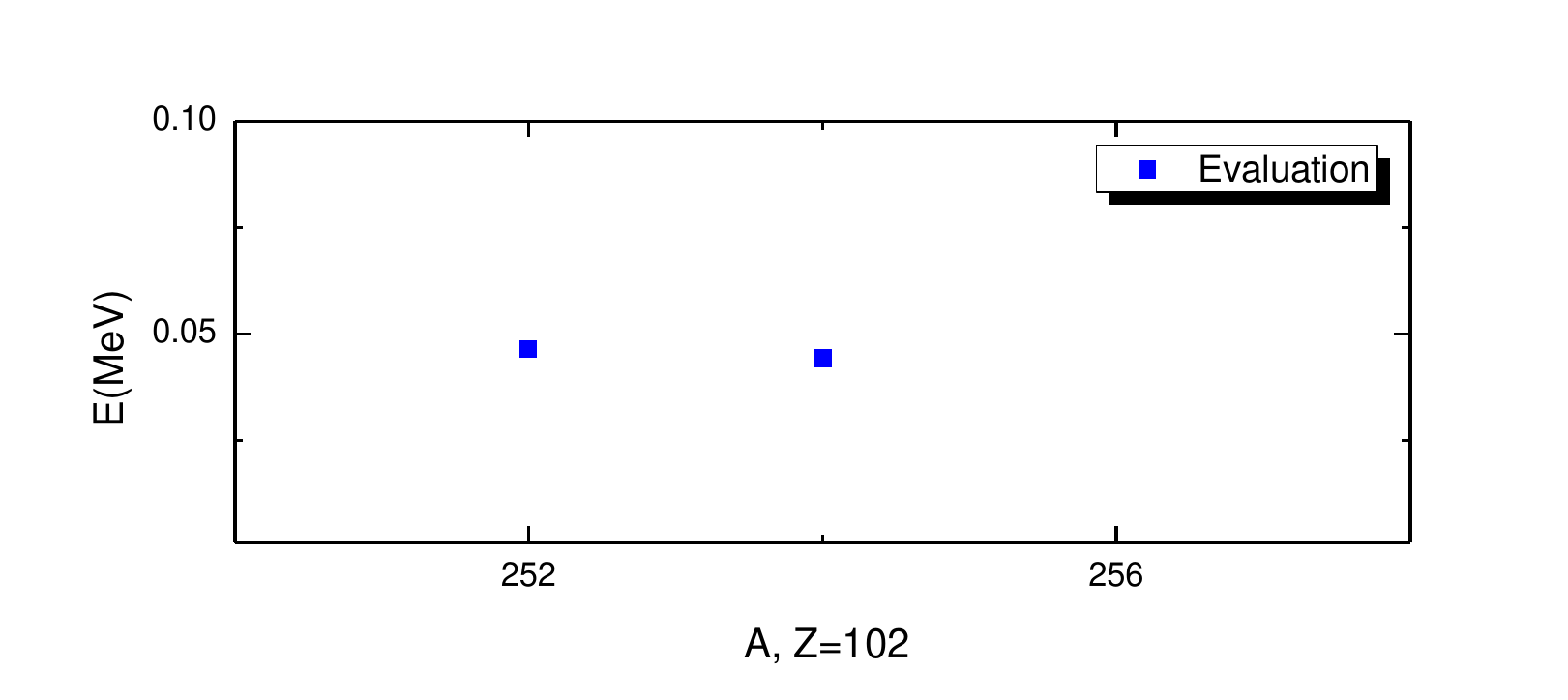}
\end{center}
\caption{Evaluated energies, E($2^{+}_{1}$)  for No nuclei.}\label{fig:graph102}
\end{Dfigures}

\begin{Dfigures}[ht!]
\begin{center}
\includegraphics[height=2.5in,width=\linewidth]{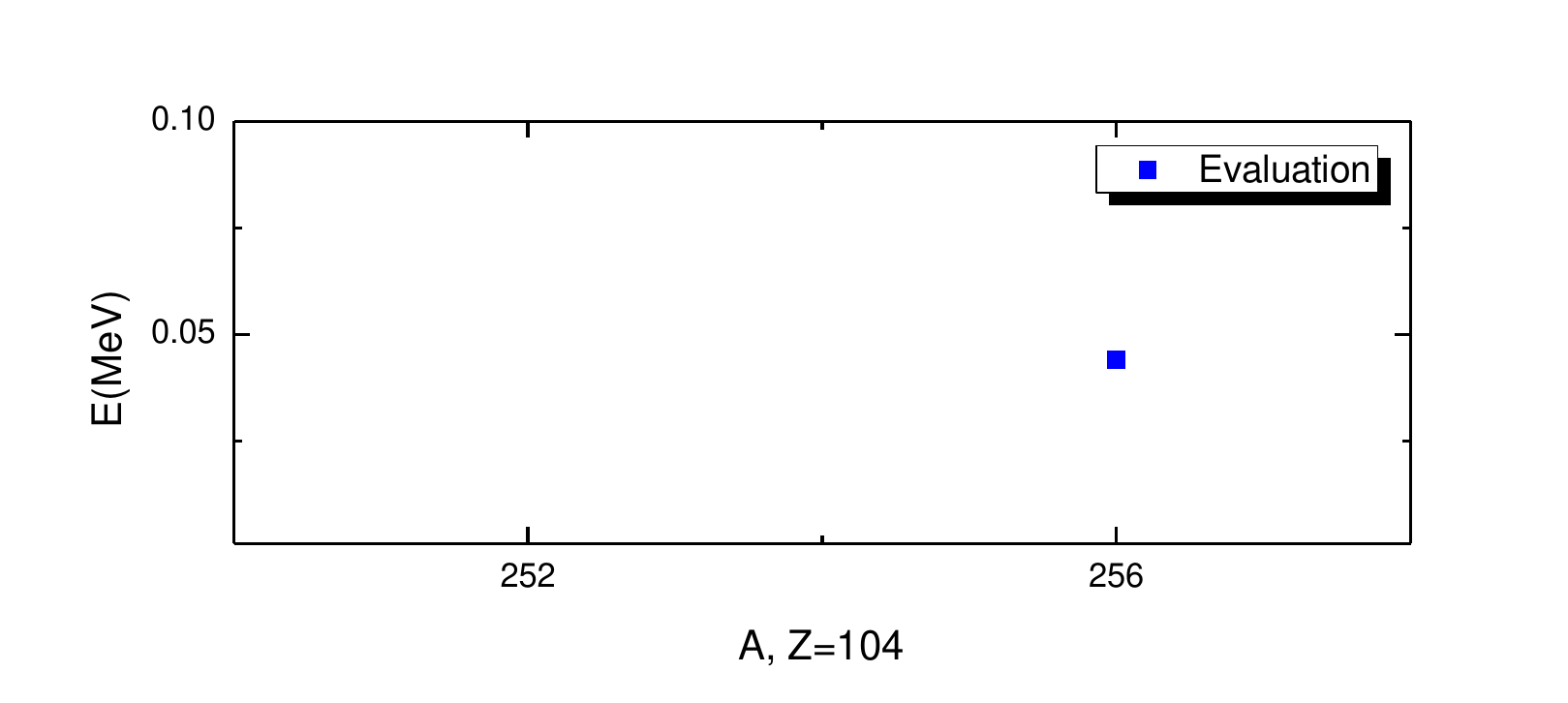}
\end{center}
\caption{Evaluated energies, E($2^{+}_{1}$)  for Rf nuclei.}\label{fig:graph104}
\end{Dfigures}
\clearpage

\end{document}